\documentclass[a4paper,hidelinks,twoside]{ociamthesis}

\usepackage[frozencache,cachedir=.]{minted}
\usepackage[inner=2.25cm,outer=1.75cm, bottom=0cm]{geometry}
\usepackage{layout}
\usepackage{geometry}
\usepackage{shadow}

\makeatletter
\renewcommand*{\cleardoublepage}{\clearpage\if@twoside \ifodd\c@page\else
\hbox{}%
\thispagestyle{empty}%
\newpage%
\if@twocolumn\hbox{}\newpage\fi\fi\fi}
\makeatother

\usepackage{titlesec}
\usepackage{setspace}

\titleformat{\section}
{\color{blue!70!black}\normalfont\Large\bfseries}
{\color{blue!70!black}\thesection}{1em}{}

\usepackage[style=numeric-comp, sorting=none, backend=biber, doi=true, isbn=false]{biblatex}
\newcommand*{\bibtitle}{References}

\usepackage{lipsum}

\usepackage{afterpage}

\usepackage{sidecap}
\usepackage[most]{tcolorbox}

\usepackage{titlesec}

\usepackage{wrapfig}
\usepackage{booktabs}
\usepackage[bottom]{footmisc}
\usepackage{physics}
\usepackage{caption}
\usepackage{mwe}
\usepackage{float}
\usepackage{marginnote}
\usepackage{caption}
\SetLipsumParListSurrounders{\colorlet{oldcolor}{.}\color{gray}}{\color{oldcolor}}
\usepackage{sidenotes}

\usepackage[T1]{fontenc}

\usepackage[T1]{fontenc}
\usepackage[font=small,labelfont=bf,tableposition=top]{caption}

\usepackage[utf8]{inputenc}
\usepackage{babel}
\usepackage{booktabs}
\usepackage{varwidth}
\newsavebox\tmpbox

\usepackage{qcircuit}

\usepackage[define-L-C-R]{nicematrix}
\usepackage{amsmath}

\usepackage{amsthm}
\usepackage{cleveref}
\usepackage[most]{tcolorbox}

\newtcbtheorem[number within=chapter]{Theorem}{}{
        enhanced,
        sharp corners,
        attach boxed title to top left={
            xshift=-1mm,
            yshift=-5mm,
            yshifttext=-1mm
        },
        top=1.5em,
        colback=white,
        colframe=blue!75!black,
        fonttitle=\bfseries,
        boxed title style={
            sharp corners,
            size=small,
            colback=blue!75!black,
            colframe=blue!75!black,
        } 
    }{thm}

\newtcbtheorem[number within=chapter]{Definition}{}{
        enhanced,
        sharp corners,
        attach boxed title to top left={
            xshift=-1mm,
            yshift=-5mm,
            yshifttext=-1mm
        },
        top=1.5em,
        colback=white,
        colframe=blue!20!black,
        fonttitle=\bfseries,
        boxed title style={
            sharp corners,
            size=small,
            colback=blue!20!black,
            colframe=blue!20!black,
        } 
    }{def}

  \usepackage{sidecap}

\usepackage{mathtools,amssymb,amsthm}

\definecolor{LightGray}{gray}{0.9}

\definecolor{codegreen}{rgb}{0,0.6,0}
\definecolor{codegray}{rgb}{0.5,0.5,0.5}
\definecolor{codepurple}{rgb}{0.58,0,0.82}
\definecolor{backcolour}{rgb}{0.95,0.95,0.92}

\usepackage{datatool}
\newcommand{\sortitem}[1]{%
  \DTLnewrow{list}
  \DTLnewdbentry{list}{description}{#1}
}
\newenvironment{sortedlist}{%
  \DTLifdbexists{list}{\DTLcleardb{list}}{\DTLnewdb{list}}
}{%
  \DTLsort{description}{list}
  \begin{itemize}%
    \DTLforeach*{list}{\theDesc=description}{%
      \item \theDesc}
  \end{itemize}%
}

\usepackage{xcolor}
\usepackage{xparse}

\usepackage{lipsum}  

\newminted{python}{fontsize=\small, 
                   linenos,
                   numbersep=8pt,
                   gobble=0,
                   breaklines,
                   frame=lines,
                   bgcolor=bg,
                   framesep=3mm} 
                   
\usepackage{mathtools,amssymb,amsthm}
\usepackage{physics}
\usepackage{graphicx}
\definecolor{bg}{rgb}{0.97,0.97,0.97}

	\newcommand{\sor}[1]{\item{\textcolor{blue}{\textsf{\textbf{{#1}}}}}}
\newcommand{\bl}[1]{\textcolor{blue}{{#1}}}

\usepackage{tikz}
\usetikzlibrary{calc}

\begin{document}

\begin{singlespace}
\raggedbottom
\begin{center}
	 \normalsize{\textsc{\large{People's Democratic Republic of Algeria}}}\\
		\normalsize{\textsc{\large{Ministry of higher education and scientific research}}}\\
		\normalsize{\textsc{\large{University saad dahleb of blida 1}}}\\
		\normalsize{\textsc{\large{Faculty of sciences}}}\\
		\normalsize{\textsc{Physics department}}\\
	\end{center}
	\begin{center}
	\vspace*{0.59cm}
		\includegraphics[width=3.5cm,height=3cm]{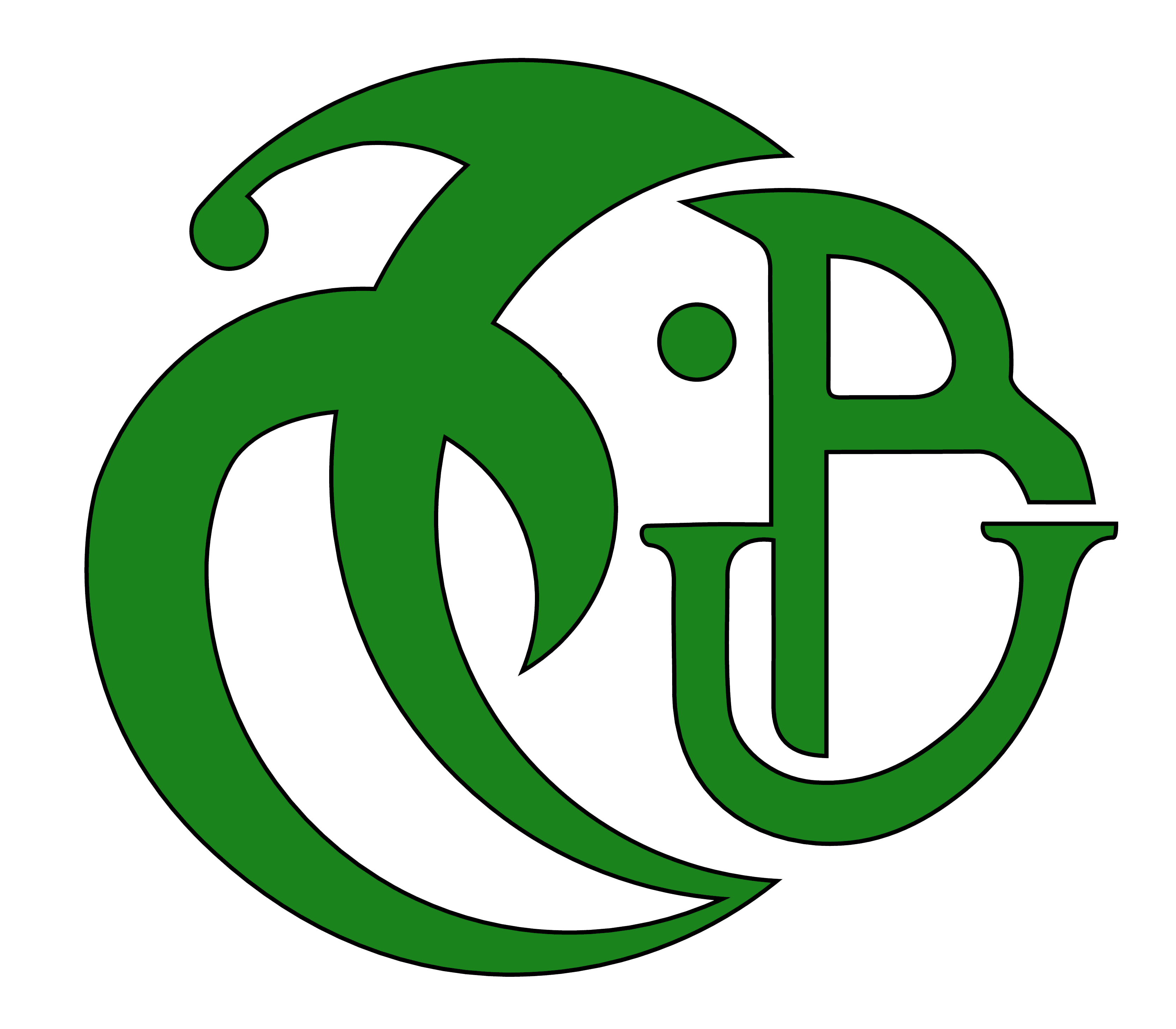}
	\end{center}
	
	\begin{center}
		\Huge{\textsc{Master Thesis}}\\
				\textit{}\\
		\large{\textbf{\textsc{Theoretical Physics}}}\\
		\textbf{}\\
	\end{center}
	\vspace*{0.05cm}
		\begin{minipage}{0.9\textwidth}
			\begin{center}
				\quad \Large{\textsc{ dealing with quantum computer readout\\ noise through high energy physics\\ unfolding methods}}
			\end{center}
		\end{minipage}
	\vspace*{0.5cm}
\thispagestyle{empty}
	
	\begin{table}[h]
		\center
		\textbf{Prepared by:}\\
		\begin{tabular}{p{8cm}p{8cm}}
			\\
		\qquad	Hacene Rabah BENAISSA  \qquad & \qquad \qquad \quad Imene OUADAH
		\end{tabular}
	\end{table}
	
	\vspace*{0.5cm}
	
	\begin{table}[h]
	\centering
		\begin{tabular}{p{14cm}p{14cm}}
			\textbf{Defended on September 20, 2021 before the jury composed of:}\\
			  \\
		Dr. K. A. BOUTELDJA \hspace{1.7em}  \qquad MAA \qquad USDB1 \hspace{3em} President \\
		Dr. A. MOUZALI \hspace{4.6em} \qquad MCB \hspace{1.75em}  USDB1 \hspace{3em} Examiner \\
		Dr. S. A. YAHIAOUI \hspace{3.1em} \qquad MCB \hspace{1.75em} USDB1 \hspace{3em} Examiner \\
		Dr. A. YANALLAH \hspace{5.75em} MCB \hspace{1.75em} USDB1 \hspace{3em} Supervisor \\
		Dr. N. BOUAYED \hspace{6.3em} MCA \hspace{1.75em} USDB1 \hspace{3em} Co-Supervisor 
		
		\end{tabular}
	\end{table}
	\begin{center}
		\vspace*{1cm}
		\textsc{{Blida1, 2020/2021}}
	\end{center}
\begin{titlepage}
\begin{romanpages}


# Dedication

إلى أمي وأبي الغاليين لو ملأت كل الدفاتر و الأوراق شكرا و تقديرا لكما لن أوفي ولو القليل من حقكما. الكون على اتساعه لا يضاهي سعة قلبكما. شكرا لكما على زرع الطموح في نفسي منذ الصغر، شكرا لكما على عطائكما و دعائكما الغير المنقطع. رضاكما وابتسامتكما نعمة لا تقدر بثمن.

إلى اخوتي سندي في الحياة لا أتذكر كم مرة طلبت منكم شراء الأوراق البيضاء لي. لم تترددوا يوما على الخروج لإحضارها. شكرا لكم على دعمكم و تشجيعاتكم لي.

إلى أختي الصغيرة التي لا تفارقني أينما ذهبت. اعلم انه كان من الصعب عليك أن تتركيني أعمل لساعات طوال دون قضاء الوقت معا كما اعتدنا سابقا. شكرا لك على كل زياراتك المشجعة لي. يمكننا أخيرا اللعب معا بعد التخرج.

أهدي لكم جميعا هذا العمل الذي استمتعت كثيرا بتحضيره. شكرا لكم لوجودكم في حياتي.

<div style="text-align: right">Imene OUADAH</div>

To my mother and my father who have been so generous to me.
To my family, who provided me with a suitable space-time to achieve this work.
To over 100 teachers who taught me,

from أ ، ب ، ت ، ... to $\alpha$, $\beta$, $\gamma$, ..., and from $1+1$ to $|1\rangle \otimes |1\rangle$.

All this brought me right here... and **beyond**!

<div style="text-align: right">Hacene Rabah BENAISSA</div>

# ACKNOWLEDGMENTS

# Abstract

Quantum computers have the potential to solve problems that are intractable to classical computers, nevertheless they have high error rates. One significant kind of errors is known as "*Readout Errors*". Current methods, as the *matrix inversion* and *least-squares*, are used to unfold (correct) readout errors. But these methods present many problems like oscillatory behavior and unphysical outcomes.

In 2020 BENJAMIN NACHMAN et al. suggested a technique currently used in HEP, to correct detector effects. This method is known as the "*Iterative Bayesian Unfolding*" (IBU), and they have proven its effectiveness in mitigating readout errors, avoiding problems of the mentioned methods.

Therefore, the main objective of our thesis is to mitigate readout noise of quantum computers, using this powerful unfolding method. For this purpose we generated a *uniform distribution* in the YORKTOWN IBM Q MACHINE, for 5 Qubits, in order to unfold it by IBU after being distorted by noise. Then we repeated the same experiment with a *Gaussian distribution*. Very satisfactory results and consistent with those of B. NACHMAN et al., were obtained. After that, we took a second purpose to explore unfolding in a larger qubit system, where we succeed to unfold a *uniform distribution* for 7 Qubits, distorted by noise from the MELBOURNE IBM Q MACHINE. In this case, the IBU method showed much better results than other techniques.

**Keywords** : Quantum computing, Qubit, Noise, Readout errors, Unfolding, High energy physics, Iterative Bayesian Unfolding, IBM Yorktown quantum computer, IBM Melbourne quantum computer.

# ملخص

تمتلك الحواسيب الكمومية القدرة على حل المشكلات المستعصية على الحواسيب الكلاسيكية ، غير أن لديها معدلات أخطاء عالية. يعرف أحد أهم أنواع هذه الأخطاء بـ "أخطاء القراءة". يتم استخدام طرائق حالية ، مثل طريقة *matrix inversion* و *least-squares* ، لبسط (تصحيح) أخطاء القراءة هذه ، لكن هذه الأساليب تقدم العديد من المشاكل مثل السلوك التذبذبي والنتائج الغير فيزيائية.

في عام 2020 إقترح بنجامين ناشمان وآخرون تقنيّة مستخدمة حاليا في HEP ، لتصحيح تأثيرات الكاشف. تعرف هذه الطريقة باسم "*Iterative Bayesian Unfolding*" (IBU) ، وقد أثبتوا فعاليتها في التخفيف من أخطاء القراءة ، وتجنب مشاكل الطرق المذكورة.

لذلك ، فإن الهدف الرئيسي من أطروحتنا هو التخفيف من ضوضاء القراءة الصادرة من أجهزة الكمبيوتر الكمومية ، باستخدام هذه الطريقة القوية للبسط. لهذا الغرض ، أنشأنا توزيعا منتظما في آلة YORKTOWN IBM Q ، لـ 5 Qubits ، من أجل بسطها بواسطة IBU بعد تشويهها بالضوضاء ، ثم كررنا نفس التجربة بتوزيع غاوسي. تم الحصول على نتائج مرضية للغاية ومتسقة مع نتائج ب. ناشمان وآخرون. بعد ذلك ، اتخذنا هدفا ثانيا لاستكشاف البسط في نظام كيوبتي أكبر ، حيث بسطنا توزيعا منتظما لـ 7 Qubits ، مشوه بالضجيج الصادر عن آلة MELBOURNE IBM Q. في هذه الحالة ، أظهرت طريقة IBU نتائج أفضل بكثير من التي أظهرتها التقنيات الأخرى.

**الكلمات المفتاحية**: الحاسوب الكمومي ، الكيوبت ، الضوضاء ، أخطاء القراءة ، البسط ، فيزياء الطاقة العالية ، IBU ، حاسوب IBM Yorktown الكمي ، حاسوب IBM Melbourne الكمي.



\dominitoc 
\flushbottom
\tableofcontents
\listoffigures
\mtcaddchapter


\end{romanpages}
\end{titlepage}

\setlength{\baselineskip}{20pt} 
\chapter*{Introduction} 
\adjustmtc
\addcontentsline{toc}{chapter}{Introduction}

We are in a new scientific revolution. A mystery box is open every day to enlighten a new era of science.  Interdisciplinary research with collaboration and a spirit of a lifelong learner are the keys to gain a more well-developed perspective and solving problems whose solutions transcend the scope of a single discipline. 

Forty years on, the theoretical physicist Richard Feynman published a paper: \textit{simulating physics with computers} \cite{Feynman}. He suggested that to simulate the quantum behavior of a physical system, we need to use quantum computers.  One of the significant advantages of quantum computing is the ability to massively parallel computations,  using quantum superposition.  In the last years, simulations of chemistry and condensed materials have developed as one of the dominant applications of quantum computing, offering an exponential speedup of solutions. However, quantum computers are noisy in consequence the results of calculations will have errors. One of the most important classes of errors are the readout errors. 

For years several methods have been proposed in order to minimize these readout errors and to approach as close as possible to the true value after the correction of the measurement. This problem is known as the inverse problem and a well-known method is the Matrix inversion. This method leads to unphysical entries (negative values) and has oscillatory behavior. For this, we need a new method that will give us better results to reduce readout errors.

In 2020, 	Ben. Nachman et al. published a paper where he explained how they use a  well-known method used in HEP and applied to quantum computing. The study focuses on a technique called “iterative Bayesian unfolding” (IBU). In their study, they simulated a quantum computer to compare the performance of three different error-correction: Matrix inversion, ignis, and IBU. The iterative Bayesian unfolding method have been proven very successful compared to the two other methods. 

In this thesis, we want to explore more about this method by unfolding errors from a real quantum computer. That means we want to use measured data from a quantum computer and correct them with the three unfolding methods. 

 It is interesting to see the effect of noise in a real quantum computer on our data.  For this, we will use two distributions as true data. The first one will be the uniform distribution, and the second one will be the Gaussian distribution. After the measurement, we will see how a noisy quantum computer will affect our data. Finally, we will use the unfolding methods to correct our data, and then, we will compare the three methods.  
\pagebreak
\clearpage
\thispagestyle{empty}
  In summary, the plan of this thesis to reach our objective is as follows: 
 \begin{itemize}
 \item \textbf{Chapter 1}

In the first chapter, we will present an overview of quantum computing and quantum information. We start with fundamental concepts of quantum mechanics: superposition and entanglement. Then, we will explain what is the difference between quantum and classical bit. Next,  we will see the most famous quantum gates and examples of quantum circuits. After that, we briefly review examples of quantum Algorithms. 
We finish this chapter by explaining the concepts of error detection and correction. 

 \item \textbf{Chapter 2}

This chapter is an introduction to unfolding for high-energy physics.  First, we explain the concept of the inverse problem and unfolding methods. Then, we introduce two methods, the first one is the matrix inversion and the second one is the iterative Bayesian unfolding. The IBU method is what interests us the most. 

 \item \textbf{Chapter 3}
 
After having all the needed tools, the last chapter is about applying unfolding methods from High Energy Physics to quantum information science (QIS).  We start by explaining what is a readout error and the methodology to construct a calibration matrix of a real quantum computer. Then, we will see how to use unfolding methods from HEP to Quantum computing. Next, we will see how to simulate the unfolding process with python. Finally, we apply unfolding techniques to mitigate the readout noise of the 5 Yorktown IBM Q machine with 5 qubits and the 16 Melbourne IBM Q machine with 7 qubits.

Finally, we draw our conclusion and, additional details are submitted in appendices.   
\newpage
 \end{itemize}
\chapter{Overview of Quantum Computing and Quantum Information}
\label{chap1}
\minitoc
Quantum information science QIS is a new field that combines \textit{quantum physics} with \textit{classical information science}. 
The foundations of \textit{quantum information theory} were established in the late 1980s by \textsc{Charles Bennett} and others \cite{Marinescu}.

In this chapter, we introduce the basic elements of QIS: \textit{qubit, quantum logic gates, quantum circuit, quantum algorithms, quantum error correction}. But before tackling any of these topics, we first provide a glimpse of their classical analogs.

This chapter represents a reference to all that will follow  because it provides us with all the necessary tools to achieve the goal of this thesis.

The quantum mechanics concepts on which QIS stands are \textit{superposition} and \textit{entanglement}, we, therefore, begin the chapter by recalling them.
\section{Superposition and Entanglement}
In this section, we recall the two concepts of quantum mechanics \textit{superposition} and \textit{entanglement}, then, we present \textit{Bell states}.
\subsection{Quantum Superposition Principle}
The principle of superposition of states is a fundamental principle in quantum mechanics, which represents a property that every quantum object has. If a quantum object can be either in a state $\psi_{1}$ or in another state $\psi_{2}$, then the superposition principle announce that this object is partially in $\psi_{1}$ and partially in $\psi_{2}$, and the general description of its state must be viewed as the result of a \textit{superposition} of the two original states

\begin{equation}
\Psi=c_{1}\psi_{1}+c_{2}\psi_{2}
\end{equation}

\begin{flushleft}
where $c_{1},c_{2}$ are complex coefficients.
\end{flushleft}

The new state $\Psi$ is completely defined by the original states $\psi_{1}$ and $\psi_{2}$ and has  \textit{intermediate properties} which are closer to the state with the greatest \textit{probability amplitude} ($c_{1},c_{2}$) \cite{Dirac}. \\
 
We can cite illustrative examples of the superposition of states such as:
\begin{itemize}
\item An electron can be in a state of \textit{spin-up $|\uparrow\:\rangle$ \emph{and} spin-down $|\downarrow\:\rangle$}, \textit{simultaneously}.
\[|\Phi\rangle_{\mathcal{S}}=a\,|\uparrow\:\rangle+b\,|\downarrow\:\rangle\]
\item An atom that can be both \textit{excited} $|\phi^{(1)}\rangle$ or \textit{not-excited} $|\phi^{(0)}\rangle$.
\[|\Phi\rangle_{\mathcal{E}}=a\,|\phi^{(0)}\rangle+b\,|\phi^{(1)}\rangle\]
\item A photon which can be in a superposition of \textit{linear polarization} states; polarized \textit{horizontally $|\!\leftrightarrow\rangle$ \emph{and} vertically $|\updownarrow\:\rangle$}, at the same time.
\[|\Phi\rangle_{\mathcal{P}}=a\,|\!\leftrightarrow\rangle+b\,|\updownarrow\:\rangle\]
\end{itemize}

We can generalize this concept to a number $n$ ($n\in\,\mathbb{N}$) of states $\{\psi_{_{1}},\,\psi_{_{2}},...,\,\psi_{_{n}}\}$ that can a system occupy:

\begin{equation}
\Psi\,=\sum_{n}\,c_{n}\:\psi_{n}\qquad;\quad c_{n}\in\,\mathbb{C}
\end{equation}

It expresses the general state $\Psi$ as a \textit{linear combination} of the \textit{discrete} original states $\psi_n$.\\

Although the \textit{superposition principle} exists in \textit{classical mechanics}, it is fundamentally different from any classical property, \textsc{Dirac} wrote about the \textit{non-classical nature} of superposition:\\

\textit{“the superposition that occurs in quantum mechanics is of an essentially different nature from any occurring in the classical theory, as is shown by the fact that the quantum superposition principle”} \cite{Dirac}.\\

\begin{figure}[H]
\centering
\includegraphics[width = 0.8\linewidth]{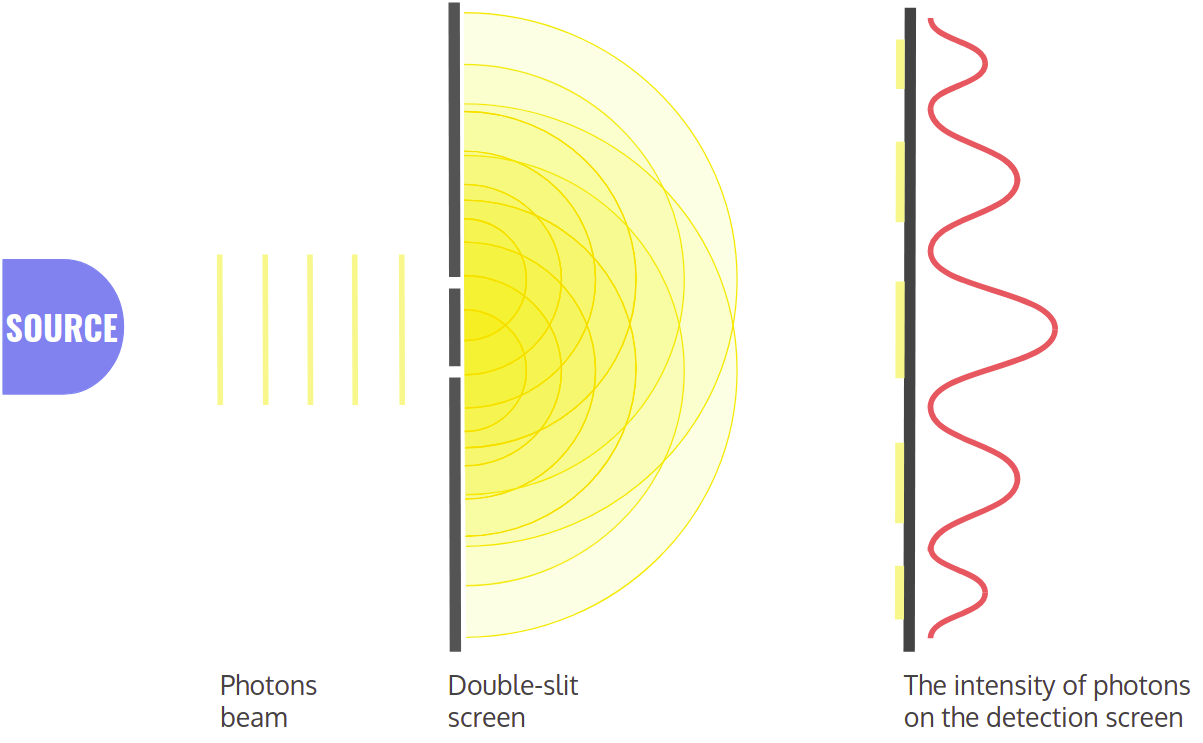}
\caption{The Double-Slit experiment.}
\label{1.1}
\end{figure}

The simplest experiment demonstrating this principle is the “\textit{double-slit experiment}”, where we emit from a source a beam of \textit{monochromatic light} to pass through an intermediate screen with two \textit{identical slits}, that must be very \textit{thin}, to finally hits a \textit{detection screen}, Figure \ref{1.1}.
When crossing the two slits, the beam splits into two components and the two components are then caused to \textit{interfere}. We need the \textit{corpuscular} aspect of light, so consider a single photon $\gamma$. Arriving at the first screen, the photon will be in a \textit{superposition} of the two paths; the photon is in a general state given by the \textit{sum} of the two states associated with the two paths. This superposition leaves marks in the form of \textit{interference fringes} in the \textit{detection screen}, when we measure the position of the photons.

The principle of superposition gives rise to an important concept, called “\textit{entanglement}”.

\subsection{Quantum Entanglement}
\label{1.1.2}
The entanglement phenomenon is a direct consequence of the superposition principle applied to composite systems, that occurs when a pair or group of particles is generated, and which are considered as an inseparable whole; where each constituent cannot be fully described without considering the other.\\
Quantum systems can become entangled through various types of interactions between subatomic particles. Experimentally there are various methods; we can cite a  well-known one, named “\emph{the  spontaneous parametric down-conversion}” (SPDC) which is an optical process that converts one photon of higher energy into a pair of entangled photons of lower energy.
In nature, a pair of entangled particles can be created for example by the decaying phenomena as the “\emph{pair production process}” \cite{Horodecki2007Feb}: \[\gamma\longrightarrow\,e^{+}+\;e^{-}\]
where the pair $(\:e^{+},\:e^{-})$ are \textit{entangled}; they form a \textit{single system}.\\

For the mathematical aspect, consider a system formed by two particles, where the first particle is in the state  $\psi \in \mathcal{H}_{1}$ and the second particle is in the state $\phi \in \mathcal{H}_{2}\,$, where $\mathcal{H}_{1}$ and $\mathcal{H}_{2}$ are respectively their \textit{Hilbert-spaces}. To represent the \textbf{general} state of the \emph{two-particle} system, we define the “\emph{Tensor product}” of the two particles states $ \psi \otimes\, \phi \,$, which is a vector in a new \textit{Hilbert-space} written $ \mathcal{H}_{1}\otimes\, \mathcal{H}_{2} $ that will carry the description of the quantum states of this composite system. This state is expressed as a \textbf{linear superposition} of the product of the two-particle states $ \psi_{i} \otimes\, \phi_{j}\:$: 
\begin{equation}\Upsilon=\sum_{i\,j}\,\alpha_{ij}\;\psi_{i} \otimes\, \phi_{j}\qquad ;\; \Upsilon\in\,\mathcal{H}_{1}\otimes\,\mathcal{H}_{2}
\end{equation}
where $\alpha_{i\,j}$ are the corresponding scalar components.\\

If $\Upsilon$ is an \textbf{entangled} state of the two particles, then it cannot be dissociated into separate states of each of them. Conversely, if one can write it as a single term $\psi_{\ast} \otimes \,\phi_{\ast}$ for some $ \psi_{\ast}\in \mathcal{H}_{1} $ and $ \phi_{\ast}\in \mathcal{H}_{2}\,$; which means that the description of the particles state in $ \Upsilon $ can be written independently (each particle in its state), hence, in the state $ \Upsilon $ the particles are \textbf{not entangled} \cite{MIT}. Otherwise, $\Upsilon$ is not entangled if and only if $\alpha_{ij}=\alpha_{i}\,\alpha_{j}\,$, which leads to:
\begin{equation}
\Upsilon=\,\Psi\,\otimes\,\Phi
\end{equation} 
where $\quad\Psi=\sum_{i}\,\alpha_{i}\,\psi_{i}\quad$ and $\quad\Phi=\sum_{j}\,\alpha_{j}\,\phi_{j}$\\
And vice versa, if $\alpha_{ij}\neq\alpha_{i}\,\alpha_{j}\,$, then the separation will be impossible.\\

We can illustrate this reasoning by taking \emph{two-dimensional Hilbert-spaces} $\mathcal{H}_{1}$ and $\mathcal{H}_{2}$. Let $( e_{1},e_{2} )$ the basis for $ \mathcal{H}_{1} $ and $( s_{1},s_{2} )$ basis for $ \mathcal{H}_{2} $. Then, the general state is written as: 
\begin{equation}
\Upsilon_{\mathcal{G}} = g_{11}\: e_{1}\otimes s_{1}+g_{12}\: e_{1}\otimes s_{2}+g_{21}\:e_{2}\otimes s_{1}+g_{22}\:e_{2}\otimes s_{2} 
\label{eq1.5}
\end{equation}
This state is encoded by a matrix of coefficients:
\begin{equation}
\mathcal{G}=\begin{pmatrix} \,g_{11} & g_{12}\, \\ \,g_{21} & g_{22}\, \end{pmatrix}
\end{equation}
If we have understood the reasoning above, the state is \emph{not entangled} if there exist arbitrary constants $\alpha_{1},\,\alpha_{2},\,\beta_{1},\,\beta_{2}$ such that:
\begin{equation}
\Upsilon_{\mathcal{G}} = (\alpha_{1}\:e_{1}+\alpha_{2}\:e_{2})\:\otimes \: (\beta_{1}\:s_{1}+\beta_{2}\:s_{2})
\label{eq1.7}
\end{equation}
By expanding the expression \eqref{eq1.7} than identifying the constants with \eqref{eq1.5}, we find the four equalities below:
\begin{equation}
g_{11}=\alpha_{1} \: \beta_{1}\qquad ;\quad g_{12} =\alpha_{1} \: \beta_{2} \qquad ;\quad g_{21} =\alpha_{2} \: \beta_{1}\qquad ;\quad g_{22} =\alpha_{2} \: \beta_{2}\quad
\end{equation}
Which therefore leads to the following condition:
\begin{equation}
\mathcal{G}=\begin{pmatrix} \,\alpha_{1} \: \beta_{1} & \alpha_{1} \: \beta_{2}  \, \\ \,\alpha_{2} \: \beta_{1} & \alpha_{2} \: \beta_{2} \, \end{pmatrix}
\quad\longrightarrow\quad \det(\,\mathcal{G}\,)=0\;
\end{equation}
So we conclude that $\Upsilon_{\mathcal{G}}$ is \textbf{not entangled} if and only if $\: \det(\,\mathcal{G}\,)=0 \;$. This result is available just for \emph{two-dimensional} vector spaces \cite{MIT}.\\

We will introduce at next the most \textbf{important} \textsc{feature} of the \textsc{entanglement phenomenon}. For this let's take the example of the \textit{neutral meson} $\eta_{\,0}$ decay, which decays into an \textit{entangled pair} of a \textit{muon} $\mu^{-}$ and an \textit{anti-muon} $\mu^{+}\,$.
\begin{equation}
\eta_{\,0}\longrightarrow\;\mu^{-}+\;\mu^{+}
\end{equation}
In this decay process, the various conservation laws are respected ( total energy, momenta, spin, \ldots). As known the $\eta_{\,0}$ is \textbf{spinless} and which decays to two particles with \emph{spin-\begin{small}(1/2)\end{small}}$\,$, Since the conservation laws require that the total \textit{initial} and \textit{final} spins must be \textit{zero}. Thus if we found after a measurement that the $1^{st}$ particle is \textit{spin-up} on some axis, then the $2^{\,nd}$ particle is found to be \textit{spin-down} when measured on the same axis \cite{MIT}. \\
We can check this mathematically by firstly describing the entangled pair $(\mu^{+},\mu^{-})$ state of total spin zero as follow:
\begin{equation}
\vert\psi\rangle=\dfrac{1}{\sqrt{2}}\:(\:\vert+\rangle\otimes\vert-\rangle-\:\vert-\rangle\otimes\vert+\rangle\:)
\end{equation}
where $\vert+\rangle\equiv|\!\uparrow\,\rangle$ and $\vert-\rangle\equiv|\!\downarrow\,\rangle$ are respectively, the \textit{spin-up \emph {and} spin-down} states.
Also, $S_{n}^{T}\:\vert\psi\rangle=0$, where $\,S_{n}^{T}$ is the \emph{total n-component of spin angular momentum}, given by $\:S_{n}^{T}=S_{n}\otimes 1\!\!1+1\!\!1\otimes S_{n}\:$ and $n$ an arbitrary direction.\\

If we measure $\vert\psi\rangle$ along the basis $\lbrace\vert+\rangle ,\vert-\rangle\rbrace$, then the probability $\mathbb{P}$ of $\vert\psi\rangle$ to be in the states $\vert\pm\pm\,\rangle\,,\:\vert\pm\mp\,\rangle$ and $\vert\psi\rangle$ after measurement are respectively:
\begin{eqnarray}
\mathbb{P_{\pm\pm}}&=&\vert \langle\,\pm\pm\vert\psi\rangle \vert^{\,2}=0\qquad\,;\qquad \vert\psi\rangle\longrightarrow\vert\psi\rangle_{\pm\pm}=0\\
\mathbb{P_{\pm\mp}}&=&\vert \langle\,\pm\mp\vert\psi\rangle \vert^{\,2}=\tfrac{1}{2}\qquad;\qquad\vert\psi\rangle\longrightarrow\vert\psi\rangle_{\pm\mp}=\pm\:\vert\pm\,\mp\:\rangle
\end{eqnarray}
These results allow us to say that the two \textit{entangled} particles have always \textit{opposite} spins. Hence, the measurement of one of them will affect the other, regardless of how much further they are apart$\,$! Hence the \textsc{entangled} particles are \textsc{perfectly correlated}\footnote{\hspace{0.1cm} This \textsc{entanglement feature} seems to violate the \textit{local realism} which gave rise in 1935 to what has been known as \textsc{the Einstein-Podolsky-Rosen (EPR) paradox} \cite{Einstein}.}.\\ 

\textsc{Alain Aspect} describes the difference between classical and quantum correlations in bipartite systems as follows:\\ 

\emph{“If we have a pair of identical twins we do not know what their blood type is before testing them, but if we determine the type of one, we know for sure that the other is the same type. We explain this by the fact that they were born with, and still carry along, the same specific chromosomes
that determine their blood type. Two entangled photons are not two distinct systems carrying identical copies of the same parameter. A pair of entangled photons must instead be considered as a single, inseparable system, described by a global wave function that cannot be factorized into single-photon states”} \cite{Jaeger}. 


\subsubsection{Bell States}
Bell states are defined as a \textit{set} of \textit{maximally entangled basis states}. Named after the Irish physicist \textsc{John Stewart Bell} \cite{MIT}.\\ 

We aim in what follows, to construct these states in the particular case of a \emph{two-particles} system of \emph{spin-\begin{small}(1/2)\end{small}}, which are described in the \textit{Hilbert-space} ($\mathcal{H}_1\otimes\mathcal{H}_2$), where $ \mathcal{H}_1$ and $ \mathcal{H}_2$ are both \emph{two-dimensional} \textit{Hilbert-spaces}. Let's suppose that these two particles are produced in the specific \textit{normalized \emph{and} entangled} state below:
\begin{equation}
\vert\Phi_{_{0}}\rangle= \frac{1}{\sqrt{2}}\;(\;\vert+\rangle_{_{1}}\,\vert+\rangle_{_{2}}\:+\:\vert-\rangle_{_{1}}\,\vert-\rangle_{_{2}})
\end{equation}
We can quickly verify these two properties, respectively:
\begin{equation}
\langle\Phi_{_{0}}\vert\Phi_{_{0}}\rangle =\tfrac{1}{2}\:\bra{++}\ket{++}\:+\:\tfrac{1}{2}\:\bra{--}\ket{--}=1
\end{equation}
\begin{equation}
\mathcal{G}=\begin{pmatrix} \,\frac{1}{\sqrt{2}} & 0 \, \\ \,0 & \frac{1}{\sqrt{2}}\, \end{pmatrix} \quad\longrightarrow\quad \det(\mathcal{G})=\tfrac{1}{2} 
\end{equation}
The dimensionality of ($\mathcal{H}_1\otimes \mathcal{H}_2$) can be calculated by the following relation:
\[\dim( \mathcal{H}_1\otimes \mathcal{H}_2 )=\dim(\mathcal{H}_1)\times \dim(\mathcal{H}_2)\]
Since $\,\dim(\mathcal{H}_1\otimes \mathcal{H}_2)=4\,$, we therefore need four \textit{entangled} states to construct our \emph{Bell basis}. If we consider the state $\ket{\Phi_{_0}}$ as the first one, then the other three can be generated by applying the operators $(\,1\!\!1\otimes\,\sigma_k)$ on $\ket{\Phi_{_0}}\,$ \cite{MIT}.
\begin{equation}
\vert\Phi_{_{k}}\rangle =(\,1\!\!1 \otimes \sigma_{k}) \: \vert\Phi_{_{0}}\rangle \qquad;\;\, k=1,\,2,\,3
\end{equation}
where $\sigma_{k}$ are the \emph{Pauli matrices}, and their actions on $\ket{\pm}$ are as follows:\\
\begin{equation}
\sigma_1\ket{\pm}=\ket{\mp}\quad;\quad\sigma_2\ket{\pm}=\pm\,i\ket{\mp}\quad;\quad\sigma_3\ket{\pm}=\pm\ket{\pm}
\end{equation}
where $\:\sigma_1=\sigma_x\,$, $\:\sigma_2=\sigma_y\,$, $\:\sigma_3=\sigma_z\,$.\\

\begin{flushleft}
Applying this, we get the full list of \textit{Bell states}, which are explicitly written:
\end{flushleft}
\begin{eqnarray}
\vert\Phi_{_{0}}\rangle =\:(\,1\!\!1 \otimes\! \:1\!\!1\,\,) \: \vert\Phi_{_{0}}\rangle \;=\:\dfrac{1}{\sqrt{2}}\;(\:\vert+\rangle_{_{1}}\,\vert+\rangle_{_{2}}\:+\:\vert-\rangle_{_{1}}\,\vert-\rangle_{_{2}}\:)\\
\vert\Phi_{_{1}}\rangle =\:(\,1\!\!1 \otimes \sigma_{1} )\: \vert\Phi_{_{0}}\rangle \;=\:\dfrac{1}{\sqrt{2}}\;(\:\vert+\rangle_{_{1}}\,\vert-\rangle_{_{2}}\:+\:\vert-\rangle_{_{1}}\,\vert+\rangle_{_{2}}\:)\\
\vert\Phi_{_{2}}\rangle =\:(\,1\!\!1 \otimes \sigma_{2} )\: \vert\Phi_{_{0}}\rangle \;=\:\dfrac{i}{\sqrt{2}}\;(\:\vert+\rangle_{_{1}}\,\vert-\rangle_{_{2}}\:-\:\vert-\rangle_{_{1}}\,\vert+\rangle_{_{2}}\:)\\
\vert\Phi_{_{3}}\rangle =\:(\,1\!\!1 \otimes \sigma_{3} )\: \vert\Phi_{_{0}}\rangle \;=\:\dfrac{1}{\sqrt{2}}\;(\:\vert+\rangle_{_{1}}\,\vert+\rangle_{_{2}}\:-\:\vert-\rangle_{_{1}}\,\vert-\rangle_{_{2}}\:)
\end{eqnarray}
By a simple calculation it will be clear that they form an \emph{orthonormal} basis 
\begin{equation}
\langle\Phi_{\mu}\vert\Phi_{\nu}\rangle = \,\delta_{\mu\nu}
\end{equation}
where 
\begin{equation}
\vert\Phi_{\mu}\rangle =(\,1\!\!1 \otimes \sigma_{\mu}) \: \vert\Phi_{_{0}}\rangle \qquad;\;\, \mu=0,\,1,\,2,\,3
\end{equation}
and $\:\sigma_0=1\!\!1\,$.\\
Finally, we have an \textit{orthonormal basis} of \textit{entangled states} which defines the \emph{\textbf{Bell basis}}.
\pagebreak
\section{The Quantum Bit (Qubit)}
\label{sec 1.2}
In this section, we introduce the \textit{basic information unit} of QIS the “qubit”, then we explore its \textit{physical implementation}. But before that, let's take a look at its \textit{classical equivalent}, the “bit”.
\subsection{Classical Bit}

In our daily life, we use decimal numbers for almost everything: weight, height, phone number, etc.
These numbers are written in \textbf{Base-ten}. For this, we use ten digits: from 0 to 9, all of the numbers use some combination of these ten digits; each number can be decomposed into successive powers of 10. For example:  
\begin{equation}
2021 = 2\times10^3 + 0\times10^2 +2\times10^1+ 1\times10^0.
\end{equation}
The number $9$ is the last digit in base-ten. To go beyond $9$, we need to change the row. If the row of units is complete, we start the row of tens, and we put the units back to zero. We continue the counting until $99$, when we run out of digits, we add another $0$ and one on the left $(100)$ and so on.

In decimal, at every change of powers of 10, one passes from the units ($10^0$) to the tens ($10^1$), then to the hundreds ($10^2$). It is the same thing for \textbf{base-2}, but they are the powers of $2$. 

For example, 
\begin{equation}
(1011)_2 = 1\times2^3+0\times2^2+1\times2^1+1\times2^0  = (11)_{10}.
\end{equation}
A computer works with electric current, and to communicate with an electronic device, we must send electric signals to it.
A computer's processor is made up of transistors; we can imagine these transistors as small switches that can be turned \textsc{on} or \textsc{off}; they let the current pass or not like a light switch. The processor interprets and executes the instructions by turning millions of connected transistors \textsc{on} and \textsc{off}.

In the language of a computer, \textsc{on} and \textsc{off} are represented by the numbers $1$ and $0$. This language is called the \textbf{binary language}; the computer does not work with decimal numbers but uses the binary system. \textit{A single binary value is called a \textbf{“bit”}}.

 A bit (binary digit) is the most basic unit of information in classical computing. To store the value of the bit, we use a capacitor inside a memory device. The value is stored as either above or below a specific level of electric charge in a capacitor; it can hold only one of two values: $0$ or $1$.

Any number from our decimal system can be represented in binary; $0$ in decimal is $0$ in binary. Similarly, $1$ in decimal is $1$ in binary. $2$ in decimal becomes $10$ in binary\footnote{Create a new column and start afresh, using $1$ and $0$.}.

The table below shows how decimal numbers convert to binary.

\vspace{0.5cm}
\begin{tabular}{|c|c|c|c|c|c|c|c|c|c|c|c|c|c|c|c|c|}
\hline 
Decimal & 0 & 1 & 2 & 3 & 4 & 5 & 6 & 7 & 8 & ... & 15 \\ 
\hline 
Binary & 0000 & 0001 & 0010 & 0011 & 0100 & 0101 & 0110 & 0111 & 1000 & ... & 1111 \\ 
\hline 
\end{tabular}
\newline

Bits are so small, it is uncommon to work with only one. Bits are usually put together in a series of eight-bit, to form a \textbf{“byte”}. A  byte contains sufficient information to store a single $ASCII$ character\footnote{Stands for "\textit{American Standard Code for Information Interchange}" $ASCII$ is a character encoding that uses numeric codes to represent characters.}, such as “$O$”: 01001111 and “$B$”: 01000010.

There is a unique binary number for each letter in upper and lower case. By knowing the binary number, we can write complete words with a series of $1s$ and $0s$. For example:

\begin{itemize}
\item Benaissa: \small{01000010 01100101 01101110 01100001 01101001 01110011 01110011 01100001}
\item Ouadah: \small{01001111 01110101 01100001 01100100 01100001 01101000}
\end{itemize}



\subsubsection*{Physical Implementation of Bit}

Computers represent data as voltage or current signals at the hardware level. They could be analog, digital, or hybrid. Computers
that use analog representations exclusively are uncommon. It is challenging to control physical signals with a high degree of accuracy.

In a digital computer, the input is converted and transmitted as electrical pulses and represented by two individual states, \textsc{on} and \textsc{off}. The sequence of \textsc{on} and \textsc{off} forms the electrical signals that the computer can understand.

\begin{figure}[H]
\centering
\includegraphics[width=0.7\textwidth]{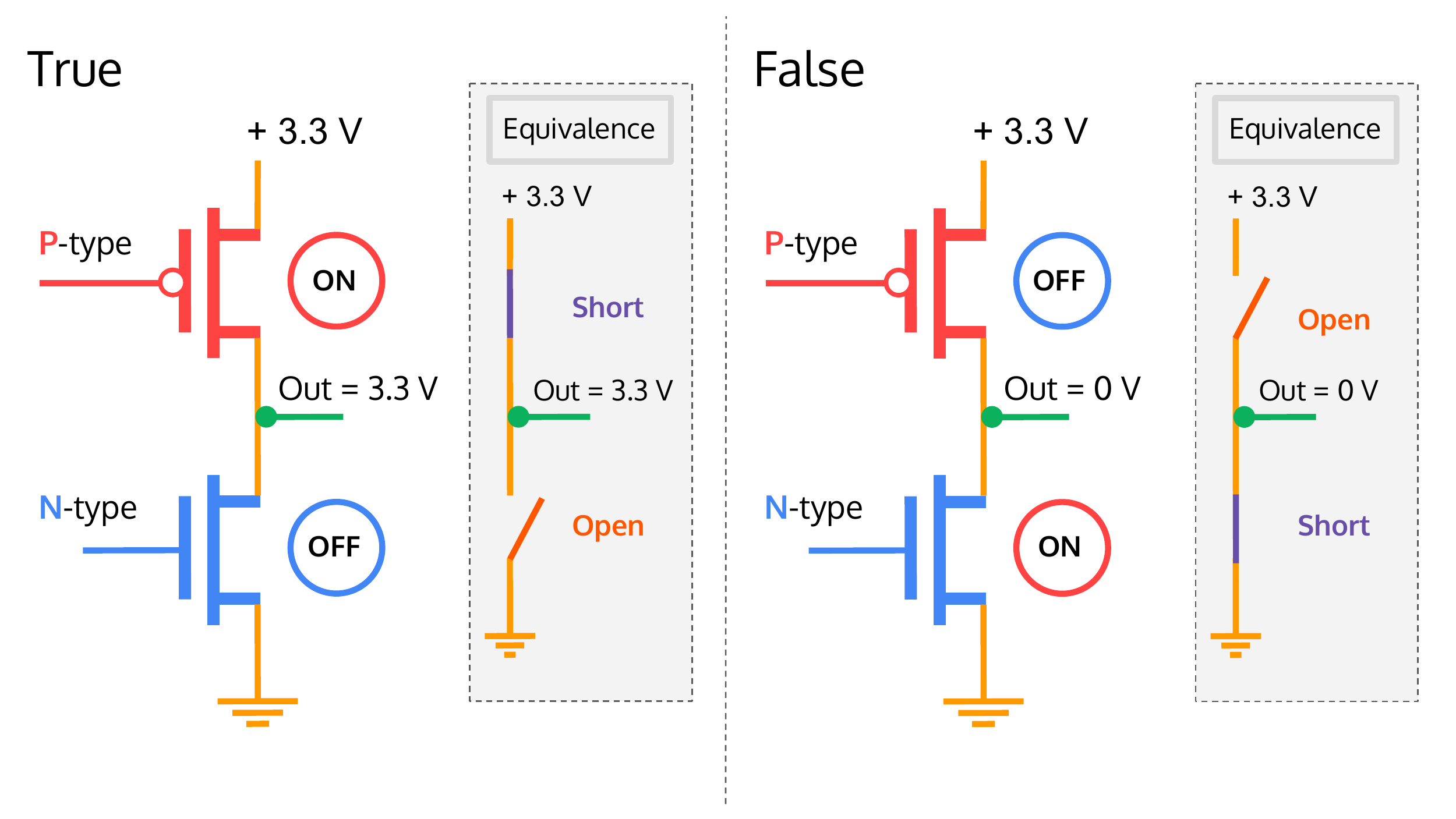}
\caption{A binary bit is true if a voltage is present and false if the voltage is absent.}
\label{1.2}
\end{figure}

In positive logic, the presence of a voltage is called: \textit{“1”}, true or high state. If the voltage is absent then this is called the \textit{“0”}, false or low state. The Figure \ref{1.2} shows the output of a typical \textit{complementary metal-oxide-semiconductor} (CMOS) circuit. The left side shows the condition with a true bit, and the right side shows a false. The output of each digital circuit consists of a P-type transistor “on top of” an N-type transistor.

In digital circuits, each transistor is essentially \textsc{on} or \textsc{off}. If the transistor is \textsc{on}, it is equivalent to a short circuit between its two output pins. If the transistor is \textsc{off}, it is equivalent to an open circuit between its outputs pins.
\subsection{Single Qubit}
\label{1.2.2}
We saw that a classical bit is the elementary unit of information, which can be in one of two defined states 0 or 1. Similarly, a quantum bit or “$\,\mathfrak{qubit}\:$” is the elementary unit of quantum information in quantum computers, and which can be unlike the classical bit, in infinite states, but gives only an observed output of 0 or 1 \cite{Dancing}. \\
A qubit is a \textit{two-level} system that can be expressed using two orthonormal basis states:
\begin{equation}
\vert0\rangle=\begin{pmatrix}
 1\\0 \end{pmatrix}\qquad;\qquad \vert1\rangle=\begin{pmatrix} 0\\1 \end{pmatrix},
\end{equation}
where $\vert0\rangle$ is the qubit state which is certain to give 0 when measured, and $\vert1\rangle$ the qubit state that give 1 with certainty. The states $\vert0\rangle$ and $\vert1\rangle $ are called the “\textit{computational basis states}”\footnote{\hspace{0.1cm}We can assign $|0\rangle$ and $|1\rangle$, respectively, to the \textit{spin-up} $|\uparrow\,\rangle$ and \textit{spin-down} $|\downarrow\,\rangle$ states of an \textit{electron}, or the \textit{horizontal polarization} $|\!\leftrightarrow\,\rangle$ and the \textit{vertical polarization} $|\updownarrow\:\rangle$ of a \textit{photon}.}.\\
In the quantum realm, the qubit is in a coherent \textbf{superposition} of these two states; hence the state describing it can be written as a linear combination of the vectors $\vert0\rangle,\vert1\rangle\!\in\!\mathbb{C}^{2}$ \cite{Dancing}:
\begin{equation}
\vert\psi\rangle= \,a\,\vert0\rangle \,+\, b\,\vert1\rangle \qquad;\quad a,b\,\in\, \mathbb{C}
\end{equation} 
When performing a measurement, the state $\vert\psi\rangle$ collapses irreversibly to either $\vert0\rangle$ or $\vert1\rangle$, with corresponding probabilities $|a|^{2}$ and $|b|^{2}$, respectively, where $a$ and $b$ represent the \textit{probability amplitudes}. Furthermore, they achieve the relation $|a|^{2}+|b|^{2}=1\,$, from the \textit{normalization condition}, and which describes the \textit{probability conservation}.\\ 
The fact that the qubit can be in infinite states, leads us to look for a representation, that is able to show the enormous \textit{flexibility} of the qubit, in comparison with a classical bit.\\
Firstly, we can express the complex numbers ($a, b$) in \textit{polar coordinates}; $a=|a|\,e^{i\phi_{a}}$ and $b=|b|\,e^{i\phi_{b}}$, then $\vert\psi\rangle$ becomes:
\begin{align*}
\vert\psi\rangle&=\;|a|\,e^{i\phi_{a}}\,\vert0\rangle+\,|b|\,e^{i\phi_{b}}\,\vert1\rangle\\
&= \,e^{i\phi_{a}}\,(|a|\,\vert0\rangle+|b|\,e^{i\phi}\,\vert1\rangle)\:
=\,e^{i\phi_{a}}\,\vert\chi\rangle
\end{align*}
where $\phi=\phi_{b}-\phi_{a}\,$.\\
If we calculate the probability of measuring $\vert\psi\rangle$ in the state $|\,\mathfrak{q}\,\rangle;\:\mathfrak{q}=0,1\,$:
\begin{equation}
\mathbb{P}_{\mathfrak{q}}=\,|\langle \mathfrak{q}|\psi\rangle|^{2}=\:|e^{i\phi_{a}}\langle\mathfrak{q}|\chi\rangle\,|^{2}=\:|\langle\mathfrak{q}|\chi\rangle|^{2}
\end{equation}
where $|e^{i\phi_{k}}|=1;\,k=a,b\,$.\\
From this result, we can see that the probability for the state $|\chi\rangle$ is identical to that for the state ($e^{i\phi_{a}} \,|\chi\rangle$). Hence, we conclude that the two states are equivalent $|\chi\rangle\,\equiv\,e^{i\phi_{a}}\vert\chi\rangle\,,\,\forall\,\phi_{a}\in\mathbb{R}\,$; we cannot differentiate between $|\chi\rangle$ and $e^{i\phi_{a}}\,|\chi\rangle$ when we  measure. \\ The factor $e^{i\phi_{a}}$ is called the “\textbf{\textit{global phase}}” and it's not observable. What we can measure is the phase difference ($e^{i(\phi_{b}-\phi_{a})}$) between the states $|0\rangle$ and $|1\rangle$, which is called the “\textit{\textbf{relative phase}}” \cite{Dancing}. \\ 
From this fact, we can choose the \textit{complex} coefficient $a$ to be \textit{real}, then the \textit{normalization condition} $|a|^{2}+|b|^{2}=1\,$, can be identified with the \textit{trigonometric identity} $\cos^{2}(x)+\sin^{2}(x)=1$. Since $|a|^2$ and $|b|^2$ are \textit{probabilities}, their values are confined between 0 and 1, i.e.
\[0\leq\,(\,|a|=\sqrt{\mathbb{P}_{0}}\:;\,|b|=\sqrt{\mathbb{P}_{1}}\,)\,\leq1\quad\Longrightarrow\quad0\leq\,(\,\cos(x)\,;\,\sin(x)\,)\,\leq 1\]
So $0\leq x\leq\pi/2$. For later reasons we take $x=\theta/2$, thus we can write:
\[|a|=\cos(\theta/2)\qquad;\qquad |b|=\sin(\theta/2)\]
We can now describe the state of any qubit through the two variables $\phi$ and $\theta\,$:
\begin{equation}
\vert\psi\rangle=\, \cos(\theta/2)\,\vert0\rangle\,+\:e^{i\phi}\,\sin(\theta/2)\,\vert1\rangle,
\label{eq1.30}
\end{equation}
where $0\leq\theta\leq\pi\:,\;0\leq\phi<2\pi$.
\vspace{0cm} If we interpret $\theta$ and $\phi$ as \textit{spherical coordinates} (with $r=\langle\psi|\psi\rangle=1$), then we can plot any single-qubit state on the surface of a unit sphere, known as the “\textit{Bloch sphere}”. Each \textbf{point} on its surface represents a \textbf{qubit state} \cite{Dancing}.
\begin{figure}[H]
\centering
\includegraphics[width = 0.6\linewidth]{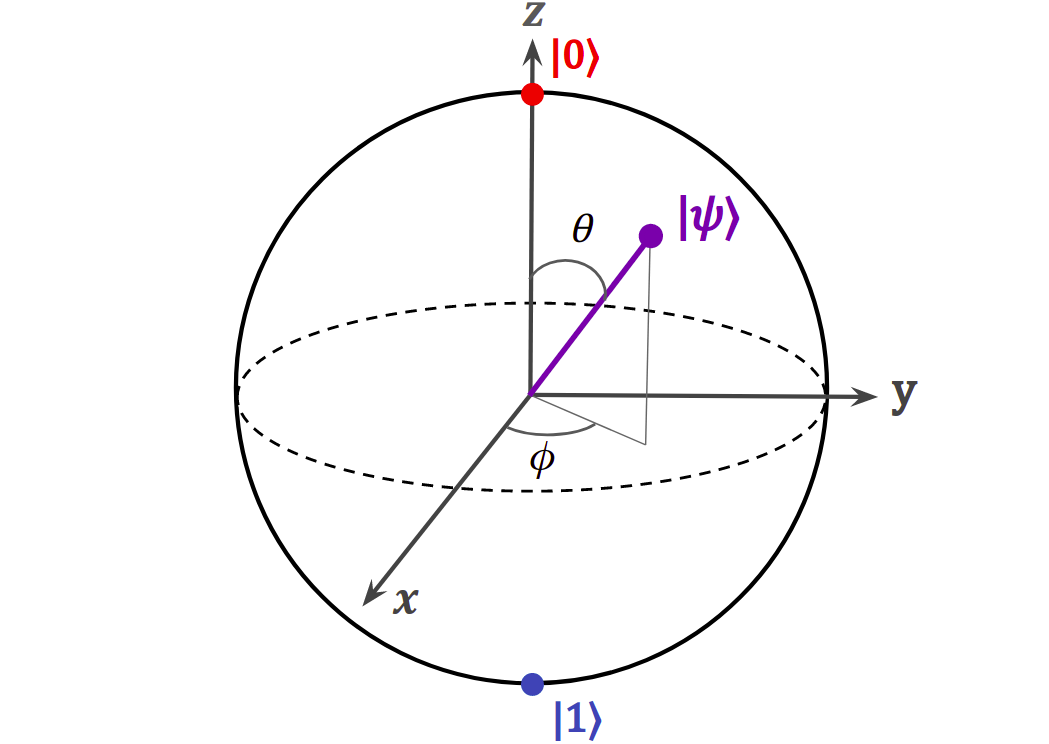}
\caption{Geometrical representation of the 2D complex state vector $|\psi\rangle$, mapped by the \textbf{Bloch vector} onto real 3D space called the \textbf{Bloch sphere}, of \textit{radius} $r=1$, \textit{colatitude angle} $\theta\,$, and \textit{azimuth angle} $\phi\,$. Each qubit state is represented by a point. The state $|0\rangle$ is at the north pole, and $\vert1\rangle$ at the south pole.\\}
\label{1.3}
\end{figure}
The two extreme cases are when the qubit is completely in either the state $|0\rangle$ or $|1\rangle$, which can be obtained respectively, for $(\theta\!=\!0)\rightarrow|\psi\rangle=|0\rangle$ and $(\theta\!=\!\pi)\rightarrow|\psi\rangle=|1\rangle$, for any angle $\phi$; since the vectors $|0\rangle$ and $|1\rangle$ are perpendicular\footnote{They are \textbf{orthogonal} in \textit{Hilbert-space}. while, in the \textit{Bloch sphere}, they are \textbf{perpendicular}, because the angles in the \textit{Bloch sphere} are \textbf{double} those in the \textit{Hilbert-space} (i.e., $\theta_{\textsc{b}}=\theta$; $\;\theta_{\textsc{h}}=\theta/2$).} to the plane ($x,y$), as shown in Figure \ref{1.4}:

\begin{figure}[H]
\centering
\includegraphics[width = 0.35\linewidth]{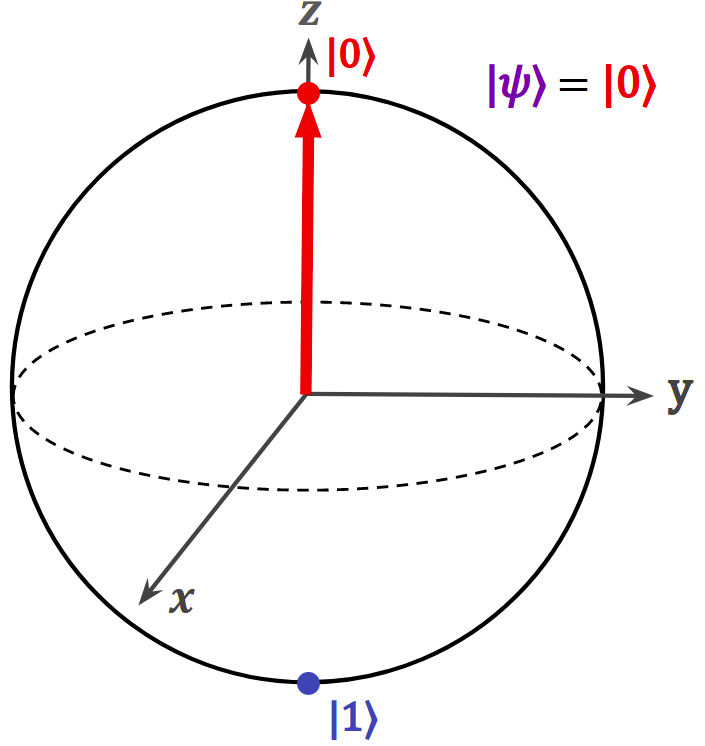}\hspace{1cm}
\includegraphics[width = 0.35\linewidth]{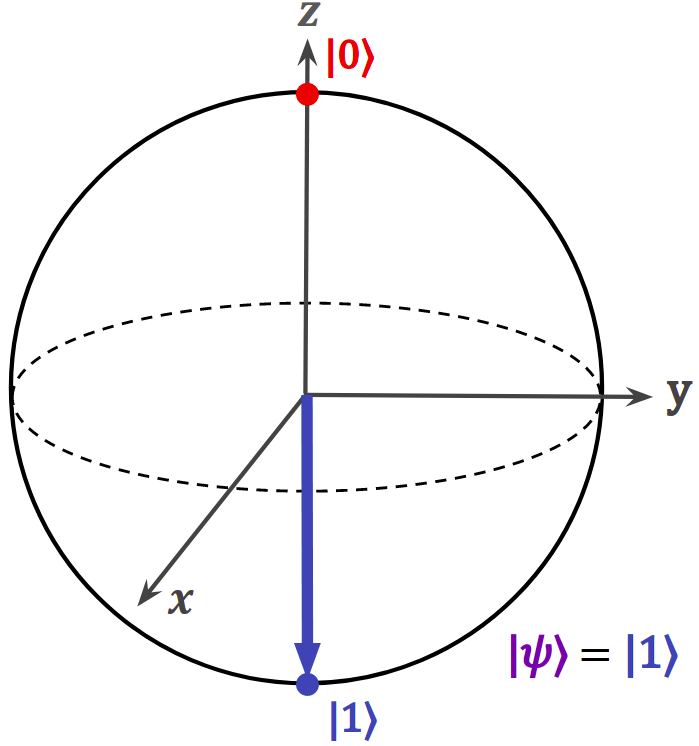}
\caption{The two extremum states of the qubit. In the left: the Bloch vector of the state $|\psi\rangle=|0\rangle$. In the right: the Bloch vector of the state $|\psi\rangle=|1\rangle$, for ($\theta=0,\,\pi;\,\forall\phi$), respectively.}
\label{1.4}
\end{figure}
These two states are those in which a \textit{classical bit} is confined. But, the qubit can be in any point of coordinates ($\theta,\phi,r=1$) on the \textit{Bloch sphere}. This clearly shows the large difference between the qubit and its classical equivalent. We can take for example $(\pi/2,0,1)$ which gives from equation \eqref{eq1.30} the state $|\psi\rangle=\frac{1}{\sqrt{2}}|0\rangle+\frac{1}{\sqrt{2}}|1\rangle$ (which is usually noted $|+\rangle$)\footnote{To avoid any confusion, the state $|+\rangle$ represents an \textbf{equal superposition} state of a qubit, which has no relation to that of \autoref{1.1.2}, which represents a \textbf{spin-up} state of a \textit{spin-\begin{small}(1/2)\end{small}} particle.}. This state is characterized by an equal probabilities $\mathbb{P}_{0,1}=\frac{1}{2}$, hence, the qubit is in perfect superpositions equilibrium of the two states $|0\rangle,\,|1\rangle$. 
\begin{figure}[H]
\centering
\includegraphics[width = 0.75\linewidth]{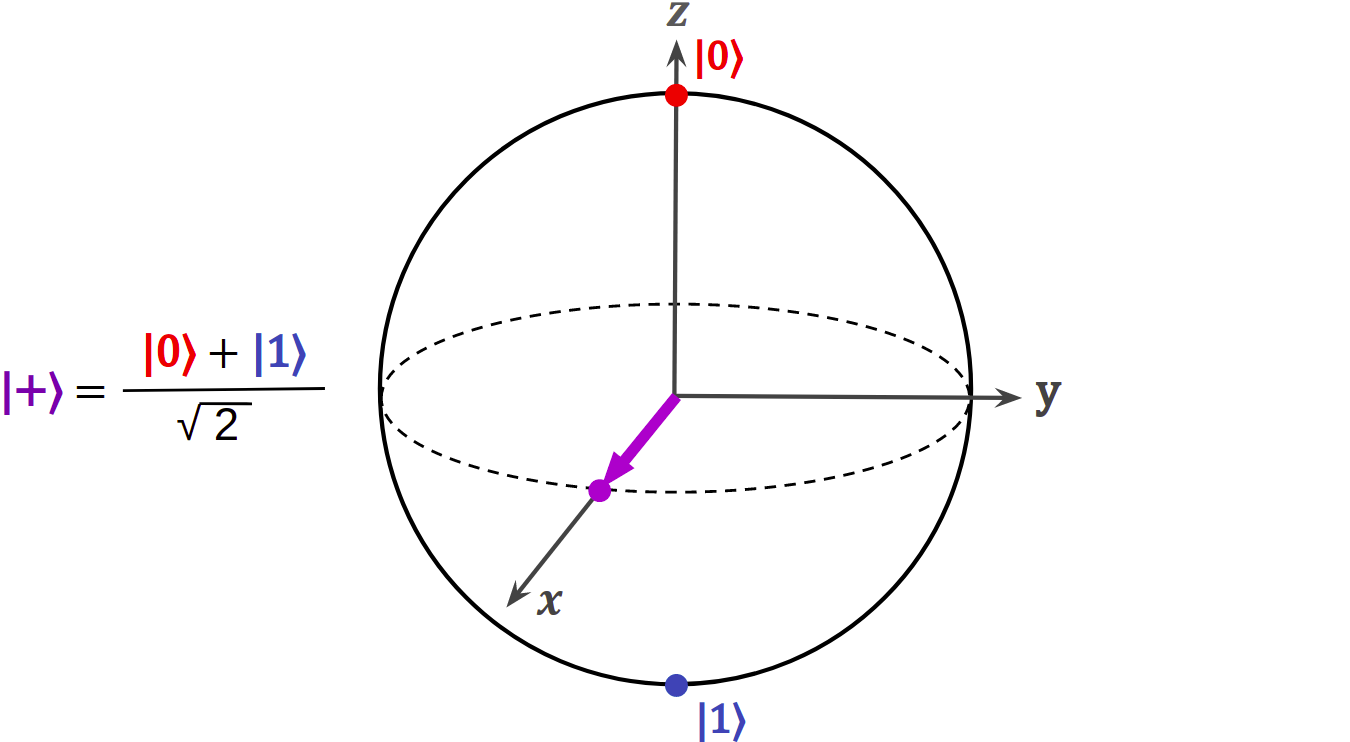}
\caption{Representation of a superposition balance state $|+\rangle$ for ($\theta=\frac{\pi}{2},\phi=0,\,r=1$).}
\label{1.5}

\end{figure}
If we measure this specific qubit state or any other one, it will collapse to either $|0\rangle$ or $|1\rangle$. One may think: “if the qubit does eventually produce only a 0 or 1 when measured, then what can bring to us more than a classical bit ?”. We say that the quantum properties of the qubit allow us to manipulate it behind the \textit{classical limits}, in ways that can only be described through \textit{quantum mechanics}; in other words, the modest 0 or 1 values, that we observe are very \textit{special}; by virtue of the \textit{remarkable quantum} manipulations that allowed us to get them. Therefore, quantum computing is the art of exploiting these quantum properties to perform tasks that are impossible by the classical equivalent.
\subsection{Multiple Qubits}
A qubit by itself is not enough to represent the \textit{useful information} needed. How much can a single \textsc{letter} tell us ? Almost nothing. But if we have two or more letters, we can then form a meaningful \textsc{word}. By grouping words, we can syntax a \textsc{sentence}, which is able to give us \textit{significant information}. With those sentences you just read, we were able to provide \textit{adequate information} to illustrate this fact. In classical information processing, one bit can give us just a 0 or 1, but by grouping bits together we can represent \textit{useful information}. The same reasoning is going on quantum information processing, where the power of quantum computing appears when considering \textit{multi-qubit} systems, in which the qubits interact with each other in a way that cannot be achieved by classical bits.\\[0.1cm]
To understand how can $\mathfrak{multi}$-$\mathfrak{qubit}$ systems make quantum computing more \textit{powerful} than their \textit{classical} equivalents, let's take a 2-$\mathfrak{qubit}$ system, where the two qubits are noted $\mathfrak{q}_{_{1}}$ and $\mathfrak{q}_{_{2}}$, with their corresponding states $|\psi\rangle_{1},\,|\psi\rangle_{2}$, and basis $\lbrace\,|0\rangle_{1},|1\rangle_{1}\rbrace\in\mathcal{H}_{1}$ and $\lbrace\,|0\rangle_{2},|1\rangle_{2}\rbrace\in\mathcal{H}_{2}$, respectively, where $\mathcal{H}_{1},\,\mathcal{H}_{2}$ are \textit{two-dimensional Hilbert spaces}. The two-qubit states are expressed:
\[\vert\psi\rangle_{_{1}}= \,a_{1}\,\vert0\rangle_{_{1}} \,+\, b_{1}\,\vert1\rangle_{_{1}} \quad;\quad\vert\psi\rangle_{_{2}}= \,a_{2}\,\vert0\rangle_{_{2}} \,+\, b_{2}\,\vert1\rangle_{_{2}} \qquad;\quad a_{1},a_{2},b_{1},b_{2}\,\in\, \mathbb{C}\]
From the normalization condition we get $|a_1|^2+|b_1|^2=1;\,|a_2|^2+|b_2|^2=1$.\\
As seen in \autoref{1.1.2}, composite systems are built by applying the “\textit{tensor product}”, where the 2-qubit system is described by the state  $|\psi\rangle_{1}\otimes\,|\psi\rangle_{2}\,$, which lives in the \textit{four-dimensional}\footnote{\hspace{0.1cm}Dimensions in a \textit{tensor product} are \textbf{multiplied}: $\dim(\,\mathcal{H}_{1}\otimes\mathcal{H}_{2})=\dim(\,\mathcal{H}_{1})\times \dim(\,\mathcal{H}_{2})$} complex vector space (\textit{Hilbert-space}) $\mathcal{H}_{1}\otimes\mathcal{H}_{2}\,$, with basis:
\[\lbrace\;|0\rangle_{1}\otimes\:|0\rangle_{2}\;\,,\;\:|0\rangle_{1}\otimes\:|1\rangle_{2}\;\,,\;\:|1\rangle_{1}\otimes\,|0\rangle_{2}\;\,,\;\:|1\rangle_{1}\otimes\:|1\rangle_{2}\:\rbrace\]
For brevity of notation we will omit the symbol “$\otimes$” between the kets $|0\rangle,\,|1\rangle$, and the subscripts $1,2$ on them:
\begin{equation}
\lbrace\;|00\rangle\,,\:|01\rangle\,,\:|10\rangle\,,\:|11\rangle\:\rbrace
\end{equation}
To explicit the form of these basis states, we compute their \textit{tensor product}. Let's calculate the first one:
\[\vert00\rangle=|0\rangle\otimes|0\rangle=\begin{pmatrix}
 1\\0 \end{pmatrix}\otimes\begin{pmatrix}
 1\\0 \end{pmatrix}=\begin{pmatrix}1\times\begin{pmatrix}
 1\\0 \end{pmatrix}\\0\times\begin{pmatrix}
 1\\0 \end{pmatrix} \end{pmatrix}=\begin{pmatrix}1\\0\\0\\0 \end{pmatrix},\]
By analogous calculations, we obtain the explicit form of the four \textit{orthonormal} basis states:
\begin{equation}
\vert00\rangle=\begin{pmatrix}1\\0\\0\\0 \end{pmatrix}\quad;\quad
\vert01\rangle=\begin{pmatrix}0\\1\\0\\0 \end{pmatrix}\quad;\quad
\vert10\rangle=\begin{pmatrix}0\\0\\1\\0 \end{pmatrix}\quad;\quad
\vert11\rangle=\begin{pmatrix}0\\0\\0\\1 \end{pmatrix},\quad
\end{equation}
The general state of the 2-qubit system can be written as:
\begin{eqnarray}
\vert\Upsilon\rangle & = &|\psi\rangle_{1}\otimes\;\vert\psi\rangle_{2}\\
& = &\,a_1\,a_2\:|00\rangle+a_1b_2\:|01\rangle+b_1\,a_2\:|10\rangle+b_1\,b_2\:|11\rangle,
\end{eqnarray}
Similar to the single-qubit state, the coefficients ($a_1a_2,\,a_1b_2,\,b_1a_2,\,b_1b_2$) represent the \textit{probability amplitudes}.\\ 
The state $|\Upsilon\rangle$ is of course normalized, which lead to the relation below:
\begin{equation}
\langle\Upsilon\vert\Upsilon\rangle=|a_1\,a_2|^2+|a_1\,b_2|^2+|b_1\,a_2|^2+|b_1\,b_2|^2=1,
\end{equation}
which can be seen as probability conservation.\\

When performing a measure, this state will collapse through projection to one of the states $|00\rangle\,,\:|01\rangle\,,\:|10\rangle\,,\:|11\rangle$, with respective probabilities 
\[\mathbb{P}_{00}=|a_1\,a_2|^2\quad;\quad\mathbb{P}_{01}=|a_1\,b_2|^2\quad;\quad\mathbb{P}_{10}=|b_1\,a_2|^2\quad;\quad\mathbb{P}_{11}=|b_1\,b_2|^2.\] 

Unlike classical bits, the qubits have a \textit{remarkable feature}, which is the \textbf{entanglement}; where the qubits are \textbf{tightly correlated} in a way that we need all of the constituents to have the \textit{complete information} of the whole system. \\
We know from the \autoref{1.1.2} that a state is not \textit{entangled} if we can separate it, like the example below:
\[|\Upsilon\rangle=\tfrac{1}{\sqrt{2}}\,|00\rangle+\tfrac{1}{\sqrt{2}}\:|10\rangle=\tfrac{1}{\sqrt{2}}\:(|0\rangle+|1\rangle)\otimes|0\rangle=|+\rangle\otimes|0\rangle,\]
where the state of the qubit $\mathfrak{q}_{1}$ is $|\psi\rangle_{1}=|+\rangle$ and $\mathfrak{q}_{2}$ is $|\psi\rangle_{_2}=|0\rangle$.\\
Contrary to an \textit{entangled} state, which is a \textit{mixed} state and cannot be factorized into \textit{pur} states: 
\begin{equation}
|\Upsilon\rangle=\tfrac{1}{\sqrt{2}}\,|00\rangle+\tfrac{1}{\sqrt{2}}\,|11\rangle
\end{equation}
If we measure the First qubit and find it to be in the state $|0\rangle$, then the second one will be in $|0\rangle$ with certainty.
\[|\Upsilon\rangle\,\xrightarrow{\scriptstyle{measure}}\,|00\rangle\]
the same can be said for the state $|1\rangle$.\\
This state is one of four \textit{entangled orthonormal} states known as the “\textit{\textbf{Bell states}}”.
\begin{eqnarray}
|\Phi^+\rangle=\tfrac{1}{\sqrt{2}}\,(\,|00\rangle+\,|11\rangle\,)\qquad;\qquad
|\Phi^-\rangle=\tfrac{1}{\sqrt{2}}\,(\,|00\rangle-\,|11\rangle\,)\,\!\\
|\Psi^+\rangle=\tfrac{1}{\sqrt{2}}\,(\,|01\rangle+\,|10\rangle\,)\qquad;\qquad
|\Psi^-\rangle=\tfrac{1}{\sqrt{2}}\,(\,|01\rangle-\,|10\rangle\,)
\end{eqnarray}
To visualize a 2-qubit system or more, we can think in the \textit{Bloch sphere}, which was perfect to represent the state of 1-qubit system. But what for more than a single qubit?.\\
On the one hand, for a multi-qubit system, the entanglement applies to the qubits to be \textit{correlated}; we cannot describe their states \textit{separately}. On the other hand, the Bloch sphere gives a local view, where each qubit state is represented on its own Bloch sphere; therefore, we miss the important effect of \textit{correlation} between the qubits when mapping them on \textit{separate} Bloch spheres.\\
For this reason, we have to look for another model of visualization. The \textit{Q-sphere} is the \textit{primary visualization method}, that provides a \textit{global viewpoint} of multi-qubit systems\footnote{The Q-sphere is limited to a \textit{small} number of qubits.}.
\begin{figure}[H]
\centering
\includegraphics[width = 0.44\linewidth]{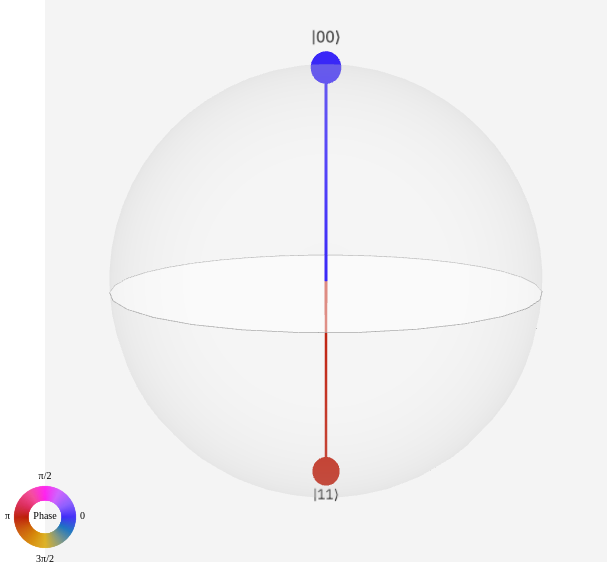}
\caption{The Q-sphere : a visualization of the entangled 2-qubit state $\ket{\Phi^{-}}$.}
\end{figure}
\label{1.6}

\begin{itemize}
\item Each \textbf{\textit{computational basis state}} is associated with a \textbf{point} on the surface of the sphere\footnote{The \textit{spherical shape} has no importance, it's just a nice way to organize the \textit{nodes}.}, where the state with all \textbf{\textit{zeros}} is at the \textit{north pole} and that with all \textbf{\textit{ones}} is at the \textit{south pole}\footnote{The basis states with the same number of \textit{zeros \emph{or} ones} that lie on a shared \textit{latitude} of the Q-sphere (e.g., $|011\rangle,|101\rangle,|110\rangle$).}.
\item At each point there is a \textbf{node}, that represents the \textit{probability amplitude}, and its \textbf{radius} is proportional to \textit{the probability} $\mathbb{P}$ of its basis states.
\item The \textbf{color} of the \textit{nodes} represents the \textit{quantum phase} of each \textbf{computational basis state}.
\end{itemize}
The Q-sphere gives us \textit{complete information} of the system state, at each given \textit{manipulation} of the qubits.

We now move from the 2-$\mathfrak{qubit}$ to the 3-$\mathfrak{qubit}$ system in order to generalize these cases to an $\mathfrak{n}$-$\mathfrak{qubit}$ system. Analogous to what we said above, the \textit{Hilbert-space} associated with the 3-qubit system is written as $\mathcal{H}_1\otimes\mathcal{H}_2\otimes\mathcal{H}_3$, with dimensionality $2\times2\times2=2^3=8$. The \textit{computational basis} of this space is
\begin{equation}
\lbrace\;|000\rangle\:;\:|001\rangle\:;\:|010\rangle\:;\:|011\rangle\:;\:|100\rangle\:;\:|101\rangle\:;\:|110\rangle\:;\:|111\rangle\:\rbrace,
\end{equation}
each of these kets is an \textit{eight-dimensional} column.\\
The general state of the 3-qubit system is expressed:
\[|\Upsilon\rangle=|\psi\rangle_{1}\otimes|\psi\rangle_{_2}\otimes|\psi\rangle_{_3}.\]
We can follow the same reasoning of separability to verify if it is entangled or not.\\
Now for the \textbf{generalization} to an $\mathfrak{n}$-$\mathfrak{qubit}$ system. Let's have $n$ qubit, where $n>1$, with $n \in\mathbb{N}$, and their general state is constructed by the tensor product of the $n$ single-qubit states; i.e.
\[|\psi\rangle_{1}\otimes|\psi\rangle_{_2}\otimes\cdots\otimes|\psi\rangle_{_{n-1}}\otimes|\psi\rangle_{_n}.\]
This $n$-qubit state lives in the \textit{Hilbert-space} $\mathcal{H}_1\otimes\mathcal{H}_2\otimes\cdots\otimes\mathcal{H}_{n-1}\otimes\mathcal{H}_n$, which can briefly be written as $\mathcal{H}\,\otimes^n\,\mathcal{H}\equiv\mathcal{H}^{\,\otimes^n}$. The \textit{dimensionality} of the $n$-qubit space is $2\times...\times2=2^{\,n}$, which is the same as the $n$-qubit state. So the dimensionality of the state vector is increasing \textit{exponentially} every time we add a qubit to the system \cite{Dancing}. The difficulty to simulate a \textit{quantum computer} appears when considering a system of a large number of qubits. A modern laptop can easily simulate a general quantum state of 20 qubits, but simulating 100 qubits is too difficult for the largest \textbf{supercomputers}!\\
The \textbf{essence power} of \textit{quantum computers} lies in the $\mathfrak{superposition}$ feature of the qubits and which gives rise to the $\mathfrak{entanglement}$ phenomenon, when considering multi-qubit systems.
\subsection{Physical Implementation of Qubit }
In quantum information theory, \textit{the qubit is often described abstractly as a two-state
(or two-level) quantum mechanical system}, in the beginning, we should not worry too much about the
characteristics of these states or what their energies are. We assume that they are
eigenstates of the Hamiltonian system with known eigenvalues. This approach allows us to
concentrate on the fundamental properties of the system without getting lost in the details.

In 1994, the mathematician \textsc{Peter Shor} presented two results \cite{Shor1994Nov} \cite{Sanders}: \textit{the
first, showing that a quantum computer would provide exponential speedup for number factorization. The second, showing that quantum error correction could overcome
decoherence.} After that, the scientific community started to come up with ideas to
build a quantum computer.

In this part, we introduce the notion of a \textbf{physical qubit} and distinguishes its relation with the \textbf{logical qubit} (described in \autoref{1.2.2}). We identify the essential criteria to realize quantum computing. Additional information on the technologies used to perform the physical implementation of qubits are provides in the  \autoref{ap c}. 


\subsubsection*{Physical and Logical Qubit}
 \begin{wrapfigure}{r}{0.35\textwidth}
\centering
\includegraphics[width=0.35\textwidth]{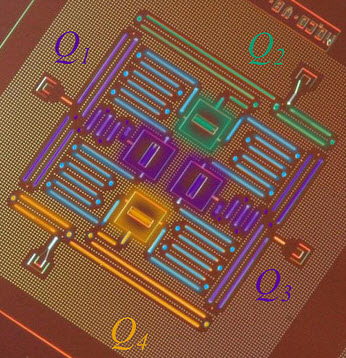}
\caption[Layout of IBM’s four superconducting Qubit device]{ \small{Layout of IBM’s four superconducting Qubit device\footnotemark.}}
\label{1.7}
\end{wrapfigure}

\footnotetext{\url{https://search.creativecommons.org/photos/30f43527-9f8d-4e77-859a-a85f0a0ada4c}}

 In quantum circuits and quantum algorithms, we use a \textbf{logical Qubit}. It specifies how a single Qubit should behave with quantum logical operations. We can use them indefinitely, they never lose state when they are not used, and we can apply as many gates to them as we wish.

A \textbf{physical Qubit} is a real Qubit produced by one of the implementation technologies and is physical hardware device. The Qubits used in quantum computers are constructed from microscopic degrees of freedom like the spin of electrons or nuclei, transition dipoles of atoms or ions in a vacuum. These degrees of freedom are naturally isolated from their environment and lead to slow decoherence \cite{Shaw2008Mar}.
 
 It is possible to create objects that act like logical qubits from physical, but it will require hundreds to thousands of physical qubits to make one logical Qubit. We use error-correcting schemes to have many physical qubits work together to create a virtual logical qubit. Hence, \textit{A physical qubit is the hardware implementation of logical qubits} \cite{Dancing}.

The quantum state of a physical qubit decays quickly and inevitably; for this, the main objective of researchers and engineers in quantum computing is to slow down this decay and correct the effects of the decoherence of qubits through the execution of circuits.

\subsubsection*{Criteria to Realize Quantum Computing:}

In 2018, the theoretical physicist David DiVincenzo published a brief retrospective on creating and implementing quantum computers that respected some criteria.
He laid out \textit{five requirements for the implementation of quantum computation} \cite{DiVincenzo}.

The DiVincenzo Criteria that a quantum computer implementation must satisfy are as follows:

\begin{enumerate}

\item \textbf{A scalable physical system with well-characterized qubits.} \\
We should be able to add over time, qubits in our quantum system without any restriction. With a limited number of Qubits, we cannot make useful quantum calculations.

\item \textbf{The ability to initialize the state of the qubits to a simple fiducial state.}\\
It is common for algorithms, to start with a qubit in state $\ket{0}$ but sometimes we need to use $\ket{1}$ instead $\ket{0}$. For this, we must be able to initialize the qubit to a known initial state with a very high
probability. This is called \textit{high-fidelity state preparation.}

\item \textbf{Long decoherence time}\\
A quantum superposition state has a lifetime. This time is called the coherence time and depends on the environment of the qubit. Environmental disruptions cause a quantum superposition to dissipate and the final state will no longer in a coherent superposition. If we have a long coherence time, but our gates take a long time to execute, that may be equivalent to a short coherence time but with fast gates.
Long coherence, fast enough gates, and low error rates will be the key to success with \textit{Noisy Intermediate-Scale Quantum computers} (NISQ).

\item \textbf{A universal set of quantum gates}\\
Without quantum logic gates, we cannot do quantum programming; that means, we cannot build quantum algorithms unless we have a sufficiently complete set of gates.

\item \textbf{A qubit-specific measurement capability}\\
When we measure a quantum system we force the state to take one of the basic orthogonal states. we must be able to force the qubit into one of the two orthogonal states $\ket{0}$ and $\ket{1}$. The error rate of this operation must be low enough to allow for useful quantum computation, if we get the wrong measurement answer fifty percent of the time, all the previous work on executing the circuit is lost. This is called \textit{high fidelity readout.}

\end{enumerate}

\section{Quantum Gates and Quantum Circuits}
Now that we understand what a qubit is, it's time to present the operators that manipulate its state called “\textit{quantum logic gates}”. Next, we introduce “\textit{quantum circuits}” and see some examples. To understand this, we give a glimpse of “\textit{classical gates}”.
\subsection{Classical Gates}
In numerical systems, algebra describes the arithmetic operation on variables that can take any defined value. For Boolean algebra, it is the same thing except that the variables cannot take more than two values. Those variables are called logical variables. A logical variable can be presented in several ways, for example, 0 and 1 or \textsc{true} and \textsc{false} and it indicates the state of the system like voltage in a circuit. Boolean algebra has only three fundamental operators: \textsc{NOT}, \textsc{AND}, \textsc{OR}. We need to manipulate data using operators (like in mathematics); this operator is called \textit{“gates”}. Digital technologies are based on repeatedly applying the three operators (\textsc{NOT}, \textsc{AND}, \textsc{OR}) in various combinations. 

\subsubsection*{Logic Gates}

A logic gate is an electronic circuit which is composed of transistors and resistors, and they are represented in schematic symbols. Each gate has one output that depends on its inputs; the output voltage depends on input voltage. A circuit or a digital system can be made using the three basic gates: \textsc{and}, \textsc{or}, \textsc{not} \cite{Sagawa}. Other fundamental gates can be listed: \textsc{nand}, \textsc{nor}, \textsc{xor}. These gates can be constructed using the three basic gates. To describe the output of a logical circuit, we need a truth table. We will see in this part the logic gates that are the most commonly used, their symbol, and their truth table\footnote{Truth tables list the output of a particular digital logic circuit for all the possible combinations of its inputs.}.
\begin{enumerate}
\item \textbf{NOT Gate}

\begin{minipage}[t]{.5\textwidth}\vspace{0pt}%
A \textsc{not} gate is a logic gate that inverts the digital input signal. If the input of a \textsc{not} gate is considered as $A$, then the output of the gate will be $\bar{A}$. When the input signal $A$ is controlled by the \textsc{not} operation, the output $O$ can be expressed as: $O=\bar{A}=A'$.\\
If the value of $A$ is $1$ then $\bar{A} = 0$ and in the opposite if the value of $A$ is 0 then $\bar{A} = 1$. 
\hfill
\end{minipage}%
\hfill
\begin{minipage}[t]{.36\textwidth}\vspace{0pt}%

 \footnotesize
    \begin{tabular}[b]{cc}
    \multicolumn{2}{c}{\includegraphics[scale=0.07]{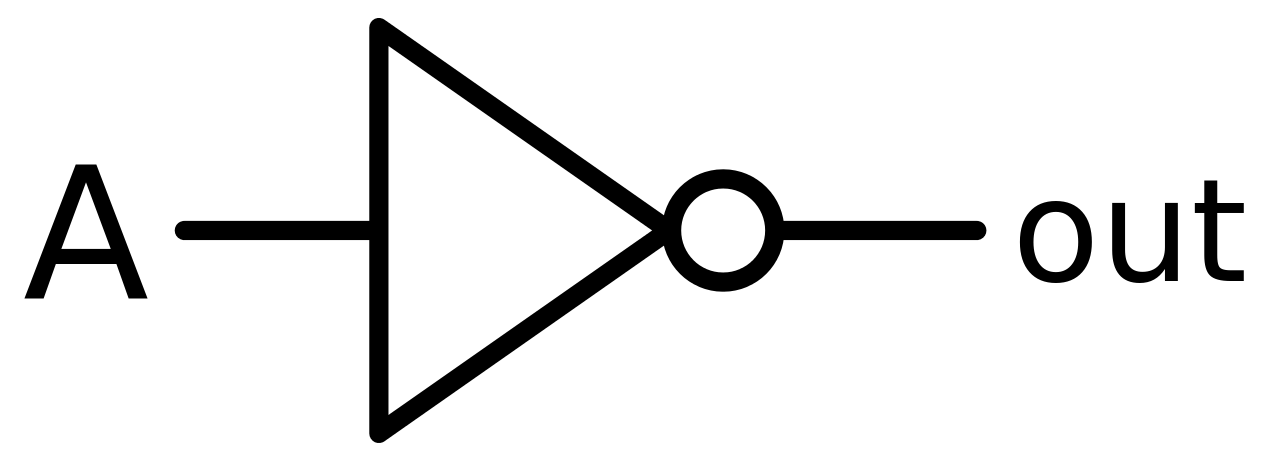}}
     \\ 
       A & Output \\ \hline
       0 & 1 \\
       1 & 0 \\ \hline  
       \multicolumn{2}{c}{\textsc{not} gate symbol and truth table} \\   
       
    \end{tabular}
\end{minipage}%
\item \textbf{AND Gate}

\begin{minipage}[t]{.5\textwidth}\vspace{0pt}%
An \textsc{and} gate is a logic gate with two or more inputs and a single output, it can have any number of inputs, but the most common is 2 and 3. The \textsc{and} gate operates on logical multiplication rules that mean the output of the gate, $O$, will be 0 unless all (both $A$ and $B$) inputs are 1. The output can be expressed as $O=A.B$, which means: output $O$ equals $A$ \textsc{and} $B$. The “.” sign stands for the \textsc{and} operation.
\hfill
\end{minipage}%
\hfill
\begin{minipage}[t]{.36\textwidth}\vspace{0pt}%
\footnotesize
    \begin{tabular}[b]{ccc}
    \multicolumn{3}{c}{\includegraphics[scale=0.045]{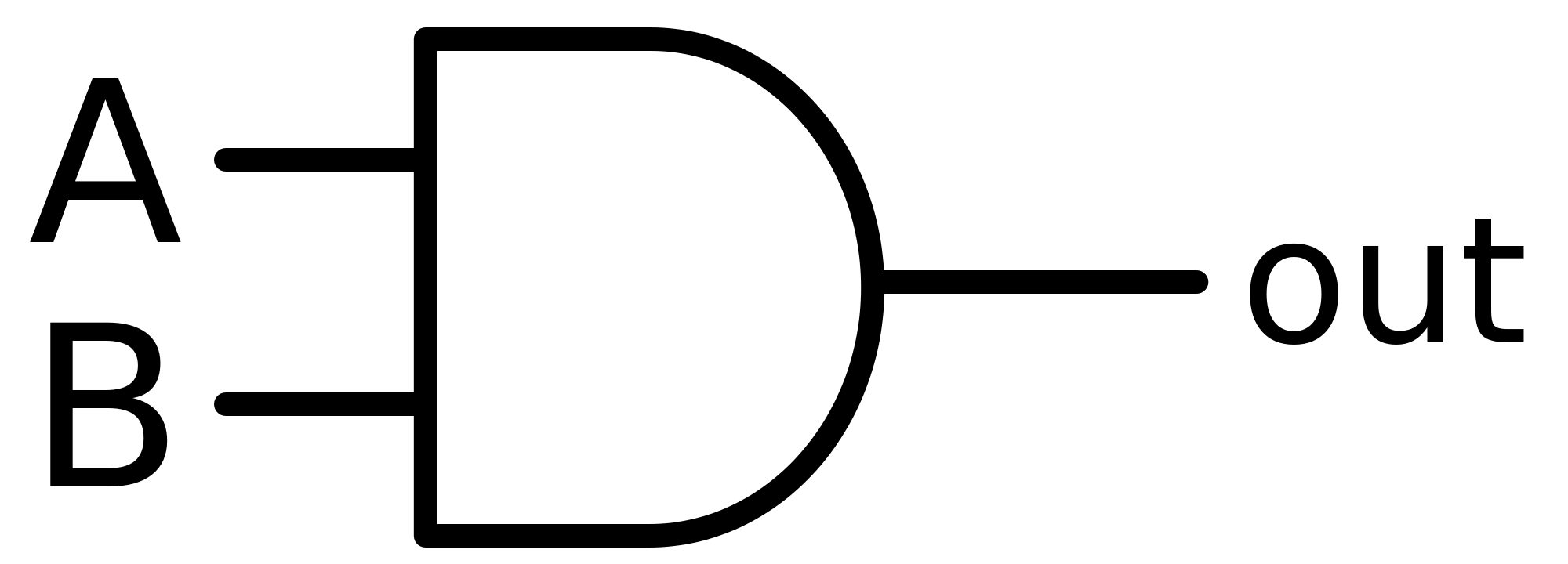}} \\ 
      A & B & Output \\ \hline
      0 & 0	&	0 \\
      0 & 1	&	0 \\
      1 & 0	&	0 \\
      1 & 1	&	1 \\ \hline   
       \multicolumn{3}{c}{\textsc{and} gate symbol and truth table} \\    
    \end{tabular}
\end{minipage}%


\item \textbf{OR Gate}

\begin{minipage}[t]{.5\textwidth}\vspace{0pt}%
An \textsc{or} gate is a logic gate that performs logical \textsc{or} operation. Just like an \textsc{and} gate, an \textsc{or} gate may have any input but only one output. When any of the \textsc{or} gate inputs is 1, the output of the \textsc{or} gate will be 1. However, the output of the \textsc{or} gate is 0 only when all the input variables are 0. The output can be expressed as $O=A+B$, which means: output $O$ equals $A$ \textsc{or} $B$. The “+” sign stands for the \textsc{or} operation, and it is not an arithmetic addition.

\end{minipage}%
\hfill
\begin{minipage}[t]{.36\textwidth}\vspace{0pt}%
\footnotesize
    \begin{tabular}[b]{ccc}
    \multicolumn{3}{c}{\includegraphics[scale=0.07]{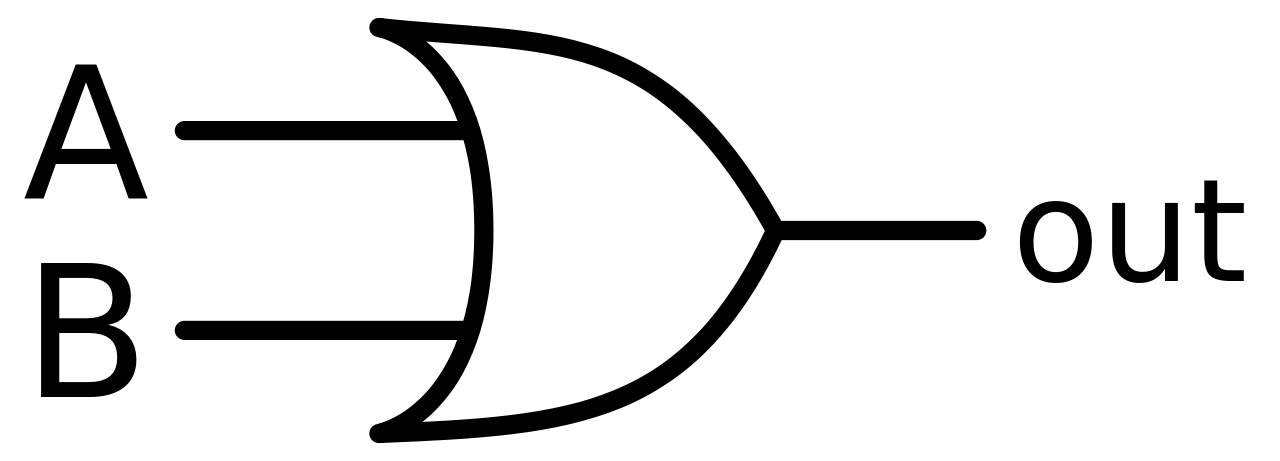}} \\ 
      A & B & Output \\ \hline
      0 & 0	&	0 \\
      0 & 1	&	1 \\
      1 & 0	&	1 \\
      1 & 1	&	1 \\ \hline 
      \multicolumn{3}{c}{\textsc{or} gate symbol and truth table} \\     
    \end{tabular}
\end{minipage}%
\pagebreak

\item \textbf{NAND Gate}

\begin{minipage}[t]{.5\textwidth}\vspace{0pt}%

The \textsc{nand} gate is the inverse of the \textsc{and} gate. Just like an \textsc{and} gate, a \textsc{nand} gate may have any number of inputs but only one output.
The output of the gate, $O$, will be 1 when at least one input is 0 ($A$ or $B$). The output can be expressed as $O = \overline{A.B}$, which means: the output $O$ equals $A.B$ bar. The small circle at the output mean the inverted operation of output.

\end{minipage}%
\hfill
\begin{minipage}[t]{.36\textwidth}\vspace{0pt}%
\footnotesize
    \begin{tabular}[b]{ccc}
    \multicolumn{3}{c}{\includegraphics[scale=0.075]{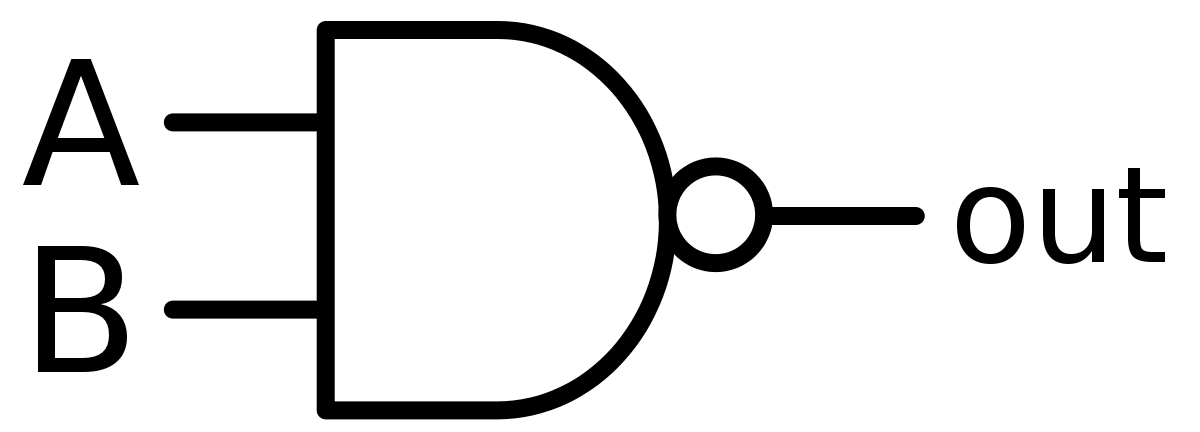}} \\ 
      A & B & Output \\ \hline
      0 & 0	&	1 \\
      0 & 1	&	1 \\
      1 & 0	&	1 \\
      1 & 1	&	0 \\ \hline 
      \multicolumn{3}{c}{\textsc{nand} gate symbol and truth table} \\     
    \end{tabular}
\end{minipage}%
\item \textbf{NOR Gate}

\begin{minipage}[t]{.5\textwidth}\vspace{0pt}%
A \textsc{nor} gate is the inverse of the \textsc{or} gate, and its circuit is produced by connecting an \textsc{or} gate to a \textsc{not} gate. The \textsc{nor} gate produces 1 only if all its inputs are 0, and 0 otherwise. The output can be expressed as $O = \overline{A+B}$, which means: the output $O$ equals $A+B$ bar. This gate has a small circle on the output. This small circle represents the inversion operation.

\end{minipage}%
\hfill
\begin{minipage}[t]{.36\textwidth}\vspace{0pt}%
\footnotesize
    \begin{tabular}[b]{ccc}
    \multicolumn{3}{c}{\includegraphics[scale=0.075]{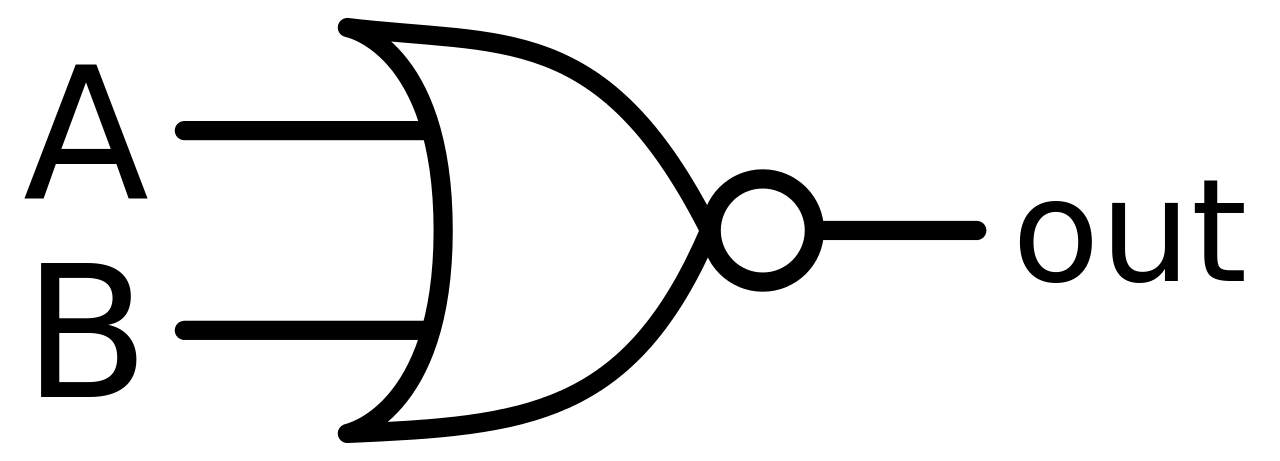}} \\ 
      A & B & Output \\ \hline
      0 & 0	&	1 \\
      0 & 1	&	0 \\
      1 & 0	&	0 \\
      1 & 1	&	0 \\ \hline 
      \multicolumn{3}{c}{\textsc{nor} gate symbol and truth table} \\     
    \end{tabular}

\end{minipage}%

\item \textbf{XOR Gate}

\begin{minipage}[t]{.5\textwidth}\vspace{0pt}%
A \textsc{xor} (pronounced as Exclusive \textsc{or}) is the logic gate that produces 1 in the output if and only if an odd number of inputs are 1. If both inputs are 1 or both inputs are 0, the output is 0. So a two-input \textsc{xor} gate is quite different from the \textsc{or} gate. \textsc{xor} represents the inequality function, which means that the output is true if the inputs are not similar. The Boolean expression for the output from a two-input \textsc{xor} gate is $O=A.\bar{B}+\bar{A}.B $
\end{minipage}%
\hfill
\begin{minipage}[t]{.36\textwidth}\vspace{0pt}%
\footnotesize
    \begin{tabular}[b]{ccc}
    \multicolumn{3}{c}{\includegraphics[scale=0.075]{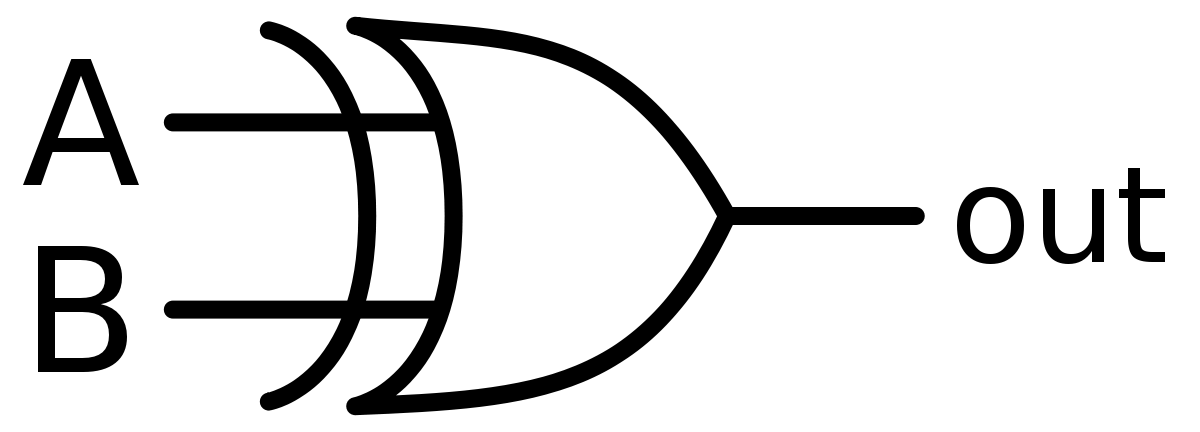}} \\ 
      A & B & Output \\ \hline
      0 & 0	&	0 \\
      0 & 1	&	1 \\
      1 & 0	&	1 \\
      1 & 1	&	0 \\ \hline 
      \multicolumn{3}{c}{\textsc{xor} gate symbol and truth table} \\     
    \end{tabular}

\end{minipage}%


\end{enumerate}

\subsection{Quantum Gates}
Now after we have seen classical gates we can have a better understanding of
quantum gates. We know that classical logic gates manipulate the classical bit values 0 or 1. Likewise,
quantum gates act on qubits and transform their states. For example by a rotation around the Bloch sphere. The main difference between classical and quantum gates is that quantum gates can manipulate multi-quantum states.
Quantum gates are represented by unitary matrices (the inverse equals the conjugate transpose) and therefore, any quantum gate is
reversible.
 When we apply a gate/operator $\mathbb{G}$ on an arbitrary state $\ket{\psi_{In}}$, the input state will be transformed as follow: $\ket{\psi_{_{Out}}} = \mathbb{G} \ket{\psi_{_{In}}}$. We can reconstruct the input state  from the output as $\ket{\psi_{_{In}}} = \mathbb{G}^{\dagger} \ket{\psi_{_{Out}}}$ \cite{Chatter}.
 
In this part, we will see almost all the quantum gates provided by IBM quantum
composer. Then, we will see some examples of quantum circuits that are very known in QIS.

\subsection*{Single-Qubit Gates}
\vspace{0.3cm}
 \textcolor{black}{\underline{\large{Identity Gate}}}
\begin{wrapfigure}{l}{0.17\textwidth}
\centering
\vspace{-11pt}
\includegraphics[width=0.09\textwidth]{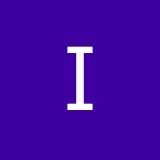}
\end{wrapfigure}
 The first gate is the \textsc{identity gate}, it is represented by the identity matrix and defined for a single qubit as follow:

\begin{equation}
\mathbb{I}=\begin{pmatrix}
1 & 0 \\
0 & 1 \\
\end{pmatrix} \quad ; \quad \mathbb{I}.\mathbb{I}=\mathbb{I}
\end{equation}
When we apply this gate to the computational basis states $\ket{0}$ and $\ket{1}$, it leaves the states unchanged.
\begin{equation}
\mathbb{I}\ket{0}=\begin{pmatrix}
1 & 0 \\
0 & 1 \\
\end{pmatrix} \begin{pmatrix}
1 \\ 0
\end{pmatrix}=  \begin{pmatrix}
1 \\ 0
\end{pmatrix}= \ket{0}
\end{equation}
\begin{equation}
\mathbb{I}\ket{1}=\begin{pmatrix}
    1 & 0 \\
    0 & 1 \\
    \end{pmatrix} \begin{pmatrix}
    0 \\ 1
    \end{pmatrix}=  \begin{pmatrix}
    0 \\ 1
    \end{pmatrix}= \ket{1}
\end{equation} 
This gate is useful for two reasons, one to prove that a gate is its own inverse. Second, it is useful to have a do-nothing operation, especially for the error models. This can be used as a simple version of decoherence of the qubit. 
\begin{figure}[H]
\centering
\includegraphics[width = 0.38\linewidth]{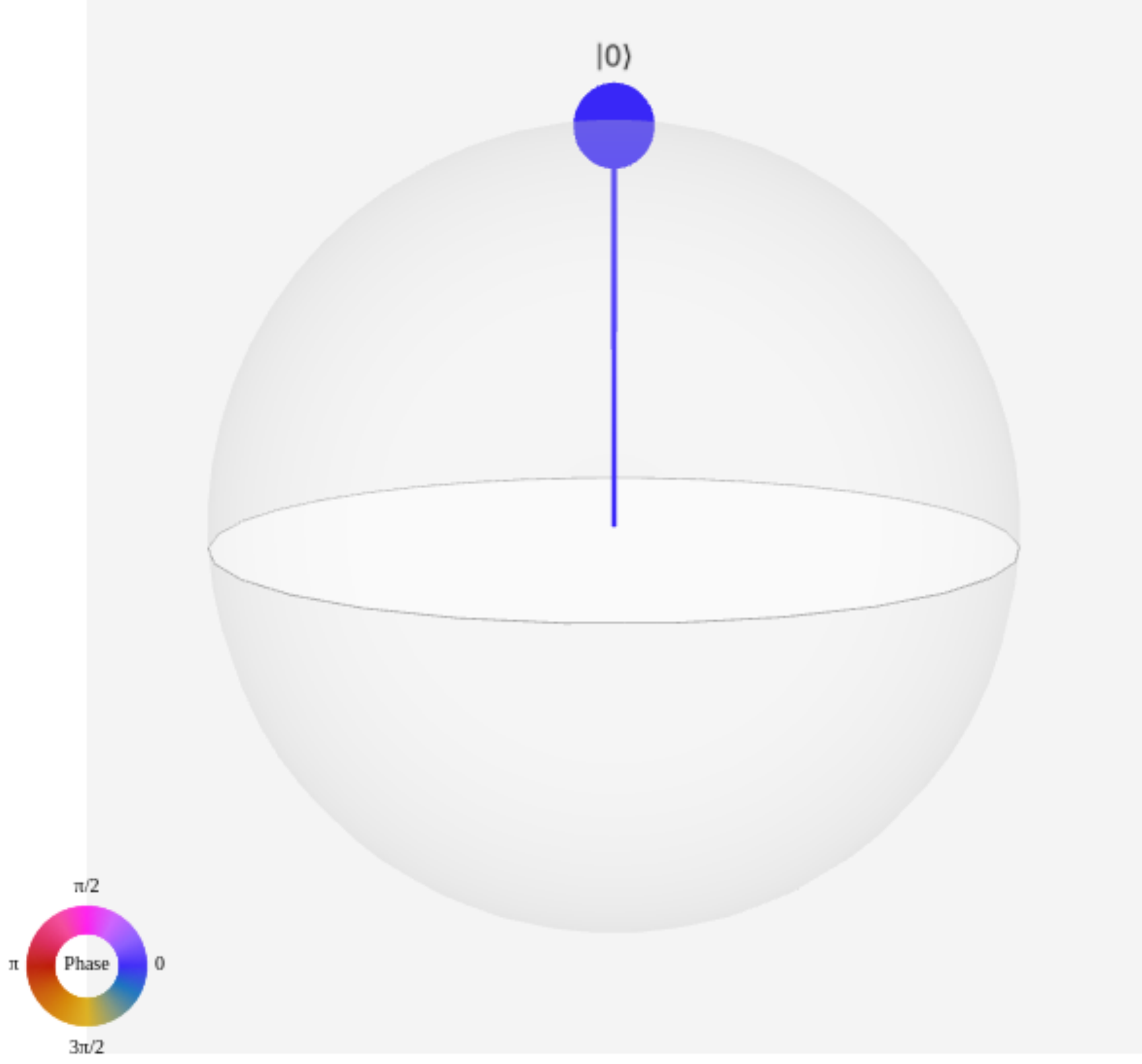}\hspace{0.4cm}
\includegraphics[width = 0.38\linewidth]{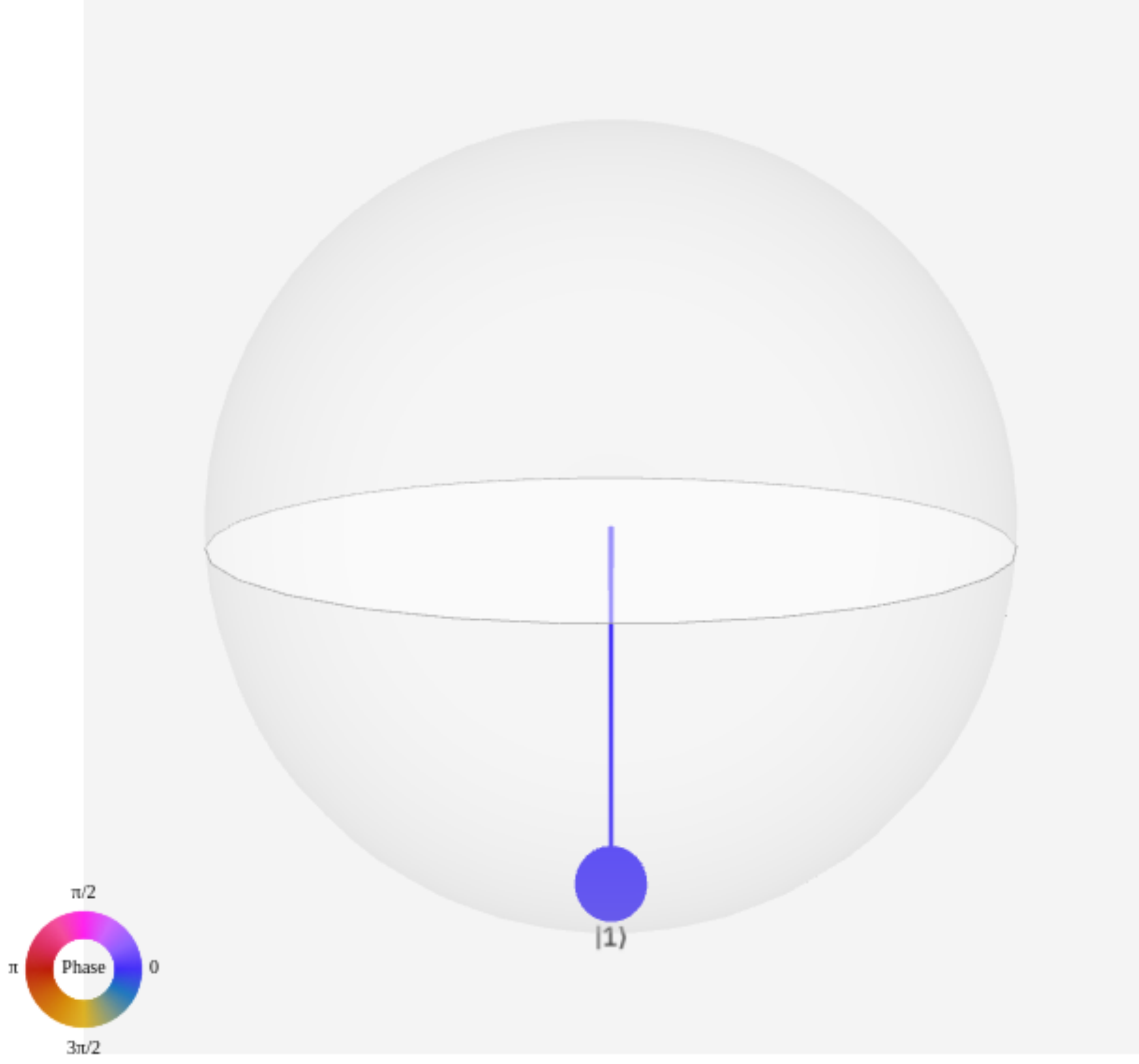}
\caption{A visualization of the resulting states from the $\mathbb{I}$ action on the states $\ket{0}$ and $\ket{1}$, respectively.}
\label{1.8}
\end{figure}
 

 \textcolor{black}{\underline{\large{The X Gate}}}
\begin{wrapfigure}{l}{0.17\textwidth}
\centering
\vspace{-11pt}
\includegraphics[width=0.09\textwidth]{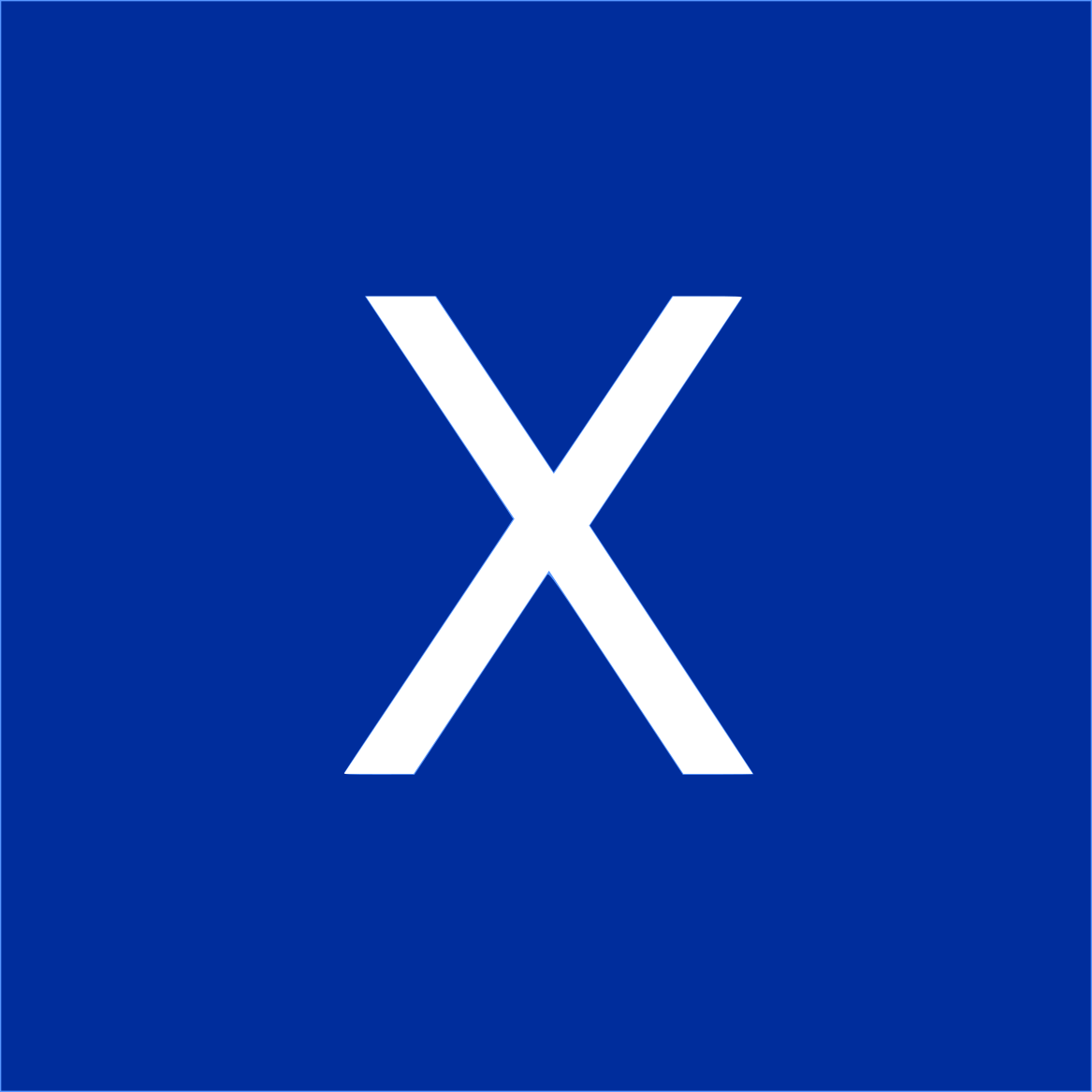}
\end{wrapfigure}
The \textsc{not gate} is the quantum equivalent of the classical \textsc{not} gate, it is also known as the \textsc{Pauli} X gate. Its matrix representation is given by: 

\begin{equation}
    \mathbb{X}=\begin{pmatrix}
    0 & 1 \\
    1 & 0 \\
    \end{pmatrix} \quad ; \quad \mathbb{X}.\mathbb{X}=\mathbb{I}
    \end{equation}

It acts in a single qubit by switching the \textbf{amplitudes} of the $\ket{0}$ and $\ket{1}$ basis. 

\begin{equation}
\mathbb{X}\ket{0}=\begin{pmatrix}
    0 & 1 \\
    1 & 0 \\
    \end{pmatrix} \begin{pmatrix}
1 \\ 0
\end{pmatrix}=  \begin{pmatrix}
0 \\ 1
\end{pmatrix}= \ket{1}
\end{equation}

\begin{equation}
\mathbb{X}\ket{1}=\begin{pmatrix}
    0 & 1 \\
    1 & 0 \\
    \end{pmatrix} \begin{pmatrix}
    0 \\ 1
    \end{pmatrix}=  \begin{pmatrix}
    1 \\ 0
    \end{pmatrix}= \ket{0}
\end{equation} 

The $\mathbb{X}$ gate rotates the qubit state by $\pi$ radians around the \textit{x-axis}. It is sometimes called \textit{bit-flip gate} as it maps $\ket{0}$ to $\ket{1}$ and $\ket{1}$ to $\ket{0}$.

\begin{figure}[H]
\centering
\includegraphics[width = 0.38\linewidth]{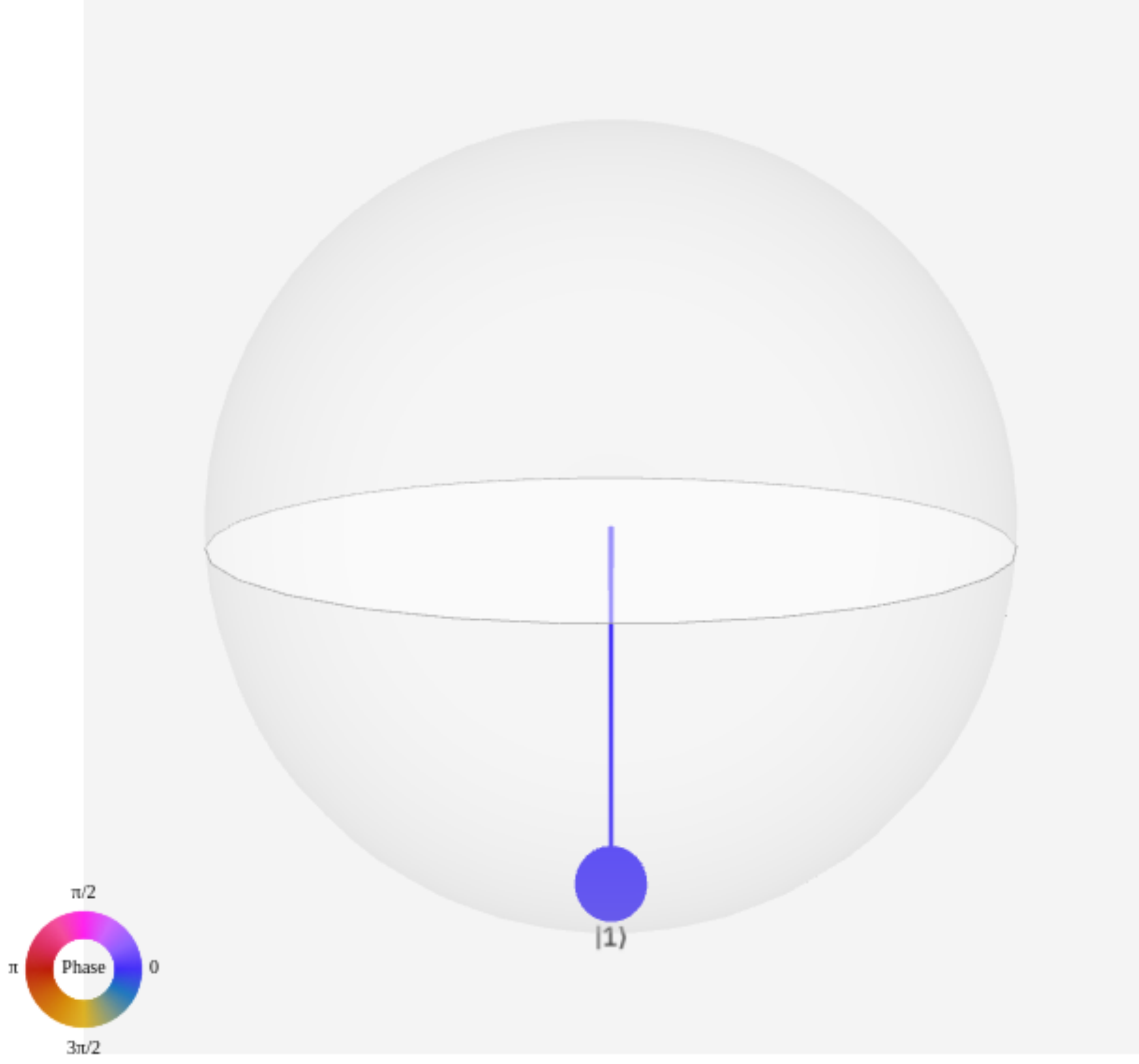}\hspace{0.4cm}
\includegraphics[width = 0.38\linewidth]{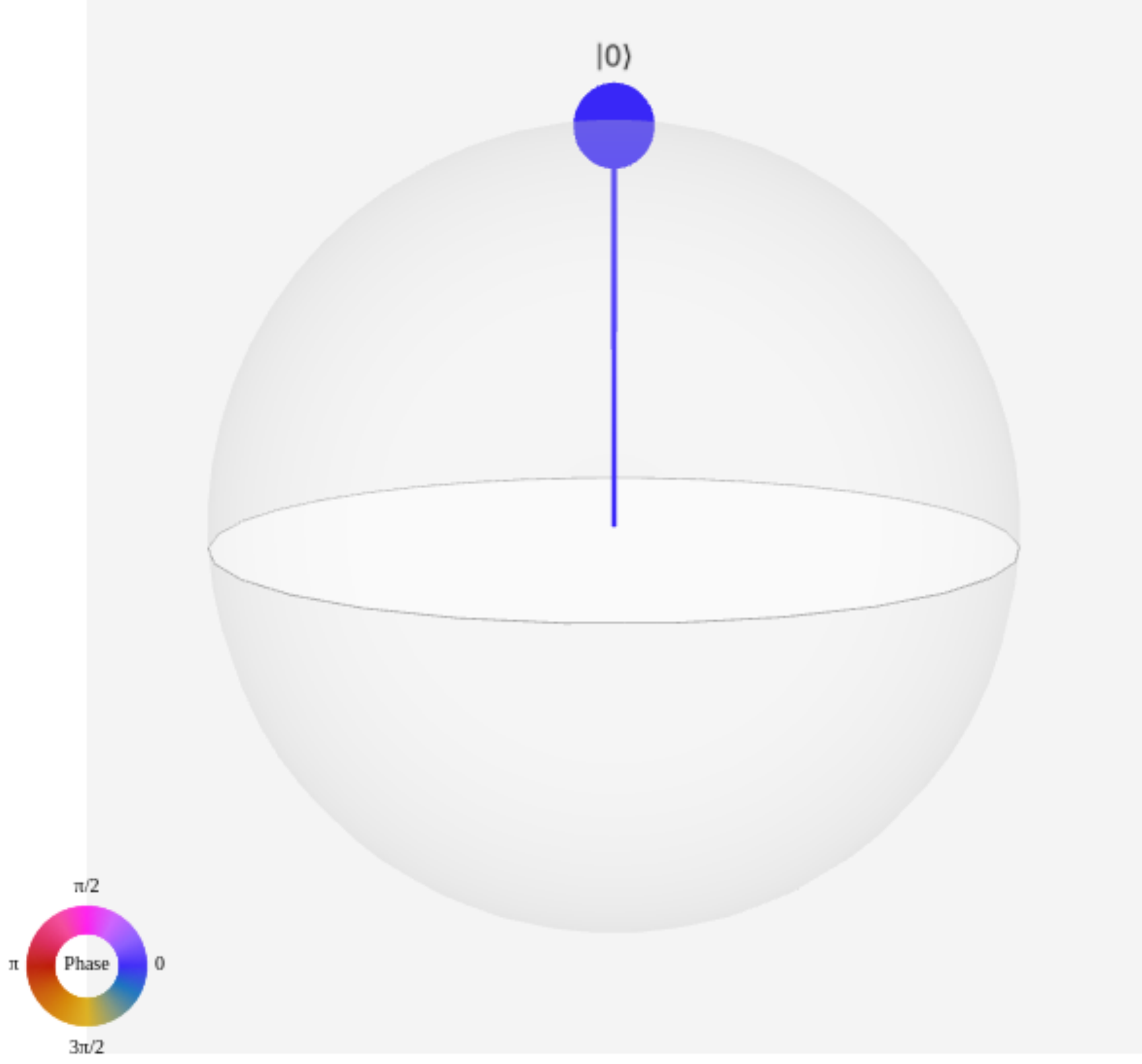}
\caption{A visualization of the resulting states from the $\mathbb{X}$ action on the states $\ket{0}$ and $\ket{1}$, respectively.}
\label{1.9}

\end{figure}


 \textcolor{black}{\underline{\large{The Y Gate}}}
\begin{wrapfigure}{l}{0.17\textwidth}
\centering
\vspace{-11pt}
\includegraphics[width=0.09\textwidth]{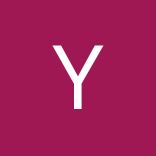}
\end{wrapfigure}
The Pauli Y gate rotates the qubit around the \textit{y-axis} by $\pi$ radians. It is represented by the following matrix:

\begin{equation}
    \mathbb{Y}=\begin{pmatrix}
    0 & -i \\
    i & 0 \\
    \end{pmatrix} \quad ; \quad \mathbb{Y}.\mathbb{Y}=\mathbb{I}
    \end{equation}

When we apply this gate to the computational basis states $\ket{0}$ and $\ket{1}$, we get:   

\begin{equation}
\mathbb{Y}\ket{0}=\begin{pmatrix}
    0 & -i \\
    i & 0 \\
    \end{pmatrix} \begin{pmatrix}
1 \\ 0
\end{pmatrix}=  \begin{pmatrix}
0 \\+ i
\end{pmatrix}= +i\ket{1}
\end{equation}

\begin{equation}
\mathbb{Y}\ket{1}=\begin{pmatrix}
    0 & -i \\
    i & 0 \\
    \end{pmatrix} \begin{pmatrix}
    0 \\ 1
    \end{pmatrix}=  \begin{pmatrix}
    -i \\ 0
    \end{pmatrix}= -i\ket{0}
\end{equation} 

This gate maps $\ket{0}$ to $i\ket{1}$ and $\ket{1}$ to $-i\ket{0}$.
\begin{figure}[H]
\centering
\includegraphics[width = 0.38\linewidth]{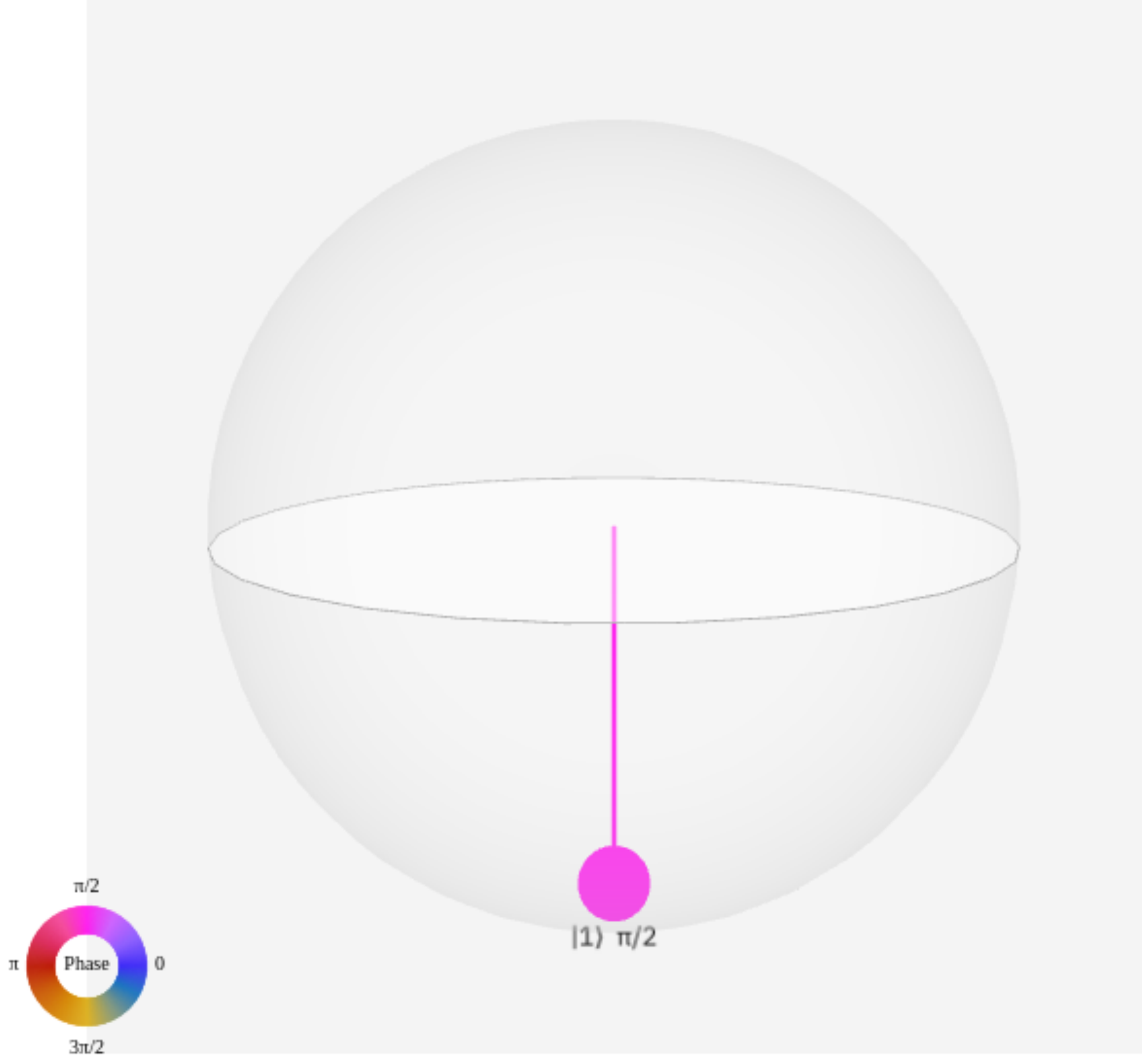}\hspace{0.4cm}
\includegraphics[width = 0.38\linewidth]{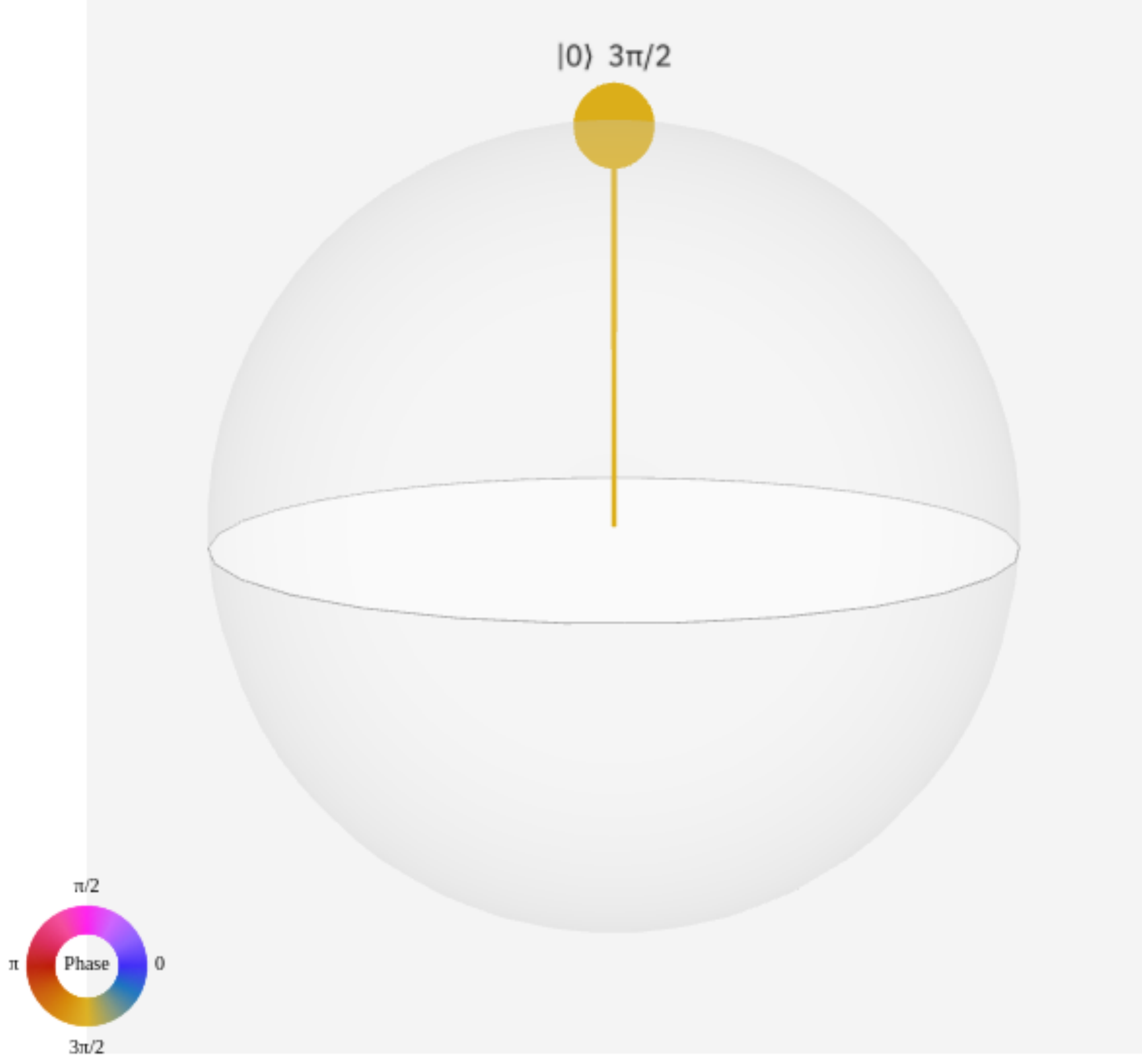}
\caption{A visualization of the resulting states from the $\mathbb{Y}$ action on the states $\ket{0}$ and $\ket{1}$, respectively.}
\label{1.10}
\end{figure}

 \textcolor{black}{\underline{\large{The Z Gate}}}

\begin{wrapfigure}{l}{0.17\textwidth}
\centering
\vspace{-11pt}
\includegraphics[width=0.09\textwidth]{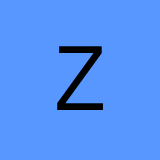}
\end{wrapfigure}

The \textsc{Pauli Z gate} is represented by the \textit{diagonal} matrix:
\begin{equation}
\mathbb{Z}=\begin{pmatrix} 1 & 0 \\ \:0 & \!\!-1\end{pmatrix}\quad;\quad\mathbb{Z.Z=I}
\end{equation}
Let's observe its action on the \textit{computational} basis states $\ket{0}$ and $\ket{1}$.
\[\mathbb{Z}\ket{0}=\begin{pmatrix} 1 & 0 \\ \:0 & \!\!-1\end{pmatrix}\begin{pmatrix} 1\\0 \end{pmatrix}=\begin{pmatrix} +1\\0 \end{pmatrix}=+\ket{0}\]
\[\mathbb{Z}\ket{1}=\begin{pmatrix} 1 & 0 \\ \:0 & \!\!-1\end{pmatrix}\begin{pmatrix} 0\\1 \end{pmatrix}=\begin{pmatrix} 0\\-1 \end{pmatrix}=-\ket{1}\]
So, this gate acts as an $\mathbb{I}$ on $\ket{0}$, and applies a phase of $\pi$ on $\ket{1}$. Moreover, its action can be seen as a rotation of the qubit state by $\pi$ around the \textit{z-axis} on the \textit{Bloch sphere}. Therefore, its action on the states $\ket{+}$ and $\ket{-}$ (recalling that $\ket{\pm}=\tfrac{1}{\sqrt{2}}\,(\,\ket{0}\,\pm\,\ket{1}\,)$)

\[\mathbb{Z}\ket{+}=\begin{pmatrix} 1 & 0 \\ \:0 & \!\!-1\end{pmatrix}\begin{pmatrix} +\tfrac{1}{\sqrt{2}}\\+\tfrac{1}{\sqrt{2}} \end{pmatrix}=\begin{pmatrix} +\tfrac{1}{\sqrt{2}}\\-\tfrac{1}{\sqrt{2}} \end{pmatrix}=\ket{-}\]
\[\mathbb{Z}\ket{-}=\begin{pmatrix} 1 & 0 \\ \:0 & \!\!-1\end{pmatrix}\begin{pmatrix} +\tfrac{1}{\sqrt{2}}\\-\tfrac{1}{\sqrt{2}} \end{pmatrix}=\begin{pmatrix} +\tfrac{1}{\sqrt{2}}\\+\tfrac{1}{\sqrt{2}} \end{pmatrix}=\ket{+}\]
The Z gate swaps these two states $\ket{+}\rightleftarrows\ket{-}$. Since it reverses the sign of the $2^{\,nd}$ \textit{amplitude}, it is also called \textit{sign-flip gate}.
\begin{figure}[H]
\centering
\includegraphics[width = 0.38\linewidth]{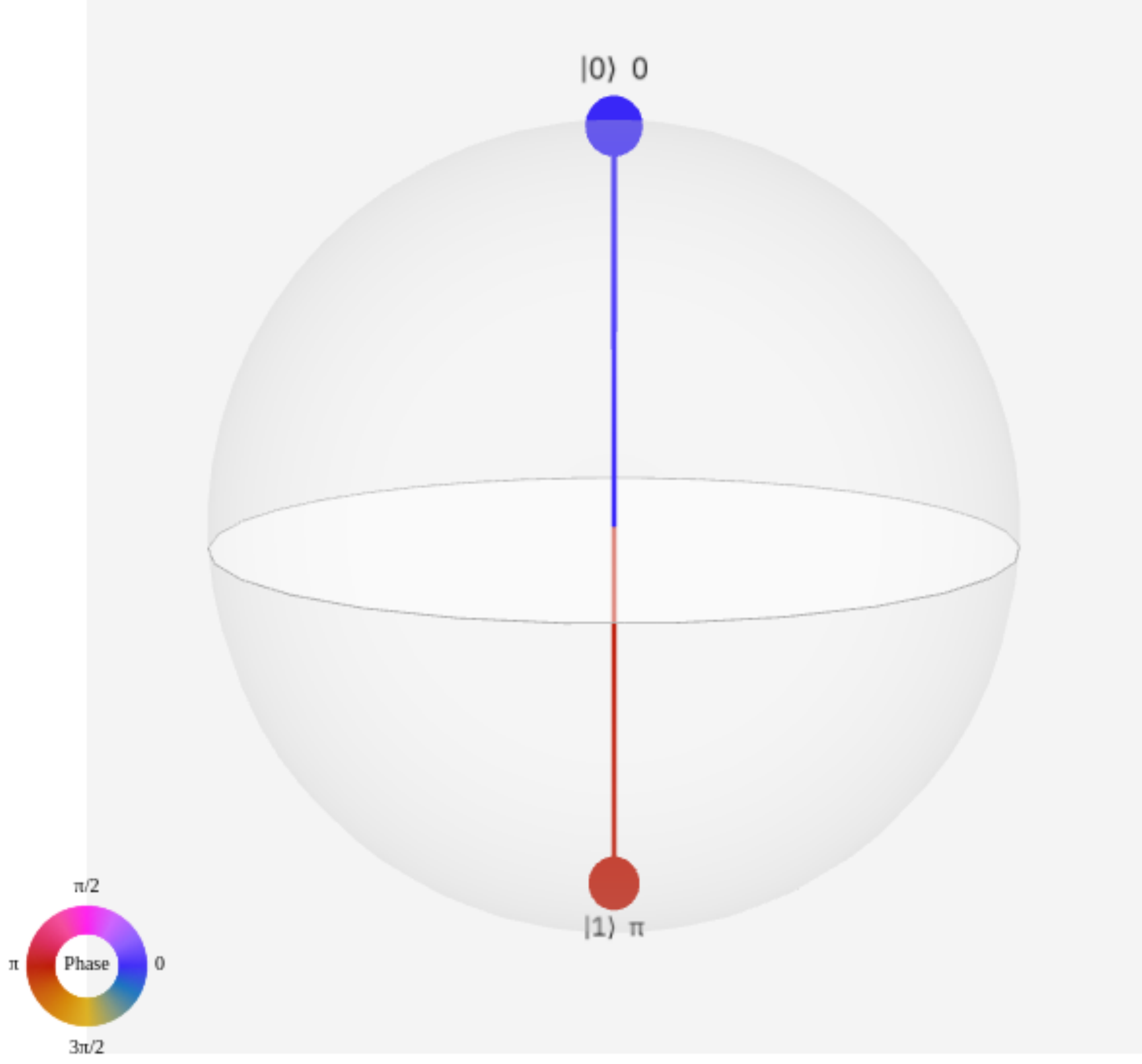}\hspace{0.4cm}
\includegraphics[width = 0.38\linewidth]{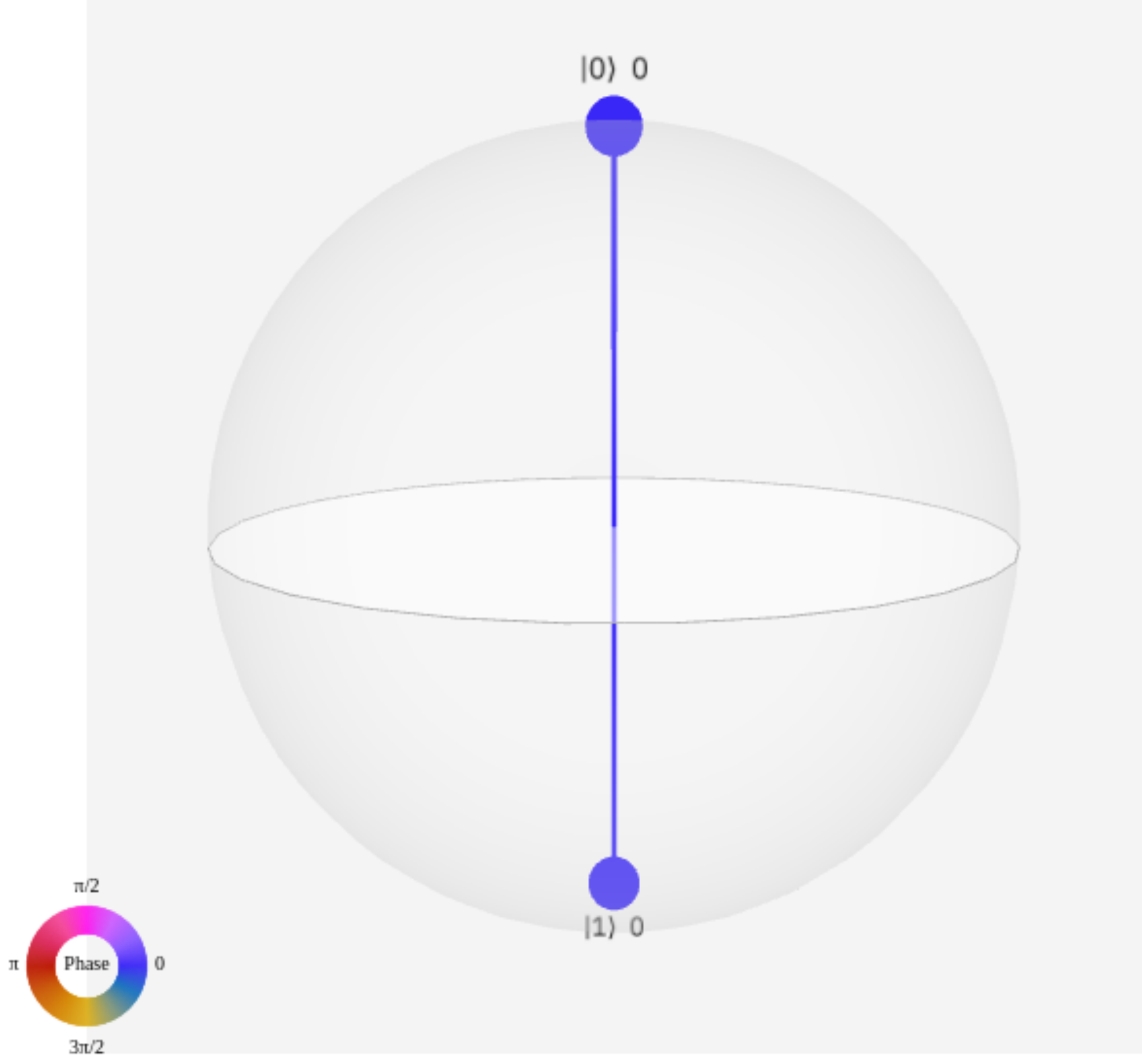}
\caption{A visualization of the resulting states from the $\mathbb{Z}$ action on the states $\ket{+}$ and $\ket{-}$, respectively.}
\label{1.11}
\end{figure}
 \textcolor{black}{\underline{\large{Hadamard Gate}}}
 
\begin{wrapfigure}{l}{0.17\textwidth}
\centering
\vspace{-11pt}
\includegraphics[width=0.09\textwidth]{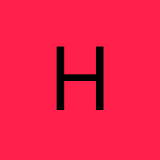}
\label{1.12}
\end{wrapfigure}
The \textsc{Hadamard gate} is represented by the matrix below:
\begin{equation}
\mathbb{H}=\frac{1}{\sqrt{2}}\begin{pmatrix}1&\,\,1\\1&\!\!-1\end{pmatrix}\quad;\quad\mathbb{H.H=I}
\end{equation}
Applied to the two computational basis states $\ket{0}, \ket{1}$ gives:
\[\mathbb{H}\ket{0}=\frac{1}{\sqrt{2}}\begin{pmatrix}1&\,\,1\\1&\!\!-1\end{pmatrix}\begin{pmatrix}1\\0\end{pmatrix}=\frac{1}{\sqrt{2}}(\ket{0}+\ket{1})=\ket{+}\]
\[\mathbb{H}\ket{1}=\frac{1}{\sqrt{2}}\begin{pmatrix}1&\,\,1\\1&\!\!-1\end{pmatrix}\begin{pmatrix}0\\1\end{pmatrix}=\frac{1}{\sqrt{2}}(\ket{0}-\ket{1})=\ket{-}\]
Therefore, the \textsc{Hadamard gate} puts the qubit in an \textbf{equal superposition}. The transformations $\ket{0}\rightarrow\ket{+}$ and $\ket{1}\rightarrow\ket{-}$ can be seen as a rotation around the \textit{Bloch vector} of coordinates (1, 0, 1) by an angle $\pi$. Without this gate, the qubit would still behave like a classical bit. Hence, implementing the \textsc{Hadamard gate} in any given classical circuit at any position will transform it into a quantum circuit.
\begin{figure}[H]
\centering
\includegraphics[width = 0.38\linewidth]{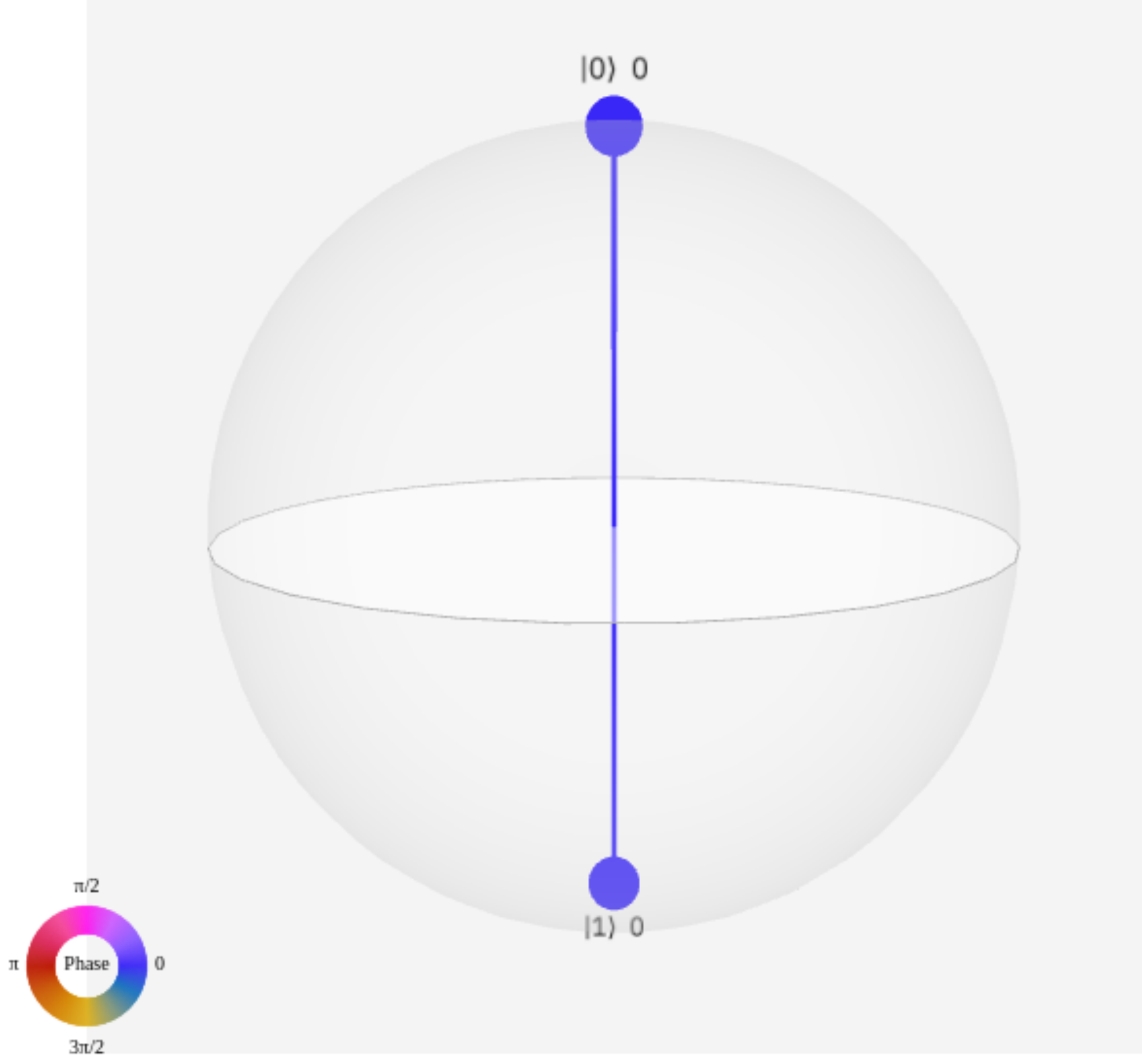}\hspace{0.4cm}
\includegraphics[width = 0.38\linewidth]{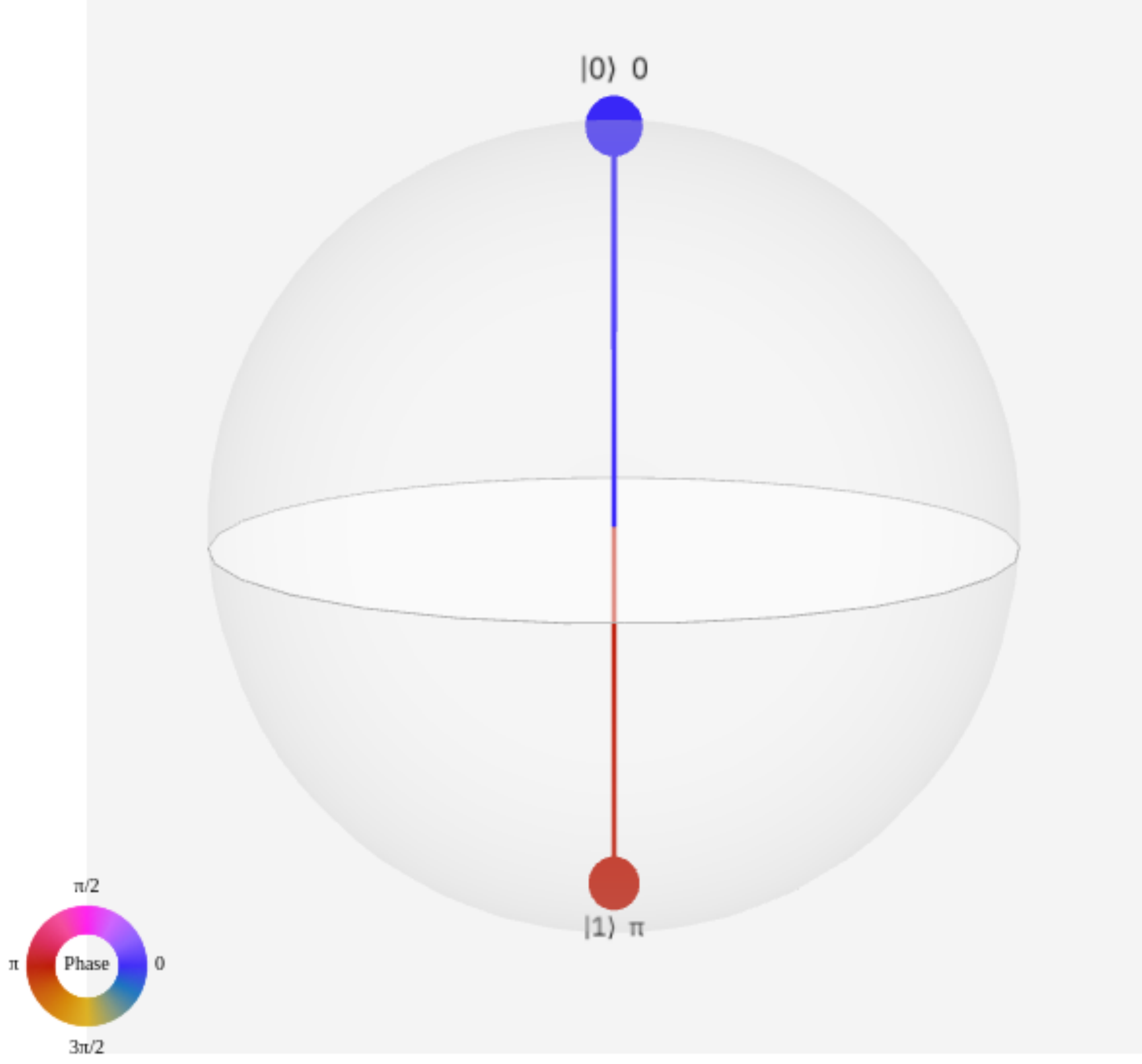}
\caption{A visualization of the resulting states from the $\mathbb{H}$ action on the states $\ket{0}$ and $\ket{1}$, respectively.}
\end{figure}
 \textcolor{black}{\underline{\large{The S Gate}}}
 
\begin{wrapfigure}{l}{0.17\textwidth}
\centering
\vspace{-11pt}
\includegraphics[width=0.09\textwidth]{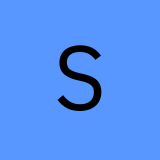}
\end{wrapfigure}
The \textsc{S gate} is represented by the following \textit{complex} matrix:
\begin{equation}
\mathbb{S}=\begin{pmatrix} 1 & 0 \\ 0 & i \end{pmatrix}
\end{equation}
When applied to the basis states $\ket{0}$ and $\ket{1}$, one obtains
\[\mathbb{S}\ket{0}=\begin{pmatrix} 1 & 0 \\ 0 & i\end{pmatrix}\begin{pmatrix} 1\\0 \end{pmatrix}=\begin{pmatrix} 1\\0 \end{pmatrix}=1\,\ket{0}\]
\[\mathbb{S}\ket{1}=\begin{pmatrix} 1 & 0 \\ 0 & i\end{pmatrix}\begin{pmatrix} 0\\1 \end{pmatrix}=\begin{pmatrix} 0\\i \end{pmatrix}=i\,\ket{1}\]
Thus, this gate induces a phase of ($\frac{\pi}{2}$) just on $\ket{1}$, which is equivalent to a ($\frac{\pi}{2}$) rotation around the \textit{z-axis}. If it is applied on the states $\ket{+}$ and $\ket{-}$, we obtain
\[\mathbb{S}\ket{+}=\begin{pmatrix} 1&0\\0&i \end{pmatrix}\begin{pmatrix} \tfrac{1}{\sqrt{2}}\\\tfrac{1}{\sqrt{2}} \end{pmatrix}=\begin{pmatrix} \tfrac{1}{\sqrt{2}}\\\tfrac{i}{\sqrt{2}} \end{pmatrix}=\ket{+i}\]
\[\mathbb{S}\ket{-}=\begin{pmatrix} 1&0\\0&i \end{pmatrix}\begin{pmatrix} \tfrac{1}{\sqrt{2}}\\\tfrac{-1}{\sqrt{2}} \end{pmatrix}=\begin{pmatrix} \tfrac{1}{\sqrt{2}}\\\tfrac{-i}{\sqrt{2}} \end{pmatrix}=\ket{-i}\]
Let's calculate its square \[\mathbb{S}^{\,2}=\begin{pmatrix} 1 & 0 \\ 0 & i \end{pmatrix}\begin{pmatrix} 1 & 0 \\ 0 & i \end{pmatrix}=\begin{pmatrix} 1 & 0 \\ \:0 & \!\!-1 \end{pmatrix}=\mathbb{Z}\]
Therefore, the $\mathbb{S}$ \textsc{gate} is not \textit{hermitian}, and its \textit{conjugate-transpose} defines the \textsc{S-dagger} or \textsc{Sdg gate}, given by the matrix:
\begin{equation}
\mathbb{S}^{\dagger}=\begin{pmatrix} 1 & 0 \\ 0 & -i \end{pmatrix}=\begin{pmatrix} 1 & 0 \\ 0 & e^{i\frac{3\pi}{2}} \end{pmatrix}\qquad;\quad \mathbb{S}^{\dagger}=\mathbb{S}^{\,3}
\end{equation}
That applies a ($\frac{3\pi}{2}$) rotation around the \textit{z-axis}.
\begin{figure}[H]
\centering
\includegraphics[width = 0.38\linewidth]{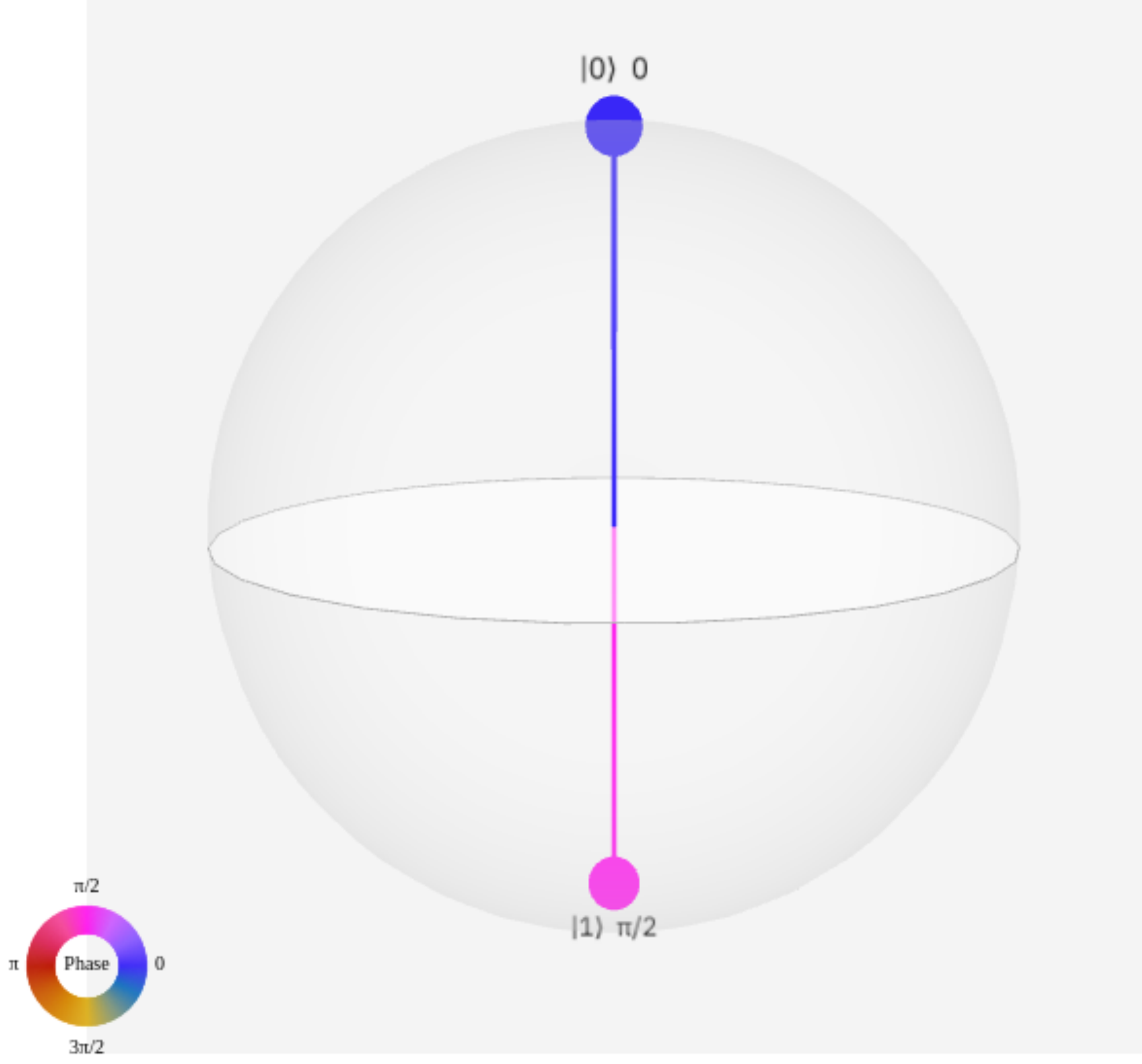}\hspace{0.4cm}
\includegraphics[width = 0.38\linewidth]{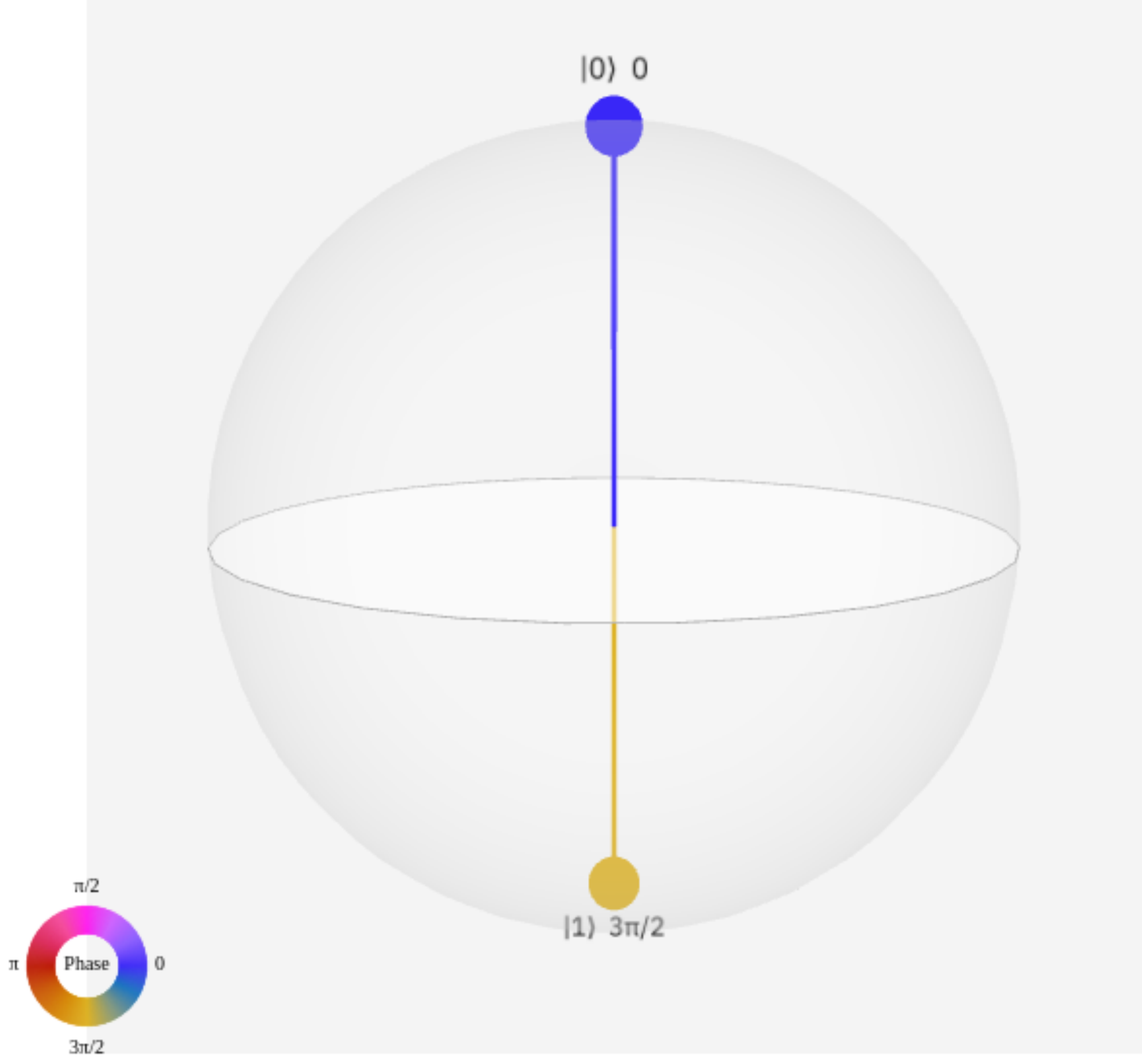}
\caption{A visualization of the resulting states from the $\mathbb{S}$ action on the states $\ket{+}$ and $\ket{-}$, respectively. (For the $\mathbb{S}^{\dagger}$ gate, the figures will be inversed).}
\label{1.13}
\end{figure}
\pagebreak
 \textcolor{black}{\underline{\large{The T Gate}}}

\begin{wrapfigure}{l}{0.17\textwidth}
\centering
\vspace{-11pt}
\includegraphics[width=0.09\textwidth]{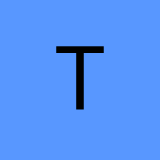}
\label{1.14}
\end{wrapfigure}
The only difference between the \textsc{S gate} and the \textsc{T gate} is the induced phase, where this gate applies a phase of ($\frac{\pi}{4}$), hence, we can write 
\begin{equation}
\mathbb{T}=\begin{pmatrix} 1 & 0 \\ \:0 & e^{i\frac{\pi}{4}} \end{pmatrix}
\end{equation}
Thus, for its action on $\ket{0}$ and $\ket{1}$, the outputs are
\[\mathbb{T}\ket{0}=\begin{pmatrix} 1 & 0 \\ 0 & e^{i\frac{\pi}{4}}\end{pmatrix}\begin{pmatrix} 1\\0 \end{pmatrix}=\begin{pmatrix} 1\\0 \end{pmatrix}=\ket{0}\quad\;\,\]
\[\mathbb{T}\ket{1}=\begin{pmatrix} 1 & 0 \\ 0 & e^{i\frac{\pi}{4}}\end{pmatrix}\begin{pmatrix} 0\\1 \end{pmatrix}=\begin{pmatrix} 0\\i \end{pmatrix}=e^{i\frac{\pi}{4}}\!\ket{1}\]
which is a rotation of ($\frac{\pi}{4}$) around the \textit{z-axis}. Applied on the states $\ket{+}$ and $\ket{-}$, gives
\[\mathbb{T}\ket{\pm}=\begin{pmatrix} 1&0\\0& e^{i\frac{\pi}{4}}\end{pmatrix}\begin{pmatrix} +\tfrac{1}{\sqrt{2}}\\ \pm\tfrac{1}{\sqrt{2}} \end{pmatrix}=\tfrac{1}{\sqrt{2}}\:(\:\ket{0}\pm e^{i\frac{\pi}{4}}\ket{1}\,)\]
\begin{figure}[H]
\centering
\includegraphics[width = 0.38\linewidth]{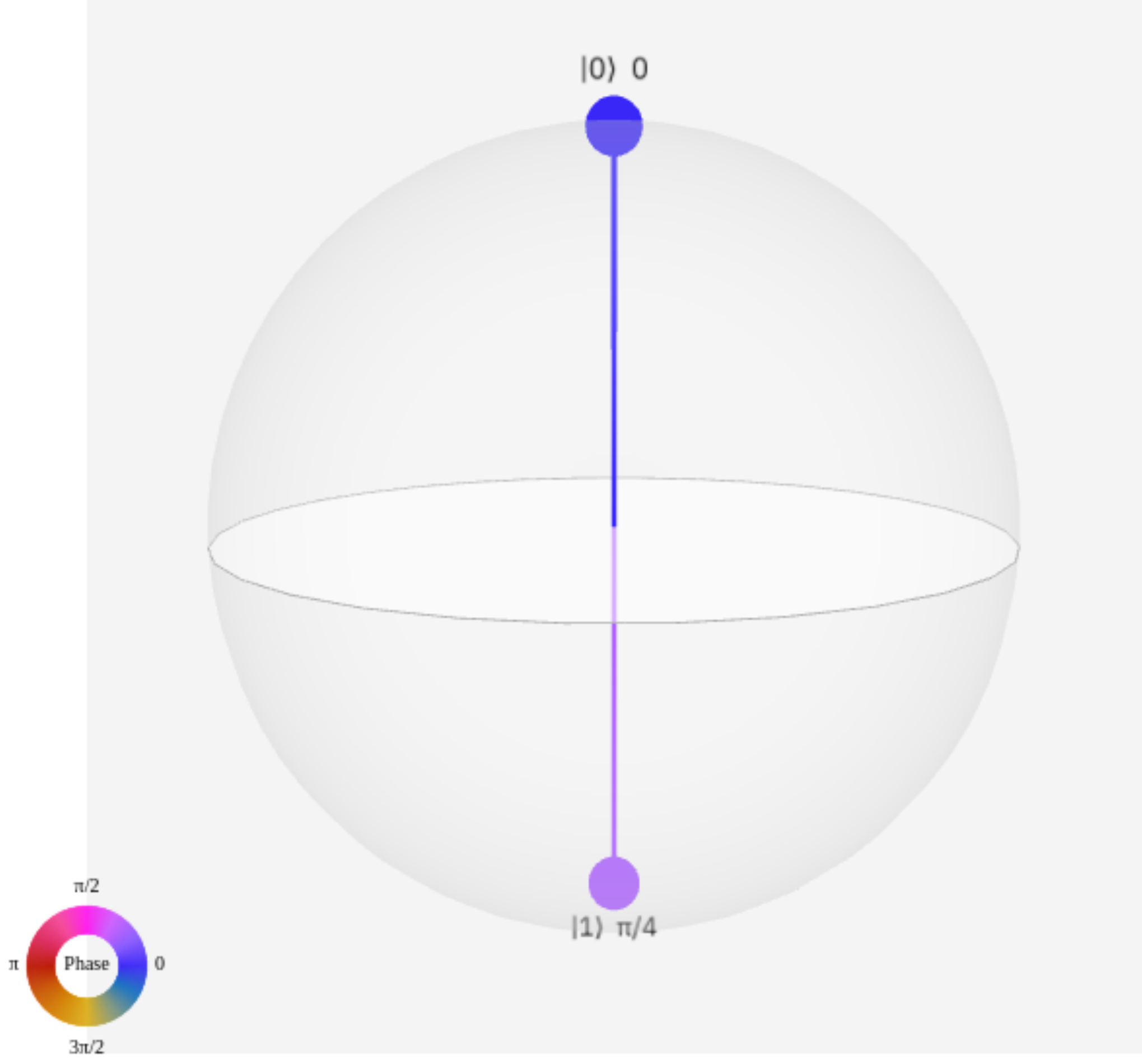}\hspace{0.4cm}
\includegraphics[width = 0.38\linewidth]{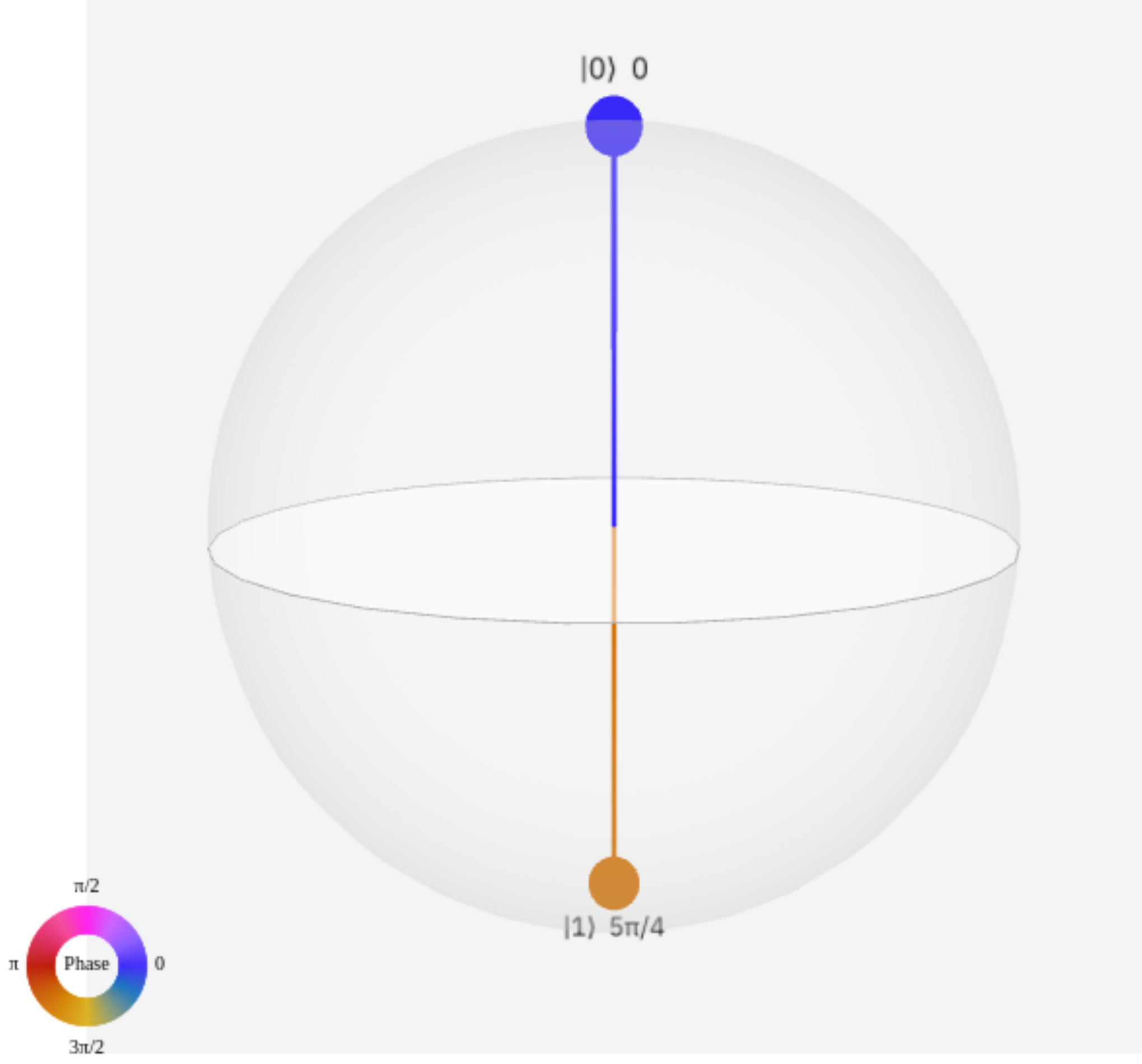}
\caption{A visualization of the resulting states from the $\mathbb{T}$ action on the states $\ket{+}$ and $\ket{-}$, respectively.}
\end{figure}
When calculating its square, we get \[\mathbb{T}^{\,2}=\begin{pmatrix} 1 & 0 \\ 0 &  e^{i\frac{\pi}{4}} \end{pmatrix}\begin{pmatrix} 1 & 0 \\ 0 &  e^{i\frac{\pi}{4}} \end{pmatrix}=\begin{pmatrix} 1 & 0 \\ 0 &  e^{i\frac{\pi}{2}} \end{pmatrix}=\mathbb{S}\]
This means that the $\mathbb{T}$ \textsc{gate} is not \textit{hermitian}, and its \textit{conjugate-transpose} defines the  \textsc{T-dagger} or \textsc{Tdg gate}, given by the \textit{adjoint} of $\mathbb{T}$
\begin{equation}
\mathbb{T}^{\dagger}=\begin{pmatrix} 1 & 0 \\ 0 & e^{-i\frac{\pi}{4}} \end{pmatrix}=\begin{pmatrix} 1 & 0 \\ 0 & e^{i\frac{7\pi}{4}} \end{pmatrix}\qquad;\quad \mathbb{T}^{\dagger}=\mathbb{T}^{\,7}
\end{equation}
Which rotates the qubit state by ($\frac{7\pi}{4}$) around the \textit{z-axis} on the \textit{Bloch sphere}.\\
\begin{figure}[H]
\centering
\includegraphics[width = 0.38\linewidth]{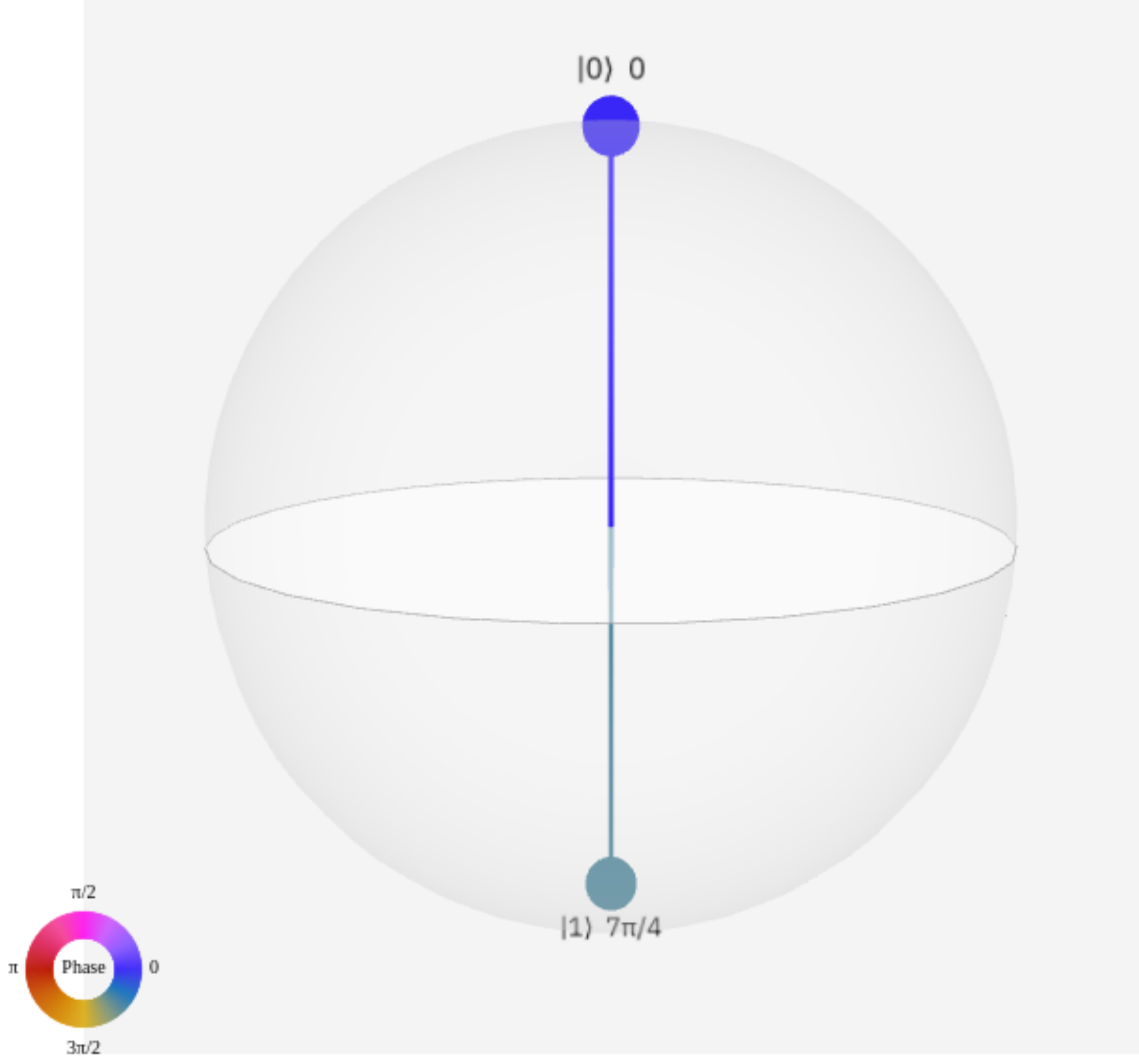}\hspace{0.4cm}
\includegraphics[width = 0.38\linewidth]{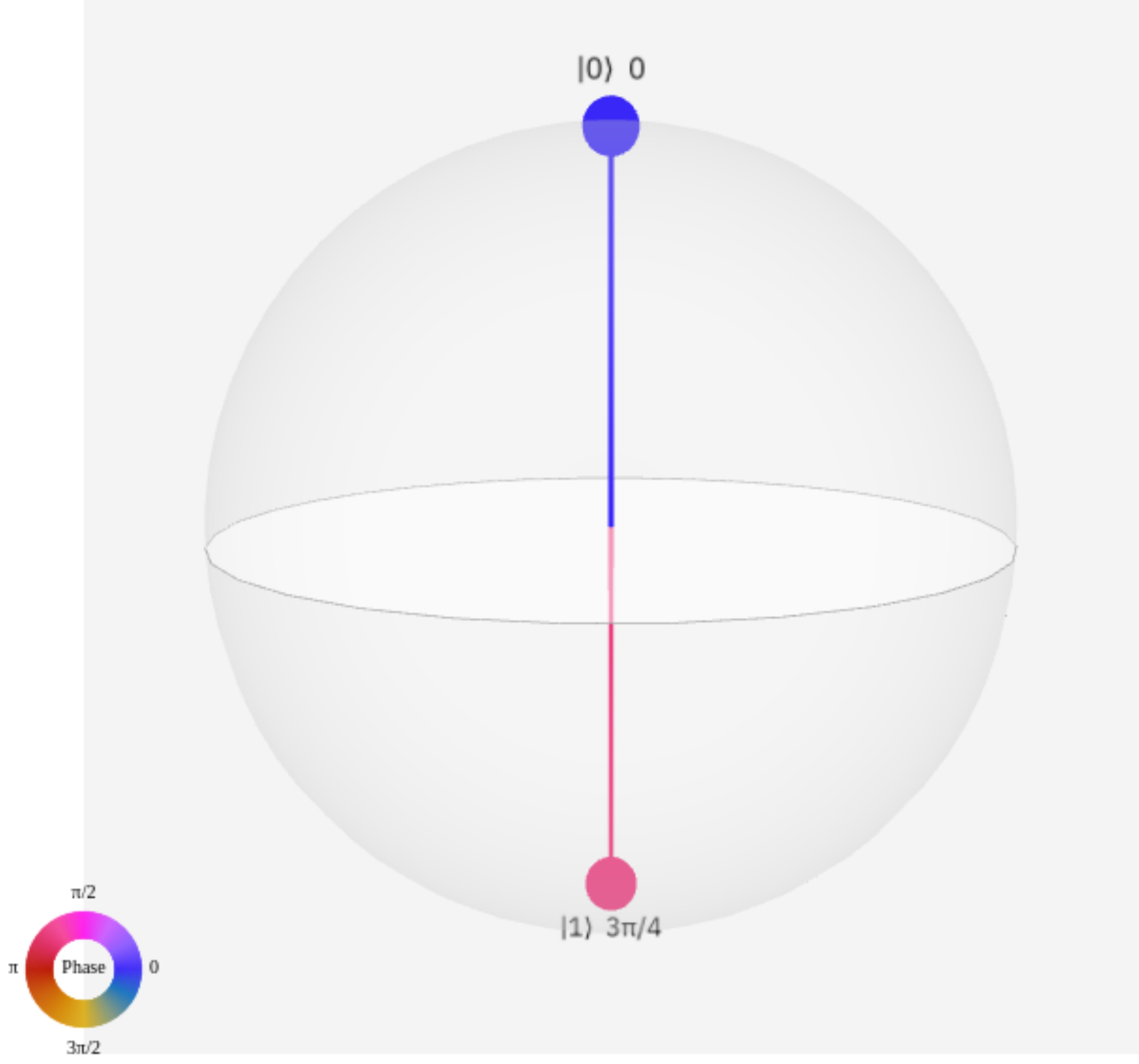}
\caption{A visualization of the resulting states from the $\mathbb{T}^{\dagger}$ action on the states $\ket{+}$ and $\ket{-}$, respectively.}
\end{figure}
 \textcolor{black}{\underline{\large{The P Gate}}}
 
\begin{wrapfigure}{l}{0.17\textwidth}
\centering
\vspace{-11pt}
\includegraphics[width=0.09\textwidth]{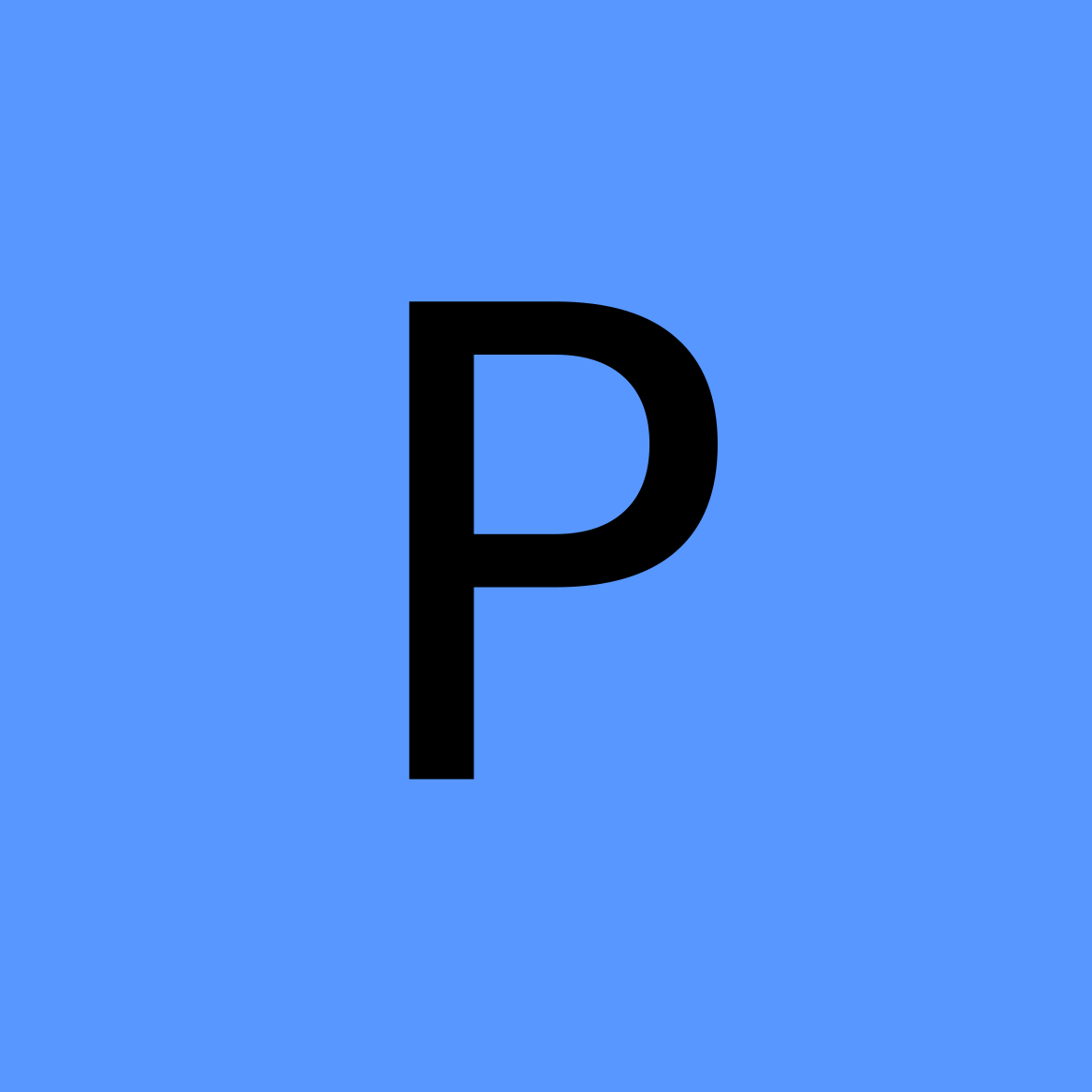}
\end{wrapfigure}
The generalization for a gate that applies any phase ($\theta$), can be written as:
\begin{equation}
\mathbf{P}_{\theta}=\begin{pmatrix} 1 & 0 \\ 0 & e^{i\theta} \end{pmatrix}\quad;\quad 0\leqslant\theta\leqslant 2\pi
\end{equation}

This gate is known as the \textsc{Phase gate}, and which leads for specific values of $\theta$ to
\[\mathbf{P}_{0,\,2\pi}=\mathbb{I}\quad;\quad\mathbf{P}_{\pi}=\mathbb{Z}\quad;\quad\mathbf{P}_{\frac{\pi}{2}}=\mathbb{S}\quad;\quad\mathbf{P}_{\frac{3\pi}{2}}=\mathbb{S}^{\dagger}\quad;\quad\mathbf{P}_{\frac{\pi}{4}}=\mathbb{T}\quad;\quad\mathbf{P}_{\frac{7\pi}{4}}=\mathbb{T}^{\dagger}\]
The \textsc{Phase gate} rotates the qubit state by $\theta$ around the \textit{z-axis} on the \textit{Bloch sphere}.
Like any quantum gate, the matrix $\mathbf{P}$ is of course \textit{unitary}  \[\mathbf{P}.\,\mathbf{P}^{\dagger}=\begin{pmatrix} 1 & 0 \\ 0 & e^{i\theta} \end{pmatrix}\begin{pmatrix} 1 & 0 \\ 0 & e^{-i\theta} \end{pmatrix}=\begin{pmatrix} 1 & 0 \\ 0& 1 \end{pmatrix}=\mathbb{I}\]

 \textcolor{black}{\underline{\large{Square-Root NOT Gate}}}
 
\begin{wrapfigure}{l}{0.17\textwidth}
\centering
\vspace{-11pt}
\includegraphics[width=0.09\textwidth]{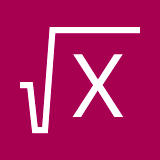}
\label{1.15}
\end{wrapfigure}
As the name suggests, it is the \textit{square root} of the \textsc{not} gate, therefore its square gives the \textsc{not} gate
\[\sqrt{\mathbb{X\,}}.\sqrt{\mathbb{X\,}}=\mathbb{X}\] 
It is represented by the matrix below :
\begin{equation}
\sqrt{\mathbb{NOT}}\equiv{\sqrt{\mathbb{X}}}=\frac{1}{2}\begin{pmatrix}1+i&1-i\\1-i&1+i\end{pmatrix} 
\end{equation}
\[\sqrt{\mathbb{X\,}}\ket{0}=\frac{1}{\sqrt{2}}(e^{i\frac{\pi}{4}}\ket{0}+e^{i\frac{7\pi}{4}}\ket{1})\]

\[\sqrt{\mathbb{X\,}}\ket{1}=\frac{1}{\sqrt{2}}(e^{i\frac{7\pi}{4}}\ket{0}+e^{i\frac{\pi}{4}}\ket{1})\]
It creates an \textbf{equal superposition} state, but the only difference with the \textsc{Hadamard gate}, is that it applies a different \textit{relative phase}. This gate is \textit{natively} implemented in some \textsc{hardware}.
\begin{figure}[H]
\centering
\includegraphics[width = 0.38\linewidth]{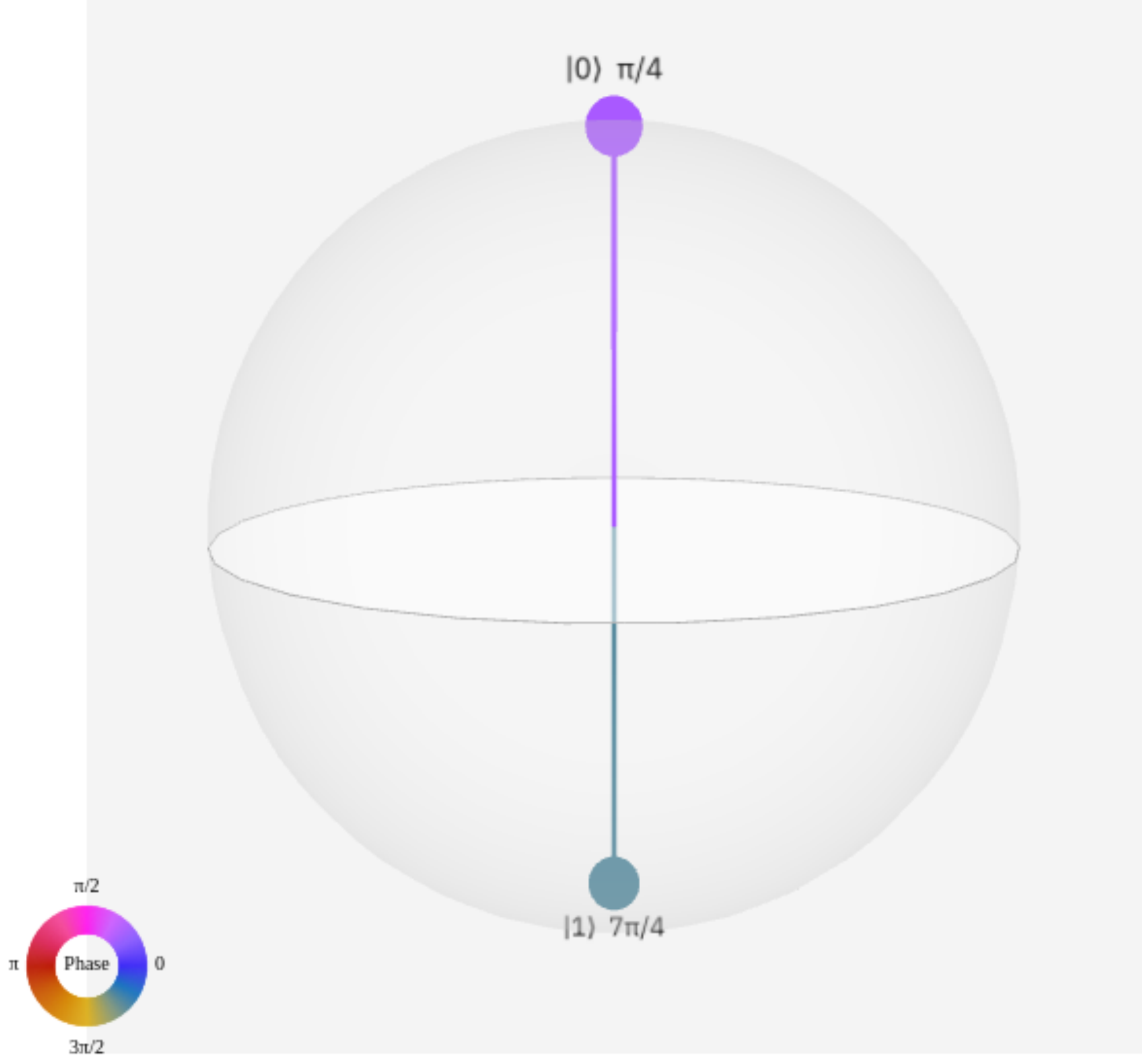}\hspace{0.4cm}
\includegraphics[width = 0.38\linewidth]{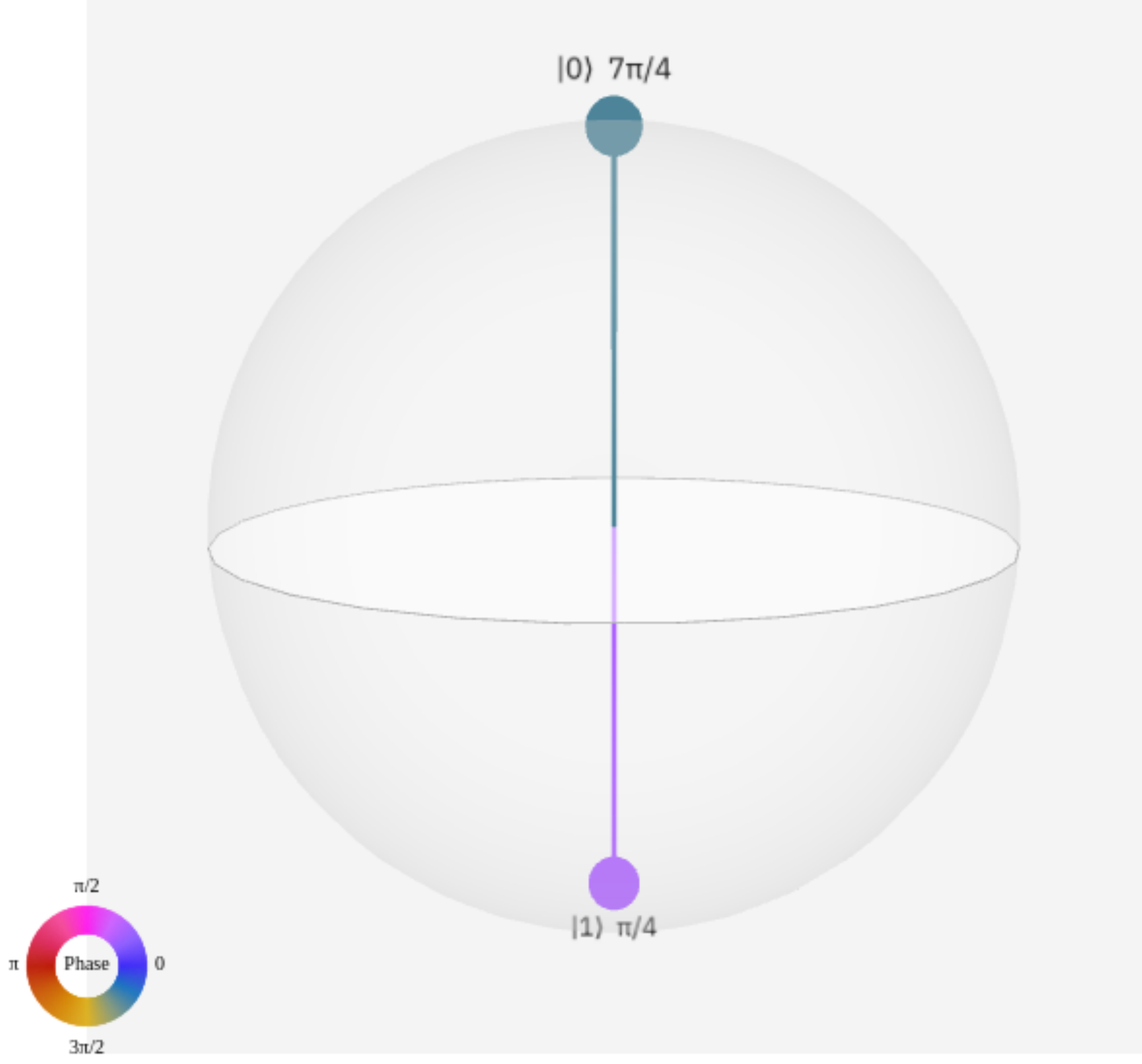}
\caption{A visualization of the resulting states from the $\sqrt{\mathbb{X}}$ action on the states $\ket{0}$ and $\ket{1}$, respectively.}
\label{1.16}
\end{figure}
 \textcolor{black}{\underline{\large{Square-Root NOT-Dagger Gate}}}
 
\begin{wrapfigure}{l}{0.17\textwidth}
\centering
\vspace{-11pt}
\includegraphics[width=0.09\textwidth]{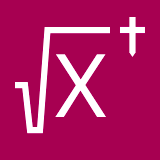}
\end{wrapfigure}

This gate is the \textit{inverse} of $\sqrt{\mathbb{X}}$ gate, so $\sqrt{\mathbb{X}}^\dagger.\sqrt{\mathbb{X}}=\sqrt{\mathbb{X}}.\sqrt{\mathbb{X}}^\dagger=\mathbb{I}$ and its square also gives the \textsc{not} gate $\sqrt{\mathbb{X\,}}^\dagger.\sqrt{\mathbb{X\,}}^\dagger=\mathbb{X}$. Its matrix representation is given below:
\begin{equation}
\sqrt{\mathbb{NOT}}^\dagger\equiv{\sqrt{\mathbb{X}}}^\dagger=\frac{1}{2}\begin{pmatrix}1-i&1+i\\1+i&1-i\end{pmatrix}
\end{equation}

\[\sqrt{\mathbb{X\,}}^\dagger\ket{0}=\frac{1}{\sqrt{2}}(e^{i\frac{7\pi}{4}}\ket{0}+e^{i\frac{\pi}{4}}\ket{1})\]

\[\sqrt{\mathbb{X\,}}^\dagger\ket{1}=\frac{1}{\sqrt{2}}(e^{i\frac{\pi}{4}}\ket{0}+e^{i\frac{7\pi}{4}}\ket{1})\]
We see that it also creates an \textbf{equal superposition} state, but with \textit{relative phases} reversed with respect to $\sqrt{\mathbb{X}}$. Therefore, the Q-spheres of $\sqrt{\mathbb{X}}^\dagger$ are also reversed.

\subsection*{Two-Qubit Gates}

 \textcolor{black}{\underline{\large{SWAP Gate}}}

\begin{wrapfigure}{l}{0.17\textwidth}
\centering
\vspace{-11pt}
\includegraphics[width=0.09\textwidth]{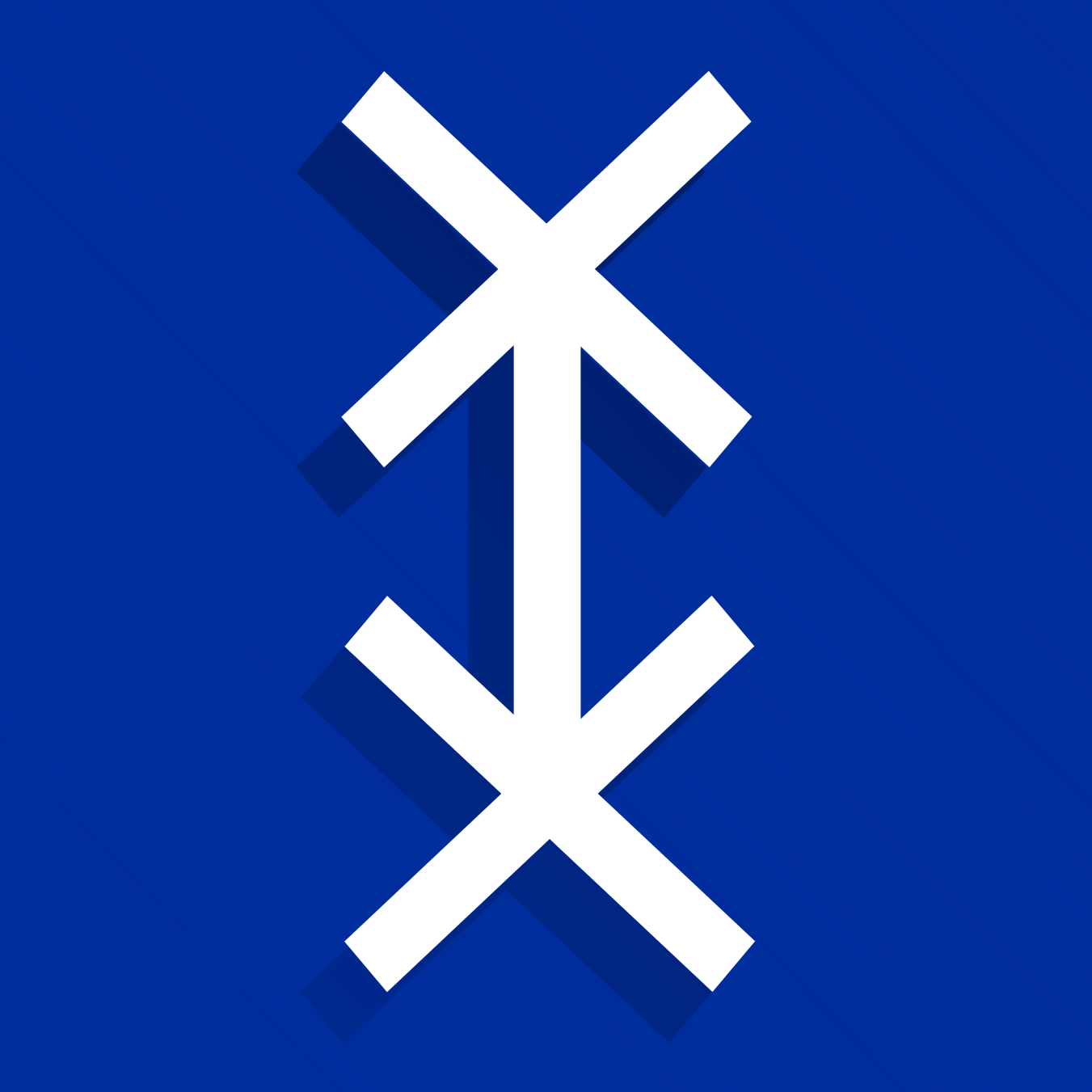}
\end{wrapfigure}
The \textsc{Swap gate} is a two-qubit gate, so it is represented by a $4\times4$ \textit{dimensional unitary matrix}, given by
\begin{equation}
\mathbb{SWAP}=\begin{pmatrix} 1&0&0&0\\0&0&1&0\\0&1&0&0\\0&0&0&1 \end{pmatrix}\qquad;\quad \mathbb{(SWAP)}^2=\mathbb{I}
\end{equation}
Let's observe its action on a two-qubit state \[\ket{\psi}=\ket{\phi}_{_0}\otimes\,\ket{\phi}_{_1}=c_{_{00}}\ket{00}+c_{_{01}}\ket{01}+c_{_{10}}\ket{10}+c_{_{11}}\ket{11}\quad;\: c_{_{kl}}=a_{_{k}}b_{_{l}}\,(k,l=0,1)\]
where $\ket{\phi}_{_0}$ and $\ket{\phi}_{_1}$ are the states of the qubits $\mathfrak{q}_0$ and $\mathfrak{q}_1$, respectively.
\begin{align*}
\mathbb{(SWAP)}\ket{\psi}&=\begin{pmatrix}1&0&0&0\\0&0&1&0\\0&1&0&0\\0&0&0&1\\\end{pmatrix}\begin{pmatrix}a_{_{0}}\,b_{_{0}}\\a_{_{0}}\,b_{_{1}}\\a_{_{1}}\,b_{_{0}}\\a_{_{1}}\,b_{_{1}}\\\end{pmatrix}=\begin{pmatrix}a_{_{0}}\,b_{_{0}}\\a_{_{1}}\,b_{_{0}}\\a_{_{0}}\,b_{_{1}}\\a_{_{1}}\,b_{_{1}}\\\end{pmatrix}\\&=c_{_{00}}\ket{00}+c_{_{10}}\ket{01}+c_{_{01}}\ket{10}+c_{_{11}}\ket{11}=\ket{\phi}_{_1}\otimes\,\ket{\phi}_{_0}
\end{align*}
So this gate \textit{switches} the states of two qubits
\[(\;\ket{\phi}_{_0}\!\otimes\,\ket{\phi}_{_1})\,\xrightarrow{\mathbb{SWAP}}\,(\;\ket{\phi}_{_1}\!\otimes\,\ket{\phi}_{_0})\]
\begin{figure}[H]
\centering
\includegraphics[width = 0.42\linewidth]{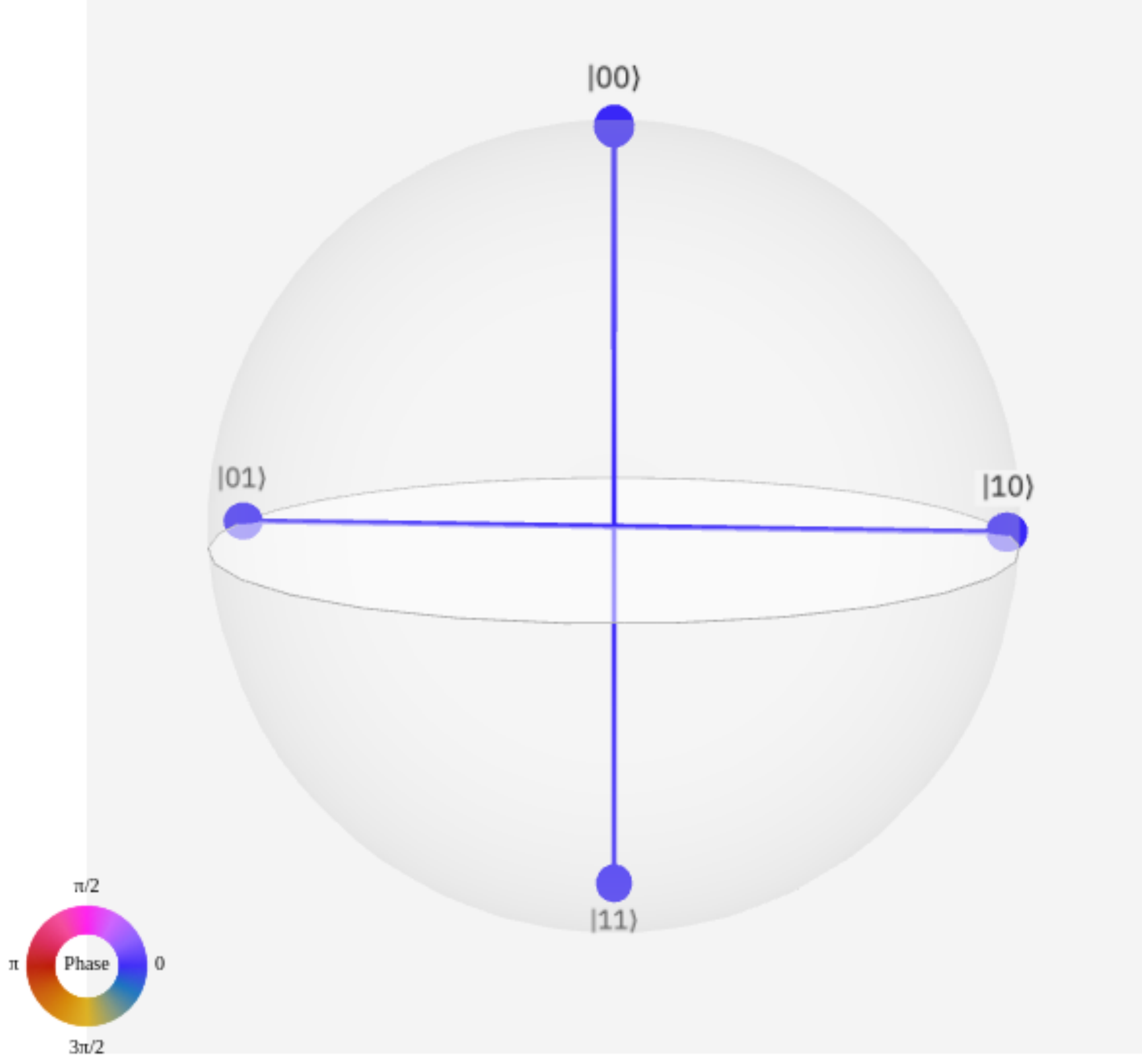}
\caption{A visualization of all 4 possible resulting states from the $\mathbb{SWAP}$ action on the 2-qubit state $\ket{\alpha\,\beta};\,\alpha,\,\beta=0,\,1$.}
\label{1.17}
\end{figure}

 \textcolor{black}{\underline{\large{Controlled-NOT Gate}}}

\begin{wrapfigure}{l}{0.17\textwidth}
\centering
\vspace{-11pt}
\includegraphics[width=0.09\textwidth]{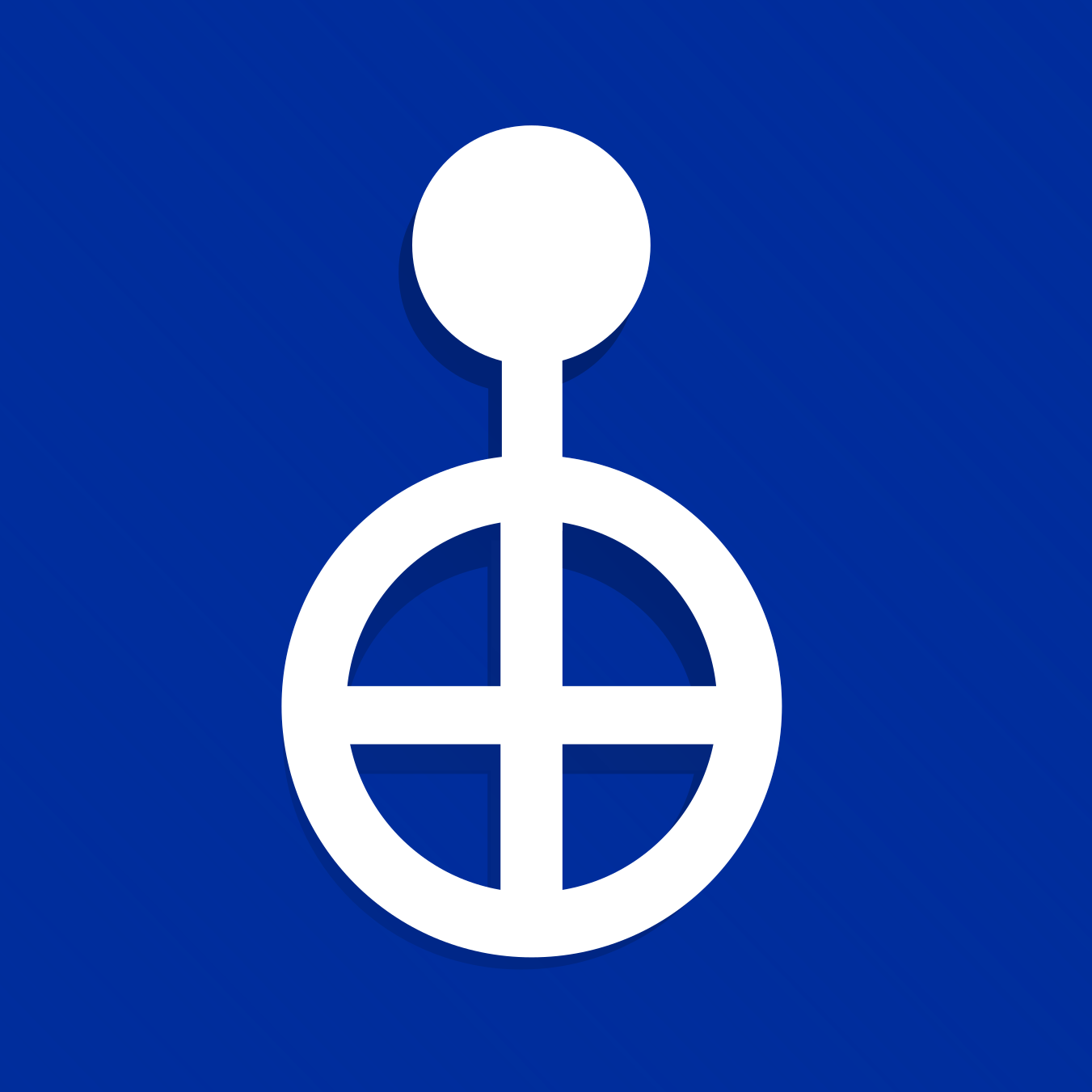}
\end{wrapfigure}

The \textsc{cnot} or \textsc{cx} gate is a two-qubit gate, acting in the $2^2$ dimensional \textit{Hilbert-space}.
The \textsc{cnot gate} is represented by the matrix of dimension $4\times4$ below:
\begin{equation}
\mathbb{CNOT}\equiv\mathbb{CX}=\begin{pmatrix}1&0&0&0\\0&1&0&0\\0&0&0&1\\0&0&1&0\\\end{pmatrix}\quad;\quad(\mathbb{CX})^{2}=\mathbb{I}
\end{equation}
When the 2-qubit state $\ket{\psi}$  pass through this gate it becomes:
\[\mathbb{CX}\ket{\psi}=\begin{pmatrix}1&0&0&0\\0&1&0&0\\0&0&0&1\\0&0&1&0\\\end{pmatrix}\begin{pmatrix}a_{_{0}}\,b_{_{0}}\\a_{_{0}}\,b_{_{1}}\\a_{_{1}}\,b_{_{0}}\\a_{_{1}}\,b_{_{1}}\\\end{pmatrix}=\begin{pmatrix}a_{_{0}}\,b_{_{0}}\\a_{_{0}}\,b_{_{1}}\\a_{_{1}}\,b_{_{1}}\\a_{_{1}}\,b_{_{0}}\\\end{pmatrix}\]
We see that this gate swaps the two \textit{amplitudes} of $\mathfrak{q}_1$ in the states $\ket{10}$ and $\ket{11}$ which is a \textsc{not} operation on $\mathfrak{q}_1$. Moreover, it's applied only when $\mathfrak{q}_0$ is in $\ket{1}$. Therefore the \textsc{cnot} gate is a \textbf{conditional} gate, where $\mathfrak{q}_0$ and $\mathfrak{q}_1$ are called \textbf{control} and \textbf{target}, respectively. If the control qubit is in the state $\ket{1}$ the \textsc{cnot} gate executes a \textsc{not} on the target qubit, otherwise it acts as an $\mathbb{I}$ on the target.\\ 
If we turn the roles ($control\leftrightarrow target$), then this gate will become what we call a \textsc{reverse cnot} gate and its matrix is given by : 
\[\mathbb{RCNOT}\equiv\mathbb{RCX}=\begin{pmatrix}1&0&0&0\\0&0&0&1\\0&0&1&0\\0&1&0&0\\\end{pmatrix} \]
The \textsc{cnot gate} is the gateway that \textbf{entangles} 2-qubits, only when the control qubit is in a \textbf{superposition}.\\
\begin{figure}[H]
\centering
\includegraphics[width = 0.4\linewidth]{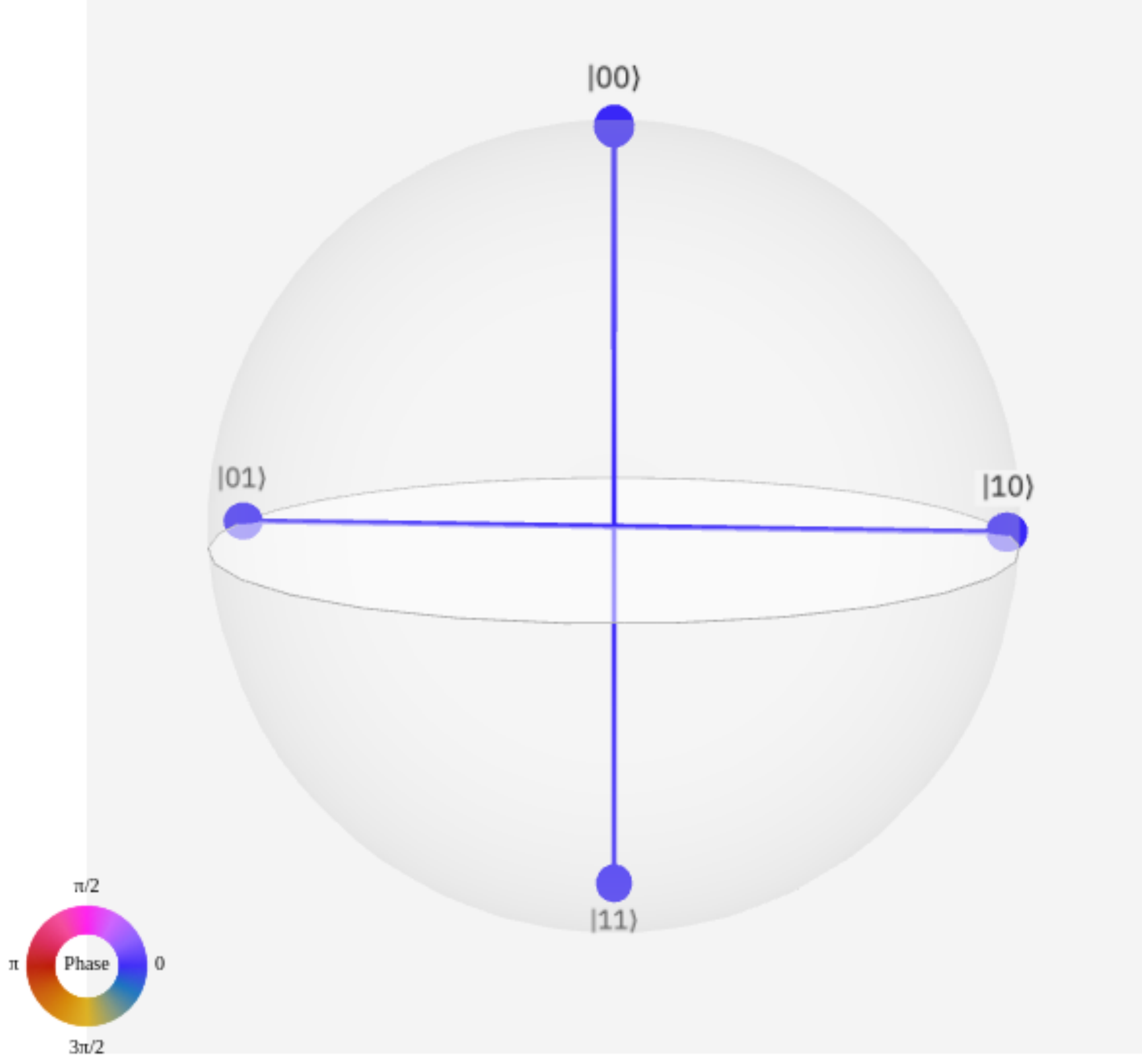}
\caption{A visualization of all 4 possible resulting states from the $\mathbb{CX}$ action on the 2-qubit state $\ket{\alpha\,\beta};\,\alpha,\,\beta=0,\,1$.}
\label{1.18}

\end{figure}
\subsection*{Three-Qubit Gates}
 \textcolor{black}{\underline{\large{Controlled-Controlled-NOT Gate}}}
 
\begin{wrapfigure}{l}{0.17\textwidth}
\centering
\includegraphics[width=0.09\textwidth]{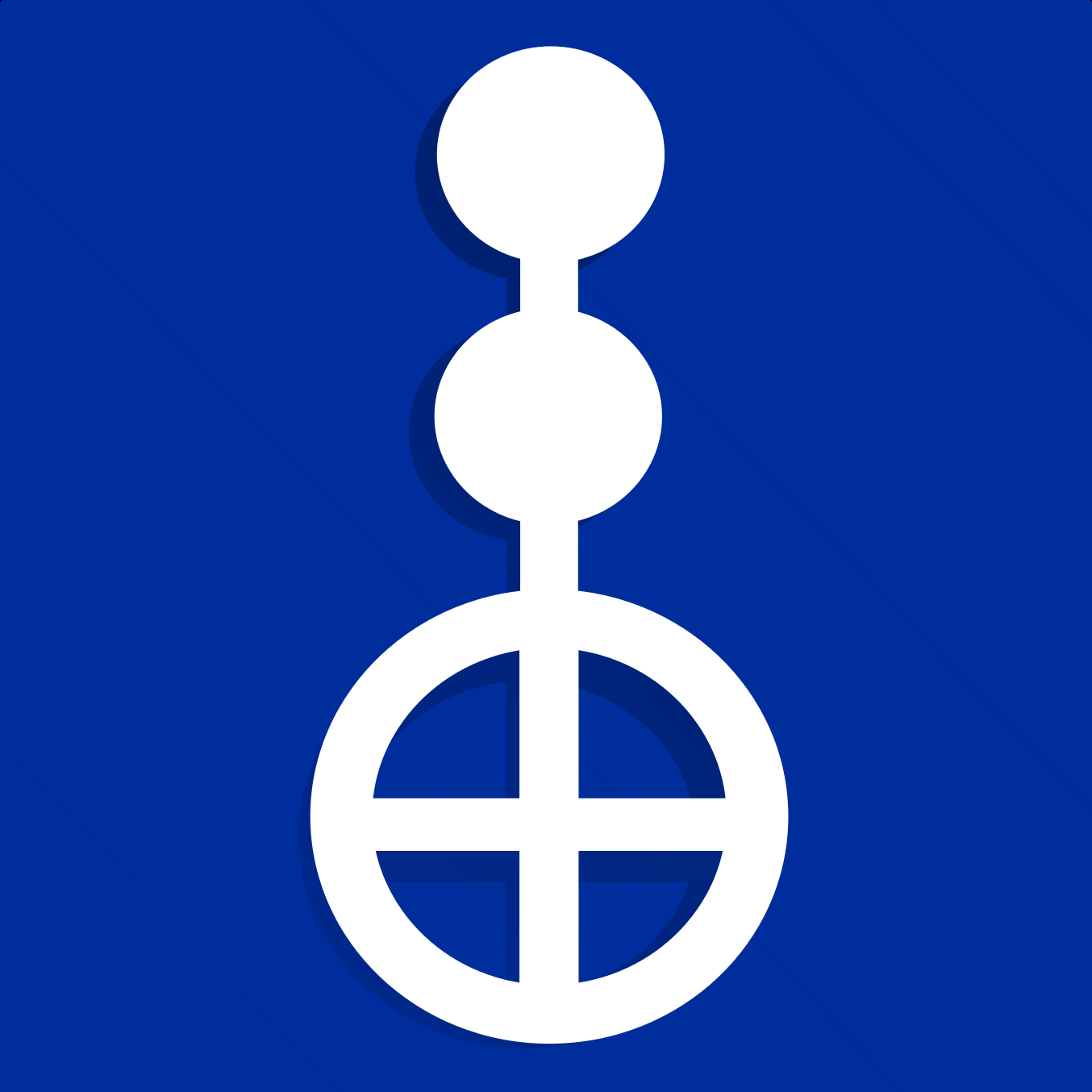}
\end{wrapfigure}

The \textsc{ccnot} or \textsc{ccx} is a \textbf{conditional} three-qubit gate, with two \textit{control} qubits and one \textit{target}. It performs a \textsc{not} to the target if and only if the two control qubits are in $\ket{1}$, otherwise it acts as an $\mathbb{I}$ on the target.
It is represented by the matrix of dimension $8\times8$ below:
\begin{equation}
\mathbb{CCNOT}\equiv\mathbb{CCX}=\begin{pmatrix}1&0&0&0&0&0&0&0\\0&1&0&0&0&0&0&0\\0&0&1&0&0&0&0&0\\0&0&0&1&0&0&0&0\\0&0&0&0&1&0&0&0\\0&0&0&0&0&1&0&0\\0&0&0&0&0&0&0&1\\0&0&0&0&0&0&1&0\end{pmatrix}\quad;\quad(\mathbb{CCX})^2=\mathbb{I}
\end{equation}
The \textsc{ccnot} is a \textit{reversible} version of classical computing's \textsc{and} and \textsc{nand} gates.
This is a \textbf{universal gate} for \textit{classical computation}, meaning that any program can be built from many instances of this gate.\\
\begin{figure}[H]
\centering
\includegraphics[width = 0.42\linewidth]{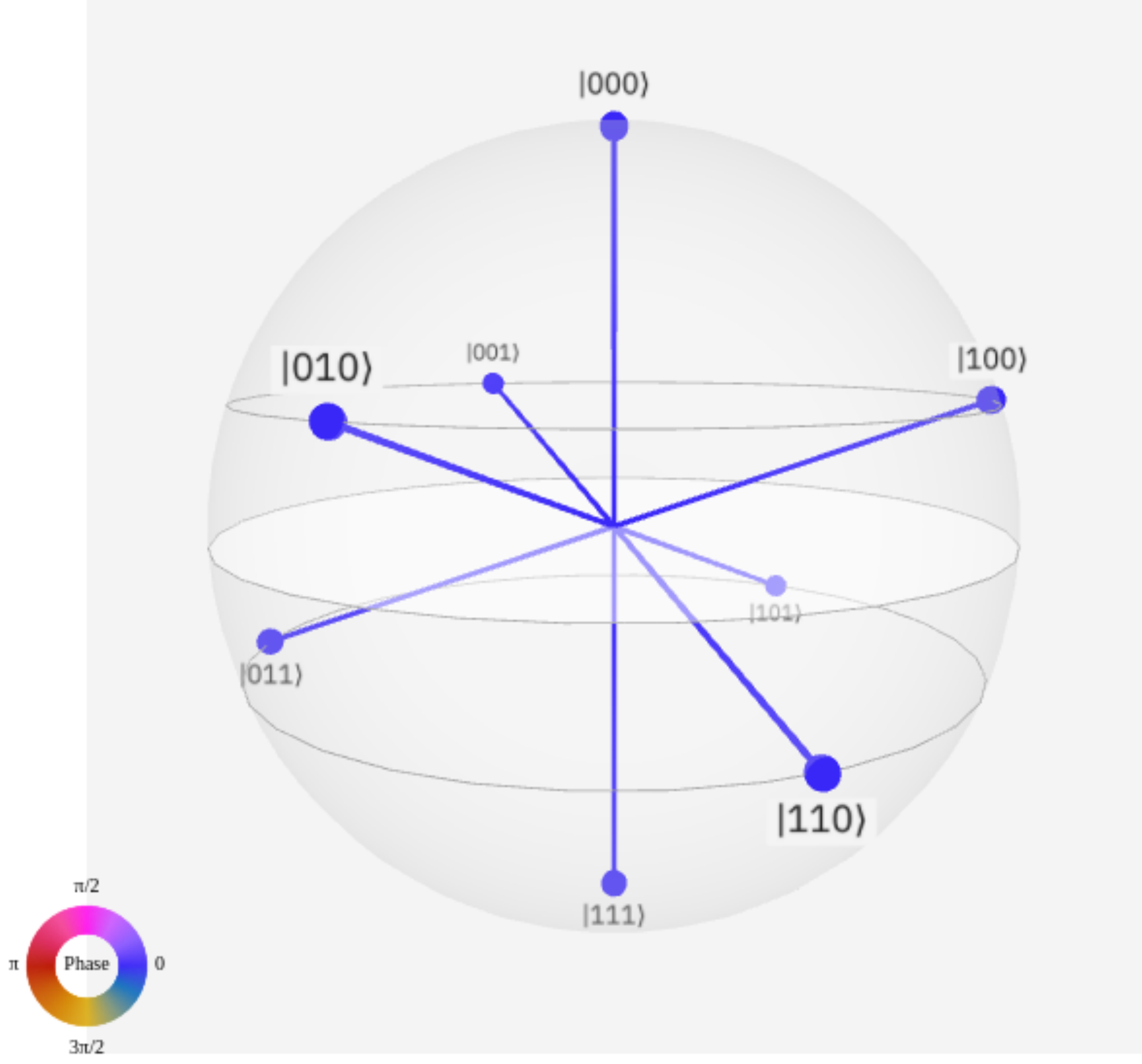}
\caption{A visualization of all 8 possible resulting states from the $\mathbb{CCX}$ action on the 3-qubit state $\ket{\alpha\,\beta\,\gamma};\,\alpha,\,\beta,\,\gamma=0,\,1$.}
\label{1.19}

\end{figure}

\subsection{Quantum Circuit Diagrams}
\label{1.3.3}
A quantum circuit is a sequence of gates applied to one qubit or more. The collection of qubits in a circuit is called a \textit{quantum register}. So, to perform calculations in quantum computers, we need to manipulate qubits (inside a quantum register) by applying a serie of gates. All qubits in the quantum register are initialized in the state $\ket{0}$ and by convention, we number the qubits in the register as $q_0$, $q_1$, $q_2$, $...$,$q_n$. 
To visualize quantum circuits, we need a \textit{quantum circuit diagram}. In what follows, we will use the interactive interface of the IBM \textsc{quantum composer}\footnote{\url{https://quantum-computing.ibm.com/composer/files/new}} to represent these quantum circuits.\\
The following figure show an “\textit{empty}” quantum circuit; we haven't added any gates yet.

\begin{figure}[!ht]
\centering
\includegraphics[scale=0.4]{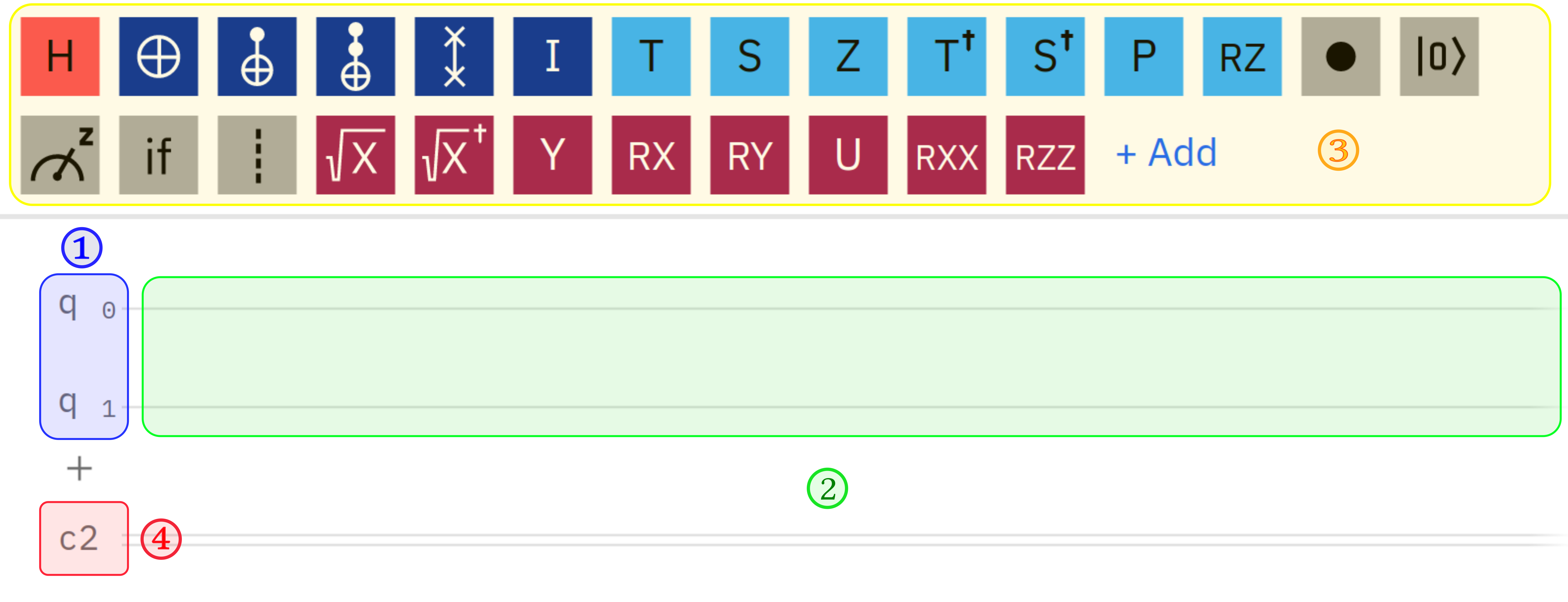}
\caption{The interactive interface of IBM quantum composer.}
\label{1.20}

\end{figure}
\pagebreak

Let us describe each part of this empty circuit.
\begin{enumerate}
 \item First, on the left-hand side, we see two qubits, the numbering starts from zero: $q_0$, $q_1$. Each of these qubits is initialized to be in the $\ket{0}$ state.
 \item In the right-hand side, we have two wires. In these wires we can place gates and measurements. 
 \item At the top, we see some of the quantum gates that we have already discussed.
 \item On the bottom we see a $C_2$, indicating that it is a classical register, a $C_4$ will indicate that the register contains four bits all initialized to zero, and a wire with a line through it will indicate that the register contains multiple bits. When we do the measurement a line will be drawn down to the classical register, indicating that the measurement is stored there.
\end{enumerate}

Now let us see our quantum circuit with some gates on it. 

\begin{figure}[!ht]
\centering
\includegraphics[scale=0.38]{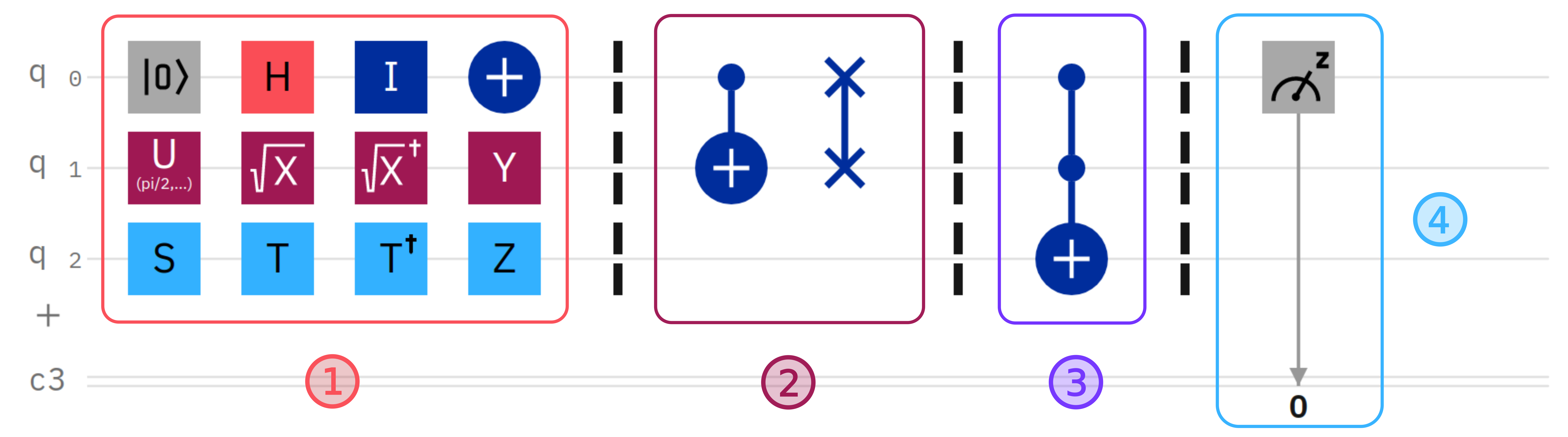}
\caption{Types of quantum gates.}
\label{1.21}

\end{figure}

We have divided the circuit into four blocks.

\begin{itemize}
\item The first block includes single-qubit gates. These gates are applied from the left to the right, we apply first the closest gate to the qubit. On the other hand from the algebra perspective, we apply gates from the right to left. For example, the first line in the first block reads as $\mathbb{XIH}\ket{0}$ which means we apply the \textsc{Hadamard} first then the \textsc{identity} gate, and finally the \textsc{not} gate.  The first operation from the left  \fbox{$\ket{0}$} returns a qubit to state $\ket{0}$, independently of its state before the operation was applied. It is not a reversible operation, so it is not a gate. The \textsc{not} gate  is drawn graphically as ($\bigoplus$).

\item The second block contains, two-qubit gates that span \textbf{two wires}. From the right to the left the \textsc{cnot} and \textsc{swap} gate. The \textsc{cnot} gate graphic contains the sign ($\bullet$) denoting the \textbf{control qubit} and the sign ($\bigoplus$) denoting the \textbf{target qubit}. If we draw it upside down, we will get the \textsc{reverse cnot}. 
\item The third block contains, the \textsc{ccnot} gate it operates on three qubits. If the first two qubits are $\ket{1}$ then it flips the third, otherwise, it does nothing. The only difference between \textsc{cnot} and \textsc{ccnot} gate is that there are two \textbf{control qubits} ($\bullet\,\bullet$).
\item Finally, the measurement operation returns a $\ket{0}$ or $\ket{1}$. The \textsc{Measurement} operation must always be at the \textbf{end} of any quantum circuit.
\end{itemize}

In quantum computing, all operations between \textit{initialization} and \textit{measurement} are \textbf{reversible}. The following table shows the gates and it inverses.

\begin{center}
\begin{large}
\begin{tabular}{|c|c|c|c|c|c|c|c|c|c|}
\hline 
Gate & $\mathbb{I}$ & $\mathbb{X}$ & $\mathbb{Y}$ & $\mathbb{Z}$ & $\mathbb{H}$ & $\mathbb{S}$ &$ \mathbb{S}^{\dagger}$ & $\mathbb{T}$ & $\mathbb{T}^{\dagger}$   \\ 
\hline 
Reverse & $\mathbb{I}$ & $\mathbb{X}$ & $\mathbb{Y}$ & $\mathbb{Z}$ & $\mathbb{H}$ & $\mathbb{S}^{\dagger}$ & $\mathbb{S}$ & $\mathbb{T}^{\dagger}$& $\mathbb{T}$  \\ 
\hline 
\end{tabular} 
\end{large}

\end{center}
\subsection*{Examples of Quantum Circuits} 
To explain how quantum circuits work, let's take two examples:\\
\begin{itemize}

\item The first example illustrate the \textsc{succession} of operations:
\begin{figure}[H]
\centering
\includegraphics[width = 0.65\linewidth]{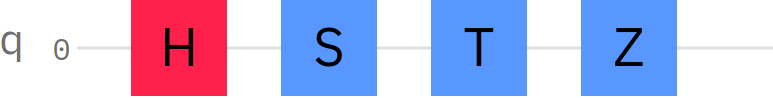}
\end{figure}
A qubit is initialized at the state $\ket{\psi}=\ket{0}$. Then each gate will act on its state and transform it according to its functionality. Mathematically we have:
\begin{align*}
\ket{0}\rightarrow\mathbb{Z\,T\,S\,H}\ket{0}&=\mathbb{Z\,T\,S}\ket{+}=\mathbb{Z\,T}\ket{+i\,}\\&=\tfrac{1}{\sqrt{2}}\mathbb{\,Z\,}(\ket{0}+i\,e^{i\frac{\pi}{4}}\ket{1})=\tfrac{1}{\sqrt{2}}\,(\ket{0}+e^{i\frac{7\pi}{4}}\ket{1})
\end{align*}

\item The second example illustrate the \textsc{equivalence}:

\end{itemize}

This circuit swaps two qubits. That means if we started in the $\ket{01}$ state, we would end up in the $\ket{10}$ state and vice versa. Inputs of $\ket{00}$ or $\ket{11}$ would be unchanged as swapping would have no effect. Note that for the first and third CNOTs, the control qubit is the first qubit and the target qubit is the second, while for the second CNOT, the control qubit is the second and the target qubit is the first.

\begin{figure}[!ht]
\centering
\includegraphics[scale=0.27]{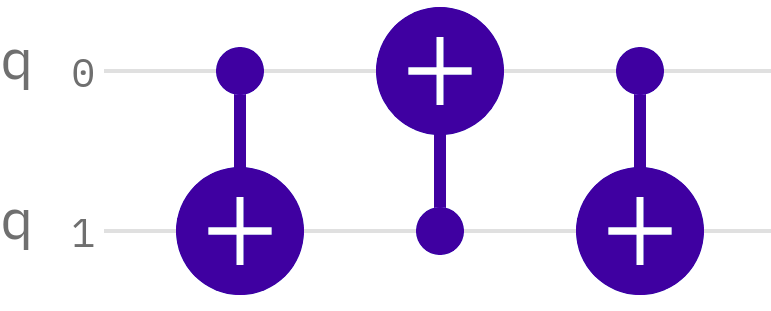}
\end{figure}

 Let's trace it out step by step:
 
\vspace*{1cm}
\begin{center}
\begin{tabular}{|c|c|c|c|}
\hline 
Input & After first CNOT & After second CNOT & After the third CNOT (output) \\ 
\hline 
00 & 00 & 00 & 00 \\ 
\hline 
01 & 01 & 11 & 10 \\ 
\hline 
10 & 11 & 01 & 01 \\ 
\hline 
11 & 10 & 10 & 11 \\ 
\hline 
\end{tabular} 
\end{center}
If we calculate the matrix product, we will found certainly the $\mathbb{SWAP}$ matrix.
\[\mathbb{CX.RCX.CX}=\begin{pmatrix}1&0&0&0\\0&1&0&0\\0&0&0&1\\0&0&1&0\end{pmatrix}
\begin{pmatrix}1&0&0&0\\0&0&0&1\\0&0&1&0\\0&1&0&0\end{pmatrix}
\begin{pmatrix}1&0&0&0\\0&1&0&0\\0&0&0&1\\0&0&1&0\end{pmatrix}=
\begin{pmatrix}1&0&0&0\\0&0&1&0\\0&1&0&0\\0&0&0&1\end{pmatrix}=\mathbb{SWAP}\]
Therefore 
\[\mathbb{SWAP=CX\,.\,RCX\,.\,CX}\] 







\subsection*{Bell States Circuits}

The central object of this part will be to construct \textbf{quantum circuits} that generate the \textbf{four Bell states}: $|\Phi^+\rangle,|\Phi^-\rangle,|\Psi^+\rangle,|\Psi^-\rangle$, which describe an entangled pair of qubits. We have seen previously simple manipulations on a single or two qubits system. But creating $\mathfrak{entanglement}$ between two qubits, seems difficult at first glance, and one may think that we will use a varied assortment of $\mathfrak{quantum\:gates}$. But in fact, it is \textit{extraordinary} $\mathfrak{simple}$.\\

Firstly, let's have two qubits, noted $\mathfrak{q}_{0}$ and $\mathfrak{q}_{1}$, which are initialized in the state $|0\rangle$, so we have $\mathfrak{q}_0$ in the state $|\psi\rangle_{_0}=|0\rangle_{_0}$ and $\mathfrak{q}_1$ in $|\psi\rangle_{_1}=|0\rangle_{_1}$. The initial 2-qubit state is of course \textit{separable} $|\psi\rangle_{_0}\otimes|\psi\rangle_{_1}=\ket{0}_{_0}\otimes\ket{0}_{_1}=|00\rangle$, and we want to construct the specific Bell state $|\Phi^+\rangle$, written
\begin{equation}
|\Phi^+\rangle=\tfrac{1}{\sqrt{2}}\,(\,|00\rangle+|11\rangle\,)
\end{equation} 
which is \textit{inseparable}. But we can think of applying $\mathfrak{operators/gates}$ to \textit{separate} it into two distinct qubit states, and vis versa; by acting gates on the \textit{initial sparable} state, it will be transformed into an \textit{inseparable} (entangled) state, i.e.

\begin{equation}
|\Phi^+\rangle=\mathbb{\hat G}\:|00\rangle\quad\rightarrow\quad|00\rangle=\mathbb{\hat G}^{-1}|\Phi^+\rangle
\end{equation}
where $\mathbb{\hat G}$ is a \textit{combination} of quantum gates, and $\mathbb{\hat G}^{-1}$ its \textit{inverse}.\\

We saw previously a gate that \textit{links} two qubits, through a \textit{condition}, it is the \textsc{cnot gate} (noted $\mathbb{CNOT\equiv CX}$), so it is a good candidate for the \textit{linking \emph{and the} separation}. \\

For the $1^{st}$ example, we will try the \textit{inverse} way; going from the \textit{inseparable} state $|\Phi^+\rangle$ to the \textit{initial separate} state $|00\rangle$, thus we apply the inverse of $\mathbb{CX}$. But we know that it is its own \textit{inverse} $\mathbb{CX}\!=\!(\mathbb{CX})^{-1}$. 
\begin{equation}
(\mathbb{CX})^{-1}\,\ket{\Phi^+}\equiv\mathbb{CX}\,|\Phi^+\rangle=\begin{pmatrix}
1 & 0 & 0 & 0 \\ 0 &1 &0 &0 \\ 0 &0 &0 &1 \\0 &0 &1 &0 \end{pmatrix}\begin{pmatrix} \tfrac{1}{\sqrt{2}}\\0\\0\\\tfrac{1}{\sqrt{2}}
\end{pmatrix}=\begin{pmatrix} \tfrac{1}{\sqrt{2}}\\0\\\tfrac{1}{\sqrt{2}}\\0\end{pmatrix}=\tfrac{1}{\sqrt{2}}\,(\,|00\rangle+|10\rangle\,),
\end{equation}
which is clrearly separable
\begin{equation}
\mathbb{CX}\,|\Phi^+\rangle=\tfrac{1}{\sqrt{2}}\,(\,|0\rangle+|1\rangle\,)\otimes|0\rangle=|+\rangle\otimes|0\rangle.
\end{equation}
We can rewrite it in terms of the two \textit{initial separate} states:
\begin{align*}
|\Phi^+\rangle &=\mathbb{CX}\,(\,|+\rangle\otimes|0\rangle\,)\\
&=\mathbb{CX}\,(\,\mathbb{H}\,|0\rangle\otimes\mathbb{I}\,|0\rangle\,)\\&=\mathbb{CX}\:(\,\mathbb{H}\otimes\mathbb{I}\:)\,(\,|0\rangle\otimes|0\rangle),
\end{align*}
where $\mathbb{H}\,|0\rangle\!=\!|+\rangle$.

So by acting the \textsc{Hadamard gate} $\mathbb{H}$, only on the $\mathfrak{q}_0$ state $|0\rangle_{_0}$, it becomes $|+\rangle_{_0}$. Then by acting $\mathbb{CX}$ on the new state $\ket{+}\otimes\ket{0}$, we obtain the \textit{entangled} state $|\Phi^+\rangle$.\\
We can now apply this reasoning into a quantum circuit:
\vspace{0.5cm}
\begin{figure}[H]
\centering
\includegraphics[width = 0.43\linewidth]{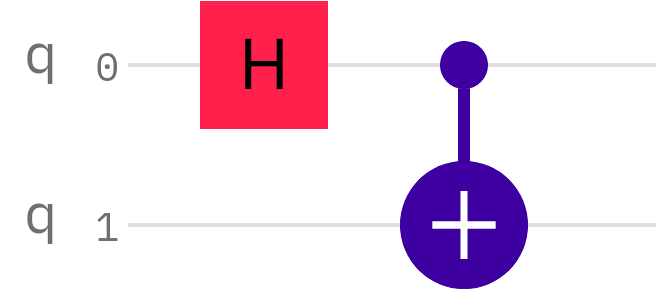}
\caption{The quantum circuit that generates the Bell state $|\Phi^+\rangle$.}
\label{1.22}

\end{figure}
For the \textit{interpretation}, we need some \textit{philosophy}. We have a qubit $\mathfrak{q}_0$ in the state $\ket{0}_{_0}$ subjected to the action of the \textsc{Hadamard gate}, which will make it in a \textit{superposition balance}. The second secret of the entanglement process lies in the \textsc{cnot gate}, which leaves $\mathfrak{q}_1$ in $\ket{0}_{_1}$ if $\mathfrak{q}_0$ is in $\ket{0}_{_0}$ and turns $\mathfrak{q}_1$ to $\ket{1}_{_1}$ if $\mathfrak{q}_0$ is $\ket{1}_{_0}$. But we know that $\mathfrak{q}_0$ is \textbf{both} in $\ket{0}_{_0}$ and $\ket{1}_{_0}$; therefore, the \textsc{cnot gate} \textbf{simultaneously} turn and don't turn the $\mathfrak{q}_1$ state!, which makes $\mathfrak{q}_0$ also in a \textit{superposition balance}; that is to say, the \textsc{cnot gate} transmits the \textit{power of superposition} from $\mathfrak{q}_0$ to $\mathfrak{q}_1$. Since $\mathfrak{q}_1$ \textit{depends directly} on $\mathfrak{q}_0$, they are therefore \textbf{\textit{linked}}, which makes them \textbf{entangled}.\\
\pagebreak

The steps followed to obtain $|\Phi^+\rangle$, begins by acting $\mathbb{H}$ on $|0\rangle_{_0}$. But what if, instead, we act $\mathbb{H}$ on $|1\rangle_{_0}$, what will be the resulting state?

For the $2^{nd}$ example, we will take the \textit{direct} way. Firstly, we can write $\mathbb{H}\,|1\rangle=\mathbb{H\,(X}\,|0\rangle)\,$. Since the operation must act just on $\mathfrak{q}_0$, therefore, our operator takes the form ($\mathbb{H\,X\otimes I}$).
\begin{equation}
|00\rangle\longrightarrow(\mathbb{H\,X\otimes I})\:|00\rangle=\mathbb{H\,X}\,|0\rangle \otimes \mathbb{I}\,|0\rangle=\mathbb{H}\,|1\rangle \otimes|0\rangle=|-\rangle\otimes|0\rangle
\end{equation}
\begin{equation}
\mathbb{CX}\,|\!-0\,\rangle=\begin{pmatrix}
1 & 0 & 0 & 0 \\ 0 &1 &0 &0 \\ 0 &0 &0 &1 \\0 &0 &1 &0 \end{pmatrix}\begin{pmatrix} \tfrac{1}{\sqrt{2}}\\0\\\tfrac{-1}{\sqrt{2}}\\0
\end{pmatrix}=\begin{pmatrix} \tfrac{1}{\sqrt{2}}\\0\\0\\\tfrac{-1}{\sqrt{2}}\end{pmatrix}
\end{equation}
which gives $|\Phi^-\rangle$
\begin{equation}
\mathbb{CX}\,|\!-0\,\rangle=\tfrac{1}{\sqrt{2}}\,(\,|00\rangle-|11\rangle\,)=|\Phi^-\rangle
\end{equation}
Therefore, the gates that take us from $|00\rangle$ to $|\Phi^-\rangle$ are:
\begin{equation}
|\Phi^-\rangle=\mathbb{CX}\:(\,\mathbb{H\,X}\otimes\mathbb{I}\:)\,\,|00\rangle
\end{equation}
If we draw the corresponding quantum circuit of this state, then the only difference compared to that of $|\Phi^+\rangle$, is the action of a \textsc{not gate} before the \textsc{Hadamard gate}. This circuit is drawn on the left side of the figure below:
\begin{figure}[H]
\centering
\includegraphics[width = 0.43\linewidth]{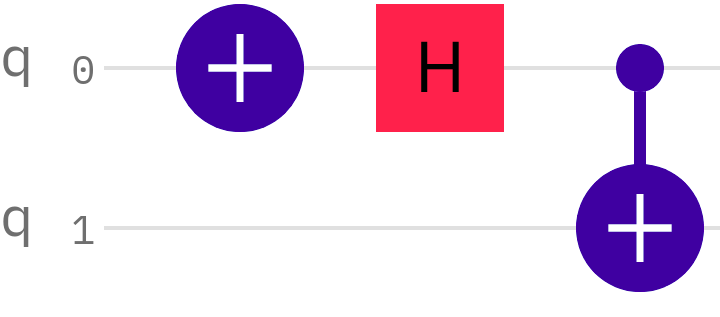}
\hspace{0.4cm}
\includegraphics[width = 0.43\linewidth]{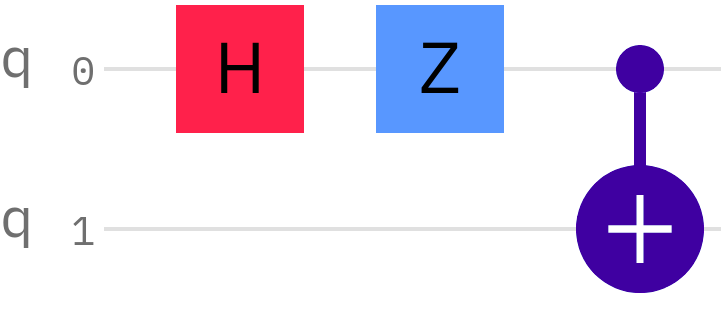}\hspace{0.2cm}
\caption{Two equivalent quantum circuits that generate the Bell state $|\Phi^-\rangle$.}
\label{1.23}
\end{figure}
We can mathematically prove that the state $|\Phi^-\rangle$, can be produced by the \textit{equivalent} circuit, shown on the right above. Firstly, we recall that $\mathbb{Z}\,|\pm\rangle=|\mp\rangle$; hence, we can write $|-\rangle=\mathbb{Z}\,|+\rangle=\mathbb{Z\,(H}\,|0\rangle)$ and we know that $|-\rangle=\mathbb{H\,(X}\,|0\rangle)$, therefore
\[\mathbb{Z\,H}=\mathbb{H\,X}\]
Then we can obtain $|\Phi^-\rangle$, by applying the transformation below on $|00\rangle$:
\begin{equation}
|\Phi^-\rangle=\mathbb{CX}\:(\,\mathbb{Z\,H\otimes I}\:)\:|00\rangle
\end{equation}
This clearly explains the circuit on the right side.

We can plot the two states $\ket{\Phi^{\pm}}$ in the Q-sphere:
\begin{figure}[H]
\centering
\includegraphics[width = 0.38\linewidth]{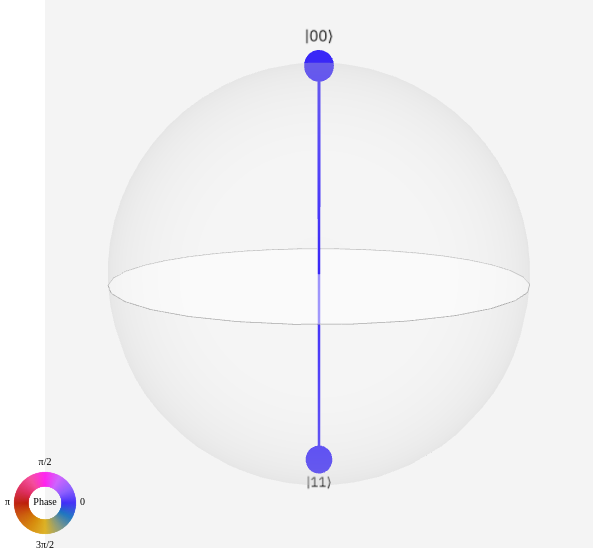}\hspace{0.4cm}
\includegraphics[width = 0.38\linewidth]{Figures/Chapter 1/Qphi-.png}
\caption{A visualization of the Bell states: $|\Phi^+\rangle$ (on the left) and $|\Phi^-\rangle$ (on the right).}
\label{1.24}
\end{figure}
In the Q-sphere, we represent the basis state with all \textbf{zeros} $|00\rangle$ at the \textit{north pole} and that with all \textbf{ones} $|11\rangle$ at the \textit{south pole}. On the left side, we see a representation of $|\Phi^+\rangle$, where the \textbf{\textit{phase}} of $|0\rangle$ and $|1\rangle$ are both $=0$. The right side concerns that of $|\Phi^-\rangle$, where the \textbf{\textit{phase}} of $|0\rangle$ is $0$ and that for $|1\rangle$ is $\pi$, which is reflected as a \textit{red color} on the $|11\rangle$ node.\\

The $3^{rd}$ and $4^{th}$ examples relate to the states $\ket{\Psi^{\pm}}$. Recalling that we have found $\ket{\Phi^{\pm}}$ by acting $\mathbb{CNOT}$ on the states $|\pm 0\,\rangle$, i.e.
 \[|\Phi^{\pm}\rangle=\mathbb{CX}\,|\pm0\,\rangle\]
What if we apply it on $|\pm 1\,\rangle\,$? 
\begin{equation}
\mathbb{CX}\,|\pm1\,\rangle=\begin{pmatrix}
1 & 0 & 0 & 0 \\ 0 &1 &0 &0 \\ 0 &0 &0 &1 \\0 &0 &1 &0 \end{pmatrix}\begin{pmatrix} 0\\\tfrac{1}{\sqrt{2}}\\0\\\tfrac{\pm1}{\sqrt{2}}
\end{pmatrix}=\begin{pmatrix} 0\\\tfrac{1}{\sqrt{2}}\\\tfrac{\pm1}{\sqrt{2}}\\0\end{pmatrix}=\tfrac{1}{\sqrt{2}}\,(\,|01\rangle\pm|10\rangle\,)
\end{equation}
which is clearly the states $\:|\Psi^{\pm}\rangle=\tfrac{1}{\sqrt{2}}\,(\,|01\rangle\pm|10\rangle\,)$.\\
We want to write it in terms of the initial states $\ket{00}$. We have two distinct states:
\begin{equation}
|\Psi^+\rangle=\mathbb{CX}\:|+1\,\rangle\quad;\quad|\Psi^-\rangle=\mathbb{CX}\:|-1\,\rangle
\end{equation}
\begin{eqnarray}
|\Psi^+\rangle =\mathbb{CX}\,(\,|+\rangle\otimes|1\rangle\,)
=\mathbb{CX}\,(\,\mathbb{H}\,|0\rangle\otimes\mathbb{X}\,|0\rangle\,)=\mathbb{CX}\:(\,\mathbb{H}\otimes\mathbb{X}\:)\;|00\rangle\;\:\:\\
|\Psi^-\rangle =\mathbb{CX}\,(\,|-\rangle\otimes|1\rangle\,)
=\mathbb{CX}\,(\,\mathbb{H}\,|1\rangle\otimes\mathbb{X}\,|0\rangle\,)=\mathbb{CX}\:(\,\mathbb{HX}\otimes\mathbb{X}\:)\;|00\rangle
\end{eqnarray}

For the first case, we apply a \textsc{Hadamard gate} $\mathbb{H}$ on $\mathfrak{q}_0$, and a \textsc{not gate} $\mathbb{X}$ on $\mathfrak{q}_1$. Then by acting $\mathbb{CX}$ on the resulting state $|+1\,\rangle$, we finally get the \textit{entangled} state $|\Psi^+\rangle$. The corresponding circuit is drawn below:
\vspace{0.5cm}
\begin{figure}[H]
\centering
\includegraphics[width = 0.42\linewidth]{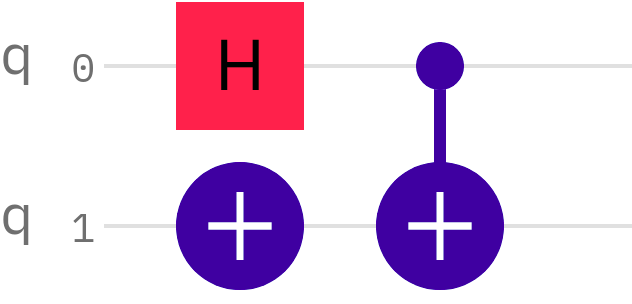}
\caption{The quantum circuit that generates the Bell state $|\Psi^+\rangle$.}
\label{1.25}

\end{figure}
In the second case, it is the same process, except that we apply a \textsc{not gate} before $\mathbb{H}$. Or like the $2^{nd}$ example, where we apply a $\mathbb{Z}$ \textsc{gate} after $\mathbb{H}$, i.e.
\[|\Psi^-\rangle=\mathbb{CX}\:(\,\mathbb{H\,X\otimes X}\:)\;|00\rangle=\mathbb{CX}\:(\,\mathbb{Z\,H\otimes X}\:)\;|00\rangle\]
The corresponding quantum circuits are drawn in the figure below:
\begin{figure}[H]
\centering
\includegraphics[width = 0.43\linewidth]{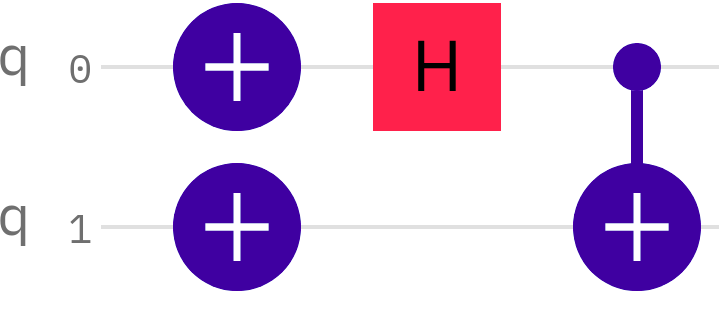}
\hspace{0.4cm}
\includegraphics[width = 0.43\linewidth]{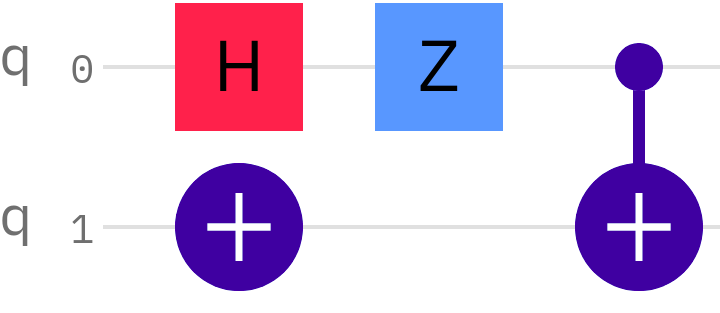}\hspace{0.2cm}
\caption{Two equivalent quantum circuits that generate the Bell state $|\Psi^-\rangle$.}
\label{1.26}
\end{figure}
We can represent the two states $\ket{\Psi^{\pm}}$ on the Q-sphere, in order to compare them with those of the states $\ket{\Phi^{\pm}}$.
\begin{figure}[H]
\centering
\includegraphics[width = 0.4\linewidth]{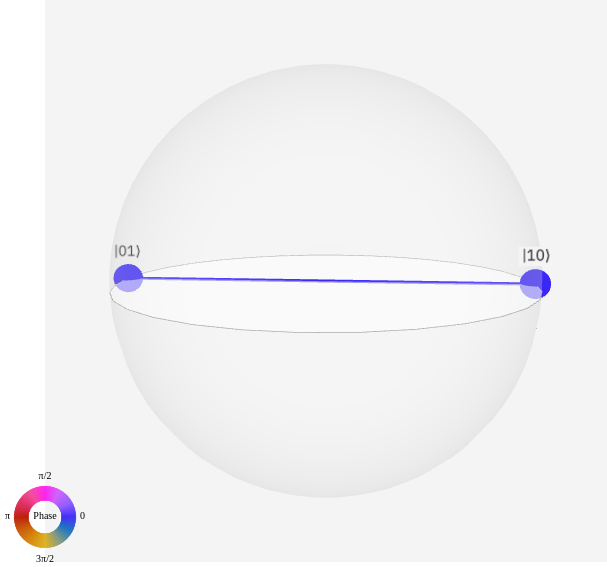}\hspace{0.4cm}
\includegraphics[width = 0.4\linewidth]{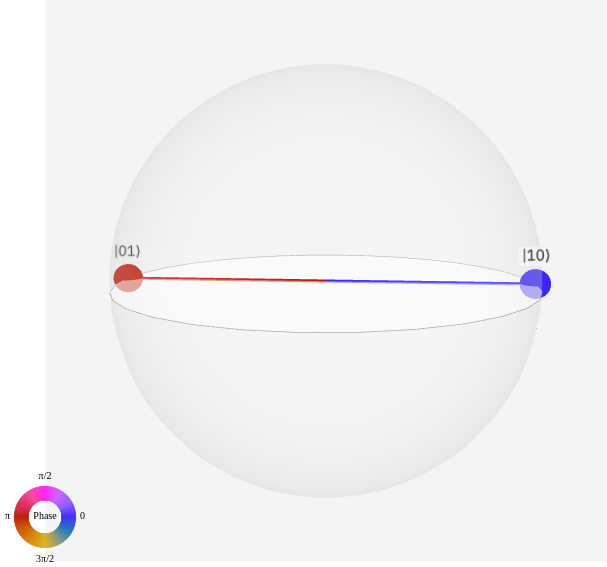}
\caption{A visualization of the Bell states: $|\Psi^+\rangle$ (on the left) and $|\Psi^-\rangle$ (on the right).}
\label{1.27}
\end{figure}
If the basis state with all \textbf{zeros} $|00\rangle$ is represented at the \textit{north pole} and that with all \textbf{ones} $|11\rangle$ at the \textit{south pole}, then the basis states with an \textbf{equal number} of $0$s and $1$s (i.e., $|01\rangle,|10\rangle$) are represented at the \textbf{equator}, which explains our situation here. Another thing is the \textbf{\textit{quantum phase}}, where in the left side, it is $0$ for both $|0\rangle, |1\rangle$ and in the right side, it's $0$ for $|0\rangle$ and $\pi$ for $|1\rangle$, which explains the difference between the two colors.\\

Finally, we built the four quantum circuits that generate the four \textit{Bell states}. As seen the two \textbf{necessary} gates to create \textbf{entanglement}, are the \textsc{Hadamard Gate} that played the role of the “\textsc{source}” of the \textbf{superposition} property, and the \textsc{Cnot Gate} that played the role of a “\textsc{key}” that \textit{close/entangle} or \textit{open/disentangle} two qubit states.\\

We have seen in this section some \textit{remarkable} quantum gates. But we are far from seeing the full picture; because there are many more gates than one can imagine, as the $\mathbb{CH,\,CZ,\,CT,\,CS,\,CSWAP}, \ldots$, which we can find in the \textsc{IBM quantum composer}.\\
\begin{figure}[H]
\centering
\includegraphics[width = 0.7\linewidth]{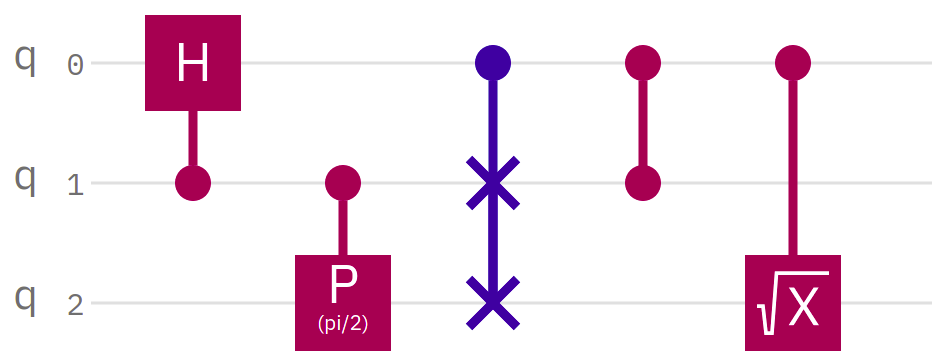}
\caption{A variety of controlled gates \small{(from the left to the right)}: $\mathbb{RCH,\,CS,\,CSWAP,\,CZ,\,C\sqrt{X}}$.}
\label{1.28}
\end{figure}
Unlike gates like $\mathbb{C\sqrt{X}^\dagger,\,CCH, \,CXHT,\ldots}$, and also gates of 4 and 5 qubits. But as shown in the \textsc{swap} gate example, we can create them through the existing basic gates. The figure\footnote{\url{https://qiskit.org/textbook/ch-gates/more-circuit-identities.html}}  below clearly explains this concept:
\begin{figure}[H]
\centering
\includegraphics[width =1\linewidth]{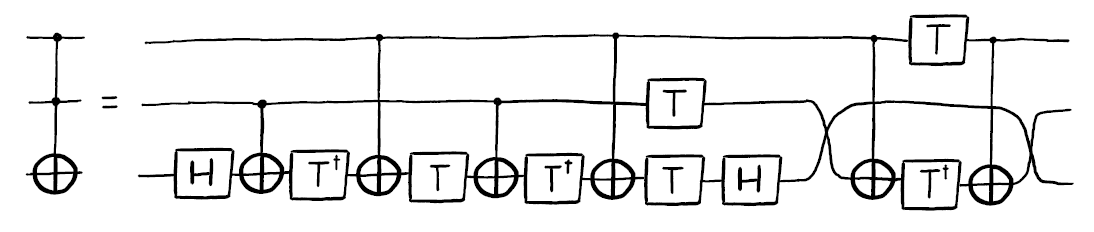}
\caption{ The Toffoli gate (on the left) and its implementation (on the right).}

\end{figure}
This figure shows the real face behind the \textsc{ccnot} or as called the \textsc{toffoli} gate. This \textit{masterpiece} was invented by \textsc{Tommaso Toffoli} in 1980\footnote{\url{https://medium.com/qiskit/ibm-quantum-challenge-2021-heres-what-to-expect-65a303753ffb}}. Therefore any specific multi-qubit gate is valuable. The challenge lies in how to combine the basic gates in order to create a desired gate?
\pagebreak

\section{Quantum Algorithms}
A classical algorithm is a procedure or a finite sequence of instructions used in problem-solving. All of these instructions can be executed on classical computers. A quantum algorithm is also a step-by-step instruction but, it is performed in quantum computers. Quantum algorithms can solve some computational issues faster than their classical counterparts. For example, \textit{integers factorization} is \textit{exponentially} faster, and an \textit{unordered search} is \textit{quadratically} faster than any classical algorithm.
\subsection{Deutsch-Jozsa Algorithm}
The Deutsch-Jozsa algorithm was the first quantum algorithm that showed that quantum computing works better than the best classical algorithm and demonstrates exponential acceleration compared to a classical deterministic algorithm \cite{Moran}. 

\vspace{2mm}
$\bullet\,$ \textbf{Deutsch-Jozsa Problem}\\
A hidden boolean function $f(x_{1},...,x_{n})$ takes as input a string of bits and returns $0$ or $1$, where  $x_{n}$ are 0 or 1. If the function returns all $0$ or all $1$ for any input, then it is a \textit{constant function}. If the function returns $0$ for exactly half of all inputs and $1$ for the other half, then it is a \textit{balanced function}. Our goal here is to determine if a given function is \textit{balanced} or \textit{constant}.
 

\vspace{2mm}
$\bullet\,$ \textbf{The Classical Solution:} 
If the task is performed with a classical algorithm on a classical computer, we can solve this problem after two calls to the function $f$. For example, if we find $0$ as output for the first combination and $1$ in the second combination, then the function $f$, is balanced. This is the best case. 
In the worst case, if we continue to see the same output for each input we try, we will have to check exactly half of all possible inputs plus one in order to be certain that $f$ is constant.

\vspace{2mm}
$\bullet\,$ \textbf{The Quantum Solution:}
We can solve this problem using quantum computing after only one call to the function $f$. This function is implemented as a quantum oracle, it maps the state from $\vert x\rangle \vert y\rangle$ to $\vert x\rangle \vert y \oplus f(x)\rangle\,$\footnote{\url{https://qiskit.org/textbook/ch-algorithms/deutsch-jozsa.html}}.
\begin{figure}[H]
\centering
\includegraphics[scale=0.06]{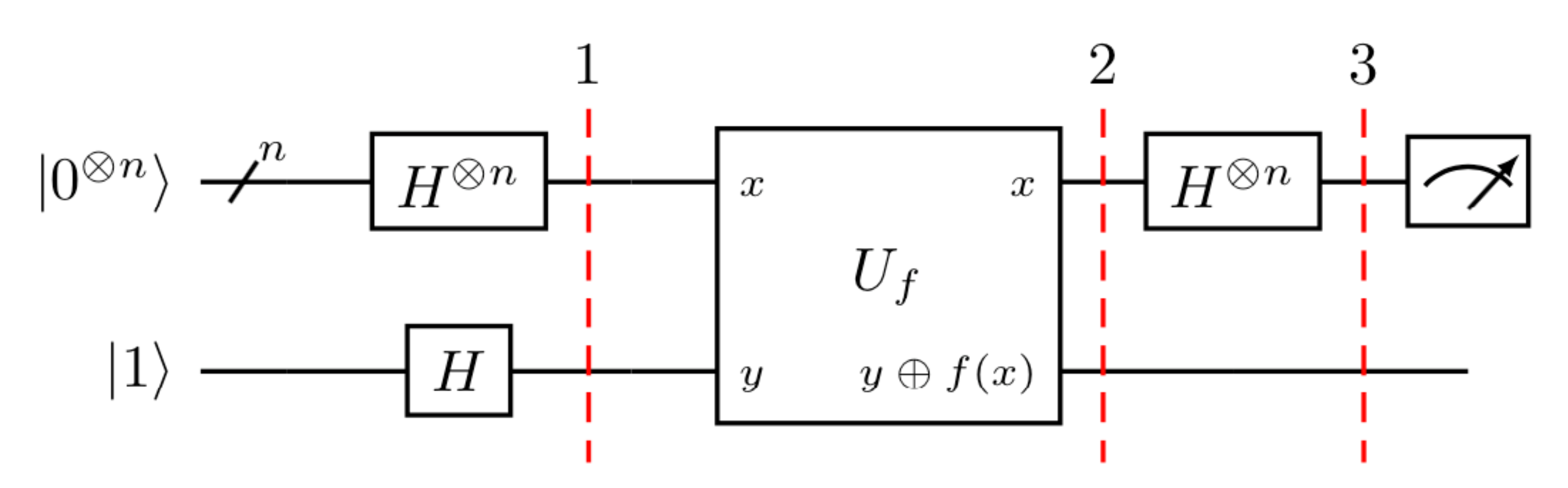}
\caption[Generic circuit for the Deutsch-Jozsa algorithm]{ \small{Generic circuit for the Deutsch-Jozsa algorithm\footnotemark.}}
\label{1.30}
\end{figure}
\footnotetext{\url{https://qiskit.org/textbook/ch-algorithms/deutsch-jozsa.html}}

Now, let's go through the steps of the algorithm:

\begin{itemize}

\item Prepare two quantum registers. The first is an $n$-qubit register initialized to $\ket{0}$, and the second is a one-qubit register initialized to $\ket{1}$
\begin{equation}
\vert \psi_0 \rangle = \vert0\rangle^{\otimes n} \vert 1\rangle
\end{equation}
\item Apply a Hadamard gate to each qubit:
\begin{align}
\ket{\psi_{1}} & =(\mathbb{H}^{\otimes n} \ket{0}^{\otimes n})(\mathbb{H}\ket{1})\\
& =  \left( \mathbb{H}\ket{0}\otimes \mathbb{H}\ket{0}\otimes \cdots \otimes \mathbb{H}\ket{0}\right) \left( \mathbb{H}\ket{1}\right)\nonumber \\
& = \left( \frac{\ket{0}+\ket{1}}{\sqrt{2}}\otimes  \frac{\ket{0}+\ket{1}}{\sqrt{2}}\otimes \cdots \otimes \frac{\ket{0}+\ket{1}}{\sqrt{2}}\right)  \left( \frac{\ket{0}-\ket{1}}{\sqrt{2}}\right) \nonumber \\
& = \left( \frac{1}{\sqrt{2^{n}}} \sum_{x=0}^{2^n-1} \vert x\rangle\right)  \left( \frac{\ket{0}-\ket{1}}{\sqrt{2}}\right) \nonumber \\
\ket{\psi_{1}} & = \frac{1}{\sqrt{2^{n+1}}} \sum_{x=0}^{2^n-1} \vert x\rangle \left(|0\rangle - |1 \rangle \right)
\end{align}

\item Apply the \textit{quantum oracle} $\vert x\rangle \vert y\rangle$ to $\vert x\rangle \vert y \oplus f(x)\rangle$:
\begin{align}
\ket{\psi_2} & = \frac{1}{\sqrt{2^{n+1}}} \sum_{x=0}^{2^n-1} \ket{x} \left( \ket{0\oplus f(x)} - \ket{1\oplus f(x)}\right) \nonumber \\
& = \frac{1}{\sqrt{2^{n+1}}}\sum_{x=0}^{2^n-1}(-1)^{f(x)}|x\rangle \left(  |0\rangle - |1\rangle \right)  
\end{align}

Since for each $x$, $f(x)$ is either 0 or 1. At this point the second single qubit register may be ignored.
 
\item Apply a Hadamard gate to each qubit in the first register:
\begin{align}
            \lvert \psi_3 \rangle 
                & = \frac{1}{2^n}\sum_{x=0}^{2^n-1}(-1)^{f(x)}
                    \left[ \sum_{y=0}^{2^n-1}(-1)^{x \cdot y} 
                    \vert y \rangle \right] \nonumber \\
                & = \frac{1}{2^n}\sum_{y=0}^{2^n-1}
                    \left[ \sum_{x=0}^{2^n-1}(-1)^{f(x)}(-1)^{x \cdot y} \right]
                    \vert y \rangle
        \end{align}
with $ \mathbb{H}\ket{x} = \sum_{y=0}^{2^n-1}  \frac{\left( -1\right)^{x.y}\ket{y}}{\sqrt{2}}$ and $ \mathbb{H}^{\otimes n}\ket{x} = \sum_{y=0}^{2^n-1}  \frac{\left( -1\right)^{xy}\ket{y}}{\sqrt{2^n}}$

Where $x \cdot y = x_0\,y_0 \oplus x_1\,y_1 \oplus \ldots \oplus x_{n-1}\,y_{n-1}$ is the sum of the \textit{bitwise product}.

\item Measure the first register. Notice that the probability of measuring $\vert 0 \rangle ^{\otimes n} = \lvert \frac{1}{2^n}\sum_{x=0}^{2^n-1}(-1)^{f(x)} \rvert^2$, which evaluates to 1 if $f(x)$ is constant and 0 if $f(x)$ is balanced.\\
\end{itemize}
\pagebreak

\underline{Example for $n=1$}

To better understand the Deutsch-Jozsa algorithm, let us take a simple example for a single input $(n=1)$. The quantum circuit that we need to use to solve the
problem is very simple:
\begin{figure}[H]
\centering
\includegraphics[width = 0.96\linewidth]{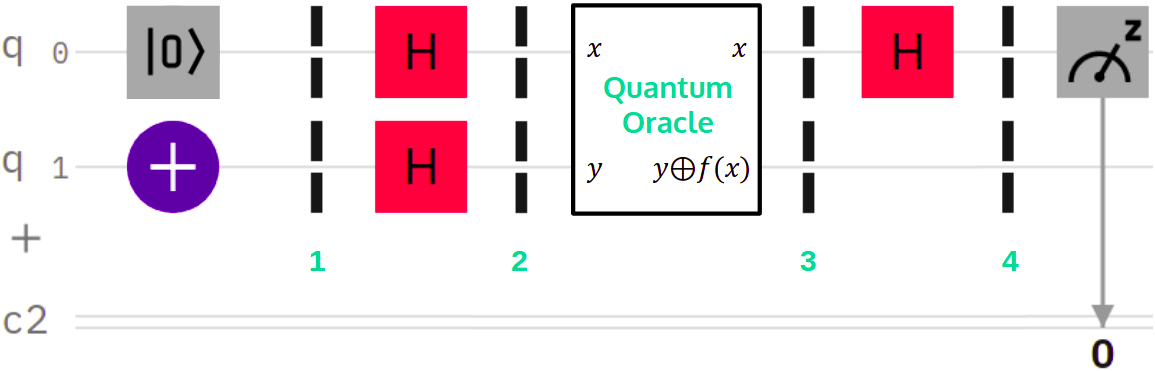}
\caption{Deutsch-Jozsa circuit example for a single input (n=1).}
\end{figure}
The algorithm is divided by four steps. Let’s look at what happens in each numbered state.\\
\textbf{Step 1:}

In this step, we prepare two quantum registers. The first is a one-qubit register initialized to $\ket{0}$, and the second is one-qubit register initialized to $\ket{1}$. The initial state is:
\begin{equation}
\ket{\psi_0}= \ket{0}\ket{1}
\end{equation}
\textbf{Step 2:}

The next step is to apply the Hadamard gates, we get the following result:
\begin{align}
\ket{\psi_1}&=\frac{(\ket{0}+\ket{1})}{\sqrt{2}} \frac{(\ket{0}-\ket{1})}{\sqrt{2}} \nonumber\\
&= \frac{(\ket{0}+\ket{1})(\ket{0}-\ket{1})}{{2}} \nonumber \\
&= \frac{\ket{0}(\ket{0}-\ket{1})}{2}+\frac{\ket{1}(\ket{0}-\ket{1})}{2}
\end{align}
\textbf{Step 3:}

The next step is to apply the oracle function, the input $(x,y)$ will become to $(x,y\bigoplus f(x))$ in the output. The state before the oracle function is 
\begin{equation}
\ket{\psi_1}=\frac{\ket{0}(\ket{0}-\ket{1})}{2}+\frac{\ket{1}(\ket{0}-\ket{1})}{2}
\end{equation}
Now let us see the effect of the oracle function. When we apply the oracle function, by linearity we obtain:
\begin{equation}
\frac{\ket{0}(\ket{0 \oplus f(0)}-\ket{1 \oplus f(0)})}{2}+ \frac{\ket{1}(\ket{0 \oplus f(1)}-\ket{1 \oplus f(1)})}{2}
\end{equation}
\begin{itemize}
\item When $f(0)=0$, we get $\ket{0 \oplus f(0)}-\ket{1 \oplus f(0)}= \ket{0}-\ket{1}$.
\item When $f(0)=1$, we have $\ket{0 \oplus f(0)}-\ket{1 \oplus f(0)}=\ket{0 \oplus 1}-\ket{1 \oplus 1}=\ket{1}-\ket{0}=-(\ket{0}-\ket{1})$.
\item For the $f(1)$ we do the same thing, so our global state can be writen as
\begin{equation}
\ket{\psi_3}=\frac{(-1)^{f(0)}\ket{0}(\ket{0}-\ket{1})}{2}+\frac{(-1)^{f(1)}\ket{1}(\ket{0}-\ket{1})}{2}.
\end{equation}
\begin{flushleft}
This expression can be also written as
\end{flushleft}
\begin{equation}
\frac{\ket{0}(\ket{0}-\ket{1})}{2}+\frac{(-1)^{f(0)+f(1)}\ket{1}(\ket{0}-\ket{1})}{2}
\end{equation}
\item If $f(0)=f(1)$, we will get
\begin{equation}
\frac{\ket{0}(\ket{0}-\ket{1})}{2}+\frac{\ket{1}(\ket{0}-\ket{1})}{2}= \frac{(\ket{0}+\ket{1})(\ket{0}-\ket{1})}{2}
\end{equation}
\end{itemize}
Lets summarize what happen when $f(x)$ is constant or balanced. 

\textbf{When constant:}

$f(0)=0$ and $f(1)=0$ $\rightarrow \frac{\ket{0}+\ket{1}}{\sqrt{2}}$ also $f(0)=1$ and $f(1)=1$  $\rightarrow$ $\frac{-\ket{0}-\ket{1}}{\sqrt{2}}$

\textbf{When balanced:}

$f(0)=0$ and $f(1)=1$ $\rightarrow$ $\frac{\ket{0}-\ket{1}}{\sqrt{2}}$ also $f(0)=1$ and $f(1)=0$ $\rightarrow$ $\frac{-\ket{0}+\ket{1}}{\sqrt{2}}$

\textbf{Step 4:}
The final step is to apply the Hadamard gate and the measure. 

When we apply the Hadamard gate, we get the state $\ket{1}$ if it is balanced and $\ket{0}$ if it is constant. We can verifier this result by the measurement.
\subsection{Shor’s Algorithm}
Shor’s Factorization Algorithm is proposed by Peter Shor. It suggests that quantum mechanics allows the factorization to be performed in \textit{polynomial time}, rather than \textit{exponential time} achieved after using classical algorithms \cite{Moran}. 
\begin{itemize}
\item \underline{The basic idea of Shor’s algorithm:}

The main problem to be solved by Shor’s algorithm is given an integer $N$, we want to find its \textit{prime factor}. In other words, we want to determine the two prime factors $p_1$ and $p_2$ which satisfy $p_1.p_2=N$ for a given large number of $N$ in polynomial time. 




Shor algorithm is a \textit{hybrid quantum-classical algorithm}. It begins with some classical processing, then uses the quantum computer to execute a particular part of the algorithm (a subroutine\footnote{A sequence of computer instructions for performing a specified task that can be used repeatedly.
}). This subroutine is difficult to perform on classical computers, but it can be solved with a fast quantum algorithm using the \textit{quantum Fourier transform} (QFT). The quantum computer essentially serves as a \textit{co-processor} to the classical computer.\\ 

In the case of Shor’s algorithm, the quantum subroutine is known as the period finding routine, which uses an important technique known as the quantum Fourier transform, or QFT, to create \textit{interference} that gives us the period. We then use a classical technique known as \textit{Euclid’s algorithm} to find the prime factors of the number. It turns out that, if we can find the period $r$ of a function $ f(x)=a^x mod\:N$ as we increment the variable $x$, then we can find the factors of $N$ with $a$ is selected prime number inferior of $N$.

\item \underline{Steps for Shor algorithm}

The algorithm consists of two parts, a classical and quantum part. The classical part reduces the factorization to a problem of finding the period of the function. This is done classically using a classical computer. Quantum part which uses a quantum computer to find the period using the Quantum Fourier Transform.
\end{itemize}
Let’s assume we’re trying to factor a number $N$. The algorithm consists of the following steps:

\begin{enumerate}
 \item Pick a random number $a$ such that $a < N$.
 \item Use Euclid’s algorithm to check if  is a factor of $N$; if so, you’re done.
 \item Otherwise, prepare your quantum computer and a program for it to execute quantum period finding.
 \item The first important quantum step is to calculate the superposition of $a^x$ for all values of $x$, all done modulo $N$. $\ket{x}$ and the remainder both appear in our quantum register.
 \item Measure the remainder (Technically, this is unnecessary).
 \item Take the quantum Fourier transform (QFT) of $\ket{x}$.
 \item Measure the register $\ket{x}$; call the result $r$, which should be a multiple of the period of the modular exponentiation function.
 \item If $r$ is even and greater than zero, calculate the numbers  $a^{(r/2)}+1$ and $a^{(r/2)}-1$; if not, repeat the quantum portion of the algorithm to find a new $r$.
 \item Use Euclid’s algorithm to calculate the GCD (Great Common Divider) of each of those numbers and $N$, and we should have two factors of $N!$.
 \end{enumerate}
 
The quantum circuit for Shor’s algorithm is represented in the following figure \cite{shors}:
\begin{center}
\includegraphics[scale=0.9]{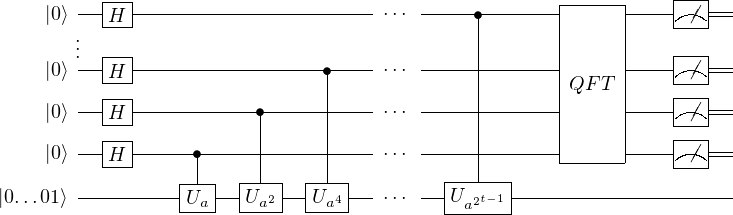}
\end{center}

\begin{itemize}

\item The first  $t$ qubits, which form the first register, will hold the exponents to which we are raising $a$. The measurement at the end will yield (with high probability) some value $y$  such that $\frac{y}{2^{t}}$ is close to a rational number $\frac{s}{r}\,$, where  is the period of $a \pmod N$. Thus increasing the number of these qubits, will simply increase the accuracy of this approximation.

\item The last $n$ qubits (initialized to be $\ket{0\cdots 01}$), which form the second register, will be where the powers $a^j \pmod N$  get stored. Hence we can take $n = \lceil \log _2 N \rceil$, the number of binary digits of $N$.
\item First, we apply $\mathbb{H}$ to each qubit in the input. The result will be that the first $t$ qubits will each become $\mathbb{H}\ket{0} = \frac{1}{\sqrt 2}(\ket 0 + \ket 1)$. The first $t$ qubits taken together will be in the state
\begin{align}
       &\frac{1}{2^{t/2}}(\ket0+\ket1)\otimes(\ket0+\ket1)\otimes\cdots\otimes(\ket0+\ket1)=\\
       &\frac{1}{2^{t/2}}(\ket{0\cdots00}+\ket{0\cdots01}+\cdots+\ket{1\cdots11})=\frac{1}{2^{t/2}}\sum_{j=0}^{2^t-1}\ket{j}.
       \end{align}
In other words, it will be in a \textit{balanced superposition} of all the basis states.

\item Then, we have a number of controlled $U_{x}$-gates. The gate $U_x$ performs multiplication by $x\pmod N$. That mean, $U_x \ket y = \ket {x\,y \bmod N}$, for $0 \leq y < N$. 

We can construct such gates efficiently since we can efficiently multiply two $n$ digit numbers, and we can compute each  $x=a^{2^k}$ simply by squaring $a \pmod N$. 

\item For example suppose $\ket j = \ket {j_{t-1}\cdots j_2\,j_1\,j_0}$ is the content of the first register. Then the controlled $U_{x}$-gates will multiply the second register by $a$ if $j_{0}=0$, then by $a^{2}$ if $j_{1}=1$, then by $a^4$ if $j_{2}=1$, and so forth. The result is that the second register will become $a^{\sum j_k2^k} = a^j.$

\item After these gates, the total state of the system will be a superposition of the form:

\begin{equation}
\frac {1}{2^{t/2}} \sum _{j=0}^{2^t-1} \ket j \ket {a^{j} \bmod N}.
\end{equation}
\end{itemize}
We can write it as:
\begin{equation}
\frac {1}{2^{t/2}} \sum _{j,b} x_{j}^{(b)} \ket j \ket b,
\end{equation}
where  $x^{(b)}_j = 1$ if $b = a^j \pmod N$, otherwise $x^{(b)}_j=0$. Thus, for any fixed $b$, the sequence $\{x_j^{(b)}\}_{j=0}^{2^t-1}$ will be periodic (in $j$) with period $r$.
\begin{itemize}
\item In this step, we input the first register into the QFT. This will send the state to 
\begin{equation}
\frac {1}{2^{t/2}} \sum _{k,b} y_k^{(b)} \ket k \ket b,
\end{equation}
where, for any fixed $b$, $\{y_k^{(b)}\}$ is the \textit{discrete Fourier transform} of $\{x_j^{(b)}\}$. Since  $\{x_j^{(b)}\}$ was periodic with period $r$,  $|y_k^{(b)}|$ will be large when  $y_k$ is approximately a multiple of $\frac{2^t}{r}$.

Let $N=2^{t}$ The quantum Fourier transform (QFT) is the linear transformation (acting on $t$  qubits) that acts on the computational basis  $\ket 0, \ket 1, \dots , \ket{N-1}$ by
\begin{equation}
\ket j \mapsto \frac {1}{\sqrt N} \sum _{k=0}^{N-1}e^{2\pi i j k/N} \ket k.
\end{equation}
This gives
\begin{equation}
\sum _{j=0}^{N-1} x_j \ket j \mapsto \sum _{k=0}^{N-1} y_k \ket k,
\end{equation}
where $\{y_k\}_{k=0}^{N-1}$ is the discrete Fourier transform of $\{x_j\}_{j=0}^{N-1}$.
\item Finally, we measure the first register. With high probability, this will give a value $k$ such that $k$ is approximately a multiple of $\frac{2^t}{r}\,$, that is, such that $\frac{k}{2^t}$ is close to a rational number with $r$ as the denominator.

Once we have this value of $k$, we compute the continued fraction expansion of $\frac{k}{2^t}$ to approximate it by a \textit{rational number} with a relatively small denominator $r$. Then we test to make sure that $a^r \equiv 1 \pmod N$.
\end{itemize}

\section{Error Detection  and Correction}
In 1948, \textsc{Claude Shannon} proved that a reliable transmission of information over \textit{noisy classical channels} is feasible. This discovery led to the search for \textit{classical error correction} (CEC) codes. Then, in the mid-1990s \textsc{Peter Shor} and \textsc{Andrew Steane} showed that \textit{quantum error correction} (QEC) is possible \cite{Marinescu}. So, in this section, we will give a glimpse of CEC and then we tackle QEC.
\subsection{Classical Error Correction}
Noise refers to any external and undesired disturbance superimposed on a useful signal, created by the operating environment and which leads to errors. In \textit{data communication}, the four most important kind of noise are: \textit{thermal noise, intermodulation noise, impulse noise \emph{and} crosstalk}. If there are defects in the hardware, either original or acquired, this can cause errors as well.\\

To correct these errors, we use \textsc{coding theory} which is critical for reliable \textit{communication}, \textit{information storage}, and \textit{processing}. Software within the storage devices can often \textbf{detect} and \textbf{correct} errors quickly using as small extra data as possible. This is called “\textit{fault-tolerance}” \cite{Dancing}.\\

\hspace{-0.3cm} The aspects of error correction process can be sited in the 4 points below:
\begin{itemize}
\item An initial information to be \textbf{sent}, \textbf{stored}, or \textbf{processed}.
\item \textbf{Encoding} the information in a larger system to protect it against noise.
\item A \textbf{noise process} that alters information.
\item \textbf{Decoding} the information, and correct it, which reduces the effects of noise.
\end{itemize}
\begin{figure}[H]
\centering
\includegraphics[width = 0.9\linewidth]{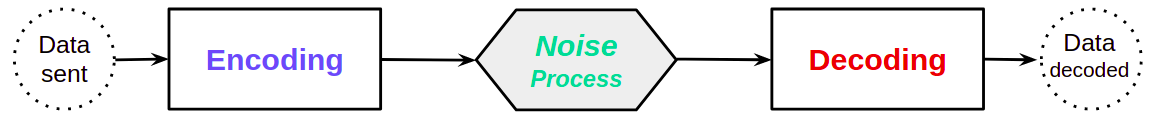}
\caption{The classical error correction process.}
\label{1.29}
\end{figure}
The trivial strategy for encoding information is to repeat it several times then the decoding is done by \textit{majority decision}. This is known as a “\textbf{\textit{repetition code}}”. For example, if we want to transmit the information 01101 through a \textit{noisy channel} that corrupt the transmission in some places, we have to repeat it in order to reduce the error rate \cite{Dancing}.
\[[\quad01101\quad01101\quad01101\quad]\rightarrow\langle\:noisy\;channel\:\rangle\rightarrow[\quad01101\quad01100\quad01101\quad]\]
An error occurred in one of The three copies, so we conclude that the original data is that occurs most often, which is $\mathbf{01101}$. We see clearly the weak point of this code; if there are one more errors than we cannot correct, but we can avoid this by more redundancy. Generally, for a repetition of length $n$, the \textit{error correction capacity} is $\frac{n-1}{2}$.\\

To detect an error there is another simple method is the use of an \textbf{even parity bit} \cite{Dancing}:
\begin{itemize}
\item If the data contains an \textbf{even} number of 1 bits, we precede it with a $\mathbf{0}$, e.g. $10100\;\rightarrow\;\mathbf{0}10100$.
\item If the data contains an \textbf{odd} number of 1 bits, we precede it with a $\mathbf{1}$, e.g. $10110\;\rightarrow\;\mathbf{1}10110$.
\end{itemize}
In the two we have an \textbf{even} number of 1s, so if there is an \textbf{odd} number of 1s after the transmission, then at least one \textit{bit flip} has occurred, anywhere. But if there is 2 or any even number of \textit{bit flips}, then we will not know.\\
There are more powerful codes that protect information. The characteristics of a good code are:
\begin{enumerate}
\item The number of repetitions in the original message must be minimal.
\item Ease of encoding and decoding information.
\end{enumerate}

The concepts of classical error detection (CED) and correction (CEC) introduced above are useless in the quantum case, because detecting errors in transmited information requires measurement. But in the quantum case, this causes a collapsing of the transmitted qubit state, so we lost any information stored in its coefficients \cite{Suter}. The philosophy of \textit{quantum error correction} is to detect and correct information without measuring it! 

\subsection{Quantum Error Correction}
Physical qubits are characterized by \textbf{decoherence}, which occurs when a state is affected by physical processes, and which cannot be reversed. For example, the effect of measures.
In the current era of \textit{quantum computing}, we are using a large number of noisy physical qubits to encode one \textbf{logical qubit}, through the process of quantum error correction (QEC) \cite{Dancing}.\\ 

The decoherence is due to the \textbf{entanglement} of the qubit with its environment, which attempts to alter the state of the system. But as we will see, the encoding of logical qubits is maintained by constantly putting the physical qubits through a \textit{highly entangling} circuit. \textsc{Peter Shor} summarizes the fundamental idea of QEC as to \textit{“fight entanglement with entanglement”} \cite{Marinescu}.\\

We have seen that to protect an information against noise requires creating multiple copies of it, in a way that decreases the probability of errors. But in quantum mechanics, the \textsc{No-Cloning Theorem} announces that a quantum state cannot be copied. \[\ket{\psi}=a\ket{0}+b\ket{1}\quad
\not\!\!\longrightarrow\quad\ket{\psi}\otimes\ket{\psi}\otimes\cdots\otimes\ket{\psi}\]
For this, a formulation of QEC was suspected that is not possible. Therefore, we need another way to employ the reasoning of redundancy of classical error correction in QEC, and it turns out that this way is \textit{entanglement}; indeed, instead of coping the state of a qubit multiple times, we can spread it onto a \textbf{highly entangled} state of multiple-qubits \cite{Dancing}\cite{Marinescu}.
The encoding in QEC combines the concept of $redundancy$ and $entanglement$, i.e.\\
\[\ket{\psi}=a\ket{0}+b\ket{1}\quad\xrightarrow{encoding}\quad\ket{\psi}=a\ket{000\cdots}+b\ket{111\cdots}\]
So we entangle one qubit carrying information, with ($n-1$) qubits of state $\ket{0}$, and create an $n$-qubit quantum state that is more resilient to errors.\\
The qubit state is prone to two types of quantum errors: \textbf{bit-flips} and \textbf{sign-flips} \cite{Aziz1}\cite{Aziz2}. 
\subsubsection{Single Bit-Flip Error Correction Code}
Bit flips are the only kind of errors that occurs in the classical case and makes the interchange of $0\rightleftarrows1$. This bit-flip can be traduced in the quantum case as the interchange of $\ket{0}\rightleftarrows\ket{1}$; which means that a state $a\ket{0}+b\ket{1}$ becomes $a\ket{1}+b\ket{0}$, where the $amplitudes$ are switched, which is a \textsc{not} action; the \textsc{X gate} leads to a \textit{bit-flip error}, and that can be corrected by applying another \textsc{X gate}, since $\mathbb{X.X=I}$ \cite{Dancing}.\\

In 1985 \textsc{Asher Peres} proposed a method called “\textit{3-qubit bit-flip code}” that corrects a single \textit{bit-flip} error. For a given transmitted state $\ket{\psi}$ this method consists of the following steps:

\begin{enumerate}
\item \textbf{Encoding} $\ket{\psi}$ to an entangled 3-qubit state.
\item The \textit{Noisy channel} cause just \textbf{bit-flip} errors.
\item \textbf{Decoding} the transmitted state.
\end{enumerate}
For the $1^{st}$ step, we have to $entangle$ the qubit $q_0\:\text{with both}\:q_1$ and $q_2$, Therefore
\[\ket{\psi\,0\,0\,}\rightarrow(\mathbb{CX}_{_{[0,2]}})(\mathbb{CX}_{_{[0,1]}})\,(\ket{\psi}_{_0}\otimes\ket{0}_{_1}\otimes\ket{0}_{_2})\]
where\[\mathbb{CX}_{_{[0,1]}}=(\mathbb{CX}\otimes\mathbb{I}\,)= (\:|0 \rangle\langle 0| \otimes \mathbb{I} \,+\, |1 \rangle\langle 1| \otimes \mathbb{X}\:)\otimes\mathbb{I} \]
\[\mathbb{CX}_{_{[0,2]}}=(\:|0 \rangle\langle 0| \otimes\mathbb{I}\otimes\mathbb{I} \,+\, |1 \rangle\langle 1| \otimes\mathbb{I}\otimes \mathbb{X}\:)\]
hence 
\[(\mathbb{CX}_{_{[0,2]}})(\mathbb{CX}\,\otimes\,\mathbb{I}\,)\,(\ket{\psi\,0\,}_{_{01}}\otimes\,\ket{0}_{_2})=(\mathbb{CX}_{_{[0,2]}})\,((a\ket{00}_{_{01}}+b\ket{11}_{_{01}})\otimes\,\ket{0}_{_2})=a\ket{000}+b\ket{111}\]
This correspond to the $1^{st}$ block in the circuit below:
\begin{figure}[H]
\centering
\includegraphics[width = 0.9\linewidth]{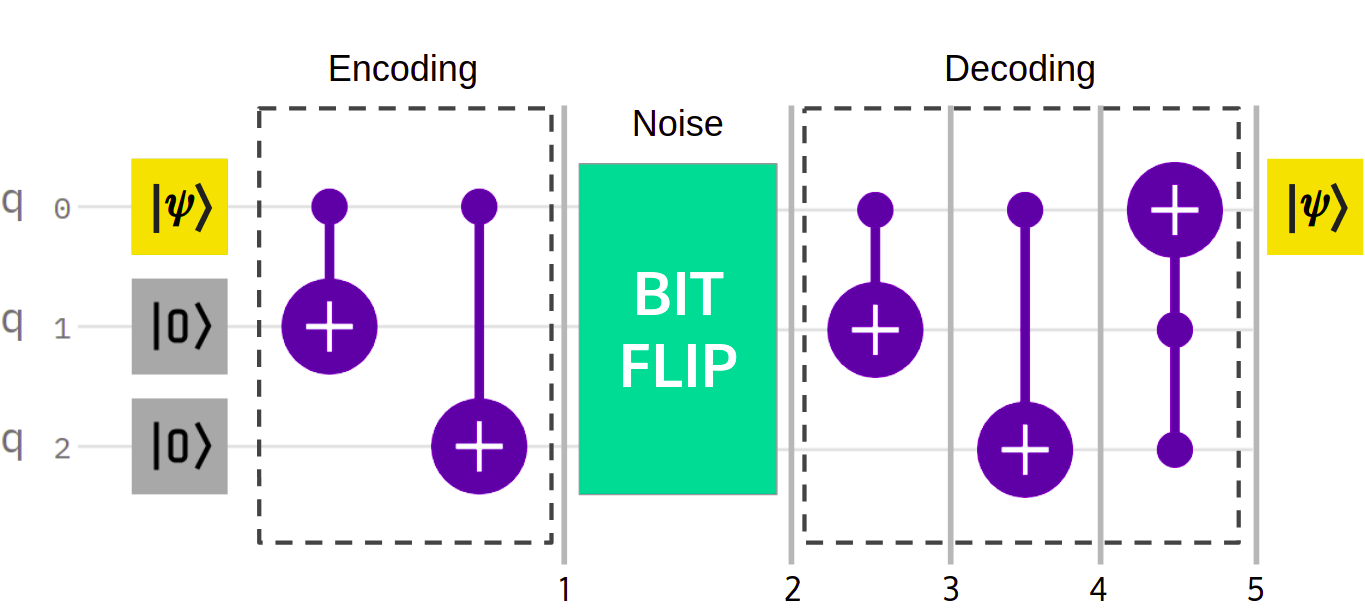}
\caption{Single bit flip error correction circuit.}
\label{1.30}
\end{figure}
After encoding one \textbf{logical qubit} as 3 \textbf{physical qubits}, this one will pass through a \textbf{noisy environment} that may cause \textit{bit-flips}. The decoding step consists of \textit{disentangling} the \textit{affected state} via the \textsc{cnot gates}. The role of \textsc{Toffoli gate} is \textbf{detection} and \textbf{correction} of errors; where the 2 $control$ parts work for $detection$ and the $target$ part is for $correction$.

\begin{center}
\begin{tabular}{||l|l|lllll||}\hline
$\ket{\psi}$&Nbr of Errors& Step 1 & Step 2 & Step 3 & Step 4 & Step 5 \\ \hline\hline
&&&&&&\\
&\textbf{0} Errors & $\ket{000}$ & $\ket{000}$ & $\ket{000}$ & $\ket{000}$ & $\ket{\mathbf{0}\,00}$ \\
&&&&&&\\
$\ket{0}$&\textbf{1} Error & $\ket{000}$ & $\ket{100}$ & $\ket{110}$ & $\ket{111}$ & $\ket{\mathbf{0}\,11}$  \\ 
&& $\ket{000}$ & $\ket{010}$ & $\ket{010}$ & $\ket{010}$ & $\ket{\mathbf{0}\,10}$\\
&& $\ket{000}$ & $\ket{001}$ & $\ket{001}$ & $\ket{001}$ & $\ket{\mathbf{0}\,01}$\\
&&&&&&\\
&\textbf{2} Errors
& $\ket{000}$ & $\ket{110}$ & $\ket{100}$ & $\ket{101}$ & $\ket{\mathbf{1}\,01}$\\
&& $\ket{000}$ & $\ket{101}$ & $\ket{111}$ & $\ket{110}$ & $\ket{\mathbf{1}\,10}$\\
&& $\ket{000}$ & $\ket{011}$ & $\ket{011}$ & $\ket{011}$ & $\ket{\mathbf{1}\,11}$\\
&&&&&&\\
&\textbf{3} Errors& $\ket{000}$ & $\ket{111}$ & $\ket{101}$ & $\ket{100}$ & $\ket{\mathbf{1}\,00}$\\
&&&&&&\\ \hline\hline
&&&&&&\\ 
&\textbf{0} Errors & $\ket{111}$ & $\ket{111}$ & $\ket{111}$ & $\ket{111}$ & $\ket{\mathbf{1}\,11}$ \\
&&&&&&\\
$\ket{1}$&\textbf{1} Error & $\ket{111}$ & $\ket{011}$ & $\ket{011}$ & $\ket{011}$ & $\ket{\mathbf{1}\,11}$  \\ 
&& $\ket{111}$ & $\ket{101}$ & $\ket{111}$ & $\ket{110}$ & $\ket{\mathbf{1}\,10}$\\
&& $\ket{111}$ & $\ket{110}$ & $\ket{100}$ & $\ket{101}$ & $\ket{\mathbf{1}\,01}$ \\
&&&&&&\\
&\textbf{2} Errors
&$\ket{111}$ & $\ket{001}$ & $\ket{001}$ & $\ket{001}$ & $\ket{\mathbf{0}\,01}$\\
&& $\ket{111}$ & $\ket{010}$ & $\ket{010}$ & $\ket{010}$ & $\ket{\mathbf{0}\,10}$\\
&& $\ket{111}$ & $\ket{100}$ & $\ket{110}$ & $\ket{111}$ & $\ket{\mathbf{0}\,11}$\\
&&&&&&\\
&\textbf{3} Errors& $\ket{111}$ & $\ket{000}$ & $\ket{000}$ & $\ket{000}$ & $\ket{\mathbf{0}\,00}$\\
&&&&&&\\ \hline
\end{tabular}\\
\end{center}

Here we can clearly understand what we have said before, about the fact that detection and correction in QEC is done without any measurement. We see that the error correction capacity of this circuit does not exceed one bit flip. Furthermore, the correction concerns just $q_0$, which is reasonable.

\subsubsection{Single Sign-Flip Error Correction Code}
Another kind of errors that can occur in quantum computers, it is the \textbf{sign-flip} or \textbf{phase-flip}, which is a $\pi$ phase error that makes the interchange $(a\ket{0}+b\ket{1})\rightleftarrows(a\ket{0}-b\ket{1})$, which describes the action of the \textsc{Z gate}.
Unlike bit-flips, this one exists only in the quantum case \cite{Dancing}. To construct the error correction circuit of sign-flips errors, we have to understand the identity obtained in \autoref{1.3.3}.
\[\mathbb{H\,X=Z\,H}\,\longrightarrow\,\mathbb{X=H\,Z\,H}\]
This means that the action of $\mathbb{X}$ (\textit{bit-flip}) on the $computational$ basis $\{\ket{0},\ket{1}\}$ is equivalent to the action of $\mathbb{Z}$ (\textit{sign-flip}) on the \textsc{Hadamard} basis $\{\ket{+},\ket{-}\}$; hence, to convert \textit{bit-flips} to \textit{sign-flips}, we have just to change the basis \cite{Dancing}.
\begin{figure}[H]
\centering
\includegraphics[width = 1\linewidth]{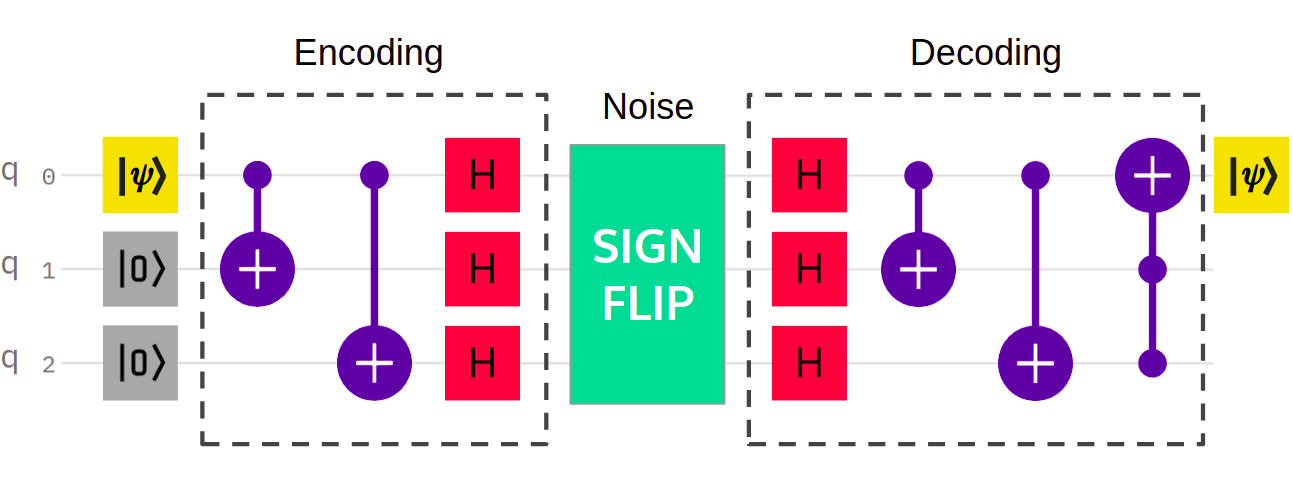}
\caption{Single sign flip error correction circuit.}
\label{1.31}
\end{figure}
The encoding used in this case consist of the following mapping:
\[\ket{\psi}=a\ket{0}+b\ket{1}\rightarrow\ket{\Psi}_{bf}=a\ket{000}+b\ket{111}\]
\[\ket{\Psi}_{bf}\rightarrow(\mathbb{H\otimes H\otimes H})\ket{\Psi}_{bf}=\ket{\Psi}_{sf}=a\ket{+++}+b\ket{---}\]
As the \textit{bit-flip} error correction circuit, this circuit corrects up to one \textit{sign-flip} error in $q_0$. 
\subsubsection{The 9-Qubit Error Correction Shor Code}
In 1995 the mathematician \textsc{Peter Shor} published work on an error correction circuit that can correct, a single $bit$-$flip$ ($\mathbb{X}$), a single $sign$-$flip$ ($\mathbb{Z}$) or both ($\mathbb{Y}$) \cite{Dancing}\cite{Marinescu}. This general error can be modeled as a \textit{unitary} transform $\mathbb{E}$, that act on the qubit, and which expresses as a $linear$ $combination$ of each error type. 
\[\mathbb{E}=\alpha\:\mathbb{I}+\beta\:\mathbb{X}+\mu\:\mathbb{Z}+\nu\:\mathbb{Y}\quad;\quad\alpha,\beta,\mu,\nu\in\mathbb{C}\]
The \textsc{Shor code}, encodes the information of $\mathfrak{one}$-$\mathfrak{qubit}$ onto a \textit{highly entangled} $\mathfrak{9}$-$\mathfrak{qubit}$ state. The followed steps are:
\begin{enumerate}
\item The qubit $q_0$ state to be transmitted is encoded by 2-qubits ($q_3$ and $q_6$) via the circuit of \textbf{Sign-flip} error correction.
\item Each of the 3 qubits ($q_0,\,q_3,\,q_6$) is encoded via the \textbf{Bit-flip} error correction code, which gives 3 $bit$-$flip$ code groups : $\{$($q_0,\,q_1,\,q_2$); ($q_3,\,q_4,\,q_5$); ($q_6,\,q_7,\,q_8$)$\}$, respectively.
\item Each of these $bit$-$flip$ groups receives its own error correction.
\item At the end, a global $sign$-$flip$ correction is carried out.
\end{enumerate}
\begin{figure}[H]
\centering
\includegraphics[width = 1.05\linewidth]{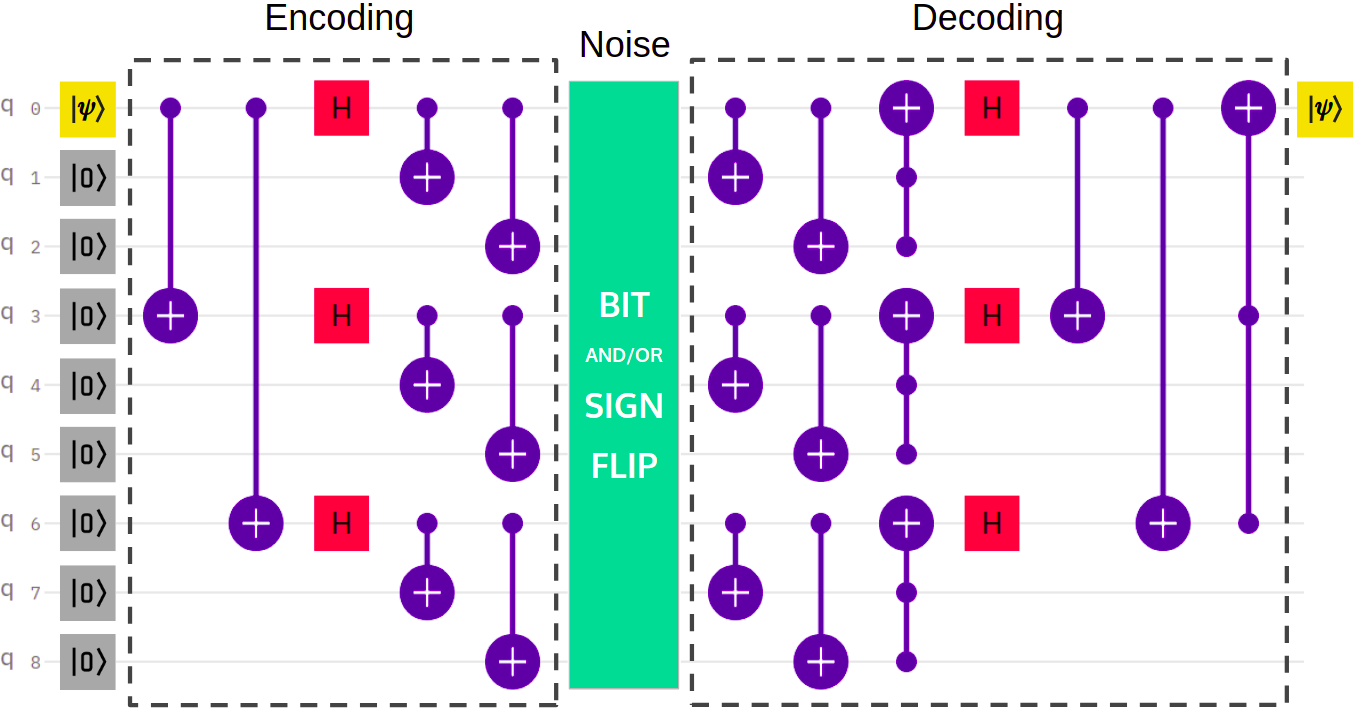}
\caption{\textsc{Shor} error correction circuit.}
\label{1.32}
\end{figure}
The \textsc{Shor code} encoding requires the following transformation:
\[\ket{0}\rightarrow\ket{0}_{\mathcal{S}hor}=\frac{1}{\sqrt{2^3}}\:(\ket{000}+\ket{111})\otimes(\ket{000}+\ket{111})\otimes(\ket{000}+\ket{111})\]
\[\ket{1}\rightarrow\ket{1}_{\mathcal{S}hor}=\frac{1}{\sqrt{2^3}}\:(\ket{000}-\ket{111})\otimes(\ket{000}-\ket{111})\otimes(\ket{000}-\ket{111})\]
The encoded qubit state is written
\[\ket{\psi}=a\ket{0}+b\ket{1}\longrightarrow\ket{\Psi}_{\mathcal{S}hor}=a\ket{0}_{\mathcal{S}hor}\!+\;b\ket{1}_{\mathcal{S}hor}\]
For this, the \textsc{9-Qubit Shor Code} is able to correct any $arbitrary$ error.\\

After understanding the QEC basic error correction codes, the problem that appears is that to correct the errors of a given qubit, we need 2 $correction$-$qubits$, then we need 2 more for each $correction$-$qubit$, and so on. Therefore, other methods are used, like the “\textit{surface codes}” based on \textsc{group theory} and \textsc{topology}\footnote{\href{https://qiskit.org/learn/intro-qc-qh}{{\color{black}Lecture Notes 5:} https://qiskit.org/learn/intro-qc-qh}}.\\

Another important kind of errors occurs when reading out data from a quantum computer, they are known as “\textsc{Readout Errors}” \cite{Nac}\cite{Leymann}. This type of error represents a \textit{central} problem in quantum computing, because it intervenes when extracting an outcome after any quantum computation, and gives us results that are not entirely correct. Readout errors and their correction methods, will be developed in depth in \autoref{chap3}.

\chapter{Unfolding for High Energy Physics }
\label{chap2}
\minitoc
In the Large Hadron Collider (LHC), beams of particles are accelerated and collide to make debris of new particles; these particles fly out from the collision point in all directions. Detectors are situated where the beams can be crossed. When a particle interacts with the detector, an electrical signal is produced and sent to a readout system. They are digitized and read out by a computer and stored on a hard disk.

The measurements of the detector are grouped into events containing all information associated with one scattering process, and events are grouped into certain regions of phase-space, called bins. The measurement of distributions $f(x)$ of some kinematical quantity $x$ is affected by the finite resolution of the particle detectors. If we have an ideal detector then we could measure this quantity in every event and could obtain the distribution by a histogram. But, with a real detector, determination of a distribution $f(x)$ is more difficult because of three factors \cite{blobel}: 

\begin{itemize}
\item \textbf{Finite resolution of the detector}:  There is only a statistical relation between the true kinematical variable $x$ and the measured quantity $y$. The measured quantity $y$ is smeared out due to the limited measurement accuracy of the detector.
\item \textbf{Limited acceptance of the detector}: Acceptance of a detector is the range of the kinematic parameters which particles are potentially detectable in it. The probability to observe a given event in a region is called acceptance. 
\item \textbf{Non-linear response}: The transformation from $x$ to $y$ can be caused by the non-linear response of a detector component. Instead of the quantity $x$ we found a different quantity $y$  in measurement.
\end{itemize} 

 To solve this problem and to estimate truth distribution (for example, cross-sections) from measurement, we use a statistical technique called unfolding methods.

\begin{figure}[!ht]
\begin{center}
\includegraphics[width=0.9\textwidth]{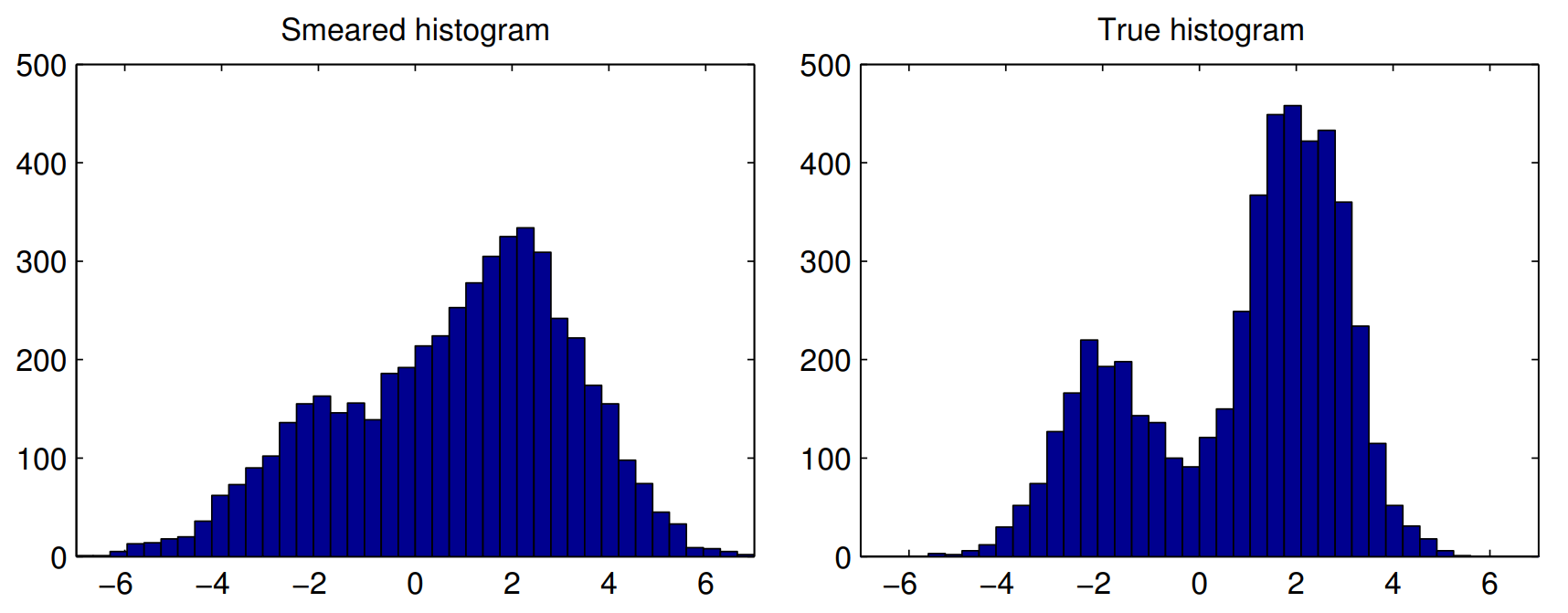}
\caption{Representation of an example of Smeared and True histogram \cite{mms}.}
\label{2.1}
\end{center}
\end{figure}



\section{Inverse Problem and Unfolding Methods}

Unfolding is the ensemble of statistical techniques used to estimate the true spectrum using the smeared observations.
The distribution $f(x)$ of the true variable $x$ is related to the measurement distribution $g(y)$ of the measured quantity $y$ by a Fredholm integral equation of the first kind:
\begin{equation}
g(y)= \int_{\omega} R(y,x)\,f(x)\,dx
\end{equation}
The problem to determine the distribution $f(x)$ from measured distribution $g(y)$ is an inverse problem and it is called unfolding. $R(y,x)$ represents the detector effect. 

The equation is usually discretized when we want to represent the probability density functions $f(x)$ and $g(y)$ in histograms. We can construct a histogram of $y$ with $M$ bins.

The integral equation becomes: 
\begin{equation}
m_i= \sum_{j=1}^M \mathcal{R}_{ij}t_j \quad i=1,...,N.
\label{eq 2.2}
\end{equation}
The response matrix $\mathcal{R}$ could be interpreted simply as a conditional probability:
\begin{equation}
\mathcal{R}_{ij}=\mathbb{P}(\text{measured in bin} \ i| \text{true value in bin}\ j)
\end{equation}
where $t=(t_1,...,t_M)$ is the expectation values for the histogram of $x$ and $m=(m_1,...,m_N)$ gives the expected number of events in bins of the observed variable $y$. The data are given as a vector of numbers $n=(n_1,...,n_N)$. 

The efficiency $\epsilon_j$ is obtained by summing $\mathcal{R}_{ij}$:

\begin{equation}
\sum^N_i \mathcal{R}_{ij}=\mathbb{P}(\text{observed anywhere} | \text{true value in bin}\ j)=\epsilon_j
\end{equation}


In the general form, the equation \eqref{eq 2.2} should include a number of background events as follow: 
\begin{equation}
m=\mathcal{R}t+\beta.
\label{2.5}
\end{equation}

Unfolding is useful when we want to compare the measurement with other experiments, or to compare the measurement with future theories. It is most often used in measurement analyses.  

\section{Unfolding with Matrix Inversion}
There are many unfolding methods and the most simple is the \textit{matrix inversion}. This is possible only if the number of bins observed is equal to the number of bins on the truth level.

We assume that the equation $m=\mathcal{R}t+\beta $ can be inverted \cite{cowan}: 
\begin{equation}
t=\mathcal{R}^{-1}(m-\beta) 
\end{equation}
The matrix inversion returns an unbiased result, because it is a simple linear transformation of the result.
\begin{figure}[H]
\begin{center}
\includegraphics[width=0.8\textwidth]{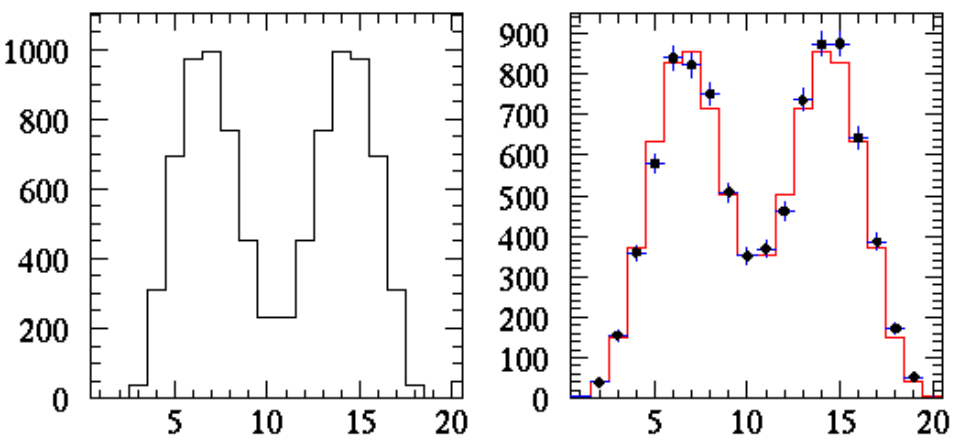}
\includegraphics[width=0.8\textwidth]{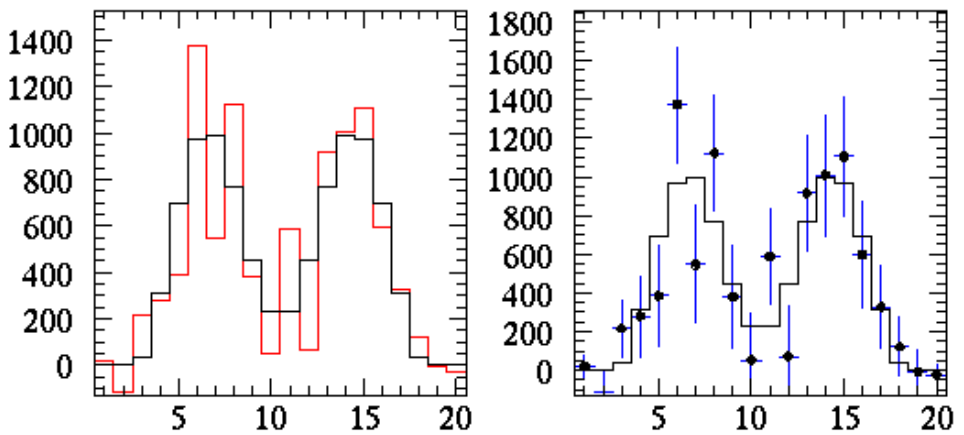}
\caption{Unfolding using matrix inversion: The true distribution (upper left), smeared distribution and points are data drawn from this distribution (upper right), unfolded data from matrix inversion with true distribution (Bottom left), error bars for the unfolded data (Bottom right) \cite{unf}.}
\label{2.3}
\end{center}
\end{figure}


\section{Iterative Bayesian Unfolding Method}
In 1994 \textsc{Giulio D'agostini} proposed an iterative method based on \textit{Bayes' theorem}, known in HEP as the \textit{Iterative Bayesian Unfolding} (IBU). This technique gives approximations very close to that of the \textit{true distributions}. The notable advantage of IBU over the \textit{matrix inversion} is that the outcome is a \textit{probability} (i.e., \textit{normalization \emph{and} non-negativeness}). This Bayes approach requires knowledge of the \textit{initial distribution}, but \textsc{D'agostini} has overcome this problem by using an \textit{iterative procedure} \cite{44} \cite{45} \cite{46}.\\

For an illustration, we start by stating the \textsc{Bayes' Theorem}, using a picture of \textit{causes \emph{and} effects}, where multiple independent \textit{causes} $c_i$ ($i=1,\,2,\,3,\,...,\,n_c$), which can produce a single \textit{effect} ($e_1$). The \textit{initial probability} of the causes $\mathbb{P}(c_i)$ is assumed to be known, as well as the \textit{conditional probability} of the $i^{th}$ cause to produce the effect $\mathbb{P}(e_1|c_i)$, then \textit{Bayes formula} is written in the form \cite{44}:
\[\mathbb{P}(c_i|e_1)=\dfrac{\mathbb{P}(e_1|c_i)\,.\:\mathbb{P}(c_i)}{\sum^{n_c}_{k}\mathbb{P}(e_1|c_k)\,.\:\mathbb{P}(c_k)}\] 
where $\mathbb{P}(c_i|e_1)$ represents the probability that a single effect $e_1$ has been due to the $i^{th}$ cause $c_i$.\\
Moreover, many possible effects  $e_j$ ($j=1,2,...,n_e$) can be created from a given cause $c_i$, then the above formula can be generalized by the change $e_1\rightarrow e_j$.\\

The causes correspond to the \textbf{true bin counts} before the \textit{distortion \emph{from} detector effects}, noted ($t_i$), and effects correspond to the \textbf{measured bin counts}, noted ($m_j$). Then the previous formula becomes  
\begin{equation}
\mathbb{P}(t_i|m_j)=\dfrac{\mathbb{P}(m_j|t_i)\,.\:\mathbb{P}(t_i)}{\sum^{n_t}_{k}\mathbb{P}(m_j|t_k)\,.\:\mathbb{P}(t_k)}
\end{equation}
From the \textit{law of total probability}, we can write $\mathbb{P}(t_i)$ as 
\[\mathbb{P}(t_i)=\sum_{j}\,\mathbb{P}(t_i\cap m_j)=\sum_{j}\,\mathbb{P}(t_i|m_j)\,.\:\mathbb{P}(m_j)\]
where we used the definition of \textit{conditional probability} : $\mathbb{P}(a|b)\overset{def}=\dfrac{\mathbb{P}(a\cap b)}{\:\mathbb{P}(b)}\;;\,\mathbb{P}(b)\neq0$.\\
Therefore, if we replace the expression of $\,\mathbb{P}(t_i|m_j)$ in the one above, we will have 
\begin{equation}
\mathbb{P}(t_i)=\sum_{j}\left(\dfrac{\mathbb{P}(m_j|t_i)\,.\:\mathbb{P}(t_i)}{\sum^{n_t}_{k}\mathbb{P}(m_j|t_k)\,.\:\mathbb{P}(t_k)}\right)\times\mathbb{P}(m_j)
\end{equation}
The probabilities $\mathbb{P}(m_j|t_i)$ must be calculated or estimated with \textsc{Monte Carlo} methods \cite{44}. The desired solution $\mathbb{P}(t_i)$, is also \textbf{unknown}, so we have to use an estimate of it, therefore we can write 
\begin{equation}
\mathbb{P}^{\,n+1}(t_i)=\sum_{j}\left(\dfrac{\mathbb{P}(m_j|t_i)\,.\:\mathbb{P}^{\,n}(t_i)}{\sum^{n_t}_{k}\mathbb{P}(m_j|t_k)\,.\:\mathbb{P}^{\,n}(t_k)}\right)\times\mathbb{P}(m_j)\quad;\quad n\in\mathbb{N}
\end{equation}
This leads to an \textit{iterative procedure}, where the choice commonly used to estimate the initial probability $\mathbb{P}^{\,0}(t_i)$ is the \textbf{uniform distribution}, so  $\mathbb{P}^{\,0}(t_i)=1/n_t\,$\cite{44} \cite{46}\cite{Nac}.\\
Recalling that the \textit{response matrix} is written as $\mathcal{R}_{kl}=\mathbb{P}(m_k|t_l)$, and with $t_l=\mathbb{P}(t_l)$, $m_l=\mathbb{P}(m_l)$, then we can write :
\begin{equation}
t^{n+1}_{i}=\,\sum_{j}\,\left(\dfrac{\mathcal{R}_{ji}\:t^{\,n}_{i}}{\sum_{k}\,\mathcal{R}_{jk}\,t^{\,n}_{k}}\right)\times m_j
\end{equation}
where $n$ is the number of iterations. The unfolding procedure starts by choosing a prior \textit{truth spectrum} $t^0_i=\mathbb{P}^{\,0}(t_i)$ from the best knowledge of the process under study, otherwise, one can take $\mathbb{P}^{\,0}(t_i)=1/n_t$. The number of iterations needed to converge depends on the desired precision. 
It is important to highlight that despite of the fact that the outcome is a probability, the unfolded outcome can have \textit{negative entries}, because the \textit{measured bin counts} may have \textit{negative entries}, which results from \textit{background subtraction}, so the result is \textit{positive} only when $m\geqslant0\,$ \cite{Nac}.\\

\textsc{D'agostini} mentioned in his paper 10 advantages of the IBU, in comparison with other unfolding methods \cite{44}:
\begin{enumerate}
\item It is theoretically well-grounded.
\item Can be applied to \textit{multidimensional} problems.
\item It can use cells of different sizes for the distribution of the true and the $experimental$ $values$.
\item The domain of the definition of the \textit{experimental values} can differ from that of the \textit{true values}.
\item It can take into account \textit{any kind of smearing \emph{and} migration} from the \textit{true values \emph{to the} measured ones}.
\item In terms of its ability to reproduce the \textit{true distribution}, it gives the \textit{best} results if one makes a realistic guess about the distribution that the true values follow. But, in case of total ignorance, satisfactory results are obtained even starting from a \textit{uniform distribution}.
\item Can take different sources of \textit{background} into account.
\item Does not require \textit{matrix inversion}.
\item Provides the \textit{correlation matrix} of the results.
\item Can be implemented in a \textit{short, simple \emph{and} fast} program, which deals directly with distributions and not with \textit{individual events}.\\
\end{enumerate}
These advantages made the IBU a widely used method in HEP, as well as \textit{statistical astronomy} in rectification and deconvolution problems \cite{45}. The next step will be to implement it in \textit{quantum information science} (QIS).

\chapter{Unfolding Quantum Computer Readout Noise}
\label{chap3}
\definecolor{bg}{rgb}{0.97,0.97,0.97}
\minitoc

Quantum computers use the principle of superposition and entanglement to solve complex problems. But today's quantum computers are noisy. Their are two sources of this noise “\textit{qubit decoherence}” and “\textit{gate infidelity}”. One important type of errors that arise from \textit{qubit decoherence} are known as “\textbf{\textit{Readout Errors}}”. These errors are similar to those caused by detector effects in HEP.

In this chapter we will introduce the so-called “\textit{readout errors}”, where their effects can be represented by a matrix called “\textit{calibration matrix}”. Then we try to use the unfolding methods from HEP to QIS. We also provide a simulation of an \textit{unfolding process}, Finally, we mitigate readout noise through HEP unfolding methods in order to achieve the objective of this thesis.

\section{Readout Errors}
After performing any quantum computation we have to get an outcome, so we have to measure our system of qubits, which cause a collapse of the system state to either 0 or 1. But we know that the measurement operation will take a certain time. On the other hand, qubits have a \textit{coherence time}, and which is \textit{negligible} compared to the \textit{measurement times}; this means that during a measurement a qubit in the state $\ket{0}$ can decay to $\ket{1}$, and vice versa (i.e., a \textit{bit-flip error}). In addition, the probability distributions of measured \textit{physical quantities} corresponding to the states $\ket{0}$ and $\ket{1}$ have an \textit{overlapping support} and there is a low probability of measuring the opposite value. This leads to biased outcomes,  which therefore change our interpretation of these results \cite{Nac}\cite{Leymann}. We can investigate this fact directly, through the circuit below: 
\begin{figure}[H]
\centering
\includegraphics[width = 0.5\linewidth]{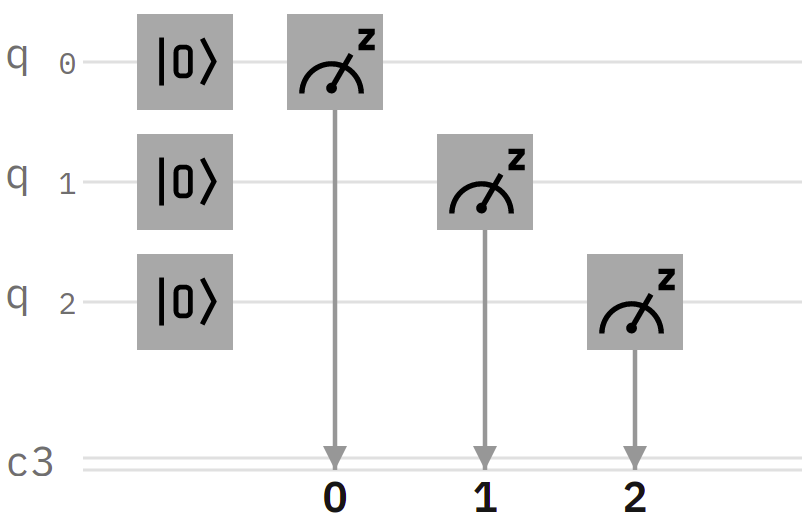}
\caption{Measuring a 3-Qubit system of state $\ket{000}$.}
\label{3.1}
\end{figure}
In this circuit we have a system of 3-qubits initialized in the state $\ket{000}$, when performing a measurement, we expect to find them 100\% in this state. But for the reasoning mentioned above, we will get the following values:
\begin{figure}[H]
\centering
\includegraphics[width = 0.8\linewidth]{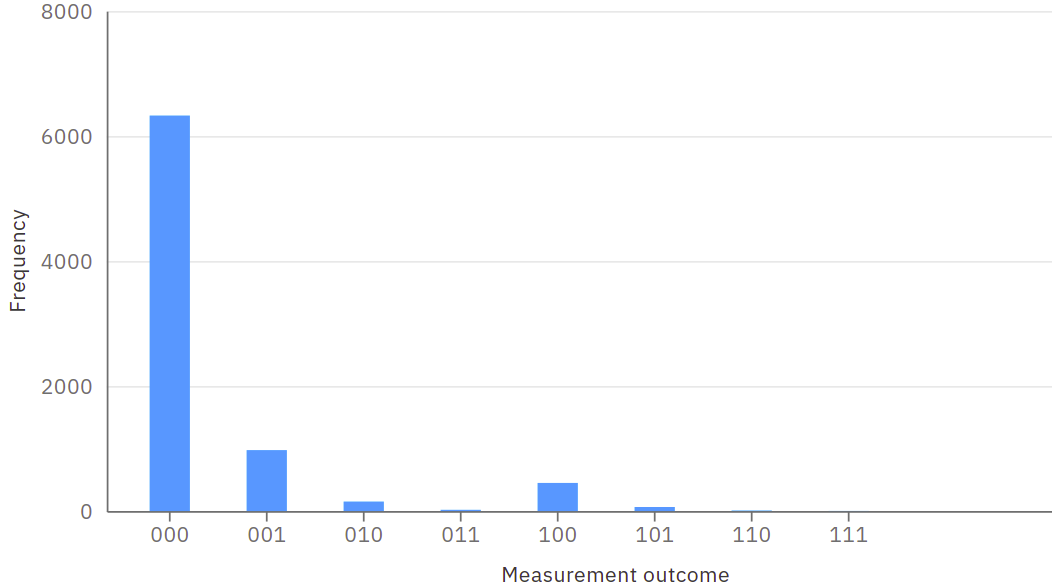}
\caption{The Frequency of measuring each basis state from the IBM Q 5 \textsc{Yorktown Machine}, for 8000 shots.}
\label{3.2}
\end{figure}
We executed the circuit above in the IBM Q 5 \textsc{Yorktown Machine} that runs it 8000 times (the number of shots is proportional to the precision of the output probability distribution). We see in this histogram the frequency $\mathcal{F}$ of measuring each basis state, where $\mathcal{F}$ is proportional to the $probability$, which achieves the following relation  \[\mathbb{P}=\frac{\mathcal{F}}{\mathcal{N}_{shot}}\]
where $\mathcal{N}_{shot}$ is the total number of shots\footnote{The number of shots defines how many times a quantum circuit is executed in a quantum
computer.}.

Due to the effects of noise, instead of having a probability of 1 to measure the state $\ket{000}$, we get $\mathbb{P}_{_{000}}\simeq0.7909$. This means that \textit{bit-flips} have occurred during the measurement. For a \textit{single bit-flip} on the state $\ket{000}$, we can get $\ket{001},\,\ket{010},\,\ket{100}$, with respective probabilities\footnote{The probabilities are not equal because the qubits are not \textbf{identical}; they have different \textit{decoherence times}.}\[\mathbb{P}_{_{001}}\simeq0.1219\,,\quad\mathbb{P}_{_{010}}\simeq0.019\,,\quad\mathbb{P}_{_{100}}\simeq0.0563\]
If \textit{two bit-flips} occurs, we get $\ket{011},\,\ket{101},\,\ket{110}$, with respective probabilities:\[\mathbb{P}_{_{011}}\simeq0.0025\,,\quad\mathbb{P}_{_{101}}\simeq0.0081\,,\quad\mathbb{P}_{_{110}}\simeq0.0001\]
If \textit{three bit-flips} occurs, we get $\ket{111}$, with the corresponding probability:\[\mathbb{P}_{_{111}}\simeq25\times10^{-5}\]
The probability that more than a single \textit{bit-flip} will occur is very low. For simple manipulations it is not significant, but for more complex manipulations (where we are in total ignorance of the results) it represents a challenge for interpreting results of experiments.

The methods of correcting readout errors will be discussed in \autoref{section 3.3}. But before that, let's explore how to construct the central object of the unfolding process, the so-called “\textit{calibration matrix}”.
\section{Constructing the Calibration Matrix $\mathcal{R}$}

The calibration matrix is the essential tool in the mitigation of readout errors, which transforms true data into noisy data, so it is equivalent to the response matrix in HEP. This matrix is specific for each quantum computer; it depends directly on the hardware. Furthermore, for a given hardware, its elements changes every time due to \textit{calibration drift}.
The method often used to measure this matrix consist of constructing a set of quantum circuits called “\textit{calibration circuits}”, where for an n-qubit system, we must construct $2^n$ calibration circuit that generates all the corresponding computational basis states of the system, where this configurations are produced via the application of the \textsc{X gate} \cite{Nac}\cite{Leymann}.

To illustrate how the construction of the calibration matrix works we will take two instances. In the reste of this section, we will use the \textsc{5 Yorktown IBM Q Machine}, and a number of 8000 of total shots.

\subsection{One Qubit Calibration Matrix}
Let's have a system of one qubit which is initialized in the state $\ket{0}$, then we perform a measurement on it, as shown in the circuit below:
\begin{figure}[H]
\centering
\includegraphics[width = 0.28\linewidth]{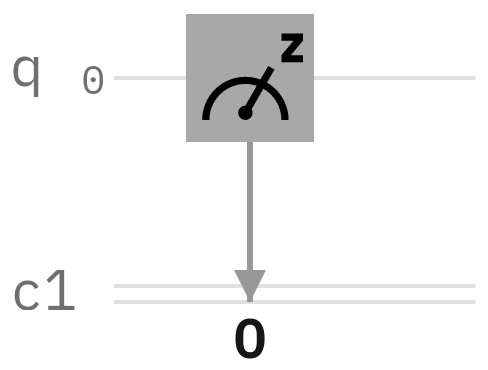}
\caption{Measuring a qubit in the state $\ket{0}$.}
\label{3.3}
\end{figure}
After running this circuit in real hardware, we get the following outcomes:
\begin{figure}[H]
\centering
\includegraphics[width=0.4\linewidth]{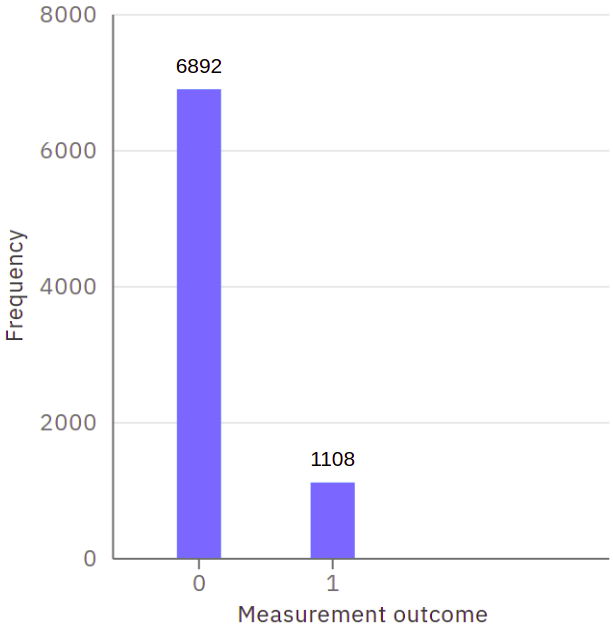}
\end{figure}
As we said in the previous section, this qubit can decay to the state $\ket{1}$, during a measurement.\\

The calibration matrix $\mathcal{R}$ is interpreted as a \textit{conditional probability}, where the coefficient $\mathcal{R}_{ij}$ represents the probability of measuring the value $i$ under the condition that the true value is $j$
\[\mathcal{R}_{ij}=\mathbb{P}\,(\text{measured\:is :\:}i\:|\text{true \:is :\:}j\,)=\mathbb{P}\,(m=i|t=j)\] Therefore, the above results can be translated into a column vector \[\mathcal{R}_{i0}=\begin{pmatrix}\mathcal{R}_{00}\\ \\\mathcal{R}_{10}\end{pmatrix}=\begin{pmatrix}\mathbb{P}(0|0)\\ \\\mathbb{P}(1|0)\end{pmatrix}=\begin{pmatrix}\frac{6892}{8000}\\ \\\frac{1108}{8000}\end{pmatrix}=\begin{pmatrix}0.861500\\ \\0.138500
\end{pmatrix}\]

The second step consists of building the circuit generating the other basis state $\ket{1}$, and which can be done by applying an X gate, before measuring the one-qubit system, as shown below:
\begin{figure}[H]
\centering
\includegraphics[width = 0.65\linewidth]{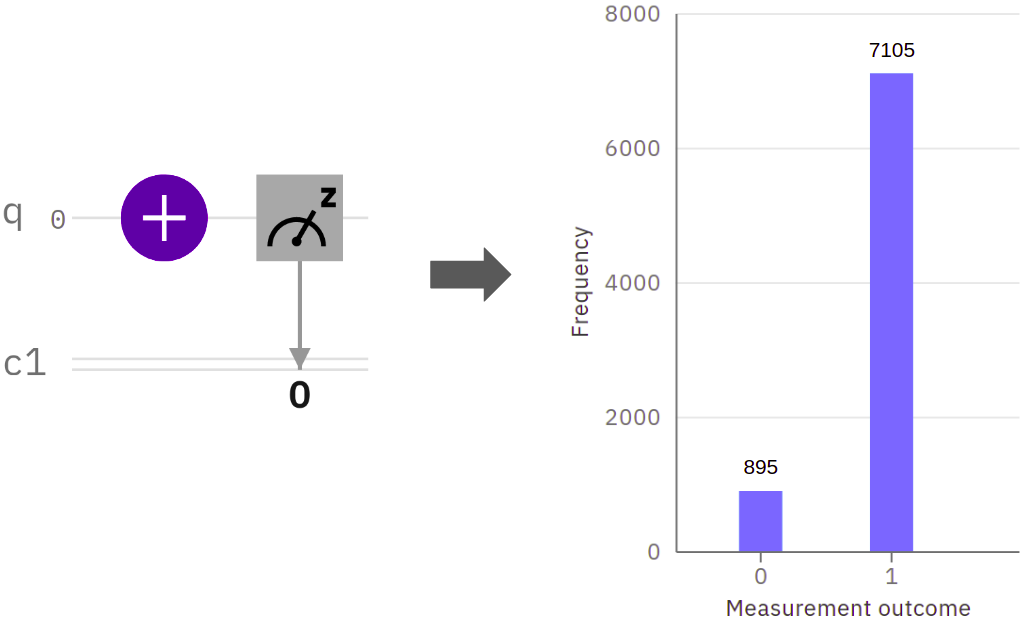}
\end{figure}
These results represent the second column of the calibration matrix  \[\mathcal{R}_{i1}=\begin{pmatrix}\mathcal{R}_{01}\\ \\\mathcal{R}_{11}\end{pmatrix}=\begin{pmatrix}\mathbb{P}(0|1)\\ \\\mathbb{P}(1|1)\end{pmatrix}=\begin{pmatrix}\frac{895}{8000}\\ \\\frac{7105}{8000}\end{pmatrix}=\begin{pmatrix}0.111875\\ \\0.888125 \end{pmatrix}\]

Finally, with these two columns, we can now construct the $2\times2$ calibration matrix of a single qubit system:
\begin{equation}\mathcal{R}_{_{1\text{Q}}}=\begin{pmatrix}\mathcal{R}_{00}&\mathcal{R}_{01}\\ \\\mathcal{R}_{10}&\mathcal{R}_{11}\end{pmatrix}=\begin{pmatrix}\;0.86150\,&\,0.111875\;\\&\\\;0.13850\,&\,0.888125\;\end{pmatrix}
\end{equation}
If there was no noise, then we will have : $\,\mathbb{P}(0|0)=\mathbb{P}(1|1)=1$ and $\mathbb{P}(1|0)=\mathbb{P}(0|1)=0$, which leads to an identity matrix.\[\mathcal{R}_{_{1\text{Q}}}=\begin{pmatrix}1&0\\0&1\end{pmatrix}=1\!\!1\] 
\subsection{Two Qubit Calibration Matrix}
Let's have two qubits that are initialized in the state $\ket{00}$, when we measure them, instead of having a probability of 1 for the state $\ket{00}$, this probability will be distributed to the $2^2$ possible computational basis states ($\ket{00}$, $\ket{01}$, $\ket{10}$, $\ket{11}$). To construct the calibration matrix we have to build $2^2$ calibration circuits, where we use the X gate to prepare these 4 qubit configurations. After the readout, we get the corresponding column of each calibration circuits.\\

It is important to note that in the IBM \textsc{quantum composer}, the convention used of the qubit order is that  qubits start from the left to the right, which is the opposite of our convention, i.e.,$\,\ket{\:\mathfrak{q}_0\:\mathfrak{q}_1\:}\:\longmapsto\,\ket{\:\mathfrak{q}_1\:\mathfrak{q}_0\:}_{_{\text{IBM}}}$. For brevity of notation, we will write the \textit{binary} numbers $i,\,j$ of the matrix elements $\mathcal{R}_{ij}$, in the \textit{decimal} base, (i.e., $(00)_{_{2}}=(0)_{_{10}}\,,\;(01)_{_{2}}=(1)_{_{10}}\,,\;(10)_{_{2}}=(2)_{_{10}}\,,\;(11)_{_{2}}=(3)_{_{10}}$).\\

The figure below shows the calibration circuits and their measurement outcomes which correspond respectively to the computational basis states $\ket{00}$, $\ket{01}$, $\ket{10}$, $\ket{11}$:
\begin{figure}[H]
\centering
\includegraphics[width=0.85\linewidth]{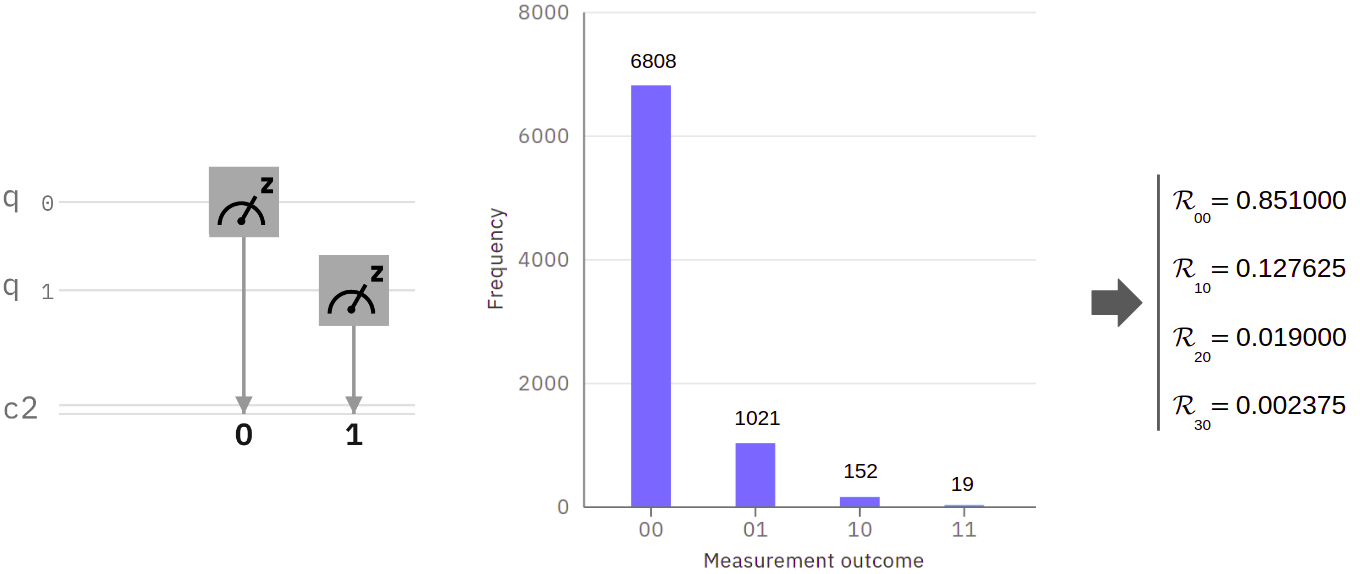}
\includegraphics[width=0.85\linewidth]{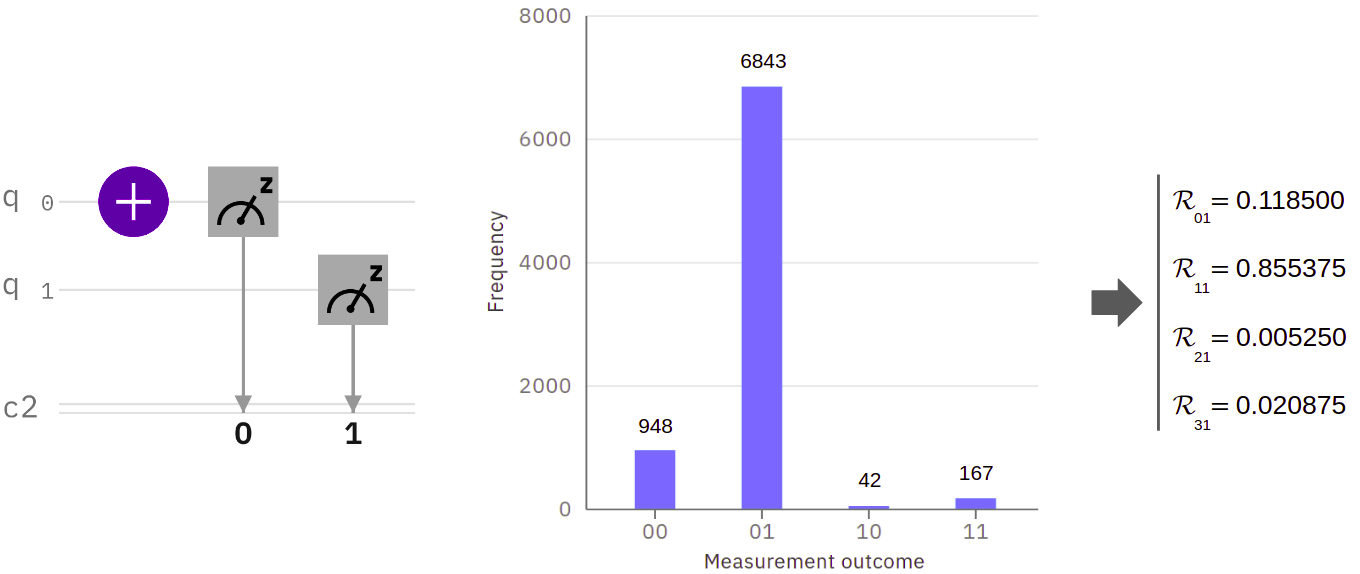}
\includegraphics[width=0.85\linewidth]{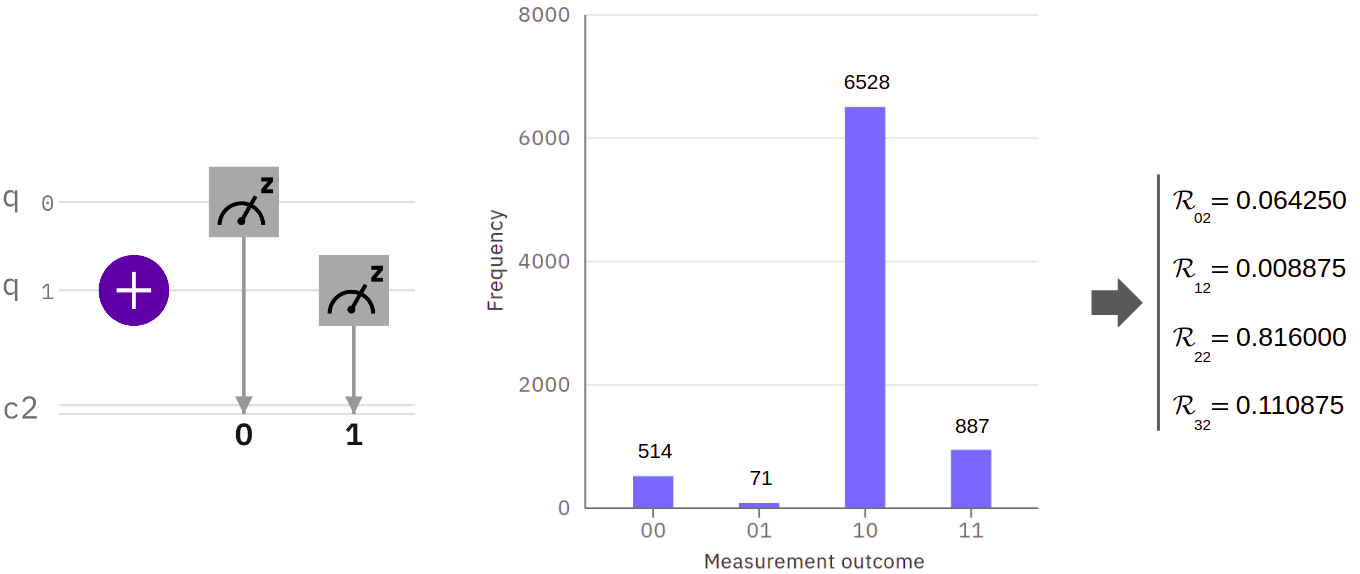}
\includegraphics[width=0.85\linewidth]{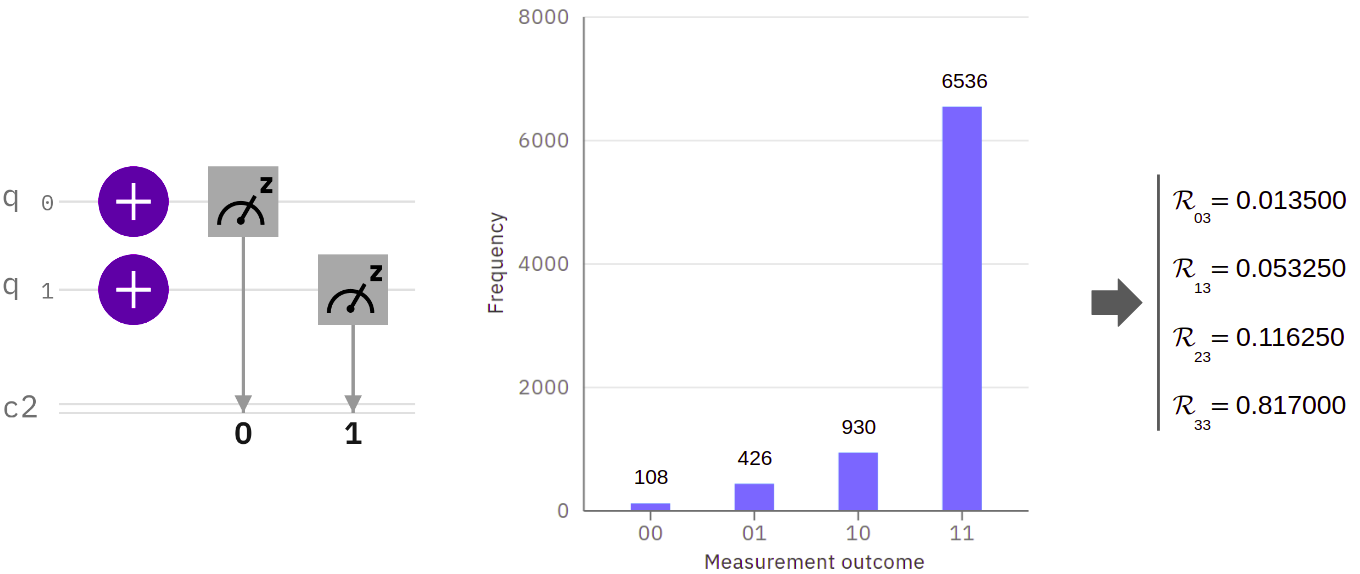}
\end{figure}
\begin{equation}
\mathcal{R}_{_{2\text{Q}}}=\begin{pmatrix}
\; \mathbf{0.851000} \, & \, 0.118500 \, & \, 0.064250 \, & \, 0.013500 \;  \\ & \\
\; 0.127625 \, & \, \mathbf{0.855375} \, & \, 0.008875 \, & \, 0.053250 \;  \\ & \\
\; 0.019000 \, & \, 0.005250 \, & \, \mathbf{0.816000} \, & \, 0.116250 \;  \\ & \\
\; 0.002375 \, & \, 0.020875 \, & \, 0.110875 \, & \, \mathbf{0.817000} \;  
\end{pmatrix}
\end{equation}
\linebreak

Finally, we get the $2^2\times2^2$ calibration matrix, where we see that the correct results (located in the diagonal) have the greatest probability.
As the number of qubits is fixed, the calibration matrix is always \textit{square}. Moreover, it is reduced to an identity matrix $\mathcal{R}_{ij}=\delta_{ij}$, in the absence of noise.\\

In the next section, we will discuss how to correct readout errors (“\textit{Unfolding}”), and show how it is done through the \textit{calibration matrix}.
\section{Unfolding Methods for QIS}
\label{section 3.3}
Unfolding is a method of determining \textit{approximately} the \textit{true distribution} from the \textit{measured disturbed distribution} (after distortions from readout noise). Several methods are used to achieve this approximation \cite{Brenner}, and which are mainly based on knowledge of the calibration matrix (a.k.a., correction matrix). The most basic and commonly used method to correct readout errors is matrix inversion. In 2020, physicist \textsc{Benjamin Nachman} et al. investigated various unfolding methods currently used in HEP, in order to connect them with current methods used in the field of \textit{quantum information science} (QIS), because current QIS methods have pathologies that can be avoided with HEP techniques. The \textit{iterative Bayesian unfolding} method, introduced in \autoref{chap2}, is shown to avoid \textit{pathologies} from the \textit{matrix Inversion} method \cite{Nac}.\\

By way of illustration, we will apply in what follows these two unfolding methods (MI, IBU), in a specific case. As we have built the \textsc{Yorktown} device-specific response matrix, we will continue to use it.\footnote{It is clear that each device has its specific response matrix; it makes no sense to unfold readout errors of the \textsc{Santiago} device (for example) using the \textsc{Yorktown} device response matrix.}
\subsection{Unfolding with Matrix Inversion Method}
The Matrix Inversion is the most naive unfolding method, it assumes that the true bin counts ($t$) and its corresponding measured counts ($m$) are related by a calibration matrix $\mathcal{R}$ as $\,m=\mathcal{R}\,t\,$. The additional assumption of this method is that $\mathcal{R}$ is \textit{regular}. Hence, it can be inverted to give us an \textit{approximation} of the true counts, denoted by $\,\hat{t}_{_{\textsf{MI}}}$. 
\begin{equation}
\hat{t}_{_{\textsf{MI}}} =\mathcal{R}^{\text{\,-}1}\:m
\end{equation}
To illustrate the unfolding procedure through this method, let's take the following example, where we want to mitigate the readout errors of two-qubit system generated in the specific Bell state $\ket{\Psi^{+}}$, written as:
\[\ket{\Psi^{+}}=\frac{1}{\sqrt{2}}\ket{01}+\frac{1}{\sqrt{2}}\ket{10}\]
The probability of finding the two qubits in different states (i.e., $\ket{01}$, $\ket{10}$), and that of finding them in the same states (i.e., $\ket{00}$, $\ket{11}$), are given respectively:
\[\mathbb{P}_{01}=\mathbb{P}_{10}=0.5\qquad;\quad\mathbb{P}_{00}=\mathbb{P}_{11}=0\]
\begin{figure}[H]
\centering
\includegraphics[width = 0.45\linewidth]{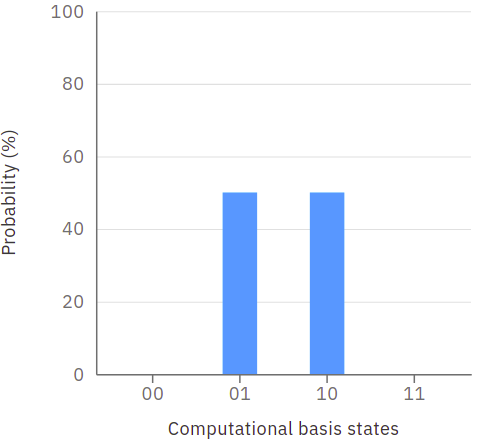}\hspace{1cm}
\end{figure}
Hence, the vector representing the true bin counts can be expressed as: 
\[t=\begin{pmatrix}0\\0.5\\0.5\\0\end{pmatrix} \]
After building the circuit of $\ket{\Psi^{+}}$ and running it on the \textsc{5 Yorktown IBM Q Machine} for $8000$ shots, the following results are obtained:
\begin{figure}[H]
\centering
\includegraphics[width = 1\linewidth]{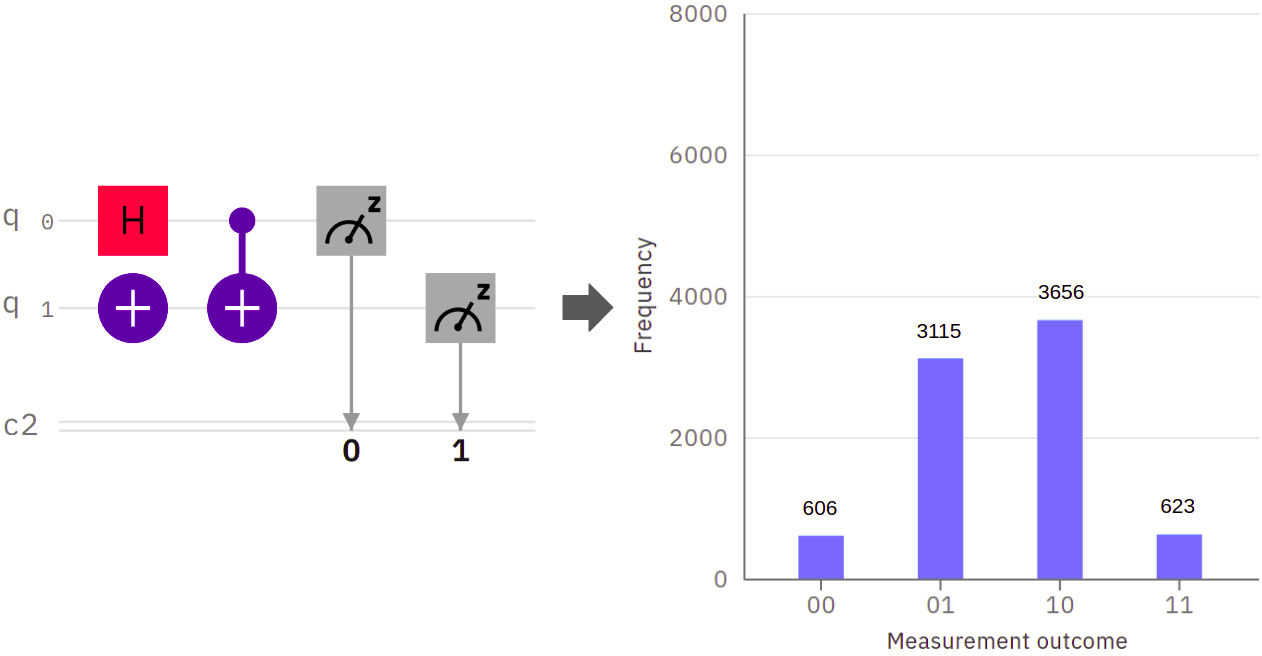}
\end{figure}
As before, these measured bin counts can be translated into a column vector. But for a better observation of the changes, we express its element in terms of \textit{frequency} instead of \textit{probability}, the same thing for the true bin counts.
\[m=\begin{pmatrix} 606\\3115\\3656\\623\end{pmatrix}\qquad;\qquad t=\begin{pmatrix} 0\\4000\\4000\\0\end{pmatrix}\]
We have a two-qubit system, so, we are going to use $\mathcal{R}_{_{2\text{Q}}}$, and all we have to do is simply to invert this matrix, in order to multiply it by the measured vector $m$. But before this, we have to check if $\mathcal{R}_{_{2\text{Q}}}$ is regular by calculating its determinant: $\,\det(\mathcal{R}_{_{2\text{Q}}})\simeq0.46448\neq0\,$, it is therefore not singular, and its inverse is given below

\begin{equation}
\mathcal{R}^{-1}_{_{2\text{Q}}}=\begin{pmatrix}
\;  +1.20206606  \, & \,  -0.16605952  \, & \,  -0.09341977  \, & \,  +0.00425315 \;  \\ & \\
\;  -0.17936727  \, & \,  +1.19572556  \, & \,  +0.01152758  \, & \,  -0.07661078 \;  \\ & \\
\;  -0.02752246  \, & \,  +0.00046624  \, & \,  +1.25179634  \, & \,  -0.17769229 \;  \\ & \\
\;  +0.00482367  \, & \,  -0.03013228  \, & \,  -0.16990414  \, & \,  +1.25004992 \;
\end{pmatrix}
\end{equation}
We can now calculate the approximation of the true bin counts $t$.

\begin{equation}
\hat{t}_{_{\textsf{MI}}} =\mathcal{R}_{_{2\text{Q}}}^{\text{-1}}m=\begin{pmatrix} -\,\mathbf{0127}.\,716\\ +\,\mathbf{3610}.\,404\\ +\,\mathbf{4450}.\,638 \\   +\,\mathbf{0066}.\,672\end{pmatrix}
\end{equation}
\begin{figure}[H]
\centering
\includegraphics[width = 0.65\linewidth]{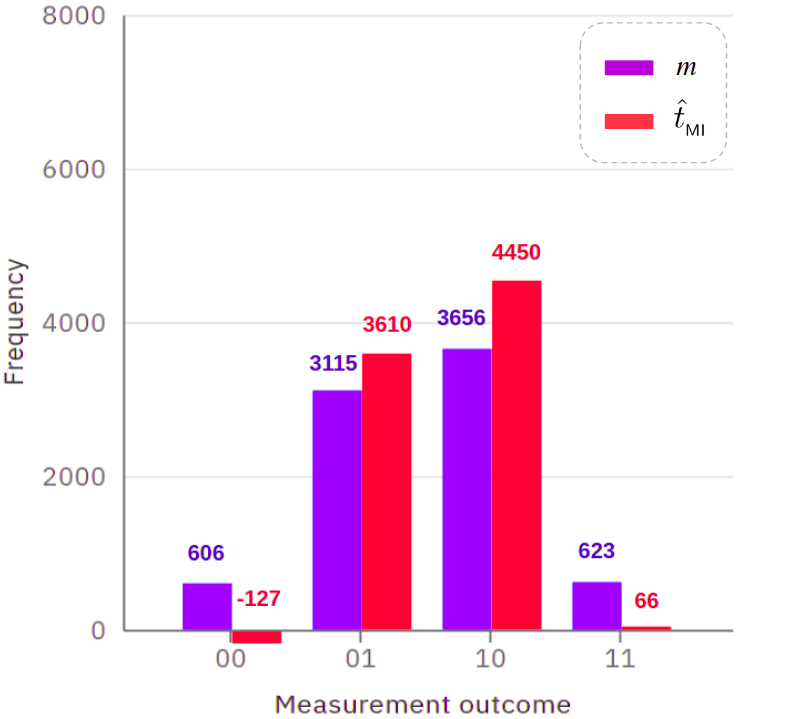}
\caption{Measured and unfolded counts via the matrix inversion method.}
\label{3.4}
\end{figure}
After unfolding the readout noise, we notice that the frequencies of the computational basis state $\ket{01},\:\ket{10}$ have increased, unlike the basis state $\ket{00},\:\ket{11}$, where their frequencies has decreased, which is suitable. But, we also notice that for the state $\ket{00}$, the mitigation causes it to decrease to a negative frequency, which is not physical. But that happens just in some cases; for example, if one gets after a measurement $m=[\,750,\,3100,\,3550,\,600\,]^\textsc{t}$, then the predicted counts will be  \[\hat{t}_{_{\textsf{IBU}}}=[\;\mathbf{57}.676\,,\: \mathbf{3567}.180\,,\: \mathbf{4318}.065\,,\:\mathbf{57}.077\;]^\textsc{t}\] which is free from \textit{unphysical outcomes}. But the fact remains that the opposite can be found.
\subsection{Unfolding with IBU}
The iterative Bayesian unfolding IBU, which was introduced in \autoref{chap2}, it is an iterative technique, where the \textit{approximate} (predicted) true counts vector $\hat{t}_{_{\textsf{IBU}}}\,$ is updated with each iteration in order to converge to the real counts $t$, recalling that the initial probability $t^0_i=\mathbb{P}^0(\text{true value is}\;i)$, is often chosen as a \textit{uniform distribution}. The IBU formula is written in the form

\begin{equation}
t^{n+1}_{i}=\,\sum_{j}\,\left(\dfrac{\mathcal{R}_{ji}\:t^{\,n}_{i}}{\sum_{k}\,\mathcal{R}_{jk}\,t^{\,n}_{k}}\right)m_j\qquad\;;\; n\in\mathbb{N}
\end{equation}
where $n$ is the number of iterations.

We will take the same example of $\ket{\Psi^{+}}$, with the same outcomes in order to compare the two methods. But before that, let's take, for simplicity, an instance of a one qubit system prepared in the state
\begin{equation}
\ket{+}=\mathbb{H}\,\ket{0}=\tfrac{1}{\sqrt{2}}\,\ket{0}+\tfrac{1}{\sqrt{2}}\ket{1}
\end{equation}
The probabilities that the qubit is in the states ($\ket{0}$, $\ket{1}$) are given respectively: \[\mathbb{P}_{0}=\mathbb{P}_{1}=0.5\]
After running the corresponding circuit of $\ket{+}$ in the same device and with the same number of shots (8000), we get $m=[\,4151,\,3849\,]^\textsc{t}$, while the true values are $t=[\,4000,\,4000\,]^\textsc{t}$. As we have a one qubit system, we will use for the unfolding the matrix
 \[\mathcal{R}_{_{1\text{Q}}}=\begin{pmatrix}\;0.86150\,&\,0.111875\;\\&\\\;0.13850\,&\,0.888125\;\end{pmatrix}\]
The IBU formula gives us the \textit{approximate components} of the true vector $t$, these components construct the predicted vector $\hat{t}_{_{\textsf{IBU}}}$, which is a \textit{two-dimensional} column vector in our case. By expanding the IBU equation for ($i,j,k=0,1$), we obtain:

\begin{equation}
t^{n+1}_{i}=\,\sum^1_{j=0}\,\left(\dfrac{\mathcal{R}_{ji}\:t^{\,n}_{i}}{\mathcal{R}_{j0}\,t^{\,n}_{0}+\mathcal{R}_{j1}\,t^{\,n}_{1}}\right) m_j
\end{equation}

\[t^{n+1}_{0}=\,\left(\dfrac{\mathcal{R}_{00}\:t^{\,n}_{0}}{\,\mathcal{R}_{00}\,t^{\,n}_{0}\,+\,\mathcal{R}_{01}\,t^{\,n}_{1}}\right)m_0\;+\;\,\left(\dfrac{\mathcal{R}_{10}\:t^{\,n}_{0}}{\,\mathcal{R}_{10}\,t^{\,n}_{0}\,+\,\mathcal{R}_{11}\,t^{\,n}_{1}}\right)m_1\]
\[t^{n+1}_{1}=\,\left(\dfrac{\mathcal{R}_{01}\:t^{\,n}_{1}}{\,\mathcal{R}_{00}\,t^{\,n}_{0}\,+\,\mathcal{R}_{01}\,t^{\,n}_{1}}\right)m_0\;+\;\,\left(\dfrac{\mathcal{R}_{11}\:t^{\,n}_{1}}{\,\mathcal{R}_{10}\,t^{\,n}_{0}\,+\,\mathcal{R}_{11}\,t^{\,n}_{1}}\right)m_1\]

\begin{flushleft}
We can schematize this iteration procedure as follows:
\end{flushleft}

\[\hat{t}^{\,0}_{_{\textsf{IBU}}}=\begin{pmatrix}t^0_0\\ \\t^0_1\end{pmatrix}\longrightarrow\hat{t}^{\,1}_{_{\textsf{IBU}}}=\begin{pmatrix}t^1_0\\\\t^1_1\end{pmatrix}\longrightarrow\hat{t}^{\,2}_{_{\textsf{IBU}}}=\begin{pmatrix}t^2_0\\\\t^2_1\end{pmatrix}\longrightarrow\cdots\longrightarrow\hat{t}^{\,n+1}_{_{\textsf{IBU}}}=\begin{pmatrix}t^{n+1}_0\\\\t^{n+1}_1\end{pmatrix}\approx t\]

Since we express the elements of vectors $t$ and $m$ in terms of frequency, the \textit{range} of the \textit{uniform distribution} will be extended from $[\,0\,,\,1\,]$ to $[\,0\:,\,8000\,]$, and any value of $t^0_i\in]\,0\:,8000\,]$ hold for the iteration procedure, so we can choose 
\[t^0_0=1\qquad;\qquad t^0_1=1\]
For $n=0\,$: \[t^1_0=4193.165\qquad;\qquad t^1_1= 3806.834\]
In the $1^{st}$ iteration these \textit{new} values will replace the \textit{old} ones $(\,t^0_0\:,\:t^0_1\,)\rightarrow(\,t^1_0\:,\:t^1_1\,)$. So, we have:\\
For $n=1\,$: \[t^2_0=4277.457     \qquad;\qquad t^2_1=3722.542\]           
For $n=2\,$:\[t^3_0=4314.417     \qquad;\qquad t^3_1=3685.582\]
and so on \[\vdots\quad\vdots\quad\vdots\qquad\qquad;\qquad\qquad\vdots\quad\vdots\quad\vdots\]
For $n=9\,$:\[t^{10}_0=4343.410\qquad;\qquad t^{10}_1=3656.589\]

We notice that the values closest to the truth $t=[4000,4000]^{\textsc{t}}$ are obtained for just \textbf{1} iteration, and the subsequent iterations diverge from $t$, where the values of $\:t^n_0$ are increasing, while those of $\:t^n_1$ are decreasing, which is the opposite of what we want.\\ 

We can now understand the unfolding example of the Bell state $\ket{\Psi^+}$ via the IBU technique. In this case, the vectors $m$, $t$ and $\hat{t}_{_{\textsf{IBU}}}$ are \textit{four-dimensional}, so we are going to use an \textit{algorithm}, written with $\mathsf{python}$ language. In doing so, we found the following results for $1,\:5\,,\:10\,,\:100\,,\:1000$ iterations, where we take the prior truth vector $\hat{t}^{\,0}_{_{\textsf{IBU}}}$ as a uniform distribution.
\[\hat{t}^{\,0}_{_{\textsf{IBU}}}\:=\begin{bmatrix}  \;001.000, & 0001.000, & 0001.000, & 0001.000\;\end{bmatrix}^\textsc{t}\]
\[\hat{t}^{\,1}_{_{\textsf{IBU}}}\:=\begin{bmatrix}\;947.003,& 2651.760,& 3255.226,& 1146.008\;\end{bmatrix}^\textsc{t}\]
\[\hat{t}^{\,5}_{_{\textsf{IBU}}}\:=\begin{bmatrix}\;187.168,& 3366.011,& 4151.338,& 0295.481\;\end{bmatrix}^\textsc{t}\]
\[\hat{t}^{10^1}_{_{\textsf{IBU}}}\,=\begin{bmatrix}\;065.849,& 3468.337,& 4302.088,& 0163.724\;\end{bmatrix}^\textsc{t}\]
\[\hat{t}^{10^2}_{_{\textsf{IBU}}}\,=\begin{bmatrix}\;000.001,& 3524.170,& 4400.027,& 0075.801\;\end{bmatrix}^\textsc{t}\]
\[\hat{t}^{10^3}_{_{\textsf{IBU}}}\,=\begin{bmatrix}\quad10^{-50},& 3524.179,& 4400.088,& 0075.731\;\end{bmatrix}^\textsc{t}\]

Note that this is the opposite of the previous example, where the $\hat{t}^{\,1}_{_{\textsf{IBU}}}$ is not the closest vector to the truth $t$. But the iterations converges it more and more to $t$. Also, note that the amount of improvement decreases with each iteration. For our situation, the values stabilize for $n\geqslant 10^{3}$, i.e., subsequent iterations will give the same outcomes. Hence, $\hat{t}^{\,10^{3}}_{_{\textsf{IBU}}}$ is the \textit{maximum approximation} of $t$.
\[t=\begin{bmatrix}\;\mathbf{000}.000, & \mathbf{4000}.000, & \mathbf{4000}.000, & \mathbf{000}.000 \;\end{bmatrix}^\textsc{t}\]

Finally, we can represent the predicted values in a histogram in order to compare the two methods MI and IBU.
\begin{figure}[H]
\centering
\includegraphics[width = 0.8\linewidth]{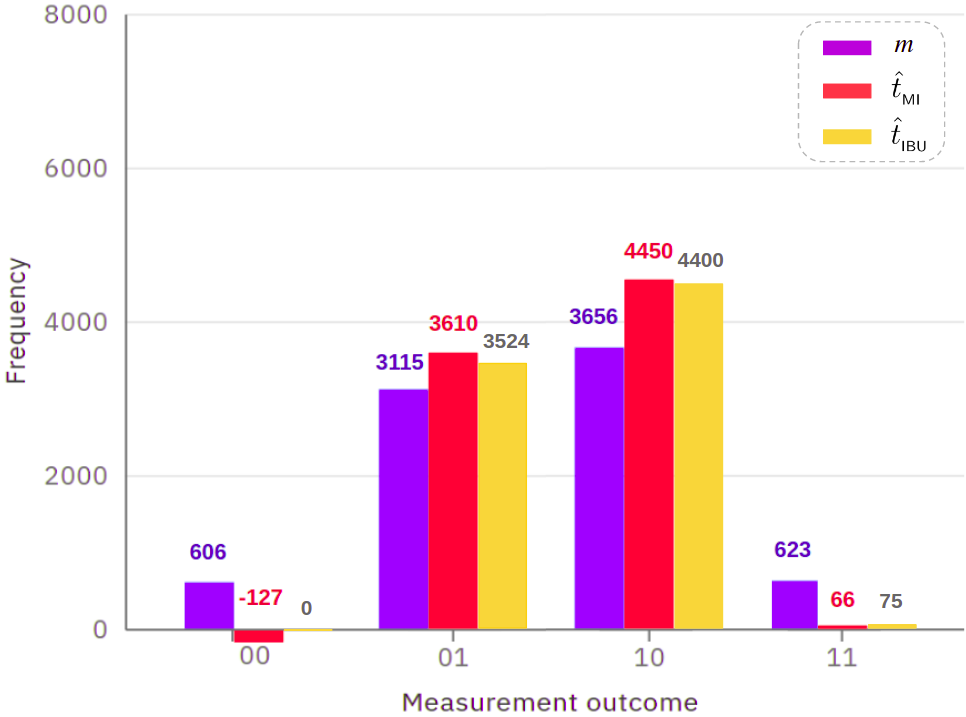}
\caption{Measured and predicted bin counts via MI and IBU for $10^3$ iterations.}
\end{figure}
We notice that in this specific example of $\ket{\Psi^+}$, the predicted values from matrix inversion method $\:\,\hat{t}_{_{\textsf{MI}}}$ and that from IBU $\:\,\hat{t}_{_{\textsf{IBU}}}$, are very close. Furthermore, no negative outcomes for IBU, if we look at state $\ket{00}$, for IBU its error attenuation has reached $100\%$. The reason of the \textit{non-negativeness} of the IBU outcomes, as we have mentioned before, is that the results are probabilities, which are always positive.\\ 

The difference in the behavior of IBU between the $1^{st}$ and $2^{nd}$ example comes from the choice of $\hat{t}^{\,0}_{_{\textsf{IBU}}}$, where in the $1^{st}$ one, both $\hat{t}^{\,0}_{_{\textsf{IBU}}}$ and $t$ where uniform distributions. But for the $2^{nd}$ example $t$ was not uniform, unlike $\hat{t}^{\,0}_{_{\textsf{IBU}}}$. If we took $\hat{t}^{\,0}_{_{\textsf{IBU}}}$ similar to $t$, e.g. $[0,\,1,\,1,\,0]^{\textsc{t}}$, then we get:
\[\hat{t}^{1}_{_{\textsf{IBU}}}\,=\begin{bmatrix}\;0,& 3598.040,& 4401.959,& 0\;\end{bmatrix}^\textsc{t}\]
\[\hat{t}^{10}_{_{\textsf{IBU}}}\,=\begin{bmatrix}\;0,& 3533.524,& 4466.475,& 0\;\end{bmatrix}^\textsc{t}\]
which shows the same behavior as in the $1^{st}$ example. We see that it gives better results than before, for one iteration.
\pagebreak

\section{Simulation of the Unfolding Process with Python }

In this section, we will explain how to simulate the unfolding process with python. To perform unfolding, we need the calibration matrix. So, in this simulation, we will assume that we know this matrix exactly and we will use it in the three unfolding methods: Matrix Inversion, IBU, and  Ignis from \hyperlink{kit}{\bl{Qiskit}}.

It is interesting to see the effect of noise in data, so we will generate some noise using a pseudo-random generator provided by python tools, this will conduct us to simulate the measured data. Then, we will see how to use an arbitrary response matrix in the unfolding methods. The matrix will be constructed with python code, but we know precisely its values. Finally, we will compare the output of the tree unfolding methods with the measured and the true data in the histograms.


\subsection{Noise Generation with Python }
The purpose of this part is to produce noisy data using python tools. The first step consists of generating \textit{pseudo-random numbers} for true data which follow the \textbf{normal law}, named \textsf{true$\_$data}. In a second step, we generate pseudo-random numbers that follow the uniform law, named as \textsf{randnoise}. We will modify the \textsf{randnoise} by using conditions on it using true data the outcome is the reconstructed noisy data. To visualize the result, we will represent it in a histogram.
\begin{center}
\textbf{Python packages}
\end{center}
Before starting, it is necessary to import some packages. We will need only two packages. The first is \textsf{numpy} which provides mathematical computations on arrays and matrices. The second is \textsf{matplotlib} for data visualization.
\definecolor{bg}{rgb}{0.97,0.97,0.97}
\begin{pythoncode}
import numpy as np
import matplotlib.pyplot as plt
\end{pythoncode}

\begin{center}
\textbf{True Data}
\end{center}

It is easier to visualize a little number of data so for the first example we will use $N=100$. The true data is generated using the function \textsf{np.random.randn(N$\_$data)*sigma}, the variable follow the normal law with \textsf{sigma =3}  the result will be clipped to the interval $[-10,10]$ using the function \textsf{np.clip()} and will be named as \textsf{true$\_$data}. The final step is to round the data to the nearest integer numbers by the function \textsf{round()}. 
\definecolor{bg}{rgb}{0.97,0.97,0.97}
\begin{pythoncode}
N_data = 100
sigma = 3
true_data = np.clip(np.random.randn(N_data)*sigma,-10.,10.)
true_data_int = [round(x) for x in true_data]
\end{pythoncode}

Now we can visualize the result by the print command as follow:
 \definecolor{bg}{rgb}{0.97,0.97,0.97}
 \begin{pythoncode}
print("true_data =\n",true_data)
\end{pythoncode}
The output for the \textsf{true$\_$data}:
\definecolor{bg}{rgb}{0.97,0.97,0.97}
\begin{pythoncode}
true_data =
 [ 5.36588542  1.30952955  0.2894924  -5.59047811 -0.83216461 -1.06427694
 -0.24822444 -1.88100203 -0.13145451 -1.43165409 -3.94159426  2.65386714
  2.64395413  5.12871919  0.15010093 -1.21403224 -1.63607984 -4.63943195
  2.9471023  -3.30320289 -3.55513958 -0.6169497   4.45844507  0.7101488
 -3.07135542 -2.1389796   1.8757349  -0.48154009 -2.30650905 -0.69009217
  2.2351688   5.92833235 -3.73236999 -1.87925073 -2.41129828 -7.25724952
 -2.77137607 -3.07162728  3.37193388 -0.3957427  -4.86985634  1.94002636
 -1.06881228 -5.22942311 -1.78994893 -1.76578314 -2.62164689  0.08914145
 -6.7447733  -0.80328559  3.03955033  2.55839352  3.3245625   3.35817197
  4.4626294  -3.35490205  2.53750022 -5.58266859 -1.80865531 -5.74341613
  3.14444254  4.00121346 -0.59224404  5.32393509 -2.02418253  0.4518506
  0.45883711 -3.19258582  1.31383983  5.81693538 -3.07479262  2.69801534
 -0.46352056  5.30888191  1.45136504  2.0286492   1.92948984  0.74726012
 -4.18729051  4.17498872 -4.11200704  0.71568958  1.84223126 -2.51373682
  0.43518964  3.50364686 -0.07231341 -2.66597225 -8.74721326 -2.91552151
 -1.77323622 -1.5492521  -2.87998854  1.1318857  -1.72412526 -0.328363
  2.0372148  -2.56631151 -0.90061822  6.47444803]
\end{pythoncode}
The output for the \textsf{true$\_$data$\_$int}: 
\definecolor{bg}{rgb}{0.97,0.97,0.97}
\begin{pythoncode}
true_data_int =
 [5, 1, 0, -6, -1, -1, 0, -2, 0, -1, -4, 3, 3, 5, 0, -1, -2, -5, 3, -3, -4, -1,
4, 1, -3, -2, 2, 0, -2, -1, 2, 6, -4, -2, -2, -7, -3, -3, 3, 0, -5, 2, -1, -5,
-2, -2, -3, 0, -7, -1, 3, 3, 3, 3, 4, -3, 3, -6, -2, -6, 3, 4, -1, 5, -2, 0, 0,
-3, 1, 6, -3, 3, 0, 5, 1, 2, 2, 1, -4, 4, -4, 1, 2, -3, 0, 4, 0, -3, -9, -3, -2,
-2, -3, 1, -2, 0, 2, -3, -1, 6]
\end{pythoncode}

\begin{center}
\textbf{Noisy Data}
\end{center}
In this part, we want to construct noisy data from true data, this will represent our measured data. Later, in the unfolding part, we will see how to correct this noisy data. So, now let's see how to generate some noise starting from the true data with \textsf{python} tools.

First, we generate $N$ pseudo-random numbers (which represent the noise) that follows the uniform law. The seed is equal to $5$ and the values of this data are in the interval $[0,1]$. 
\definecolor{bg}{rgb}{0.97,0.97,0.97}
\begin{pythoncode}
np.random.seed(5)
randnoise = np.random.uniform(0.,1.,N_data)
\end{pythoncode}
The output for the randnoise:
\definecolor{bg}{rgb}{0.97,0.97,0.97}
\begin{pythoncode}
randnoise  =
       [2.74780505e-01 6.52223125e-01 9.56449511e-01 4.35520556e-01
	7.01325051e-02 5.77314878e-02 8.28710188e-02 9.59707187e-01
	5.40760836e-01 8.37462433e-01 1.70033544e-01 2.60345073e-01
	6.91977512e-01 8.95570328e-01 3.40688484e-01 6.46731980e-02
	8.64119669e-01 2.90872446e-01 7.41082406e-01 1.58033655e-01
	6.94963435e-01 8.41419619e-01 7.27152079e-01 3.59107525e-01
	7.26689751e-01 1.39467124e-01 3.13819115e-01 4.19582757e-01
	8.77212039e-01 1.53740209e-01 8.80124790e-01 7.98964319e-01
	9.71624297e-01 3.67702983e-01 2.04939769e-01 2.40570320e-01
	8.27862801e-01 9.65228149e-01 6.98809998e-01 4.82497042e-01
	2.87049765e-01 8.33687884e-01 8.72179508e-01 9.21315918e-02
	2.15949471e-01 8.31761090e-01 8.48303897e-01 3.14652999e-01
	2.79294597e-01 4.30815022e-01 5.39446500e-01 9.55668150e-02
	8.36912139e-01 5.34734870e-01 7.74967815e-01 2.30836266e-01
	9.65293351e-01 7.51027307e-01 3.43093864e-01 9.48527647e-01
	7.00511779e-01 8.40561085e-01 4.54973059e-02 5.56415411e-02
	7.42737274e-01 3.04686433e-01 5.16784366e-01 1.56262424e-01
	9.77952410e-01 5.02751048e-01 8.29001078e-01 7.40377963e-02
	4.78915452e-01 6.22794804e-02 8.84241431e-01 4.45810179e-01
	6.85499183e-02 7.64962823e-02 5.38792658e-01 7.55664045e-02
	1.83772318e-01 4.36357084e-01 4.97782831e-01 5.83311915e-01
	6.20512681e-01 3.72811500e-01 6.18736583e-01 1.57244658e-01
	2.75508475e-01 7.98718271e-01 1.53089301e-01 2.23322973e-01
	2.42978180e-01 4.79507305e-01 7.45522042e-04 3.03113605e-02
	4.61548152e-01 1.62520685e-01 6.79501804e-01 7.95204587e-01]
\end{pythoncode}
The second step is to build new data using conditions on the real data and the random noise. That means, each value in the coordinate $i$ will be added or subtracted by $1$, this will depend on the condition between the random noise and the true value. The new reconstructed data is the measured data.
\definecolor{bg}{rgb}{0.97,0.97,0.97}
\begin{pythoncode}
reco_data_int = []
for i in range(len(true_data_int)):
    reco_data_int+=[true_data_int[i]]
    if (true_data_int[i]==-10 and randnoise[i]<0.2):
        reco_data_int[i]+=1
    elif (true_data_int[i]==10 and randnoise[i]<0.3):
        reco_data_int[i]-=1
    elif (abs(true_data_int[i]) < 10 and randnoise[i]<0.2):
        reco_data_int[i]-=1
    elif (abs(true_data_int[i]) < 10 and randnoise[i]<0.8):
        reco_data_int[i]+=1 
        pass
    pass
\end{pythoncode}

Before seeing the result let us reproduce the Reconstructed data by hand for the five values using the preview conditions.
\begin{center}
\begin{footnotesize}
\begin{tabular}{|c|c|c|c|c|c|}
\hline 
i & 0 & 1 & 2 & 3 & 4  \\ 
\hline 
True Data & 5 & 1 & 0 & -6 & -1  \\ 
\hline 
Random Noise & 2.74780505e-01 & 6.52223125e-01 & 9.56449511e-01 & 4.35520556e-01 & 7.01325051e-02  \\ 
\hline 
Reconstructed Data & 5+1=6 & 1+1=2 & 0 (pass) & -6+1=-5 &-1-1= -2 \\ 
\hline 
\end{tabular} 
\end{footnotesize}
\end{center}

The output for the Reconstructed Noisy data:
\definecolor{bg}{rgb}{0.97,0.97,0.97}
\begin{pythoncode}
reco_data_int =
[6, 2, 0, -5, -2, -2, -1, -2, 1, -1, -5, 4, 4, 5, 1, -2, -2, -4, 4, -4, -3, -1,
5, 2, -2, -3, 3, 1, -2, -2, 2, 7, -4, -1, -1, -6, -3, -3, 4, 1, -4, 2, -1, -6,
-1, -2, -3, 1, -6, 0, 4, 2, 3, 4, 5, -2, 3, -5, -1, -6, 4, 4, -2, 4, -1, 1, 1,
-4, 1, 7, -3, 2, 1, 4, 1, 3, 1, 0, -3, 3, -5, 2, 3, -2, 1, 5, 1, -4, -8, -2, -3,
-1, -2, 2, -3, -1, 3, -4, 0, 7]
\end{pythoncode}

\begin{center}
\textbf{The Plot}
\end{center}
To visualize the effect of noise and the true data together, we will represent the result in two histograms. The true data $t$ and the measured data $m$ are discretized with $15$ uniform unit width bins spanning the interval $[-10, 10]$. 

The code that we use to draw the histograms is as follow:
\definecolor{bg}{rgb}{0.97,0.97,0.97}
\begin{pythoncode}
m,b,_ = plt.hist(reco_data_int,bins=np.linspace(-10.5,10.5,16),facecolor="orange",
                 edgecolor='none',label='m')
t,b,_ = plt.hist(true_data_int,bins=np.linspace(-10.5,10.5,16),facecolor="none",
                 edgecolor='black',label="t")
plt.legend(frameon=False)
plt.xlabel('Value')
plt.ylabel('Counts')
\end{pythoncode}
\begin{figure}[H]
\centering
\includegraphics[width = 0.7\linewidth]{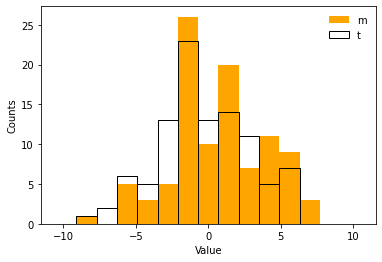}
\caption{Histogram representation of simulated true data and measured data.}
\end{figure}

We can see that the noise changed the real values to the point where it is not possible to see the pattern of the distortion in the histogram. So, we will need to solve this inverse problem by applying the unfolding technique. For the next example, we will show only the histograms\footnote{Because it is not practical to see data in dozens of pages.}. But first, let us see how to construct the response matrix with python.

\subsection{Constructing R}
We assume that we know our calibration matrix and will use the previews information about true and measured data. The dimension of $m$ is equal to the dimension of $t$ and they are related by the equation $m=R\,t$.
 
The elements of the matrix are constructed with the following python code:
\definecolor{bg}{rgb}{0.97,0.97,0.97}
\begin{pythoncode}

R = []
for i in range(len(m)):
    R += [[]]
    for j in range(len(t)):
        if (i==0):
            if (j==0):
                R[i] += [0.75]
            elif (j==1):
                R[i] += [0.25]
            else:
                R[i] += [0.]
        elif (i==len(t)-1):
            if (j==len(t)-1):
                R[i] += [0.75]
            elif (j==len(t)-2):
                R[i] += [0.25]
            else:
                R[i] += [0.]
        else:
            if (abs(i-j)==0):
                R[i] += [0.5]
            elif (abs(i-j)==1):
                R[i] += [0.25]
            else:
                R[i] += [0.] 
                    \end{pythoncode}
The index $i$ takes values from 0 to \textsf{dim(m)} and the index $j$ takes values from 0 to \textsf{dim(t)}. In our case, the calibration matrix will have dimension $ 15 \times15 $.
                    
  The output of the calibration matrix: 
\definecolor{bg}{rgb}{0.97,0.97,0.97}
 \begin{pythoncode}
[[0.75, 0.25, 0.0, 0.0, 0.0, 0.0, 0.0, 0.0, 0.0, 0.0, 0.0, 0.0, 0.0, 0.0, 0.0],
 [0.25, 0.5, 0.25, 0.0, 0.0, 0.0, 0.0, 0.0, 0.0, 0.0, 0.0, 0.0, 0.0, 0.0, 0.0],
 [0.0, 0.25, 0.5, 0.25, 0.0, 0.0, 0.0, 0.0, 0.0, 0.0, 0.0, 0.0, 0.0, 0.0, 0.0],
 [0.0, 0.0, 0.25, 0.5, 0.25, 0.0, 0.0, 0.0, 0.0, 0.0, 0.0, 0.0, 0.0, 0.0, 0.0],
 [0.0, 0.0, 0.0, 0.25, 0.5, 0.25, 0.0, 0.0, 0.0, 0.0, 0.0, 0.0, 0.0, 0.0, 0.0],
 [0.0, 0.0, 0.0, 0.0, 0.25, 0.5, 0.25, 0.0, 0.0, 0.0, 0.0, 0.0, 0.0, 0.0, 0.0],
 [0.0, 0.0, 0.0, 0.0, 0.0, 0.25, 0.5, 0.25, 0.0, 0.0, 0.0, 0.0, 0.0, 0.0, 0.0],
 [0.0, 0.0, 0.0, 0.0, 0.0, 0.0, 0.25, 0.5, 0.25, 0.0, 0.0, 0.0, 0.0, 0.0, 0.0],
 [0.0, 0.0, 0.0, 0.0, 0.0, 0.0, 0.0, 0.25, 0.5, 0.25, 0.0, 0.0, 0.0, 0.0, 0.0],
 [0.0, 0.0, 0.0, 0.0, 0.0, 0.0, 0.0, 0.0, 0.25, 0.5, 0.25, 0.0, 0.0, 0.0, 0.0],
 [0.0, 0.0, 0.0, 0.0, 0.0, 0.0, 0.0, 0.0, 0.0, 0.25, 0.5, 0.25, 0.0, 0.0, 0.0],
 [0.0, 0.0, 0.0, 0.0, 0.0, 0.0, 0.0, 0.0, 0.0, 0.0, 0.25, 0.5, 0.25, 0.0, 0.0],
 [0.0, 0.0, 0.0, 0.0, 0.0, 0.0, 0.0, 0.0, 0.0, 0.0, 0.0, 0.25, 0.5, 0.25, 0.0],
 [0.0, 0.0, 0.0, 0.0, 0.0, 0.0, 0.0, 0.0, 0.0, 0.0, 0.0, 0.0, 0.25, 0.5, 0.25],
 [0.0, 0.0, 0.0, 0.0, 0.0, 0.0, 0.0, 0.0, 0.0, 0.0, 0.0, 0.0, 0.0, 0.25, 0.75]]
\end{pythoncode}
We can also introduce each value of a calibration matrix by hand like the following example:
\definecolor{bg}{rgb}{0.97,0.97,0.97}
\begin{pythoncode}

R = [[0.75, 0.25, 0.00, 0.00],
     [0.25, 0.50, 0.25, 0.00],
     [0.00, 0.25, 0.50, 0.25],
     [0.00, 0.00, 0.25, 0.75]]
\end{pythoncode}


\subsection{Unfolding Methods}

In this part, we will use three unfolding methods to correct the measured data. Each unfolding method can be defined as a function in \textsf{python}. 
Here, we will change the number of data   $\mathsf{N\_data}$ (from 100 to 10000) and the number of bins (from 15 to 21) and also the measured data (Reconstructed Data).  All the other parameters from the previews code will remain unchanged. 
The new histograms of measured and true data are represented as follow: 

\begin{figure}[H]
\centering
\includegraphics[width = 0.7\linewidth]{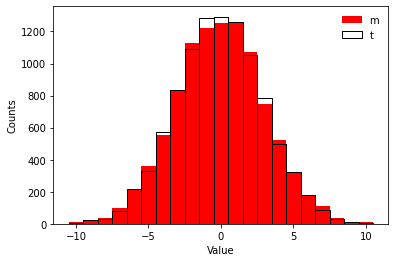}
\caption{Histogram representation of simulated true data and measured data.}
\end{figure}

\pagebreak
\begin{center}
\textbf{Matrix Inversion}
\end{center}
One of the methods to unfold the true distribution from experimental data is to calculate the inverse of the response matrix, then we will get the true distribution using the following equation $\hat{t}=R^{-1}m$.

To calculate the inverse of a matrix in \textsf{python} we use $\mathsf{np.linalg.inv()}$ function. The multiplication between the inverse of the matrix and the measured distribution vector is performed by the function: $\mathsf{np.matmul()}$. The  $\mathsf{np.ravel}$ return a contiguous flattened \textsf{1-D} array, containing all the elements of the input.
 \definecolor{bg}{rgb}{0.97,0.97,0.97}
\begin{pythoncode}
def MatrixInversion(ymes,Rin):
    return np.ravel(np.matmul(np.linalg.inv(np.matrix(Rin)),ymes))
                    \end{pythoncode}
                    
\begin{center}
\textbf{Ignis Function}
\end{center}
The ignis function uses the Sequential Least SQuares Programming (SLSQP) method. This code can be found in \textsf{qiskit-ignis}\footnote{ \url{https://github.com/Qiskit/qiskit-ignis/blob/master/qiskit/ignis/mitigation/measurement/filters.py}}. 
\definecolor{bg}{rgb}{0.97,0.97,0.97}
\begin{pythoncode}
def fun(x,ymes,Rin):
    mat_dot_x = np.ravel(np.matmul(Rin,x))
    return sum((ymes - mat_dot_x)**2)

def Ignis(ymes,Rin):
    x0 = np.random.rand(len(ymes))
    x0 = x0 / sum(x0)
    nshots = sum(ymes)
    cons = ({'type': 'eq', 'fun': lambda x: nshots - sum(x)})
    bnds = tuple((0, nshots) for x in x0)
    res = minimize(fun, x0, method='SLSQP',constraints=cons,
    bounds=bnds, tol=1e-6,args=(ymes,Rin))
 return res.x
 \end{pythoncode}

\begin{center}
\textbf{IBU Function}
\end{center}
The iterative Bayesian unfolding function is defined in this code as IBU. Its input parameters are the measured data $\mathsf{ymes}$ and the $\mathsf{t0}$ is an initial truth spectrum of the process, in our case, it is an array of ones. The third parameter is the response matrix, and finally $\mathsf{n}$ is the number of iterations. The lines from 3 to 9 are for initialization. The IBU method start in line ten.
The code for the IBU function is as follow:  
\definecolor{bg}{rgb}{0.97,0.97,0.97}
\begin{pythoncode}
def IBU(ymes,t0,Rin,n):
  
    tn = t0
    for q in range(n):
        out = []
        for j in range(len(t0)):
            mynum = 0.
            for i in range(len(ymes)):
                myden = 0.
                for k in range(len(t0)):
                    myden+=Rin[i][k]*tn[k]
                    pass
                mynum+=Rin[i][j]*tn[j]*ymes[i]/myden
                pass
            out+=[mynum]
        tn = out
        pass
    
    return tn
 \end{pythoncode}
We recall the expression of this unfolding method:

\begin{center}
\includegraphics[width = 1\linewidth]{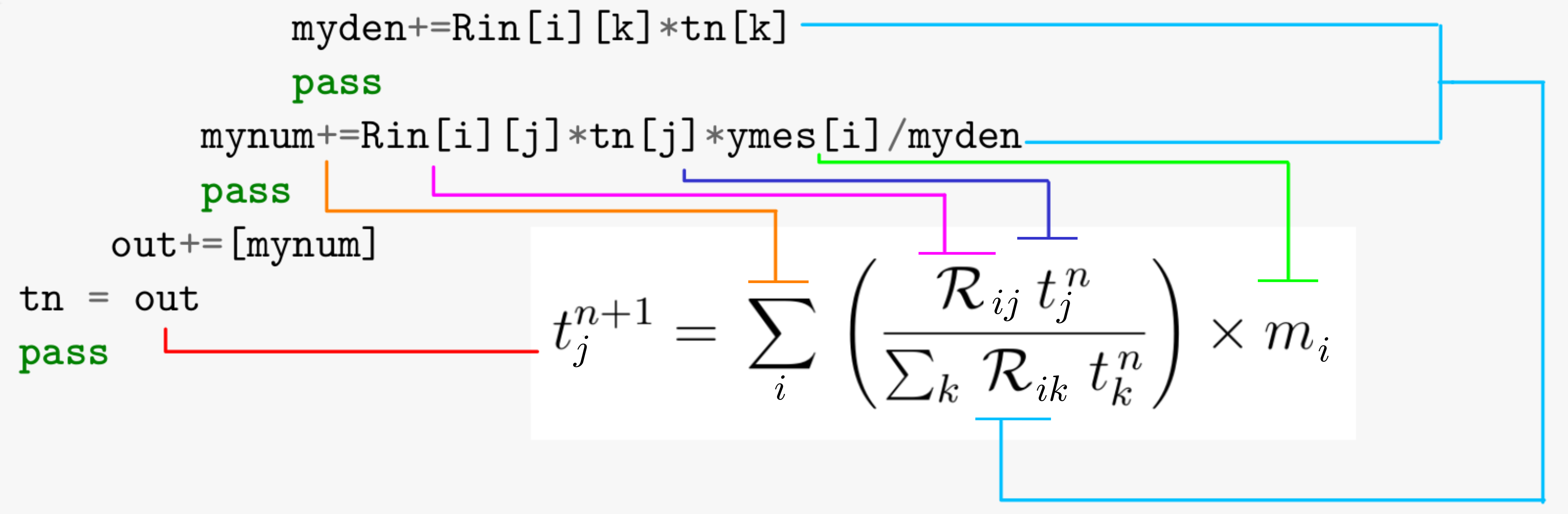}
\end{center}
Now, let us unfold by including the measured data and the calibration matrix to our functions. The matrix inversion and ignis have two parameters \textsf{m} and \textsf{R}. For  IBU it has four parameters: first the measured data $\mathsf{m}$, the initial truth data $\mathsf{t0}$ that we choose it as a vector of elements 1 (\textit{uniform distribution}) and which have the same dimension as the measured data, the response matrix, and finally the number of iterations \textsf{n=10}.
\definecolor{bg}{rgb}{0.97,0.97,0.97}
 \begin{pythoncode}
MI    = MatrixInversion(m, R)
ignis = Ignis(m, R)
ibu   = IBU(m, np.ones(len(m)), R, 10) 
 \end{pythoncode}

To see the result of the three unfolding methods, we use the \textsf{print} function.
\pagebreak

\begin{center}
Matrix Inversion
\end{center}
\definecolor{bg}{rgb}{0.97,0.97,0.97}
\begin{pythoncode} 
 print(MI)

 [2.36190476e+01 -3.48571429e+01  1.50095238e+02 -1.21333333e+02
 4.80571429e+02  1.90476190e-01  9.67047619e+02  2.89714286e+02
 1.77752381e+03  6.71238095e+02  1.77600000e+03  7.88761905e+02
 1.65447619e+03  1.82285714e+02  9.60952381e+02 -2.01904762e+01
 3.83428571e+02 -2.26666667e+01  1.01904762e+02 -3.71428571e+01
 2.83809524e+01]
\end{pythoncode}
 
\begin{center}
Ignis
\end{center}
\definecolor{bg}{rgb}{0.97,0.97,0.97}
\begin{pythoncode} 
print(ignis)

[4.6175683800  21.291404430  59.2658346200  0.             334.52132545
165.440539820  787.74328185  478.197504920  1584.36645844  864.83510512
1585.82646343  971.91899933  1481.55800967  342.012482900  816.99820065
105.680456820  277.58084116  61.4881026600  40.7416841000  0.
15.9157361800]
 \end{pythoncode} 
 \begin{center}
IBU
\end{center}
\definecolor{bg}{rgb}{0.97,0.97,0.97}
 \begin{pythoncode} 
print(ibu)

[5.259372656831409, 25.388591380196175, 33.557337461054985,
67.96709437048914, 208.8410238062819, 356.97817161684395,
521.2370635355416, 822.9011004066444, 1193.1006184446076,
1246.8670968072922, 1232.6156342943655, 1329.7638967204725,
1095.5138161773816, 719.1611893853731, 514.3011592136613,
299.016654127236, 181.16794027605107, 100.54732163607031,
23.88223196059284, 9.002561105297106, 12.930124617715322]
 
 \end{pythoncode} 
\begin{center}
The True Data 
\end{center} 
\definecolor{bg}{rgb}{0.97,0.97,0.97}
  \begin{pythoncode} 
bins=np.linspace(-10.5,10.5,22)
np.histogram(true_data,bins)

(array([   5,   23,   32,   78,  219,  329,  571,  837, 1089, 1287, 1292,
        1260, 1052,  783,  500,  324,  182,   83,   32,   14,    8]),
 array([-10.5,  -9.5,  -8.5,  -7.5,  -6.5,  -5.5,  -4.5,  -3.5,  -2.5,
         -1.5,  -0.5,   0.5,   1.5,   2.5,   3.5,   4.5,   5.5,   6.5,
          7.5,   8.5,   9.5,  10.5]))
 \end{pythoncode} 
 
 The plot of the truth and measured data with the result of the three unfolding methods.
\begin{figure}[H]
\centering
\includegraphics[width = 0.85\linewidth]{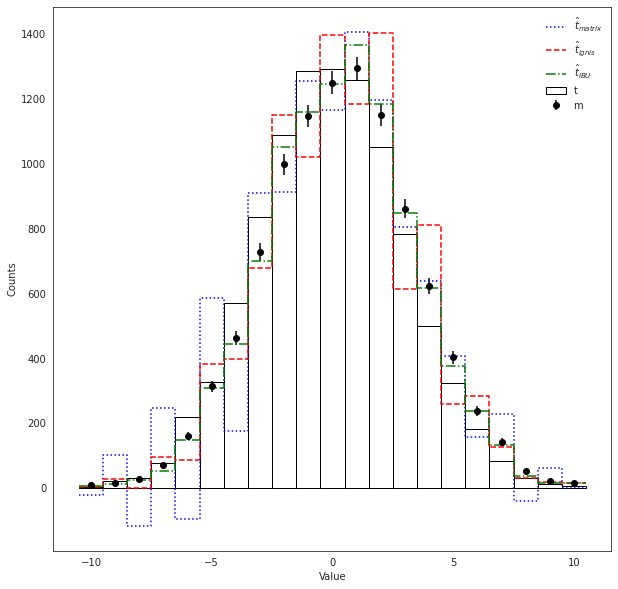}
\caption{Unfolding results of a Gaussian distribution using three methods: Matrix inversion ($\hat{t}_{matrix}$), ignis ($\hat{t}_{ignis}$), and IBU ($\hat{t}_{IBU}$) with $10$ iterations and a uniform prior truth spectrum.}
\end{figure}

We can clearly see that the IBU results are close to the real spectrum compared to matrix inversion and ignis. We can also see that the results of ignis and matrix inversion show large oscillations. We also notice that matrix inversion give negative outcomes, while IBU does not encounter this problems. Finally, we can deduce that the IBU  leads us to better results compared to the other two methods in this simulation.

\pagebreak
\section{Unfolding Readout Errors from the Yorktown IBM Q Machine}
The central objective of the article \cite{Nac} published in 2020 by \textsc{Benjamin Nachman} et al., is to provide a more powerful technique to mitigating readout errors in the field of QIS. This technique is the \textit{iterative Bayesian unfolding} IBU.
They studied three techniques: \textsf{matrix inversion, ignis}, and \textsf{IBU} in order to compare them. This study is based on \textsf{Qiskit Simulator}\footnote{A local simulation is provided by Aer. In the article \cite{Nac} they used the so-called “$\mathsf{qasm\_simulator}$”.}, to mitigate readout errors for 5 qubits generated in a gaussian distribution. The unfolding was done through a \textbf{realistic} response matrix (from the \textsc{Johannesburg IBM Q} Machine), and also for an \textbf{artificial} response matrix, and they demonstrated the advantages of the IBU over other methods.\\ 

But it is very interesting to explore the behavior of unfolding techniques, through \textbf{real noise} coming from a \textbf{real quantum computer}. The \textit{International Business Machines corporation} (IBM) provide us with 9 devices that we can use. In the following, we will be using the \textsc{Yorktown IBM Q} machine (a.k.a. IBMQX2). The reason we chose it is because it has the highest value of readout errors (average $5.826\times10^{-2}$).\\

We will start our study with a \textit{uniform distribution} as the true distribution, then we will tackle the \textit{Gaussian distribution}, both are based on 5 qubits. The number of shots taken in all experiments is fixed at $8192$\footnote{This is the maximum number allowed.}.\\

Another interesting question we are trying to answer is how do these unfolding methods behave for a system beyond 5-qubits. 
We will use for this purpose the 16 \textsc{Melbourne IBMQ} machine, which is a 15-Qubit quantum computer, we will study the unfolding a 7-qubit system, in the specific case of a \textit{uniform distribution}. In this situation, we are dealing with readout noise described by a $2^7\times2^7$ dimensional \textit{response matrix}, which represents a challenge for any \textit{unfolding technique}.\\

Our work can be schematized in 4 steps:
\begin{enumerate}
\item Constructing the response matrix $\mathcal{R}$.
\item Generating the desired distribution to be the truth.
\item Distortion of the true distribution by readout noise.
\item Unfolding readout errors via \textsf{MI, Ignis, IBU}.
\end{enumerate}

\subsection{Construction of the Response Matrix from the Yorktown IBM Q Machine}
In this part, we will explain how to extract a \textbf{calibration matrix} from a quantum computer. In our work, we will use the \textsc{IBM 5 Yorktown} system. This system contains 5 qubits, and we will use all the five qubits to build this matrix. So the calibration matrix we want to build has dimension  $ 2^{5}\times 2^{5} $. The process will be done in IBM \textsc{quantum Lab}\footnote{\url{https://quantum-computing.ibm.com}}. The response matrix will be applied later to unfold distorted distributions by readout noise.

Now, let us explore how to construct the calibration matrix of the IBM Yorktown system step by step.

\begin{center}
\textbf{Step 1: Import the packages }
\end{center}

The first step as always is to import the packages that we will need in our code. When we open a new file in the IBM \textsc{quantum lab}, a list of packages will be available in the notebook.
\definecolor{bg}{rgb}{0.97,0.97,0.97}
\begin{pythoncode}
import numpy as np
# Importing standard Qiskit libraries
from qiskit import QuantumCircuit, transpile, Aer, IBMQ
from qiskit.tools.jupyter import *
from qiskit.visualization import *
from ibm_quantum_widgets import *
# Loading your IBM Quantum account(s)
IBMQ.load_account()
\end{pythoncode}

We will add additional packages as we need more qiskit modules to construct the calibration matrix. The imported modules are qiskit ignis, and the measurement calibration for the complete measurement calibration. The \textsf{execute} module is to execute the circuit, \textsf{matplotlib} and \textsf{seaborn} to visualize the results.   

\definecolor{bg}{rgb}{0.97,0.97,0.97}
\begin{pythoncode}
import qiskit
from qiskit import QuantumRegister
# Execute a list of circuits or pulse schedules on a backend
from qiskit import execute 
# Measurement calibration modules.
from qiskit.ignis.mitigation.measurement import complete_meas_cal
from qiskit.ignis.mitigation.measurement import CompleteMeasFitter
# For data visualization
import matplotlib.pyplot as plt
import seaborn as sns

\end{pythoncode}
\pagebreak

Now let us see how to construct the \textit{calibration circuits} with 5-qubits for the ibmq Yorktown system.
\begin{center}
\textbf{Step 2: Construct the calibration circuits}
\end{center}
The first step after importing the packages is to access to quantum devices by a provider. All IBMQ providers are specified by a \hyperlink{hb}{\bl{hub}}, group, and project. To access a given provider we should use the \textsf{get$\_$provider()} function of the IBMQ account. In our case, our provider gives us access to the public IBM Quantum devices available to all IBM quantum users. 
\definecolor{bg}{rgb}{0.97,0.97,0.97}
\begin{pythoncode}
provider = IBMQ.get_provider(hub='ibm-q')
# List of all available providers
IBMQ.providers()
\end{pythoncode}
To see all available quantum computers, we use the following function:
\definecolor{bg}{rgb}{0.97,0.97,0.97}
\begin{pythoncode}
provider.backends(simulator = False)
\end{pythoncode}
We get the whole list of systems without simulators:
\definecolor{bg}{rgb}{0.97,0.97,0.97}
\begin{pythoncode}

[<IBMQBackend('ibmqx2') from IBMQ(hub='ibm-q', group='open', project='main')>,
 <IBMQBackend('ibmq_16_melbourne') from IBMQ(hub='ibm-q', group='open', project='main')>,
 <IBMQBackend('ibmq_armonk') from IBMQ(hub='ibm-q', group='open', project='main')>,
 <IBMQBackend('ibmq_athens') from IBMQ(hub='ibm-q', group='open', project='main')>,
 <IBMQBackend('ibmq_santiago') from IBMQ(hub='ibm-q', group='open', project='main')>,
 <IBMQBackend('ibmq_lima') from IBMQ(hub='ibm-q', group='open', project='main')>,
 <IBMQBackend('ibmq_belem') from IBMQ(hub='ibm-q', group='open', project='main')>,
 <IBMQBackend('ibmq_quito') from IBMQ(hub='ibm-q', group='open', project='main')>,
 <IBMQBackend('ibmq_manila') from IBMQ(hub='ibm-q', group='open', project='main')>]
\end{pythoncode}


Now we will generate a list of measurement calibration circuits for the full Hilbert space. Each circuit creates a basis state. For 5 qubits, we get $2^5$ calibration circuits.
\definecolor{bg}{rgb}{0.97,0.97,0.97}
\begin{pythoncode}
qr = qiskit.QuantumRegister(5)
qubit_list = [0,1,2,3,4]
meas_calibs, state_labels = complete_meas_cal(qubit_list, qr, circlabel='mcal')
\end{pythoncode}
 \textsf{ibmq$\_$5$\_$yorktown} system (\textsf{ibmqx2}) contains five qubits, and we want to use all the qubits to build our calibration matrix. So, we need five quantum registers. The qubit list is a list of qubits to perform the measurement on, and the last function return a list of objects containing the calibration circuits, and a list of the calibration state labels.

\begin{center}
\textbf{Step 3: Calculate the Calibration Matrix}
\end{center}
Now, we can execute the calibration circuits, we choose the \textsf{ibmq\_5\_yorktown} backend to run our circuit, and we  specify the maximum number of shots\footnote{How many times an algorithm is run to get a probability distribution of results.}. 
\definecolor{bg}{rgb}{0.97,0.97,0.97}
\begin{pythoncode}
backend = provider.get_backend('ibmq_5_yorktown')
job = qiskit.execute(meas_calibs, backend, shots=8192)
\end{pythoncode}
To get the result from the \hyperlink{jb}{\bl{job}}, we use the \textsf{result()} function:
\definecolor{bg}{rgb}{0.97,0.97,0.97}
\begin{pythoncode}
 cal_results=job.result()
 counts=cal_results.get_counts()  
\end{pythoncode}
After we run the calibration circuits and obtain the results \textsf{cal\_results}, we can construct the calibration matrix and it will be ordered according to the \textsf{state\_labels}.
\definecolor{bg}{rgb}{0.97,0.97,0.97}
\begin{pythoncode}
meas_fitter = CompleteMeasFitter(cal_results, state_labels, circlabel='mcal')
R= meas_fitter.cal_matrix
print(R)
\end{pythoncode}
Finally, our response matrix for the IBM Yorktown quantum computer is as follows:
\definecolor{bg}{rgb}{0.97,0.97,0.97}
\begin{pythoncode}
array([[7.23632812e-01, 1.22558594e-01, 3.18603516e-02, ...,
        2.44140625e-04, 0.00000000e+00, 0.00000000e+00],
       [7.58056641e-02, 7.46582031e-01, 3.05175781e-03, ...,
        1.46484375e-03, 0.00000000e+00, 1.22070312e-04],
       [3.51562500e-02, 6.22558594e-03, 7.75512695e-01, ...,
        0.00000000e+00, 9.76562500e-04, 0.00000000e+00],
       ...,
       [0.00000000e+00, 0.00000000e+00, 0.00000000e+00, ...,
        6.40136719e-01, 4.63867188e-03, 3.78417969e-02],
       [0.00000000e+00, 0.00000000e+00, 0.00000000e+00, ...,
        3.54003906e-03, 6.21093750e-01, 8.39843750e-02],
       [0.00000000e+00, 0.00000000e+00, 0.00000000e+00, ...,
        1.48925781e-02, 8.98437500e-02, 6.16088867e-01]])
\end{pythoncode}
To better visualize the result we will use the code provided with the article to represent the matrix in the form of a heatmap.
\begin{center}
\textbf{Step 4: Heatmap Representation of the Response Matrix}
\end{center}
First, we need some functions that will help us to represent the heatmap.

The following code allows us to construct the state labels for the heatmap. Let us print the output to better understand what this function is doing.
\definecolor{bg}{rgb}{0.97,0.97,0.97}
\begin{pythoncode}
nqbits = 5
def mybin(input,q,r=1):
  if (r==0):
    return bin(input).split('b')[1].zfill(q) 
  else:
    return bin(input).split('b')[1].zfill(q)[::-1]  
xvals = []
xlabs = []
for i in range(2**nqbits):
  xvals+=[i+0.5]
  xlabs+=[r'$|'+mybin(i,nqbits)+r'\rangle$']
  pass
\end{pythoncode} 
The \textsf{bin()} function returns the binary version of a specified integer. The result will always start with the prefix \textsf{0b}. For example, if \textsf{input=7} the result will be as follow:
\definecolor{bg}{rgb}{0.97,0.97,0.97}
\begin{pythoncode}
'0b111'
\end{pythoncode} 
We do not need the prefix \textsf{0b} so we will divide the result and we will retain  the second part in position 1 (position 0 is for the first part \textsf{0b}) 
\definecolor{bg}{rgb}{0.97,0.97,0.97}
\begin{pythoncode}
bin(7).split('b')[1]
\end{pythoncode}
And the result is as follows: 
\definecolor{bg}{rgb}{0.97,0.97,0.97}
\begin{pythoncode}
'111'
\end{pythoncode}
And the \textsf{zfill} will fill the string with zeros until it is \textsf{(q)} characters long, for example, \textsf{zfill(5)} will give us:
\definecolor{bg}{rgb}{0.97,0.97,0.97}
\begin{pythoncode}
bin(7).split('b')[1].zfill(5)
\end{pythoncode}
That will give us: 
\definecolor{bg}{rgb}{0.97,0.97,0.97}
\begin{pythoncode}
'00111'
\end{pythoncode}
Finally, let us see how the \textsf{mybin()} function works. For \textsf{input=7}, \textsf{q=5}, and \textsf{r=1} and \textsf{r=0}:
\definecolor{bg}{rgb}{0.97,0.97,0.97}
\begin{pythoncode}
mybin(7,5,0)
mybin(7,5,1)
\end{pythoncode}
We get the following result:
\definecolor{bg}{rgb}{0.97,0.97,0.97}
\begin{pythoncode}
'11100'
'00111'
\end{pythoncode}
The following code allows us to generate labels for the heatmap:
\definecolor{bg}{rgb}{0.97,0.97,0.97}
\begin{pythoncode}
response_normalized = np.zeros((2**nqbits,2**nqbits))
response_normalized_reversed = np.zeros((2**nqbits,2**nqbits))
response_labels = np.zeros((2**nqbits,2**nqbits))
for i in range(2**nqbits):
  mysum = 0.
  for j in range(2**nqbits):
    mysum+=meas_fitter.cal_matrix[j,i]
    pass
  for j in range(2**nqbits):
    response_labels[j][i] = '0'
    if (mysum > 0):
      response_normalized[j,i] = 100.*meas_fitter.cal_matrix[j,i]/mysum
      response_normalized_reversed[i,j] = response_normalized[j,i]
      response_labels[j][i] = '%1.0f' % (response_normalized[j,i])
      pass
    pass
  pass
\end{pythoncode}
Finally, we used this code to represent the response matrix for the Yorktown quantum computer in heatmap representation.
\definecolor{bg}{rgb}{0.97,0.97,0.97}
\begin{pythoncode}
f = plt.figure(figsize=(20, 20))
sns.set(font_scale=1.5)
ax = sns.heatmap(R,annot=response_labels,annot_kws={"size": 15}
,cmap="rocket_r")

ax.figure.axes[-1].yaxis.label.set_size(18)
cbar = ax.collections[0].colorbar
cbar.set_label('Pr(Measured | True) [%]', labelpad=30)

plt.xticks(xvals,xlabs,rotation='vertical',fontsize=15)
plt.yticks(xvals,xlabs,rotation='horizontal',fontsize=15)
plt.xlabel('True',fontsize=18)
plt.ylabel('Measured',fontsize=18)

plt.subplots_adjust(bottom=0.18,left=0.18)
plt.savefig("York_Calibration.pdf",bbox_inches='tight')
\end{pythoncode}

\begin{figure}[H]
\centering
\includegraphics[width = 1\linewidth]{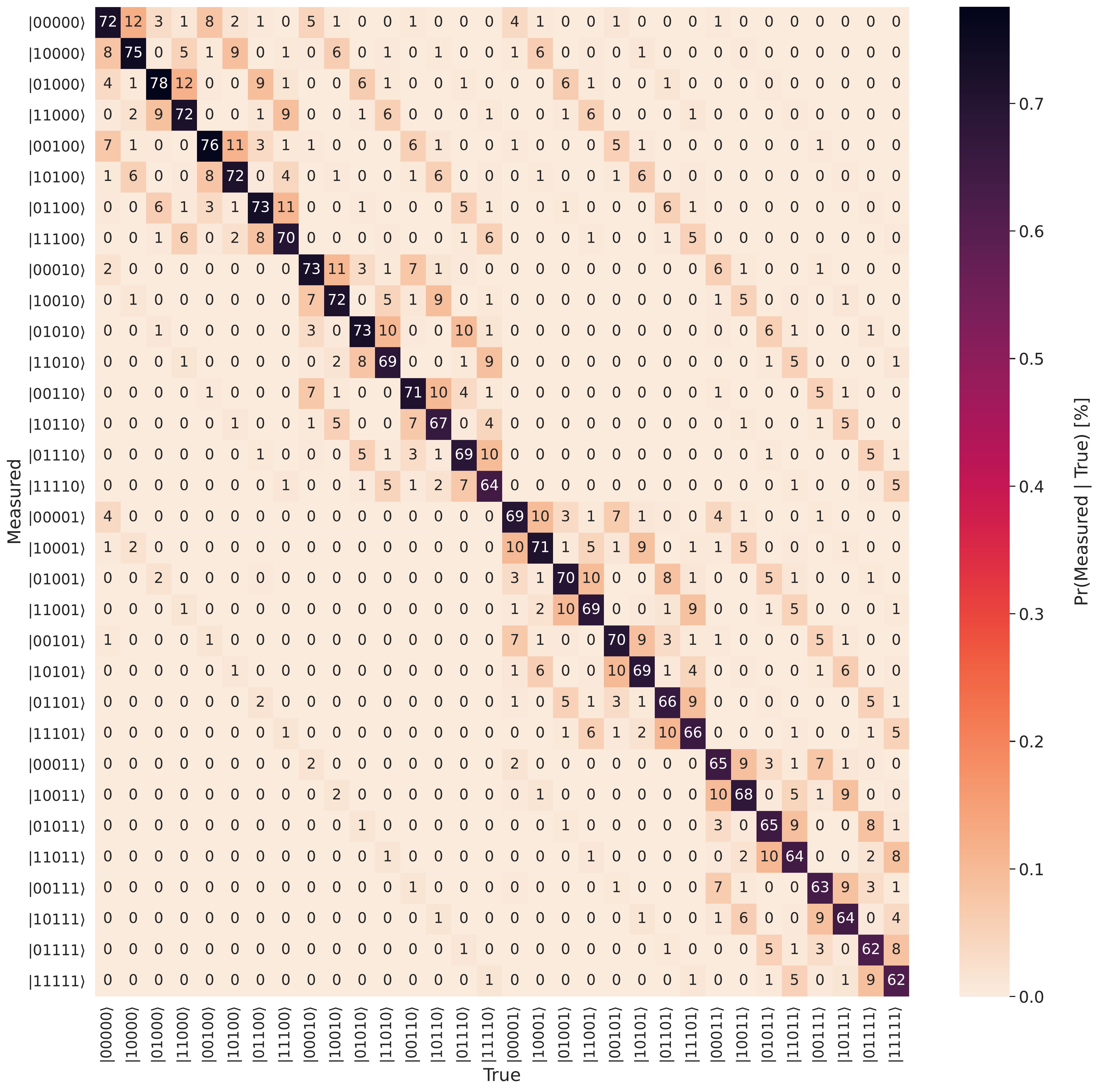}
\caption{Response matrix for $5$ qubits from the IBM Q Yorktown machine.}
\end{figure}

This is the result of our calibration matrix from the Yorktown machine for $5$ qubits. The dimension of this matrix is $2^5 \times 2^5$. First, we can see that the probability is high in the diagonal component, when the measured and true states are the same. Secondly, we observe that the diagonal stripes with transition probabilities of about $4-11 \%$  are the result of the qubit flipping from $0$ to $1$ and vice versa. For a number of flips equal to $2$ or more, the probability is almost zero in this transition matrix.
\subsection{Unfolding a Uniform Distribution}
In this part, we will use the unfolding methods on measured data that is obtained from the Yorktown quantum computer. First, we construct a quantum circuit, theoretically, this circuit will generate a uniform distribution (but quantum computer is noisy so it will destroy the distribution). Then, the circuit will be launched in the IBM Yorktown computer, and after the data collection, we will use our calibration matrix. Finally we will use the code provided by Ben. Nachman et al., to unfold our data by the three techniques: Matrix inversion, Ignis and IBU.
\begin{center}
\textbf{ Packages  }
\end{center}
First, we complete our needed packages, some of them are needed to build our quantum circuit and other packages are for data visualization.

\definecolor{bg}{rgb}{0.97,0.97,0.97}
\begin{pythoncode}
from qiskit import QuantumCircuit, QuantumRegister
from qiskit.circuit import ClassicalRegister
#---------------------------------------------------------------
# Packages for ploting 
from matplotlib.gridspec import GridSpec
from mpl_toolkits.axes_grid1 import make_axes_locatable
import matplotlib.lines as mlines
#---------------------------------------------------------------
from scipy.optimize import minimize

\end{pythoncode}

\begin{center}
\textbf{ Generation of a uniform distribution with a quantum circuit  }
\end{center}
The quantum circuit that we want to construct is five \textsc{Hadamard} gates in parallel. Initially, all the five qubits are in the $\ket{0}$ state. When we apply a \textsc{Hadamard} gate on the first qubit it will give us an \textit{equal superposition} state $\ket{+}=\frac{(\ket{0}+\ket{1})}{\sqrt{2}}$. For five \textsc{Hadamard} gates in parallel acting on $\ket{0}^{\otimes 5}$, we have the tensor product $\mathbb{H}^{\otimes 5}\ket{0}^{\otimes 5}= \frac{1}{\sqrt{2^{5}}} \sum_{x}\ket{x}$.\\

The code below is for creating this quantum circuit. We need five quantum registers and five classical registers. We first apply a \textsc{Hadamard} gate in each quantum register. Then we measure each qubit and we store the result in the classical registers.   
\definecolor{bg}{rgb}{0.97,0.97,0.97}
\begin{pythoncode}
qreg_q = QuantumRegister(5, 'q')
creg_c = ClassicalRegister(5, 'c')
circuit = QuantumCircuit(qreg_q, creg_c)
circuit.h(qreg_q[0])
circuit.h(qreg_q[1])
circuit.h(qreg_q[2])
circuit.h(qreg_q[3])
circuit.h(qreg_q[4])
circuit.measure(qreg_q[0], creg_c[0])
circuit.measure(qreg_q[1], creg_c[1])
circuit.measure(qreg_q[2], creg_c[2])
circuit.measure(qreg_q[3], creg_c[3])
circuit.measure(qreg_q[4], creg_c[4])

circuit.draw(output='mpl')
\end{pythoncode}
The circuit diagram of five Hadamard gates in parallel is displayed as follow: 
\begin{figure}[H]
\centering
\includegraphics[width = 0.6\linewidth]{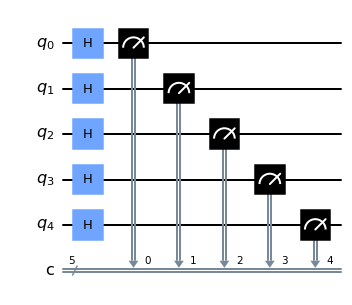}
\caption{A circuit that generates a uniform distribution based on $5$ qubits.}
\end{figure}

This circuit will create a uniform distribution, where the amplitude of each basis state will be $a=\frac{1}{\sqrt{2^{5}}}\approx 0.177$ and the equal probability is $\mathbb{P}= |a|^2 \approx 0.031.$ 

\begin{center}
\textbf{ Executing the circuit in the IBM Yorktown QC }
\end{center}
We will now run the circuit in the Yorktown quantum computer with 8192 shots. If the computer has no noise, we will have equal probabilities. That means each state among the $ 2^5 = 32 $ states will have the same number of counts and it is equal to 8192/32 = 256. In a real quantum computer, this is not the case because it is very noisy so the result will not be what we expect. This is why unfolding is so important to get closer to the true data.
The following code allows us to execute the circuit in the IBM Yorktown system.
\definecolor{bg}{rgb}{0.97,0.97,0.97}
\begin{pythoncode}
# We use the Yorktown quantum computer
backend = provider.get_backend('ibmqx2')
# Number of shots to run the program (experiment); maximum is 8192 shots.
shots = 8192
job = execute(circuit, backend, shots)
\end{pythoncode}
\begin{center}
\textbf{The result of the measurement }
\end{center}
The results do not appear directly. We have to wait in a \textbf{chain} and then wait for the execution time (Sometimes it takes 5 minutes and sometimes it takes hours). The total time differs depending on the number of people on the waiting list and the speed of the quantum computer. 
We use the following code to collect the measurement.

\definecolor{bg}{rgb}{0.97,0.97,0.97}
\begin{pythoncode}
m=job.result().get_counts(circuit)
\end{pythoncode}
When we print  $\mathsf{m}$ we get our result for a uniform distribution. We can see that the counts are not equal because of readout noise from the quantum device, so they will not be uniformly distributed.  
\definecolor{bg}{rgb}{0.97,0.97,0.97}
\begin{pythoncode}
{'00000': 324, '00001': 220, '10000': 264, '10001': 240,
 '10010': 265, '10011': 274, '10100': 281, '10101': 268,
 '10110': 299, '10111': 289, '11000': 220, '11001': 211,
 '11010': 224, '11011': 216, '11100': 242, '11101': 196,
 '11110': 222, '11111': 221, '00010': 306, '00011': 262,
 '00100': 309, '00101': 294, '00110': 342, '00111': 311,
 '01000': 243, '01001': 188, '01010': 236, '01011': 215,
 '01100': 283, '01101': 218, '01110': 267, '01111': 242}
\end{pythoncode}
\begin{center}
\textbf{Data Organization by label }
\end{center}
The labeled measures are not arranged according to the binary system, so we will use the following code to arrange our measurement. 
\definecolor{bg}{rgb}{0.97,0.97,0.97}
\begin{pythoncode}
M = {val[0] : val[1] for val in sorted(m.items(), key = lambda x: (x[0], x[1]))}
\end{pythoncode}
After the arrangement we get: 
\definecolor{bg}{rgb}{0.97,0.97,0.97}
\begin{pythoncode}
{'00000': 324, '00001': 220, '00010': 306, '00011': 262,
 '00100': 309, '00101': 294, '00110': 342, '00111': 311,
 '01000': 243, '01001': 188, '01010': 236, '01011': 215,
 '01100': 283, '01101': 218, '01110': 267, '01111': 242,
 '10000': 264, '10001': 240, '10010': 265, '10011': 274,
 '10100': 281, '10101': 268, '10110': 299, '10111': 289,
 '11000': 220, '11001': 211, '11010': 224, '11011': 216,
 '11100': 242, '11101': 196, '11110': 222, '11111': 221}
\end{pythoncode}

\begin{center}
\textbf{ Counts extraction from the measurement }
\end{center}
Now, we will extract the counts from the arranged data, but we will keep the same order. The following code will allow us to do this.
\definecolor{bg}{rgb}{0.97,0.97,0.97}
\begin{pythoncode}
m_u =[]
for key  in  M.keys()
m_u.append(M[key])
\end{pythoncode}
Finally, our measures are ready to use in the unfolding methods.
\definecolor{bg}{rgb}{0.97,0.97,0.97}
\begin{pythoncode}
[324, 220, 306, 262, 309, 294, 342, 311, 243, 188, 236,
 215, 283, 218, 267, 242, 264, 240, 265, 274, 281, 268,
 299, 289, 220, 211, 224, 216, 242, 196, 222, 221]
\end{pythoncode}

Now, let us prepare our true data for a uniform distribution.
\begin{center}
\textbf{The true count }
\end{center}
The true data is a uniform dist. That means for a number of shots equal to 8192, every state will have $8192/2^5=256$ of counts.
\definecolor{bg}{rgb}{0.97,0.97,0.97}
\begin{pythoncode}
t_u = [8192/32 for i in range(32)]

\end{pythoncode}
The result:
\definecolor{bg}{rgb}{0.97,0.97,0.97}
\begin{pythoncode}
[256.0, 256.0, 256.0, 256.0, 256.0, 256.0, 256.0, 256.0, 256.0, 256.0, 256.0,
 256.0, 256.0, 256.0, 256.0, 256.0, 256.0, 256.0, 256.0, 256.0, 256.0, 256.0,
 256.0, 256.0, 256.0, 256.0, 256.0, 256.0, 256.0, 256.0, 256.0, 256.0]
\end{pythoncode}
If we want to represent the true counts in a histogram, we need to label each state:
\definecolor{bg}{rgb}{0.97,0.97,0.97}
\begin{pythoncode}
T_u = {
 '00000': 256.0, '00001': 256.0, '00010': 256.0, '00011': 256.0, '00100': 256.0,
 '00101': 256.0, '00110': 256.0, '00111': 256.0, '01000': 256.0, '01001': 256.0, 
 '01010': 256.0, '01011': 256.0, '01100': 256.0, '01101': 256.0, '01110': 256.0,
 '01111': 256.0, '10000': 256.0, '10001': 256.0, '10010': 256.0, '10011': 256.0,
 '10100': 256.0, '10101': 256.0, '10110': 256.0, '10111': 256.0, '11000': 256.0, 
 '11001': 256.0, '11010': 256.0, '11011': 256.0, '11100': 256.0, '11101': 256.0; 
 '11110': 256.0, '11111': 256.0}
\end{pythoncode}
We can see the representation of our data in the histogram  by the $\mathsf{plot\_histogram()}$ function:
\definecolor{bg}{rgb}{0.97,0.97,0.97}
\begin{pythoncode}
plot_histogram(T_u,figsize=(35,7))
\end{pythoncode}

\begin{center}
\includegraphics[width = 1\linewidth]{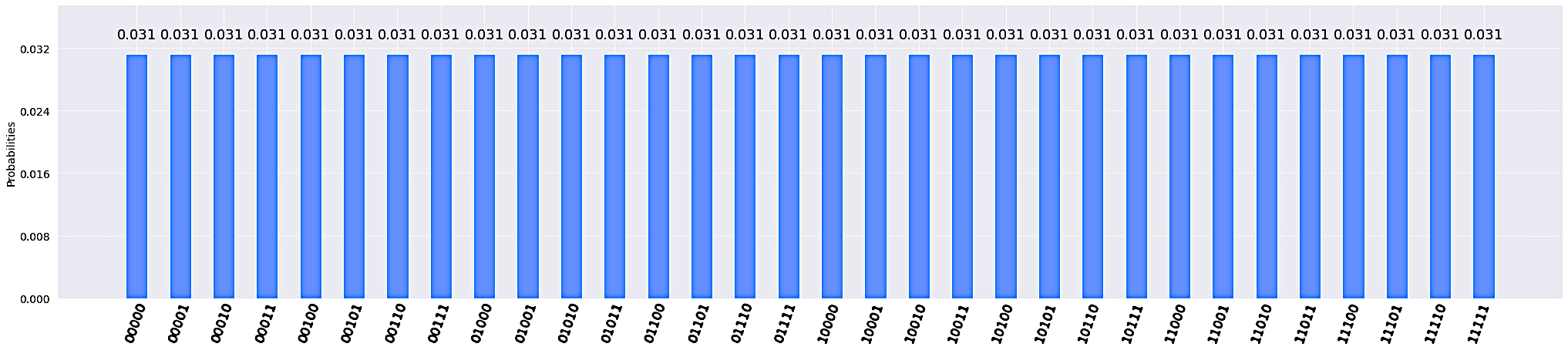}
\end{center}
The measured data are represented also in the following histogram:
\definecolor{bg}{rgb}{0.97,0.97,0.97}
\begin{pythoncode}
plot_histogram(M,figsize=(30,6))
\end{pythoncode}

\begin{center}
\includegraphics[width = 1\linewidth]{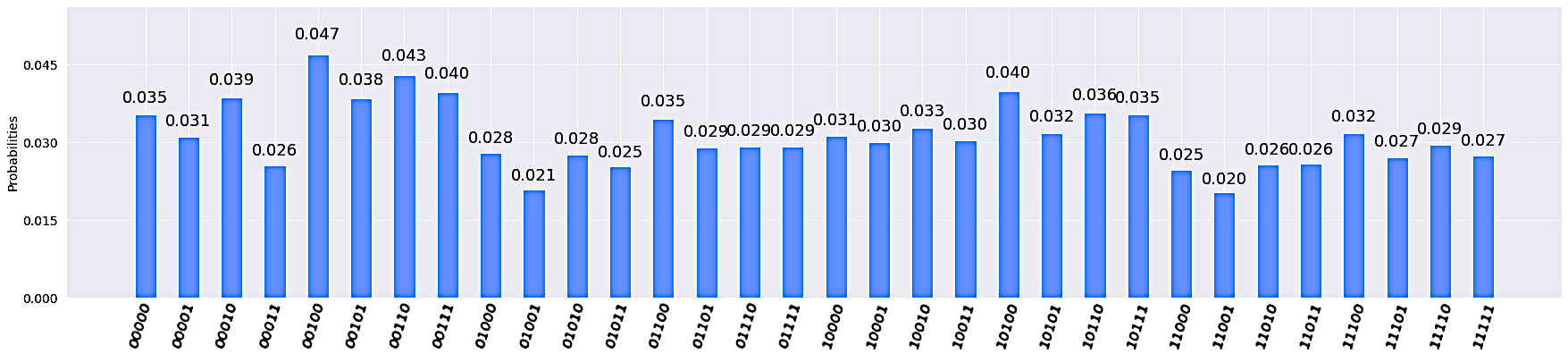}
\end{center}
We can see clearly the effect of noise from the quantum computers by comparing true and measured data.  
\begin{center}
\textbf{  Unfolding functions  }
\end{center}
Now, we will use the unfolding functions provided by Ben. Nachman et al., to unfold our measured data from a real quantum computer. This part of the code is already explained in the \textsf{python} simulation section. The matrix inversion and the ignis function have two arguments (\textsf{m$\_$u, R}). The IBU have four arguments (\textsf{m$\_$u, t0, R, n}).
At this point, we have all the ingredients for the unfolding methods.  We can choose any \textsf{t0} and any \textsf{n}. But we will fix the number of iterations to \textsf{n=1}.  The unfolding methods are defined as follow:
\definecolor{bg}{rgb}{0.97,0.97,0.97}
\begin{pythoncode}
def MatrixInversion(ymes,Rin):
    return np.ravel(np.matmul(np.linalg.inv(np.matrix(Rin)),ymes))

def fun(x,ymes,Rin):
    mat_dot_x = np.ravel(np.matmul(Rin,x))
    return sum((ymes - mat_dot_x)**2)

def Ignis(ymes,Rin):
    
    x0 = np.random.rand(len(ymes)) 
    x0 = x0 / sum(x0)
    nshots = sum(ymes)
    cons = ({'type': 'eq', 'fun': lambda x: nshots - sum(x)})
    bnds = tuple((0, nshots) for x in x0)
    res = minimize(fun, x0, method='SLSQP',constraints=cons, bounds=bnds, tol=1e-6,
     args=(ymes,Rin))
    return res.x

def IBU(ymes,t0,Rin,n):
    
    tn = t0
    for q in range(n):
        out = []
        for j in range(len(t0)):
            mynum = 0.
            for i in range(len(ymes)):
                myden = 0.
                for k in range(len(t0)):
                    myden+=Rin[i][k]*tn[k]
                    pass
                mynum+=Rin[i][j]*tn[j]*ymes[i]/myden
                pass
            out+=[mynum]
        tn = out
        pass
    return tn
\end{pythoncode}
\begin{center}
\textbf{ Unfolding a uniform distribution  }
\end{center}

Now, let us unfold our measured distribution by the three unfolding methods. For matrix inversion and Ignis we need only $\mathsf{m\_u}$ and the response matrix. For IBU, we additionally need \textsf{t0} and \textsf{n}.
These is all our ingredients in one place:

The true data  $\mathsf{t\_u}$:
\definecolor{bg}{rgb}{0.97,0.97,0.97}
\begin{pythoncode}
array([256, 256, 256, 256, 256, 256, 256, 256, 256, 256, 256, 256, 256,
       256, 256, 256, 256, 256, 256, 256, 256, 256, 256, 256, 256, 256,
       256, 256, 256, 256, 256, 256])
\end{pythoncode}
The measuresd data $\mathsf{m\_u}$:
\definecolor{bg}{rgb}{0.97,0.97,0.97}
\begin{pythoncode}

array([324, 220, 306, 262, 309, 294, 342, 311, 243, 188, 236,
       215, 283, 218, 267, 242, 264, 240, 265, 274, 281, 268,
       299, 289, 220, 211, 224, 216, 242, 196, 222, 221])
\end{pythoncode}
The response matrix \textsf{R}:
\definecolor{bg}{rgb}{0.97,0.97,0.97}
\begin{pythoncode}
array([[7.39257812e-01 1.07177734e-01 4.65087891e-02 ... 0.00000000e+00
  0.00000000e+00 0.00000000e+00]
 [7.34863281e-02 7.59277344e-01 4.39453125e-03 ... 1.09863281e-03
  0.00000000e+00 2.44140625e-04]
 [3.97949219e-02 7.32421875e-03 7.72827148e-01 ... 0.00000000e+00
  8.54492188e-04 1.22070312e-04]
 ...
 [0.00000000e+00 0.00000000e+00 0.00000000e+00 ... 6.33178711e-01
  5.24902344e-03 4.82177734e-02]
 [1.22070312e-04 0.00000000e+00 0.00000000e+00 ... 5.24902344e-03
  6.30249023e-01 8.23974609e-02]
 [0.00000000e+00 0.00000000e+00 0.00000000e+00 ... 2.61230469e-02
  9.57031250e-02 6.31103516e-01]])
\end{pythoncode}
We choose the prior truth distribution \textsf{t0}, as a uniform distribution.
\definecolor{bg}{rgb}{0.97,0.97,0.97}
\begin{pythoncode}
array([1, 1, 1, 1, 1, 1, 1, 1, 1, 1, 1, 1, 1, 1, 1, 1, 1, 1, 1, 1, 1, 1, 1, 1, 1, 1, 1, 1, 1, 1, 1, 1])
\end{pythoncode}
Let us see our output of the three unfolding methods:
\definecolor{bg}{rgb}{0.97,0.97,0.97}
\begin{pythoncode}
unfolded_Matrix_r = MatrixInversion(m_u, R)
unfolded_IBU_r = IBU(m_u, np.ones(len(m_u)), R, 1) 
unfolded_ignis_r = Ignis(m_u, R)
\end{pythoncode}

\begin{center}
\textbf{Visual representation of unfolding result}
\end{center}
In this part, we will compare the three unfolding methods by representing the results of unfolded data graphically.  

We will need the following code to label our histograms. 
\definecolor{bg}{rgb}{0.97,0.97,0.97}
\begin{pythoncode}
nqubits_r = 5

mymapping_r = {}
mymapping_inverse_r = {}
for i in range(2**nqubits_r):
    mymapping_r[i] = bin(i).split('b')[1]
    mymapping_inverse_r[bin(i).split('b')[1]] = i
    pass

xvals_r = []
xlabs_r = []
xlabs0_r = []
for i in range(2**nqubits_r):
    xvals_r+=[i]
    xlabs_r+=[r'$|'+mybin(i,nqubits_r)+r'\rangle$']
    xlabs0_r+['']
    pass
 \end{pythoncode}

The following code allow us to represent the measured data, the true data and the unfolded data via the three unfolding methods.  
 \definecolor{bg}{rgb}{0.97,0.97,0.97}
 
\begin{pythoncode}
def myre(x):
    return np.concatenate([[x[0]],x])
bincenters_r = np.array(xvals_r)+0.5
sns.set_style("white")
f = plt.figure(figsize=(8.5, 8.5))
gs = GridSpec(2, 1, width_ratios=[1], height_ratios=[3.5, 1])
ax1 = plt.subplot(gs[0])
ax1.set_xticklabels( () )

mybincenters = np.linspace(-0.5,31.5,33)
plt.step(mybincenters,myre(t_u),color='black',label=r"t")
plt.errorbar(0.5*(mybincenters[1:]+mybincenters[:-1]),m_u,yerr=np.sqrt(m_u),
label='m',marker='o',linestyle='none',color='black')
plt.step(mybincenters,myre(unfolded_Matrix_r),color='blue',linestyle=':'
,label=r"$\hat{t}_{matrix}$",linewidth=3.0)
plt.step(mybincenters,myre(unfolded_ignis_r),color='red',linestyle='--'
,label=r"$\hat{t}_{ignis}$")
plt.step(mybincenters,myre(unfolded_IBU_r),color='green',linestyle='-.'
,label=r"$\hat{t}_{IBU}$")

_=plt.xticks(xvals_r,xlabs0_r,rotation='vertical',fontsize=10)
plt.ylabel('Counts')
plt.ylim([0,2.*max(t_u)])
plt.legend(loc='upper left',fontsize=12,frameon=False)
_=plt.text(20., max(t_u)*1.8, "5 Yorktown Machine", fontsize=18)
_=plt.text(20., max(t_u)*1.7, "IBMQ Yorktown Readout Errors", fontsize=12)

ax2 = plt.subplot(gs[1])
ratio_ignis2_r = np.zeros(2**nqubits_r)
ratio_ignis_r = np.zeros(2**nqubits_r)
ratio_matrix_r = np.zeros(2**nqubits_r)
ratio_unfold_r = np.zeros(2**nqubits_r)
for i in range(len(ratio_unfold_r)):
    ratio_unfold_r[i]=1.
    ratio_matrix_r[i]=1.
    ratio_ignis_r[i]=1.
    ratio_ignis2_r[i]=1.
    if (t_u[i] > 0):
        ratio_unfold_r[i] = unfolded_IBU_r[i]/t_u[i]
        ratio_matrix_r[i] = unfolded_Matrix_r[i]/t_u[i]
        ratio_ignis_r[i] = unfolded_ignis_r[i]/t_u[i]
        pass
    pass

plt.ylim([0.8,1.2])
plt.step(mybincenters,myre(ratio_ignis_r),color='red',linestyle="--")
plt.step(mybincenters,myre(ratio_matrix_r),color='blue',linestyle=":",linewidth=3.0)
plt.step(mybincenters,myre(ratio_unfold_r),color='green')
plt.ylabel('Extracted / Truth')
xx2 = [1.,1.]
plt.plot([xvals_r[0],xvals_r[len(xvals_r)-1]],xx2,color='black',linestyle=':')
_=plt.xticks(xvals_r,xlabs_r,rotation='vertical',fontsize=12)
plt.savefig("Unfold_york.pdf",bbox_inches='tight')
\end{pythoncode}

\begin{figure}[H]
\centering
\includegraphics[width = 0.8\linewidth]{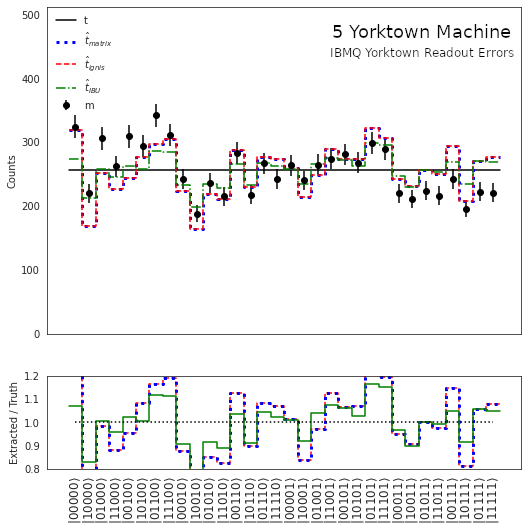}
\caption{Unfolding results of a uniform distribution using three methods: Matrix inversion, ignis, and IBU with one iteration and a uniform prior truth spectrum.}
\end{figure}

Now, let us compare the three unfolding methods.  First, we can see clearly that the values of the unfolded data with ignis and Matrix inversion are nearly identical. For IBU, the result is better than the two previous methods. where its values are near to the true data. The number of iterations that we used here is \textsf{n=1} and the test data (\textsf{t0}) is an array of 32 ones. We noticed that these parameters give better results for IBU.       

\subsection{Unfolding a Gaussian Distribution}

\definecolor{bg}{rgb}{240,248,255}
In this part we will study unfolding in the case of a Gaussian distribution (the ground state of a harmonic oscillator), the steps followed are similar to that of a uniform distribution. But we will use a more developed tool to build the circuit corresponding to the Gaussian case. The general idea is that we can initialize our 5-qubit system in any desired state vector, specifying the \textit{complex amplitudes} of the general 5-qubit state. \\
\[\ket{\psi}=\alpha_{_{00000}}\ket{00000}\,+\cdots+\,\alpha_{_{11111}}\ket{11111}\]

We will explain this concept as we go along. But first, let's define our quantum circuit with its corresponding quantum and classical registers.
\definecolor{bg}{rgb}{0.97,0.97,0.97}
\begin{pythoncode}
nq = 5

qr = QuantumRegister(nq, 'qr')
cr = ClassicalRegister(nq, 'cr')
circuit_g = QuantumCircuit(qr, cr)
\end{pythoncode}
Since we want a Gaussian distribution, we define the function: $\mathsf{HO}\!_{_{gs}}(x)=e^{-\frac{x^2}{2}}$
\definecolor{bg}{rgb}{0.97,0.97,0.97}
\begin{pythoncode}
def HO_groundstate(x):
    return np.exp(-x**2/2)
\end{pythoncode}
which represents the \textit{harmonic oscillator ground state}. The general form of the Gaussian distribution \textit{probability density function} is written as 
\[f(z)\sim e^{-\frac{1}{2}\left(\frac{z-\mu}{\sigma}\right)^2},\]

\begin{flushleft}
where $\mu$ is the mean, and $\sigma$ is the standard deviation.\\
\end{flushleft}

Instead of having $z$ continuous in our specific case it is discrete, where $z\in[0,31]$. We will plot the Gaussian in this interval, therefore, to center it we take $\mu=(31/2)$, and we can choose $\sigma=8$. So we have 
\[x=\frac{z-\mu}{\sigma}=\frac{z-(16-0.5)}{8}=2\left(\dfrac{z+0.5-2^4}{2^4}\right).\]

We now define our desired vector, as the vector of elements $e^{-\frac{1}{2}\left(\frac{z-\mu}{\sigma}\right)^2},\:z\in[\:0,\:31\:]$:
\definecolor{bg}{rgb}{0.97,0.97,0.97}
\begin{pythoncode}
desired_vector = []
for z in range(2**nq):
    desired_vector += [HO_groundstate(2.*( z+ 0.5-2**(nq-1))/2**(nq-1))]
    pass   
\end{pythoncode}
The above algorithm creates the vector
\[\mathsf{desired\_vector}=[\:e^{-\frac{1}{2}\left(\frac{0-\mu}{\sigma}\right)^2},\:e^{-\frac{1}{2}\left(\frac{1-\mu}{\sigma}\right)^2},\:e^{-\frac{1}{2}\left(\frac{2-\mu}{\sigma}\right)^2},\:\cdots,\:e^{-\frac{1}{2}\left(\frac{31-\mu}{\sigma}\right)^2}]\]
\definecolor{bg}{rgb}{0.97,0.97,0.97}
\begin{pythoncode}
print(desired_vector) 

[0.15305573773635817, 0.19348058160704673, 0.24079047427224856, 0.295022
65617444284, 0.35586525214213843, 0.4226004432231888, 0.4940699735316383
4, 0.568671053718672, 0.6443887248251953, 0.7188675697091758, 0.78952156
96607879, 0.8536763613451477, 0.9087337563610701, 0.9523447998951764, 0.
9825754689579004, 0.9980487811074755, 0.9980487811074755, 0.982575468957
9004, 0.9523447998951764, 0.9087337563610701, 0.8536763613451477, 0.7895
215696607879, 0.7188675697091758, 0.6443887248251953, 0.568671053718672,
0.49406997353163834, 0.4226004432231888, 0.35586525214213843, 0.29502265
617444284, 0.24079047427224856, 0.19348058160704673, 0.1530557377363581]
\end{pythoncode}
Since these components represent \textit{complex amplitudes}, they must check the \textit{normalization condition}:
\[|\alpha_{_{00000}}|^2\:+\:|\alpha_{_{00001}}|^2\:+\,\cdots\,+\:|\alpha_{_{11111}}|^2=1\]
Thus, we need to normalize our desired vector ($\:\textsf{desired\_vector}/\sqrt{\sum_{k=0}^{31}v_k^2}\:$).
\definecolor{bg}{rgb}{0.97,0.97,0.97}
\begin{pythoncode}    
desired_vector /= np.sqrt(np.sum([v**2 for v in desired_vector]))

print(desired_vector)
\end{pythoncode}
Output
\definecolor{bg}{rgb}{0.97,0.97,0.97}
\begin{pythoncode}
[0.04074024 0.05150049 0.06409340 0.07852887 0.09472390 0.11248742
 0.13151111 0.15136836 0.17152283 0.19134754 0.21015416 0.22723083
 0.24188595 0.25349430 0.26154108 0.26565975 0.26565975 0.26154108
 0.25349430 0.24188595 0.22723083 0.21015416 0.19134754 0.17152283
 0.15136836 0.13151111 0.11248742 0.09472390 0.07852887 0.06409340
 0.05150049 0.04074024]
\end{pythoncode}
To configure the general 5-qubit state $\ket{\psi}$ in this desired vector, we use the \textsf{initialize} instruction, which requires a vector of \textbf{complex amplitudes} (\textsf{desired\_vector}) and a \textbf{list of qubits}, which can be done by the loop below:
\definecolor{bg}{rgb}{0.97,0.97,0.97}
\begin{pythoncode}
qrs = []
for i in range(nq):
    qrs+=[qr[i]]
    pass
print(qrs)
\end{pythoncode}
Output 
\definecolor{bg}{rgb}{0.97,0.97,0.97}
\begin{pythoncode}
[Qubit(QuantumRegister(5, 'qr'), 0), Qubit(QuantumRegister(5, 'qr'), 1),
 Qubit(QuantumRegister(5, 'qr'), 2), Qubit(QuantumRegister(5, 'qr'), 3),
 Qubit(QuantumRegister(5, 'qr'), 4)]
\end{pythoncode}
We have therefore defined a list of 5 qubits $[\,q_0,\:q_1,\:q_2,\:q_3,\:q_4\,]$. We can now initialize them in a \textit{Gaussian distribution}, through the following input:
\definecolor{bg}{rgb}{0.97,0.97,0.97}
\begin{pythoncode}
circuit_g.initialize(desired_vector, qrs)
\end{pythoncode}
Once the initialization is done, the last thing to do is to measure the system. Here we are using the \textsf{measure()} function, with the parameters \textsf{(qr,cr)}, which represent a shorthand that allow us to implement a \textit{measurement operation} on each qubit. 
\definecolor{bg}{rgb}{0.97,0.97,0.97}
\begin{pythoncode}
circuit_g.measure(qr, cr)
\end{pythoncode}
Since we are curious to see this circuit, we visualize it.
\definecolor{bg}{rgb}{0.97,0.97,0.97}
\begin{pythoncode}
circuit_g.draw('latex') 
\end{pythoncode}
\begin{figure}[H]
\centering
\includegraphics[width = 1\linewidth]{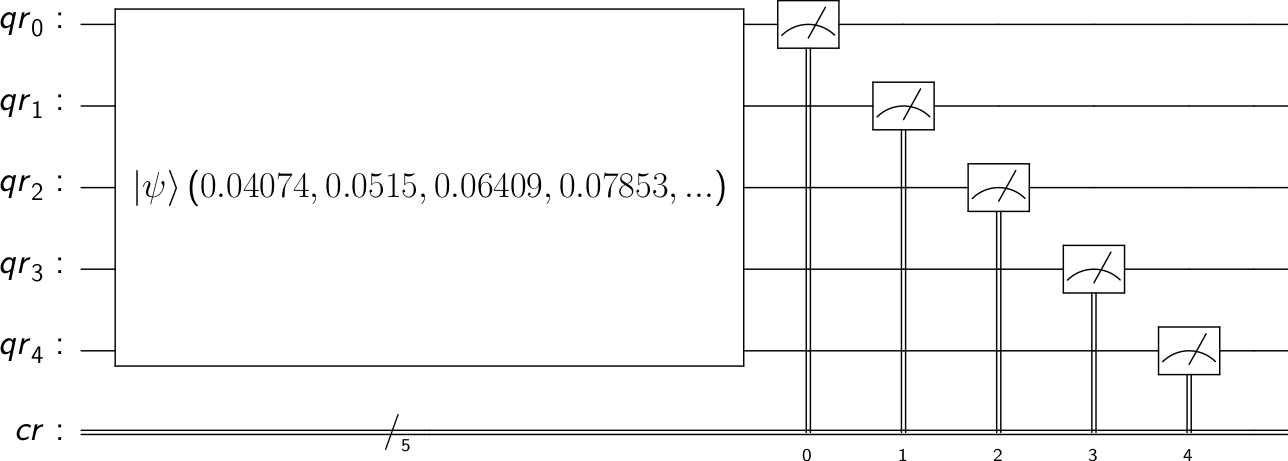}
\caption{A circuit that generates a Gaussian distribution based on $5$ qubits.}
\end{figure}
Note that \textsf{initialize} contains a \textsf{reset} instruction, which is not \textit{unitary}, therefore it is an \textbf{Instruction} and not a \textbf{Gate}.\\

Simulators are characterized by negligible readout noise, therefore, we can use one of them to get the \textit{true Gaussian distribution}, denoted by \textsf{t$\_$g}. Here we are using the simulator called “\textsf{qasm$\_$simmulator}” provided by \textsf{Aer}. To use it we enter the same commands that allow us to use a real quantum computer, with the only difference instead of the command \textsf{provider$\,$=$\,$IBMQ.load$\_$account()}, we use \textsf{Aer}.\footnote{For simulators provided by IBM Q (non-local), we use the \textsf{provider$\,$=$\,$IBMQ.load$\_$account()}.}
\definecolor{bg}{rgb}{0.97,0.97,0.97}
\begin{pythoncode}
simulator = Aer.get_backend('qasm_simulator')
result = execute(circuit_g, simulator,shots=8192).result()
count_s = result.get_counts(circuit_g)
print(count_s)
\end{pythoncode}
Output
\definecolor{bg}{rgb}{0.97,0.97,0.97}
\begin{pythoncode}
{'00000': 12, '00001': 18, '10000': 603, '10001': 563,'10010': 540,'10011
': 470, '10100': 449,'10101': 369, '10110': 324, '10111': 236,'11000': 19
0, '11001': 150, '11010': 96, '11011': 67, '11100': 50, '11101': 25, '111
10': 21, '11111': 11, '00010': 31, '00011': 51, '00100': 75, '00101': 103
, '00110': 141, '00111': 195, '01000': 240, '01001': 288, '01010': 342, 
'01011': 429, '01100': 449, '01101': 527, '01110': 566, '01111': 561} 
\end{pythoncode}
To plot this distribution, we write the following code:
\definecolor{bg}{rgb}{0.97,0.97,0.97}
\begin{pythoncode}
nq=5
xvals = []
yvals = []
for j in range(2**nq):
    xvals+=[j]
    yvals+=[count_s[mybin(j,nq)[::-1]]]
    
sns.set_style("white")
plt.errorbar(xvals,yvals,yerr=np.sqrt(yvals),linestyle='none',marker='o')
\end{pythoncode}
\begin{figure}[H]
\centering
\includegraphics[width = 0.6\linewidth]{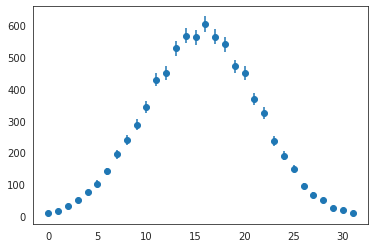}

\end{figure}
\definecolor{bg}{rgb}{0.97,0.97,0.97}
\begin{pythoncode}
plot_histogram(count_s,figsize=(25,6))
\end{pythoncode}
\begin{figure}[H]
\centering
\includegraphics[width = 1\linewidth]{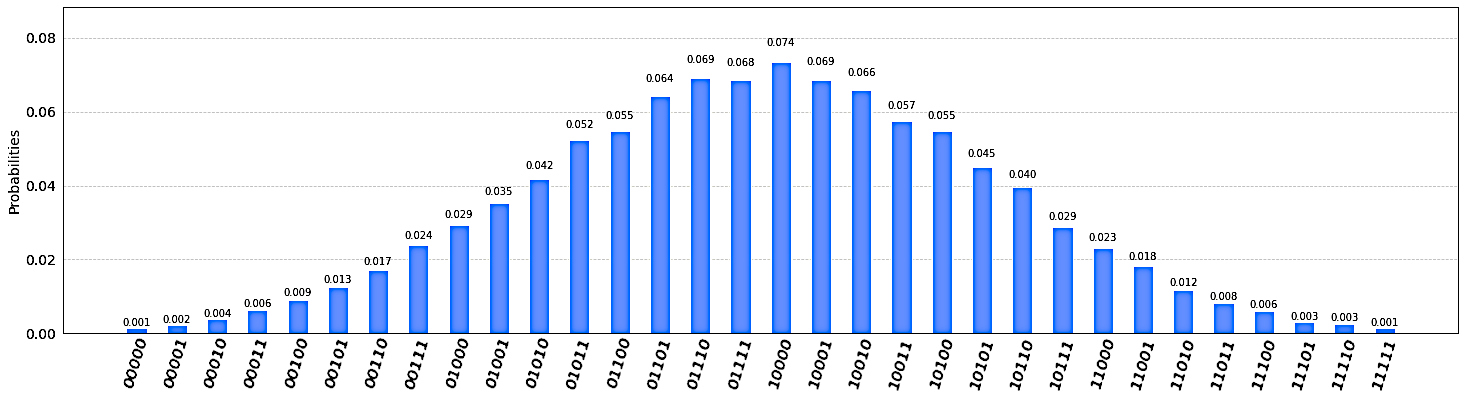}
\end{figure}
We have now confirmed that we really have a Gaussian distribution.\\

The readout noise from the Yorktown machine will distort this perfect Gaussian and gives us a disturbed distribution, denoted by \textsf{m$\_$g}. To realize this, we execute our circuit (\textsf{circuit$\_$g}) in the “\textsf{ibmq$\_$5$\_$yorktown}” machine.
\definecolor{bg}{rgb}{0.97,0.97,0.97}
\begin{pythoncode}
backend = provider.get_backend('ibmq_5_yorktown')  #or:'ibmqx2'
result = execute(circuit_g, backend,shots=8192).result()
count_y = result.get_counts(circuit_g)
print(count_y)
\end{pythoncode}
Output
\definecolor{bg}{rgb}{0.97,0.97,0.97}
\begin{pythoncode}
{'00000': 223, '00001': 327, '10000': 356, '10001': 371, '10010': 353, '
10011': 419, '10100': 282, '10101': 325, '10110': 270, '10111': 254, '11
000': 149, '11001': 120, '11010': 129, '11011': 88, '11100': 77, '11101'
: 59, '11110': 49, '11111': 43, '00010': 266, '00011': 383, '00100': 291
, '00101': 346, '00110': 281, '00111': 289, '01000': 282, '01001': 329, 
'01010': 249, '01011': 409, '01100': 298, '01101': 330, '01110': 272, '0
1111': 273}
\end{pythoncode}
Same inputs for visualization:
\definecolor{bg}{rgb}{0.97,0.97,0.97}
\begin{pythoncode}
xvals = []
yvals = []
for j in range(2**nq):
    xvals+=[j]
    yvals+=[count_y[mybin(j,nq)[::-1]]]
    
sns.set_style("white")
plt.errorbar(xvals,yvals,yerr=np.sqrt(yvals),linestyle='none',marker='o')
\end{pythoncode}
\begin{figure}[H]
\centering
\includegraphics[width = 0.6\linewidth]{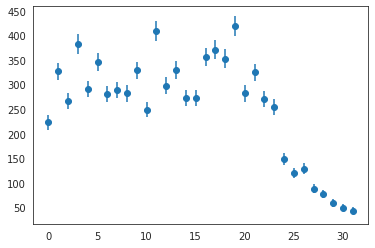}
\end{figure}
\definecolor{bg}{rgb}{0.97,0.97,0.97}
\begin{pythoncode}
plot_histogram(count_y,figsize=(25,6))
\end{pythoncode}
\begin{figure}[H]
\centering
\includegraphics[width = 1\linewidth]{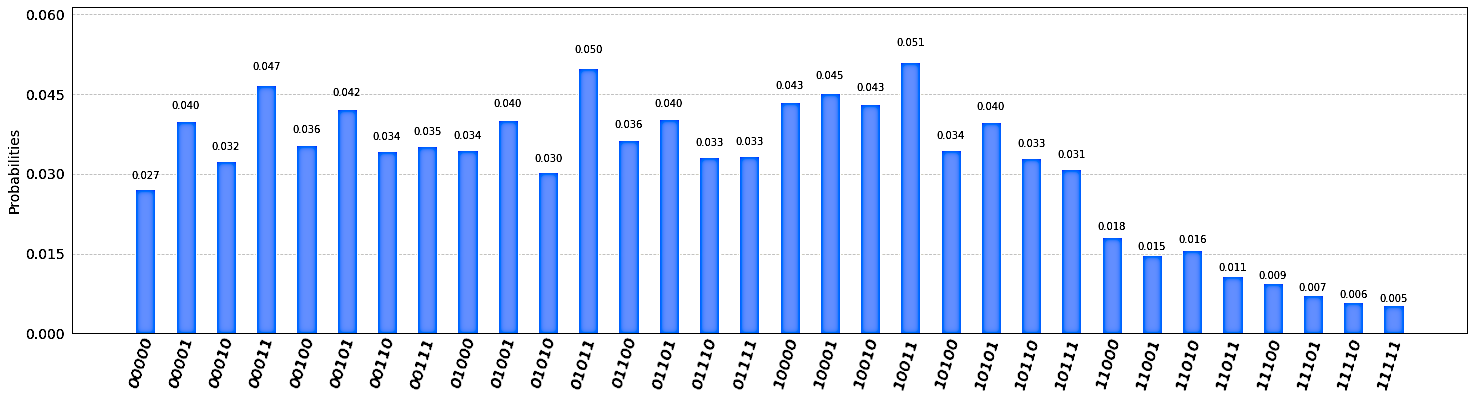}

\end{figure}
We notice that the readout noise has distorted the Gaussian (true) distribution to the point that we cannot recognize its original shape.\\ 

Now we have just to extract the \textit{frequencies} from ($\mathsf{count\_s}$) and ($\mathsf{count\_y}$), so we use the same code seen for the uniform case, in order to finally obtain the vectors $\textsf{t\_g}$, and $\textsf{m\_g}$.
\definecolor{bg}{rgb}{0.97,0.97,0.97}
\begin{pythoncode}
true_g= {val[0] : val[1] for val in sorted(count_s.items(), key = lambda 
                                           x: (x[0], x[1]))}
t_g = []
for key in true_g.keys() :
    t_g.append(true_g[key])
    pass
print(t_g)

[12, 18, 31, 51, 75, 103, 141, 195, 240, 288, 342, 429, 449, 527, 566, 
561, 603, 563, 540, 470, 449, 369, 324, 236, 190, 150, 96, 67, 50, 25,
21, 11]
\end{pythoncode}
We now extract the measured distribution vector $\mathsf{m\_g}$:
\definecolor{bg}{rgb}{0.97,0.97,0.97}
\begin{pythoncode}
meas_g= {val[0] : val[1] for val in sorted(count_y.items(), key = lambda 
                                           x: (x[0], x[1]))}
m_g = []
for key in meas_g.keys() :
    m_g.append(meas_g[key])
    pass
print(m_g)

[223, 327, 266, 383, 291, 346, 281, 289, 282, 329, 249, 409, 298, 330, 2
72, 273, 356, 371, 353, 419, 282, 325, 270, 254, 149, 120, 129, 88, 77, 
59, 49, 43]
\end{pythoncode}
Here we are at the crucial part of this experience; the reconstruction of truth “Unfolding”. For this objective, we need the \textit{Yorktown response matrix} and the result obtained from the \textit{Yorktown readout noise}. The two unfolding methods \textsf{MatrixInversion} and \textsf{Ignis} (commonly used in QIS), give us the following outputs:
\definecolor{bg}{rgb}{0.97,0.97,0.97}
\begin{pythoncode}
MatrixInversion(m_g,R)

array([151.96727863, 278.69220815, 183.01138799, 361.83956380,
       231.16886853, 318.11018472, 228.28325324, 261.47569561,
       262.90470574, 336.10572204, 204.19246932, 511.82630002,
       303.77496880, 395.06103499, 303.84402141, 307.09007413,
       383.34934568, 372.73466745, 360.07688363, 510.94824455,
       271.18915214, 358.70672324, 295.37171705, 251.92702464,
       176.83096465, 128.69790459, 163.81956139,  87.15086908,
        75.30630507,  42.56811204,  38.78854417,  35.18624349])
\end{pythoncode}
\definecolor{bg}{rgb}{0.97,0.97,0.97}
\begin{pythoncode}
Ignis(m_g,R)

array([151.96736728, 278.69210044, 183.01138211, 361.83954193,
       231.16889604, 318.11019459, 228.28326267, 261.47568405,
       262.90461210, 336.10575357, 204.19247446, 511.82626943,
       303.77501047, 395.06101942, 303.84403513, 307.09006300,
       383.34934156, 372.73459805, 360.07668674, 510.94840325,
       271.18909030, 358.70681406, 295.37191834, 251.92685430,
       176.83112506, 128.69782248, 163.81966319,  87.15077385,
        75.30624717,  42.56814870,  38.78852282,  35.18632344])
\end{pythoncode}
We notice that the unfolding results for \textsf{MatrixInversion} and \textsf{Ignis} are very close, despite the fact that these are two different methods.\\

The IBU from HEP have 4 parameters ($\mathsf{m\_g,\:t0,\:R,\:n}$), the only parameters we are free to choose are the \textit{number of iterations} (\textsf{n}) and the \textit{prior truth distribution} ($\mathsf{t0}$). We recall the $6^{th}$ advantage of the IBU mentioned by D'agostini in his article \cite{44}, where he states that satisfactory results are obtained if we start from a \textit{uniform distribution}. But it gives the best results if one makes a realistic guess about the true distribution. Therefore, we can choose $\mathsf{t0}$ for example $[\,0, 1, 2, \cdots, 2, 1, 0\,]$, which is of course a $2^5$\textit{dimensional} vector. In \textsf{python} it can be produced as follows:
\definecolor{bg}{rgb}{0.97,0.97,0.97}
\begin{pythoncode}
c =[]
for s in range(16):
    c.append(s)
    pass
t_s=[ *c , *c[::-1] ]
print(t_s)
\end{pythoncode}
Output
\definecolor{bg}{rgb}{0.97,0.97,0.97}
\begin{pythoncode}
[0, 1, 2, 3, 4, 5, 6, 7, 8, 9, 10, 11, 12, 13, 14, 15, 15, 14, 13, 12, 
11, 10, 9, 8, 7, 6, 5, 4, 3, 2, 1, 0]
\end{pythoncode}
In the algorithm above, we are adding the elements \textsf{s} to the empty vector \textsf{c} by the function \textsf{append()}. The object \textsf{[:$\,$:$\,$-1]} reverse the elements of \textsf{c}, we can then concatenate the two vectors.\\

The second parameter to set, is the number of iterations, where we just take \textsf{n=1}.
\definecolor{bg}{rgb}{0.97,0.97,0.97}
\begin{pythoncode}
print(IBU(m_g,t_s,R,1))

[0.0, 87.29209365373235, 115.43210028814272, 210.70351275475588
, 208.88814802549277, 285.4138302647153, 238.21228525158762, 30
0.7623880678843, 279.92435625976367, 353.78385668972953, 285.58
200778083545, 433.46993222855866, 351.6549239367971, 424.667582
80482253, 346.4533319644723, 406.80177950302266, 454.7858110448
58, 474.61896151747544, 430.9510312411226, 481.0646239820014, 3
29.50013439662126, 344.7721592177831, 294.3503960979849, 278.53
393415499374, 184.08290907346077, 151.61192190751058, 153.01420
394758765, 121.38914195438865, 78.25374972502975, 52.2064547630
8817, 33.82243750178114, 0.0]
\end{pythoncode}
These values are clearly different from the two previous methods. We have now to visualize the three unfolding results in ordrer to get a more accurate observation of their behavior and to compare them. We use the same algorithm as the uniform case, where the only changes to be made are the true and measured distribution $\mathsf{t\_g,\,m\_g}$, and the prior truth distribution $\mathsf{t0=ts}$.\\

\definecolor{bg}{rgb}{0.97,0.97,0.97}
\begin{pythoncode}

unfolded_Matrix = MatrixInversion(m_g, R)
unfolded_IBU = IBU(m_g, t_s, R, 1)
unfolded_ignis = Ignis(m_g, R)


bincenters_r = np.array(xvals_r)+0.5
sns.set_style("white")
f = plt.figure(figsize=(8.5, 8.5))
gs = GridSpec(2, 1, width_ratios=[1], height_ratios=[3.5, 1])
ax1 = plt.subplot(gs[0])
ax1.set_xticklabels( () )

def myre(x):
    return np.concatenate([[x[0]],x])

mybincenters = np.linspace(-0.5,31.5,33)

plt.step(mybincenters,myre(t_g),color='black',label=r"t")
plt.errorbar(0.5*(mybincenters[1:]+mybincenters[:-1]),m_g,yerr=np.sqrt(m
               _g),label='m',marker='o',linestyle ='none',color='black')
plt.step(mybincenters,myre(unfolded_Matrix),color='blue',linestyle=':',
                              label=r"$\hat{t}_{matrix}$",linewidth=3.0)
plt.step(mybincenters,myre(unfolded_ignis),color='red',linestyle='--',
                                             label=r"$\hat{t}_{ignis}$")
plt.step(mybincenters,myre(unfolded_IBU),color='green',linestyle='-.',
                                               label=r"$\hat{t}_{IBU}$")

_=plt.xticks(xvals_r,xlabs0_r,rotation='vertical',fontsize=10)
plt.ylabel('Counts')
plt.ylim([0,2.*max(t_g)])
plt.legend(loc='upper left',fontsize=12,frameon=False)
_=plt.text(20., max(t_g)*1.8, "5 Yorktown Machine", fontsize=18)
_=plt.text(20., max(t_g)*1.7, "IBMQ Yorktown Readout Errors", fontsize=12)

nqubits = 5

ax2 = plt.subplot(gs[1])
ratio_ignis2 = np.zeros(2**nqubits)
ratio_ignis = np.zeros(2**nqubits)
ratio_matrix = np.zeros(2**nqubits)
ratio_unfold = np.zeros(2**nqubits)
for i in range(len(ratio_unfold)):
    ratio_unfold[i]=1.
    ratio_matrix[i]=1.
    ratio_ignis[i] =1.
    ratio_ignis2[i]=1.
    if (t_g[i] > 0):
        ratio_unfold[i] = unfolded_IBU[i]/t_g[i]
        ratio_matrix[i] = unfolded_Matrix[i]/t_g[i]
        ratio_ignis[i] = unfolded_ignis[i]/t_g[i]
        pass
    pass

plt.ylim([0.8,1.2])
plt.step(mybincenters,myre(ratio_ignis),color='red',linestyle="--")
plt.step(mybincenters,myre(ratio_matrix),color='blue',linestyle=":"
                                                    ,linewidth=3.0)
plt.step(mybincenters,myre(ratio_unfold),color='green')
plt.ylabel('Extracted / Truth')
xx2 = [1.,1.]
plt.plot([xvals_r[0],xvals_r[len(xvals_r)-1]],xx2,color='black',lines
                                                             tyle=':')
_=plt.xticks(xvals_r,xlabs_r,rotation='vertical',fontsize=12)
\end{pythoncode}
\begin{figure}[H]
\centering
\includegraphics[width = 0.8\linewidth]{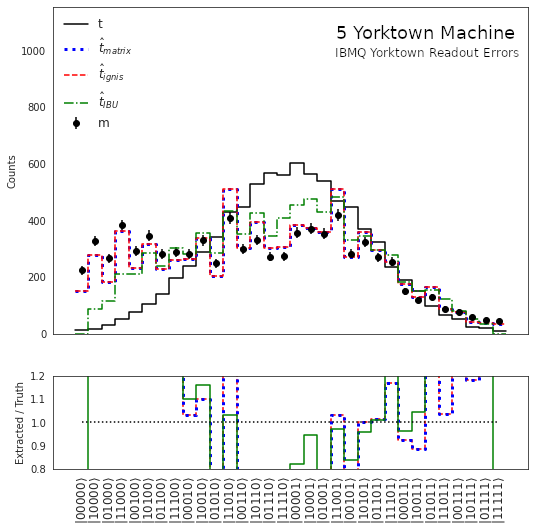}
\caption{Unfolding results of a Gaussian distribution using three methods: Matrix inversion, ignis, and IBU with one iteration and a prior truth spectrum \textsf{t0=t\_s}.}
\end{figure}
Note that the values of \textsf{MatrixInversion} and \textsf{Ignis} are identical. Moreover, the unfolding amount is very low. In the case of \textsf{IBU}, its values are closer to the true distribution and give a more important unfolding compared the other two methods, and for only 1 iteration.

In the second graph (Extracted/truth) we are dividing the components of each resulting vector by the component of the vector \textsf{t$\_$g}, so if they are equal we get 1. Hence, we are representing a Gaussian distribution as a uniform one. If we look at this graph we see that the values closest to 1 in the majority of the time are those of IBU.

The results obtained above, concern the choice \textsf{t0=t$\_$s}. But what if we tried another vector \textsf{t0}, knowing that taking the vector $\lambda.\textsf{t0}$, where $\lambda\in\mathbb{R}^*$, will not change the results, this fact can be directly verified from the formula of IBU for one iteration

\begin{equation}
t^{1}_{i}=\,\sum_{j}\,\left(\dfrac{\mathcal{R}_{ji}\:t^{\,0}_{i}}{\sum_{k}\,\mathcal{R}_{jk}\,t^{\,0}_{k}}\right)m_j\longrightarrow\,\sum_{j}\,\left(\dfrac{\mathcal{R}_{ji}\;(\lambda.t^{\,0}_{i})}{\sum_{k}\,\mathcal{R}_{jk}\:(\lambda.t^{\,0}_{k})}\right)m_j\equiv t^{1}_{i}.
\end{equation}

So we have to try a different vector, for example, a vector which takes \textit{triangular numbers}\footnote{Triangular numbers are generated by the formula: $\mathfrak{T}_{n}=\sum^{n}_{p=0}p=0+1+\cdots+n=\frac{n(n+1)}{2}$.}
as elements, where the values are increasing until the $16^{th}$ element, then decreasing, i.e.
\[\mathsf{t\_trig}=[\:0,\:1,\:3,\:6,\:\cdots,\:6,\:3,\,1,\:0\:]\]
This can be obtained by the following algorithm:
\definecolor{bg}{rgb}{0.97,0.97,0.97}
\begin{pythoncode}
b =0; p =[] 
for n in range(16):
    b = b + n
    p.append(b)
t_trig = [*p,*p[::-1]] 
print(t_trig)        
\end{pythoncode}
Output
\definecolor{bg}{rgb}{0.97,0.97,0.97}
\begin{pythoncode}
[0, 1, 3, 6, 10, 15, 21, 28, 36, 45, 55, 66, 78, 91, 105, 120, 120, 105,
 91, 78, 66, 55, 45, 36, 28, 21, 15, 10, 6, 3, 1, 0]
\end{pythoncode}
Let's try this new \textit{prior truth distribution} in the previous algorithm, where it suffices to change \textsf{t$\_$s} to \textsf{t$\_$trig}:
\begin{figure}[H]
\centering
\includegraphics[width = 0.76\linewidth]{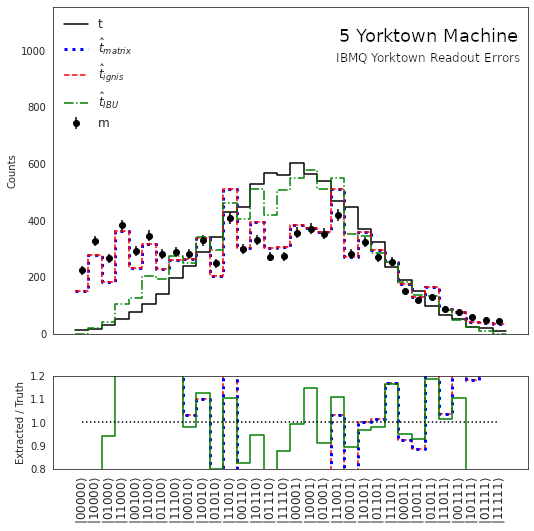}
\caption{Unfolding results of a Gaussian distribution using three methods: Matrix inversion, ignis, and IBU with one iteration and a prior truth spectrum \textsf{t0=t\_{trig}}.}
\end{figure}
With this specific initial truth spectrum (\textsf{t0}), IBU really gives the best unfolding results, the amount of unfolding is very important. From these two results, we conclude that IBU is very sensitive to the choice of $\mathsf{t0}$. To observe this fact let's take 3 possible \textsf{t0}'s:

\[\mathsf{t\_unif}=[\;1,\:1,\:1,\:\cdots,\:1,\:1,\:1\;]\]
\[\quad\;\mathsf{t\_s}=[\;0,\:1,\:2,\:\cdots,\:2,\:1,\:0\;]\]
\[\mathsf{t\_trig}=[\;0,\:1,\:3,\:\cdots,\:3,\:1,\:0\;]\]
\begin{flushleft}
We will apply unfolding via IBU with these 3 \textsf{t0}'s, and for \textsf{n=1}:
\end{flushleft}

\definecolor{bg}{rgb}{0.97,0.97,0.97}
\begin{pythoncode}
unfolded_IBU_a = IBU(m_g, t_unif, R, 1)
unfolded_IBU_b = IBU(m_g, t_s   , R, 1)
unfolded_IBU_c = IBU(m_g, t_trig, R, 1)
\end{pythoncode}

After visualizing the results using the same algorithm, we get: 
\begin{figure}[H]
\centering
\includegraphics[width = 0.8\linewidth]{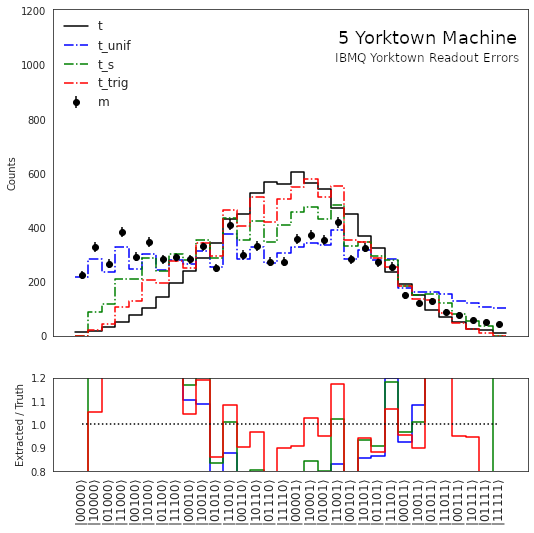}
\caption{A comparison of unfolding results for a Gaussian distribution using \textsf{IBU} with three prior truth distributions: \textsf{t\_{unif}, t\_s, t\_{trig}} and with \bl{1} iteration.}
\end{figure}
The worst unfolding results are when we take \textsf{t0} as a \textit{uniform distribution}.
In our case, we know $\mathsf{t\_g}$, but in real experiences, we ignore it, so a good idea is to take the measured distribution as the prior truth distribution i.e., $\mathsf{t0=m\_g}$ . But its effectiveness depends on the amount of noise.\\

Another property to note and which appears when the number of iterations increases, where the three prior truth distributions converge towards the same result, as seen below for 10 iterations:
\begin{figure}[H]
\centering
\includegraphics[width = 0.8\linewidth]{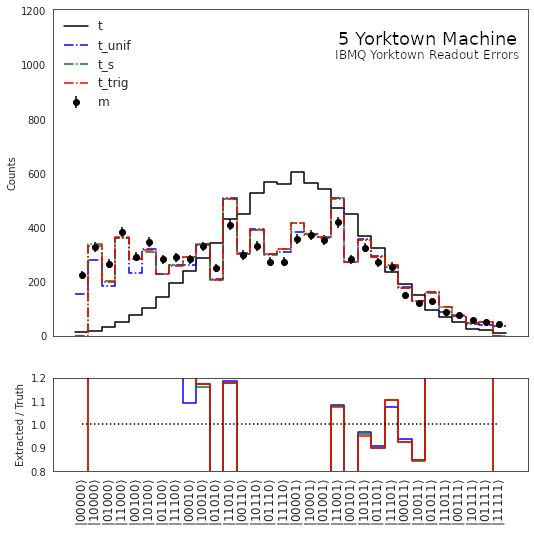}
\caption{A comparison of unfolding results for a Gaussian distribution using \textsf{IBU} with three prior truth distributions: \textsf{t\_{unif}, t\_{s}, t\_{trig}} and with \bl{10} iterations.}
\end{figure}

Now let's investigate the behavior of the IBU, depending on the number of iterations. For this purpose we first do an unfolding on a uniform distribution, then we tackle the Gaussian one, keeping the same parameters ($\mathsf{t\_u,\:m\_u,\:t\_g,\:m\_g}$), and we just vary the number of iterations, where we will take 1, 2, 3, and 10 iterations. Thus, for the first distribution, we have 
\definecolor{bg}{rgb}{0.97,0.97,0.97}
\begin{pythoncode}
t0=np.ones(len(m_u))

unfolded_IBU1  = IBU(m_u,t0,R, 1)
unfolded_IBU2  = IBU(m_u,t0,R, 2)
unfolded_IBU3  = IBU(m_u,t0,R, 3)
unfolded_IBU10 = IBU(m_u,t0,R,10)
\end{pythoncode}
The visualization algorithm: 
\definecolor{bg}{rgb}{0.97,0.97,0.97}
\begin{pythoncode}
bincenters_r = np.array(xvals_r)+0.5
sns.set_style("white")
f = plt.figure(figsize=(8.5, 8.5))
gs = GridSpec(2, 1, width_ratios=[1], height_ratios=[3.5, 1])
ax1 = plt.subplot(gs[0])
ax1.set_xticklabels( () )

mybincenters = np.linspace(-0.5,31.5,33)

plt.step(mybincenters,myre(t_u),color='black',label=r"t")
plt.errorbar(0.5*(mybincenters[1:]+mybincenters[:-1]),m_u,yerr=np.sqrt(
              m_u),label='m',marker='o',linestyle='none',color='black')
plt.step(mybincenters,myre(unfolded_IBU1),color='red',linestyle='-.',la
                                         bel=r"$\hat{t}_{IBU}\,$(n=1)")
plt.step(mybincenters,myre(unfolded_IBU2),color='blue',linestyle='-.',l
                                        abel=r"$\hat{t}_{IBU}\,$(n=2)")
plt.step(mybincenters,myre(unfolded_IBU3),color='orange',linestyle='-.'
                                      ,label=r"$\hat{t}_{IBU}\,$(n=3)")
plt.step(mybincenters,myre(unfolded_IBU10),color='green',linestyle='-.'
                                     ,label=r"$\hat{t}_{IBU}\,$(n=10)")
_=plt.xticks(xvals_r,xlabs0_r,rotation='vertical',fontsize=10)

plt.ylabel('Counts')
plt.ylim([0,2.*max(t_u)])
plt.legend(loc='upper left',fontsize=8,frameon=False)
_=plt.text(20., max(t_u)*1.8, "5 Yorktown Machine", fontsize=18)
_=plt.text(20., max(t_u)*1.7, "IBMQ Yorktown Readout Errors",fontsize=12)

ax2 = plt.subplot(gs[1])
ratio_unfold1  =  np.zeros(2**nqubits)
ratio_unfold2  =  np.zeros(2**nqubits)
ratio_unfold3  =  np.zeros(2**nqubits)
ratio_unfold10 = np.zeros(2**nqubits)

for i in range(len(ratio_unfold1)):
    ratio_unfold1[i] =1.
    ratio_unfold2[i] =1.
    ratio_unfold3[i] =1.
    ratio_unfold10[i]=1.
    if (t_u[i] > 0):
        ratio_unfold1[i]  = unfolded_IBU1[i]/t_u[i]
        ratio_unfold2[i]  = unfolded_IBU2[i]/t_u[i]
        ratio_unfold3[i]  = unfolded_IBU3[i]/t_u[i]
        ratio_unfold10[i] = unfolded_IBU10[i]/t_u[i]
        pass
    pass

plt.ylim([0.8,1.2])
plt.step(mybincenters,myre(ratio_unfold1 ),color='red' )
plt.step(mybincenters,myre(ratio_unfold2 ),color='blue')
plt.step(mybincenters,myre(ratio_unfold3 ),color='orange')
plt.step(mybincenters,myre(ratio_unfold10),color='green' )
plt.ylabel('Extracted / Truth')

xx2 = [1.,1.]
plt.plot([xvals_r[0],xvals_r[len(xvals_r)-1]],xx2,color='black'
                                                ,linestyle=':')
_=plt.xticks(xvals_r,xlabs_r,rotation='vertical',fontsize=12)
\end{pythoncode}
\begin{figure}[H]
\centering
\includegraphics[width = 0.9\linewidth]{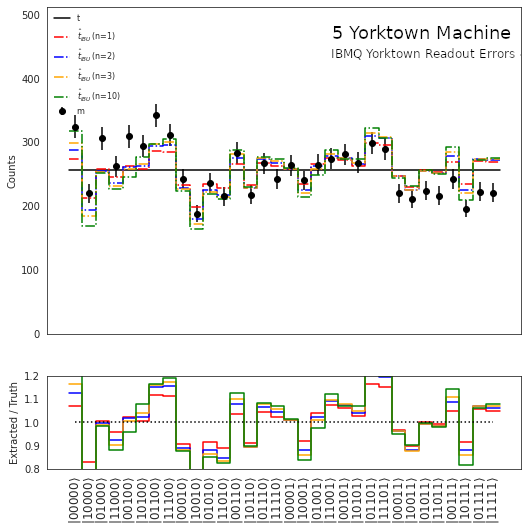}
\caption{Unfolding results for a uniform distribution using IBU with 1,2,3,10 iterations and a uniform prior truth spectrum.}
\end{figure}
\pagebreak

Now, for the Gaussian distribution:
\begin{figure}[H]
\centering
\includegraphics[width = 0.9\linewidth]{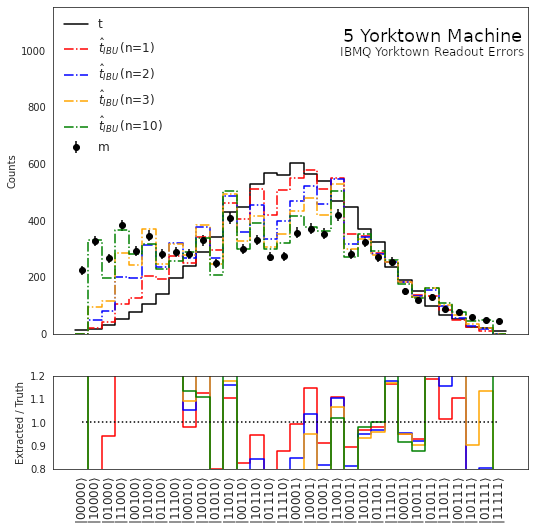}
\caption{Unfolding results for a Gaussian distribution using IBU with 1,2,3,10 iterations and a prior truth spectrum \textsf{t0=t\_{trig}}.}
\end{figure}
Note that for the two distributions, the most efficient unfolding is for $\mathbf{n=1}$, where further iterations \textit{diverge} more and more from the truth $\mathsf{t}$ until we finish with results identical to that obtained by \textsf{MI} and \textsf{Ignis}. Therefore, just \textbf{1} iteration is needed to get satisfactory results.

\pagebreak
\section{Unfolding Readout Errors from The Melbourne IBM Q Machine}
In this section, we want to explore more about unfolding methods for the IBM Melbourne quantum
computer. This quantum computer has 15 qubits, but we will construct a response matrix for 7 qubits (we can not visualize clearly the result for more then 7 qubits). Then, we will follow the same steps to unfold the readout errors from the Melbourne
quantum machine and represent the result graphically.
\begin{center}
\textbf{The Response Matrix}
\end{center}
To construct the response matrix we follow the same steps as the Yorktown QC with the only difference is that we should split ours measurement calibration circuits generated by \textsf{complete\_meas\_cal}. We do this step because the number of experiments supported by the device is $75$ which means for a number of qubits greater than $6$ we should split the experiment. ($75$ is between $2^6=64$ and $2^7=128$).
 
We split our measurement calibration circuits generated by \textsf{complete\_meas\_cal} into two batches: 
\begin{pythoncode}
device=provider.get_backend('ibmq_16_melbourne')
job1_res = qiskit.execute(meas_calibs[0:64], backend=device, shots=8192,optimization_level=0).result()
job2_res = qiskit.execute(meas_calibs[64:128], backend=device, shots=8192,optimization_level=0).result()
\end{pythoncode}
Now, we use the results to initialize a measurement correction fitter with \textsf{CompleteMeasFitter} with the first 64 circuits:
\begin{pythoncode}
meas_fitter = CompleteMeasFitter(job1_res, state_labels,circlabel='mcal')
\end{pythoncode}
In this step we use the \textsf{CompleteMeasFitter.add\_data} function to update the measurement correction fitter with the rest of calibration circuit batches.
\begin{pythoncode}
meas_fitter.add_data(new_results=job2_res)
\end{pythoncode}
We need to update the fitter with the last batch of calibration circuits to see the full calibration matrix.  If we plot the calibration matrix after initializing the measurement correction fitter with the first batch of calibration circuits, we will see a half-calibration matrix. 

Finally, we use the same code as Yorktown QC to plot the heatmap of the Melbourne response matrix, the only change is the number of the qubits. The matrix will have a dimension of $2^7 \times 2^7$.

\begin{center}
\begin{figure}[H]
\includegraphics[width = 1\linewidth]{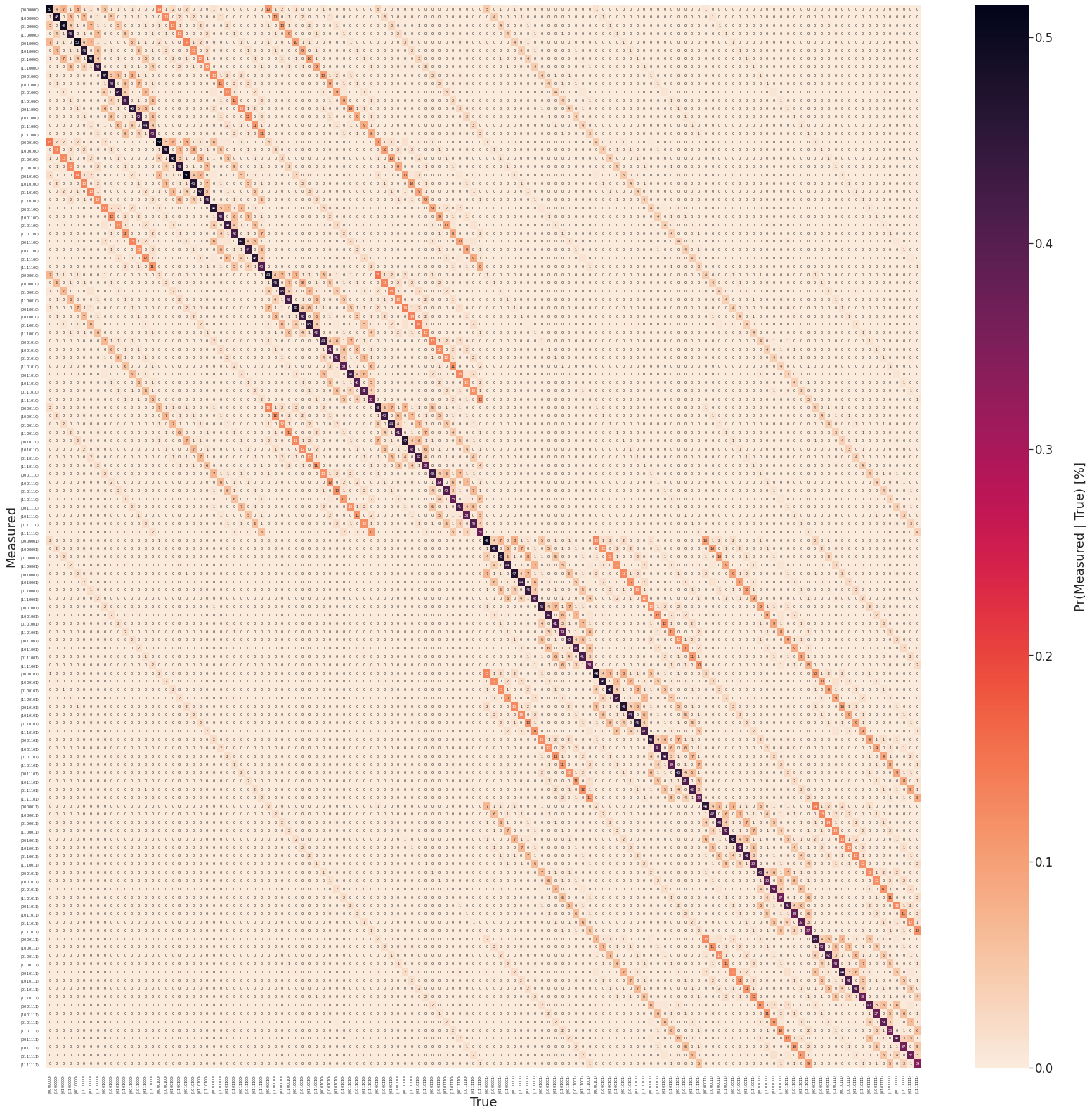}
\caption{Response matrix for $7$ qubits from the IBM Q Melbourne machine.}
\end{figure}
\end{center}

We notice that as the preview calibration matrix, there is a significant diagonal component with high probability and also many off-diagonal components with a non-zero probability of transitions.

\begin{center}
\textbf{Unfolding a Uniform Distribution}
\end{center}
First, we create a quantum circuit that generates a uniform distribution. The output will be a uniform distribution only if quantum computers was perfect, but this is not the case. So, the distribution will be distorted by the noisy Melbourne quantum machine.

The circuit for seven qubits which generate uniform distribution is represented as follow: 
\begin{figure}[H]
\centering
\includegraphics[width = 0.65\linewidth]{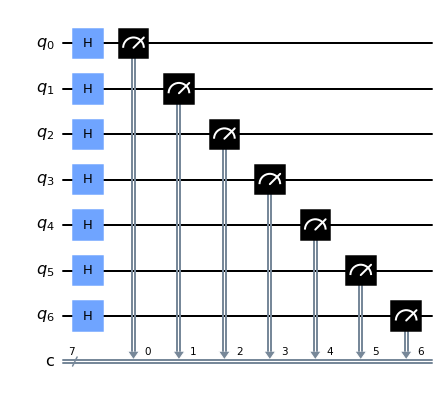}

\caption{A circuit that generates a uniform distribution based on $7$ qubits.}
\end{figure}

After the measurement, we organize the data and, extract the count. We get the following data from IBM Melbourne quantum computer: 
\begin{pythoncode}
[73, 74, 100, 84, 69, 79, 70, 72, 79, 62, 87, 68, 82, 63, 60, 62, 61, 70, 67, 76, 68, 72, 64, 74, 54, 82, 69, 65, 61, 72, 54, 64, 74, 57, 52, 66, 60, 77, 85, 75, 88, 63, 67, 50, 75, 52, 48, 46, 57, 47, 67, 52, 90, 68, 59, 80, 70, 69, 68, 82, 57, 79, 48, 57, 54, 63, 41, 66, 64, 75, 82, 73, 68, 64, 78, 62, 73, 78, 61, 61, 56, 60, 66, 58, 65, 50, 82, 75, 67, 78, 61, 67, 49, 65, 48, 59, 52, 56, 43, 60, 63, 50, 29, 63, 65, 63, 61, 48, 67, 63, 59, 50, 49, 48, 35, 47, 65, 35, 47, 41, 71, 62, 61, 60, 62, 71, 68, 67]
\end{pythoncode}
We can represent the result of the measurement in a histogram: 
\begin{pythoncode}
plot_histogram(m,figsize=(45,10))
\end{pythoncode}
The output of the plot:
\begin{figure}[H]
\centering
\includegraphics[width = 1\linewidth]{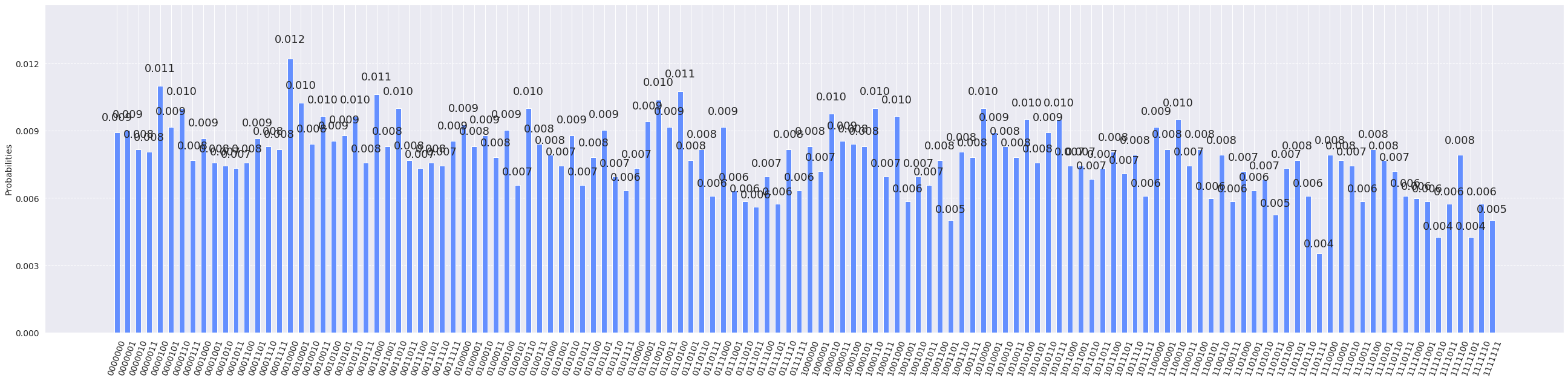}

\end{figure}
We recall that our true distribution can be generated using the following code: 
\begin{pythoncode}
t_u = [8192/128 for i in range(128)]
\end{pythoncode}
The output after using the print function:
\begin{pythoncode}
[64.0, 64.0, 64.0, 64.0, 64.0, 64.0, 64.0, 64.0, 64.0, 64.0, 64.0, 64.0, 64.0, 64.0, 64.0, 64.0, 64.0, 64.0, 64.0, 64.0, 64.0, 64.0, 64.0, 64.0, 64.0, 64.0, 64.0, 64.0, 64.0, 64.0, 64.0, 64.0, 64.0, 64.0, 64.0, 64.0, 64.0, 64.0, 64.0, 64.0, 64.0, 64.0, 64.0, 64.0, 64.0, 64.0, 64.0, 64.0, 64.0, 64.0, 64.0, 64.0, 64.0, 64.0, 64.0, 64.0, 64.0, 64.0, 64.0, 64.0, 64.0, 64.0, 64.0, 64.0, 64.0, 64.0, 64.0, 64.0, 64.0, 64.0, 64.0, 64.0, 64.0, 64.0, 64.0, 64.0, 64.0, 64.0, 64.0, 64.0, 64.0, 64.0, 64.0, 64.0, 64.0, 64.0, 64.0, 64.0, 64.0, 64.0, 64.0, 64.0, 64.0, 64.0, 64.0, 64.0, 64.0, 64.0, 64.0, 64.0, 64.0, 64.0, 64.0, 64.0, 64.0, 64.0, 64.0, 64.0, 64.0, 64.0, 64.0, 64.0, 64.0, 64.0, 64.0, 64.0, 64.0, 64.0, 64.0, 64.0, 64.0, 64.0, 64.0, 64.0, 64.0, 64.0, 64.0, 64.0]
\end{pythoncode}
Finally, after using the three unfolding methods, the result is represented in the following graphics:

\begin{figure}[H]
\centering
\includegraphics[width = 0.8\linewidth]{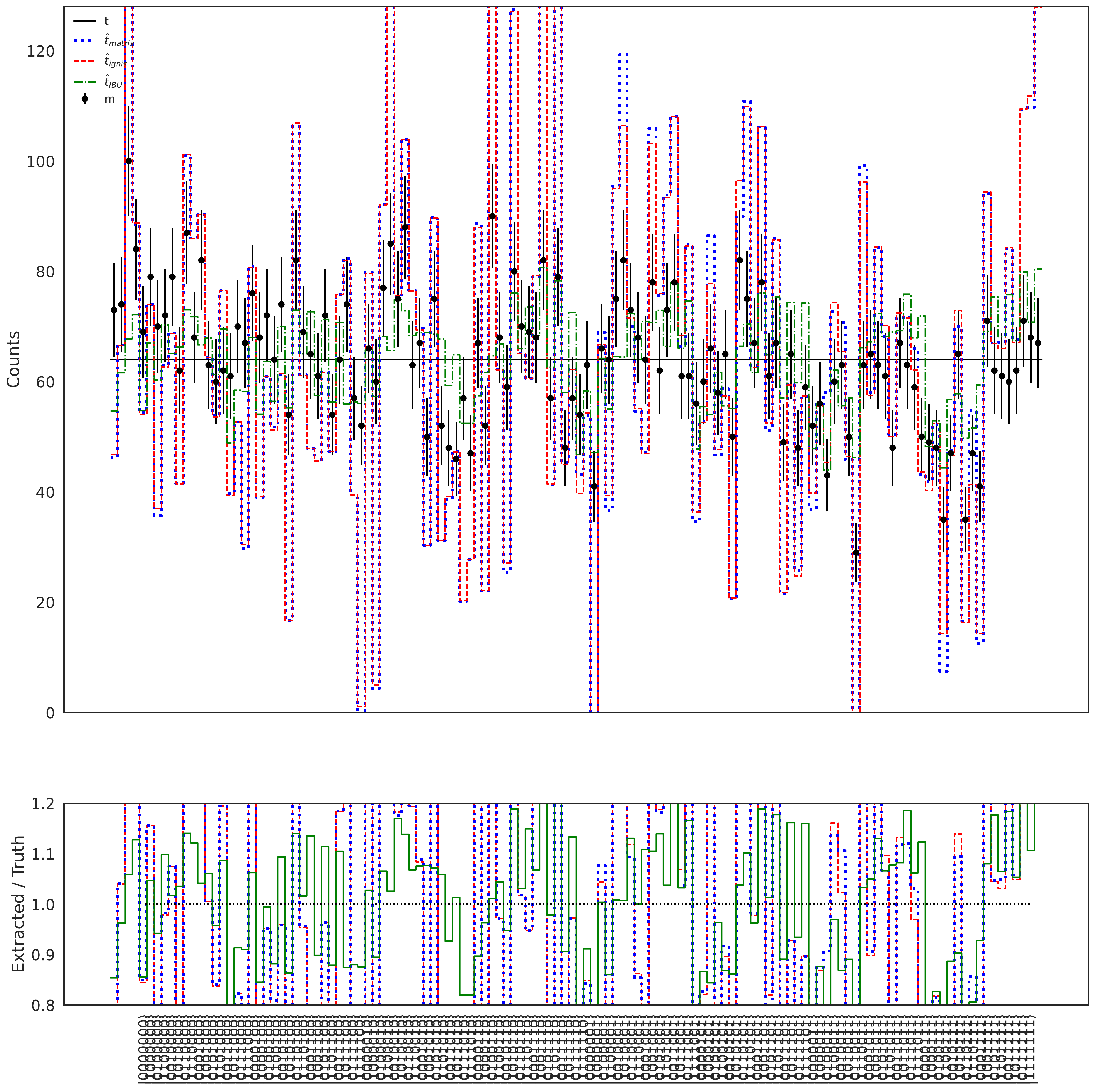}

\caption{Unfolding results of a uniform distribution based on seven-qubits using $3$ methods: Matrix inversion, ignis, and IBU with one iteration and a uniform prior truth spectrum.}
\end{figure}

This is clearly different from the previous situation of 5 Qubits. First, we notice that the uncertainties on the \textit{measured distribution} values are very high. Secondly the two unfolding methods \textsf{MI} and \textsf{ignis} show \textit{large oscillations}. Unlike the case of 5 Qubits, their values, in this case, are not completely \textit{identical}. Neither technique was able to unfold this measured distribution. The \textsf{IBU} does not exhibit large oscillations, and it was able to unfold a certain extent although the great uncertainties on \textsf{m$\_$u}. 

So, after increasing the number of qubits we notice that the first two methods have difficulties to unfold; for the IBU, its ability to unfold has decreased but compared to the two other methods, its unfolding is important.\\

After these several experiences which generally aim to deal with readout noise of quantum computers, and especially to study the \textsc{Iterative Bayesian Unfolding} method. We can recap what we have learned about this unfolding technique:
\begin{itemize}

\item IBU is very sensitive to the choice of the \textit{prior truth distribution} \textsf{t0}, as well as to the number of iterations \textsf{n}.
\item Unfolding with the vector \textsf{t$0$} or with $\mathsf{\lambda.t0}$ gives the same values, $\forall\,\lambda\in \mathbb{R^*}$.
\item If the $i^{\,th}$ component of the vector \textsf{t$^0\!\equiv\,$ t0} equal to 0 then the $i^{\,th}$ component of the vector resulting from the unfolding process is equal to 0, i.e.
\[\mathsf{t^0}=[\;\alpha,\:0,\:\beta,\:\gamma,\:0,\:0\;]\;\xrightarrow{\mathsf{IBU}}\;
\mathsf{\hat{t}^{\,n}_{\,IBU}}=[\;\alpha',\:0,\:\beta',\:\gamma',\:0,\:0\;]\]
\item If the chosen $\textit{\textsf{t}}^\mathbf{0}$ is of the same \textbf{nature} of the truth $\textit{\textsf{t}}$, then one have to take the \textbf{minimum} number of iterations, and vice versa.
\item For a higher number of iterations $\left(\textsf{\textit{n}}\gg\mathbf{1}\right)$ all chosen \textit{prior truth distributions} converge towards the same result. Moreover, this common result is identical to that obtained by \textsf{MI} and \textsf{Ignis}.

\end{itemize}
\chapter*{Conclusion}
\addcontentsline{toc}{chapter}{Conclusion}
The invention of the Quantum Computer represents a leap for human civilization, which will open new horizons and allow the emergence of a new era, not only in physics, but will contribute to the development of all sciences. However, current challenges limit its effectiveness, and one of the important limitations is the rate of errors.\\

In this master thesis, our objective was to mitigate Quantum Computer Readout Errors, using of Unfolding Methods from the field of High Energy Physics. To achieve this goal, we first had to understand the basic concepts of Classical and Quantum Information Science QIS at the same time. The final subject in this overview was Classical and Quantum Error Detection and Correction, where we explored the differences between the classical and quantum situation.\\

After this we visited the field of High Energy Physics, to understand the concept of unfolding, where we saw the simplest method called “Matrix Inversion” and a more advanced one called “Iterative Bayesian Unfolding” (IBU), which represented a promising technique for mitigating readout errors in QIS.\\

Then we investigated readout errors in the field of QIS, where we tried to understand their nature and their sources and how to construct the response matrix in order to mitigate these errors.\\

Finally, we studied three unfolding methods (MI, ignis, IBU) through a real Quantum Computer Readout Noise, for a uniform and Gaussian distribution, and we found that it gives an excellent unfolding compared to MI and ignis, which contribute to confirm the results of the researchers \textsc{Benjamin Nachman} et al. in 2020. After achieving this, we took a small step beyond these results to explore in particular the behavior of IBU in a larger system of qubits, where we found that IBU maintained its effectiveness.\\

Our work is a small step to move beyond the era of Noisy Intermediate-Scale Quantum computers. Further studies are needed to overcome the noise of quantum computers. \\

\begin{center}
\textit{“We can only see a short distance ahead, but we can see plenty there that needs to be done.”}
\end{center}
\begin{flushright}
\textsc{\textit{Alan Turing}}
\end{flushright}
\startappendices

\minitoc

\let\cleardoublepage\clearpage
\chapter{Boolean Algebra}

A boolean algebra represents the mathematical rules between two-state quantities to solve logical problems. These quantities are known as \textbf{Boolean logic variables} and can be represented by $0$ and $1$ or \textsc{true} and \textsc{false}. Boolean algebra use only three operators: \textsc{not},  \textsc{and},  \textsc{or}. The \textsc{and} and \textsc{or} operators work on two or more variable, the \textsc{not} operator operate on a single variable. These operators are represented in table \ref{TAB1} as follow \cite{DElec}:\\  

\begin{table}[ht!]
\centering
\begin{tabular}{|c|c|c|c|}
\hline 
Operator & Symbol & Usage & Spoken as \\ 
\hline 
\textsc{not} & $\bar{}$ & $\bar{A}$ & A bar , or not A \\ 
\textsc{or}  & $+$      & $A+B$     & A or B \\ 
\textsc{and} & $.$      & $A.B$     & A and B \\ 
\hline 
\end{tabular} 

\caption{Boolean variables and operators.}
\label{TAB1}
\end{table}
$\bullet$ \textbf{The NOT Operator}: A boolean variable can be 0 or 1. If we want to switch the state from 0 to 1, we need an operator that produces the complement of the variable 0. In the same way, we apply \textsc{not} to the state 1 to get 0. In the general form, every variable $A$ has an opposite value denoted as $A$-not or $A$-bar, or $A$-prime. If $A=1$ then $\bar{A}=0$ or if $A=0$ then $\bar{A}=1$.\\

$\bullet$ \textbf{The OR Operator}: The \textsc{or} operator acts on at least two inputs variables and produces only one output. The result depends on the variables; if there is at least one variable with the value $1$, the result will be $1$. If all the variables are $0$, the result will be $0$. The equation for \textsc{or} operation (for two variables) is $A+B=R$ with $A$ and $B$ the inputs, and they take the value  $0$ or $1$.  The «+» sign stands for the \textsc{or} operation and $R$ for the boolean result. We can see that this logical function acts according to these two boolean conditions: $X+1=1$ and $X+0=X$. The input $X$ takes the value $0$ or $1$ \cite{Mandal}. We will see this condition in the \textit{Boolean lows}.\\

$\bullet$ \textbf{The AND Operator}: The \textsc{and} operator works as a product of two or more variables. If all variables are $1$, the result will also be $1$. However, if at least one of the variables is $0$, the result will be $0$. The equation for \textsc{and} operation (for two variables) is $A.B=R$ with $A$, $B$, the input, and they take the value  $0$ or $1$.  The «.» sign stands for the \textsc{and} operation, and $R$ is the boolean result. The \textsc{and} function acts according to the following relation: $X.1=X$ and $X.0=0$. The variable $X$ takes the value $0$ or $1$. 
\chapter{Boolean Lows}
In this part, we will see the boolean laws that describe how operators act on boolean variables. There are two categories of boolean laws; the first describes operations on a single variable, and the second, for multivariable \cite{DElec}.

\begin{enumerate}

\item \textbf{Single variable}

This part will explain the rules that describe the logic operation for only one variable. There are three laws for a single variable: idempotent, Inverse, Involution. These lows are represented as follow: 

\begin{itemize}

\item \textbf{Idempotent}

Idem means “\textit{same}” in Latin. These laws describe the effect of an operator when the same variable goes to all inputs. Let’s take two-input \textsc{or} for example: $A+A=S$ this will give 1 if $A=1$, and 0 if $A=0$; therefore $A+A=A=S$.

\item \textbf{Inverse elements}

This law is about the effect of an operator on a variable $A$ with its complement $\bar{A}$. Let us take \textsc{and} as example, the expression is written as follow: $R=A.\bar{A}$. If $R=0$ then either $A$ or $\bar{A}$ must have $0$ because they have a complementary values. 

\item \textbf{Involution law}

When we apply the \textsc{not} for the first time on a variable $A$, we obtain $\bar{A}$, but by applying it for the second time, this will return the variable to its original state $\bar{\bar{A}}=A$.

\end{itemize}

\item \textbf{Multivariable}

Multi-variable theorems are about operations on more than one variable, and these laws are: \textit{commutative law, associative, distributive, De Morgan’s laws, Absorption}.

\begin{itemize}

\item \textbf{Commutative law}

when applying the operators \textsc{or}, \textsc{and} between two variables; the order of these variables does not matter. So $R=A.B=B.A$  and $S=A+B=B+A$.

\item \textbf{Associative law}

These laws show how the operations can be grouped, and it is the same as for conventional algebra. For example: $A+(B+C)=(A+B)+C$ and $A.(B.C)=(A.B).C$.

\item \textbf{Distributive law}

Distribution is used to expand and simplify the boolean expression. Example: $A.(B+C) = A.B+A.C$.

\item \textbf{De Morgan’s law}

This theorem is about \textit{complementation} and announces that, complementing $A.B$ is equivalent to $\bar{A}+\bar{B}$  and complementing $A+B$ is equivalent to $\bar{A}.\bar{B}$.

\item \textbf{Absorption}

These laws are used for \textit{circuit simplification} to increase the \textit{reliability} of the logic circuit and decrease the cost of manufacture. To understand the laws of Absorption, let us develop some expressions.

\begin{itemize}
\item[ ] $A + A.B = A.(1+B) = A.1=A$
\item[ ] $A.(A + B) =  A.A + A.B = A + A.B = A.(1+B)=A$
\item[ ] $A.(\bar{A} + B) = A.\bar{A} + A.B  = 0 + A.B = A.B$
\item[ ] $A.B + \bar{B} =  \bar{B}.(B.A.B+1)= \bar{B}.(B.A+1)= A+ \bar{B}$
\item[ ] $A + \bar{A}.B = (\bar{A}+A).(A+B)=1.(A+B)=A+B$
\end{itemize}
\end{itemize}
\end{enumerate}

\chapter{Qubit Design}

Due to the complexity of quantum mechanics, it is not easy to build a practical quantum computer. There is great competition and enormous effort from large companies to accelerate the development of quantum programming. Today several types of Qubits are available. We site the most common types of qubits:\\

\vspace{2mm}
$\bullet$ \textbf{Superconducting Loops}\\
Superconducting quantum computing is an implementation of a quantum computer in superconducting electronic circuits.
Superconducting quantum circuits use Josephson junction to create a \textit{non-linear} inductance which allows a design of \textit{anharmonic oscillators}. A quantum harmonic oscillator cannot be used as a qubit, as there is no way to address only two of its states. A microwave pulse is injected which excites the current into a superposition of states. This design is used by IBM.

\vspace{2mm}
$\bullet$ \textbf{Trapped Ions}\\
Ion trap uses a combination of electric or magnetic fields to capture ions or charged atomic particles in a system (isolated from the external environment). Qubits are stored in stable electronic states of each ion, and quantum information can be transferred through interacting by the Coulomb force. To induce coupling between the qubit states, we apply lasers for single-qubit operations or entanglement between qubits.

\vspace{2mm}
$\bullet$ \textbf{Silicon Quantum Dots}\\
Quantum dots are semiconductor devices at the nanoscale. Electrons are vertically confined to the ground state of quantum \textit{gallium arsenide} (GaAs) and form a two-dimensional electron gas (2DEG). This gas is free to move in two dimensions but tightly confined in the third. This confinement leads to quantized energy levels for movement in the third direction.

\vspace{2mm}
$\bullet$ \textbf{Diamond Vacancies}\\
This multiqubit system is based on a \textit{nitrogen-vacancy} (NV) center in diamond.
The NV center’s electron spin constitutes the central qubit, while the nuclear spins of the nitrogen and surrounding carbon atoms form the other qubits. 
Electron spin responds optically, so the quantum state can be programmed and read out quickly using a laser, the electron spin is used as a bus, that can be coupled through magnetic interaction. For quantum memory and processing, we use the nuclear spin, which does not respond optically and has longer coherence times.
\section*{Superconducting Qubit}
\label{ap c}
Qubit design significantly impacts creating devices with high performance, like long coherence time and high controllability. There are many quantum computing platforms, but the most famous one is based on \textbf{superconducting Qubits}. Because of its advantages, superconducting Qubit became the leading candidate for a scalable quantum processor. Like any quantum mechanical object, superconducting qubits can be controlled, placed into quantum superposition states, exhibit quantum interference effects, and become entangled \cite{Materials}.

\vspace{2mm}
$\bullet$ \textbf{What is superconducting quantum computing?}\\
A superconducting quantum computer is a connected artificial atom (rather than natural atoms). Each artificial atom is represented as a non-linear inductor-capacitor circuit. The inductor-capacitor circuits are harmonic and the energy levels are equally spaced if we add a non-linear element in the circuit; therefore, the energy separation between pairs of levels differ from each other, therefore the transitions between pairs of energy levels can be separately addressed \cite{Superconducting}.

\begin{figure}[H]
\centering
\includegraphics[width=13cm]{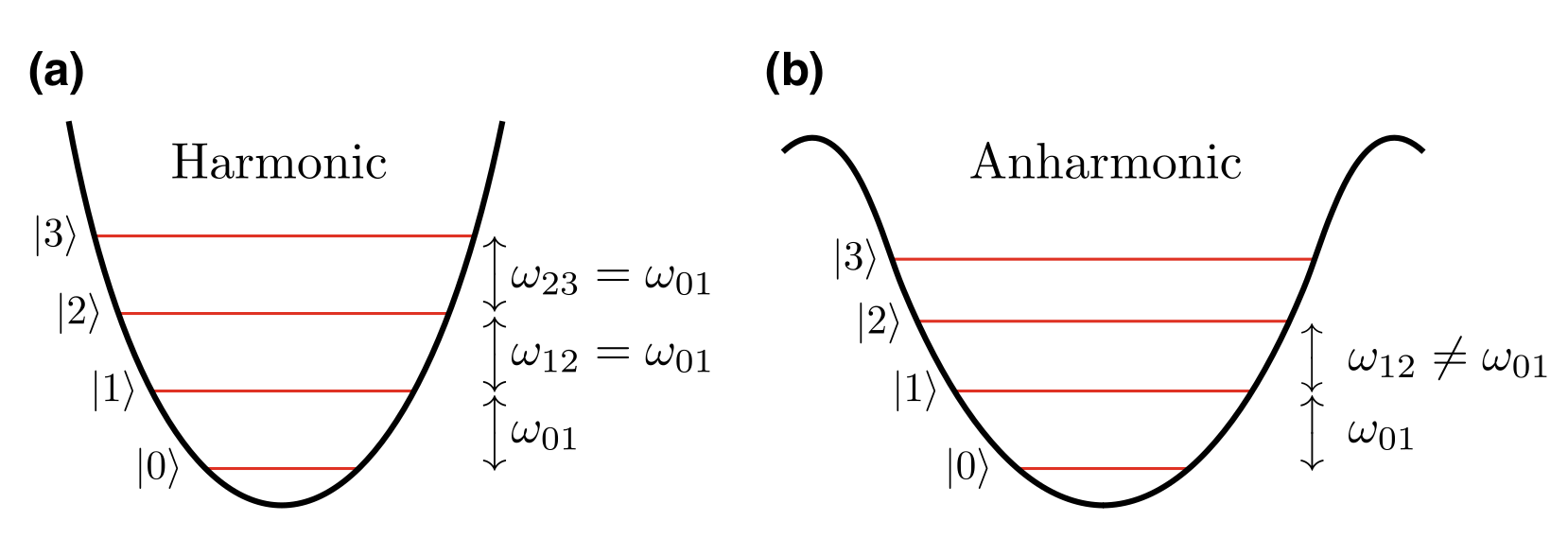}
\caption{Harmonic and anharmonic systems and their suitability as qubits \cite{Kockum2019Aug}.}
\end{figure}

\textbf{(a)} In the quadratic potential (black curve) of a harmonic system, the energy levels (red lines) are equally spaced, i.e. $\omega_{j, j+1} = \omega_{01}$, where $\omega_{jk}$ is the transition frequency between energy levels $j$ and $k$. A signal at frequency $\omega_{01}$ will thus not only transfer population from $\ket{0}$ to $\ket{1}$, but also from $\ket{1}$ to $\ket{2}$, etc.

\textbf{(b)} In the potential of an anharmonic system, e.g., the cosine potential characteristic of a Josephson
junction, $\omega_{01} \neq \omega_{12}$ . A signal at frequency $\omega_{01}$ will thus only drive transitions between $\ket{0}$ to $\ket{1}$
and not affect any other levels in the system (provided that the signal is not too strong). This limits the dynamics to the two-level system formed by $\ket{0}$ and $\ket{1}$, which can be interpreted as a qubit \cite{Superconducting}.

\vspace{2mm}
$\bullet$ \textbf{Why we use superconductors?}\\
 A superconductor enables an electric current in a circuit loop to circulate without
resistance. Depends on the 
inductance and the capacitance of the electric circuit, the electric field oscillates with a specific frequency and the electric current experiences no resistance during its oscillation and is called a \textit{supercurrent}. If the supercurrent is counter-clockwise then it generates an upward magnetic flux, and a clockwise supercurrent generates a downward magnetic flux. Superposing
clockwise and counter-clockwise supercurrents creates a superposition of up and down magnetic fields. For a small circuit, the magnetic flux is quantized and equally spaced which does not help make a two-level qubit. To make the spacing unequal, a Josephson junction is placed into the circuit.

\vspace{2mm}
$\bullet$\textbf{ What is Josephson junction?}\\  
A Josephson junction contains two superconducting materials weakly linked via another medium, such as an insulator. The supercurrent in a loop faces an energy barrier at this weak link, which modifies the \textit{harmonic supercurrent oscillation} to an \textit{anharmonic dynamic}. This results in unequally spaced magnetic flux levels. For sufficiently high anharmonicity, just the two lowest levels play a role in the dynamics so we have an effective two-level atom that functions as a flux qubit. The downward magnetic flux serves as $\ket{0}$ and the upward flux as $\ket{1}$ \cite{Sanders2017Nov}.

     
\section*{The Basic Josephson-Junction Qubits}

There are three basic designs for Josephson-junction qubits. The three are known as a charge qubit, a flux qubit, and a phase qubit. The charge qubit is a box for charge, controlled by an external voltage $V_g$; the flux qubit is a loop controlled by an external magnetic flux $\phi_{ext}$; and the phase qubit is a Josephson junction biased by a current $I_b$ \cite{Kockum2019Aug}.

\begin{figure}[H]
\centering
\includegraphics[scale=0.25]{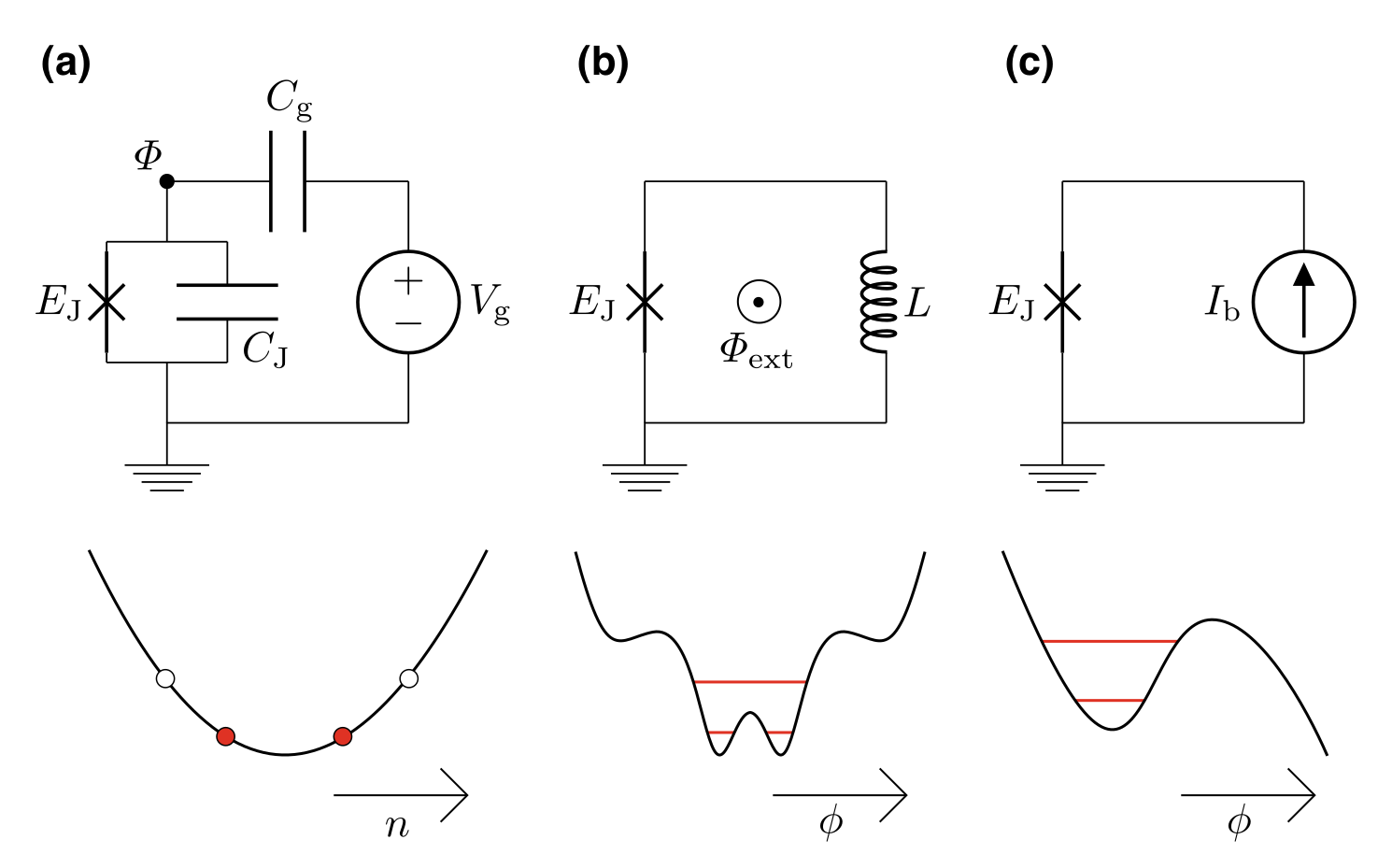}
\caption{The three basic \textbf{Josephson-junction} qubit circuits and their potential-energy landscapes, with the two lowest energy levels marked in red. \textbf{(a) Charge qubit}. \textbf{(b) Flux qubit}. \textbf{(c) Phase qubit}. For simplicity, the capacitance $C_J$ is only shown in panel \textbf{(a)}, although it is also present in the circuits in panels \textbf{(b)} and \textbf{(c)} \cite{Kockum2019Aug}.}
\end{figure}

\section*{Quantum Processor}

IBM’s quantum chips consist of a \textbf{silicon wafer} that contains qubits and resonators composed of \textbf{aluminum} and \textbf{niobium}. A resonator is connected to each qubit to deliver the control pulses that govern the qubit’s state. 

An entangled qubit’s state cannot be described independently of its paired qubit. If the state of the first qubit is measured, measuring the state of the second qubit is correlated to that of the first. Practical quantum computations often involve executing conditional logic operations in which one qubit’s state depends on that of another qubit.

To determine the result of a qubit’s operation, a measurement signal is sent as a \textbf{microwave pulse} to the qubit’s resonator. The measurement destroys the qubit’s superposition and collapses the qubit’s state to 0 or 1. The reflected signal travels back up the \textbf{dilution refrigerator}, passing through \textit{isolators} that protect the qubits from noise as well as \textit{amplifiers} that boost the signal’s strength enough to read the qubit’s state accurately. The signal then passes back through the stack of \textbf{microwave electronics}, this time to be converted from \textbf{microwave frequency} to a digital signal that is sent over the internet to the local computer to display the result of the quantum calculation\footnote{\url{https://electronics360.globalspec.com/article/13553/how-quantum-computers-work}}.

\begin{figure}[!ht]
\centering
\includegraphics[scale=0.4]{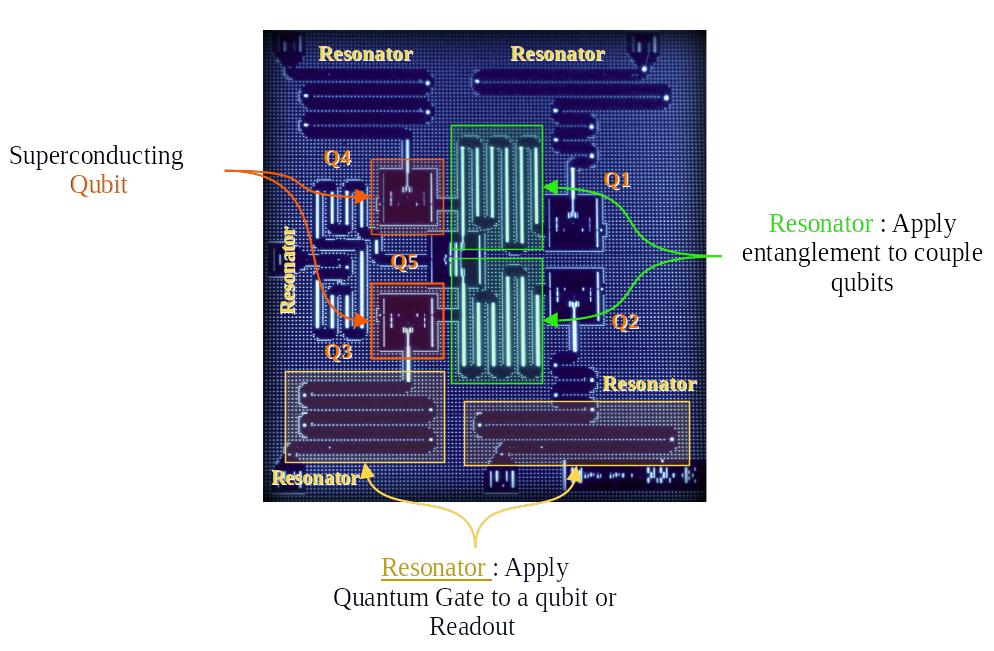}
\end{figure}


\chapter{Python and Qiskit toolkit}
In this part, we will see the most used python and qiskit functions.
\section*{Qiskit packages}
\begin{sortedlist}
\sortitem{{\color{blue}{QuantumRegister}}: Implement a quantum register.} 
\sortitem{{\color{blue}{ClassicalRegister}}: Implement a classical register.}
\sortitem{{\color{blue}{QuantumCircuit}}: Holds quantum operations and instructions for the quantum system.}
\sortitem{{\color{blue}{Aer}}: is an element that provides quantum computing simulators with realistic noise models.}
\sortitem{{\color{blue}{qiskit.providers.aer}}: This module contains classes and functions to build a noise model for simulating a Qiskit quantum circuit in the presence of errors.}
\sortitem{{\color{blue}{readout$\_$error}}: Readout error class for Qiskit Aer noise model.}
\sortitem{{\color{blue}{qiskit.ignis.mitigation.measurement}}: The Qiskit Ignis Measurement Calibration module and mitigation is used to calibrate measurement errors via a series of basis state measurements and then apply that calibration to correct the average results of another experiment of interest.}
\end{sortedlist}
\section*{Python tools}
\begin{sortedlist}
\sortitem{{\color{blue}{NumPy}}: which stands for Numerical Python, is a library consisting of multidimensional array objects and a collection of routines for processing those arrays. Using NumPy, mathematical and logical operations on arrays can be performed.}
\sortitem{{\color{blue}{matplotlib}}: is a library for creating static, animated, and interactive visualizations in Python.}
\sortitem{{\color{blue}{matplotlib.pyplot}}: is a collection of command style functions that make Matplotlib work like MATLAB. Each Pyplot function makes some change to a figure. For example, a function creates a figure, a plotting area in a figure, plots some lines in a plotting area, decorates the plot with labels, etc.}
\sortitem{{\color{blue}{matplotlib.gridspec}}: is used to specify the grid’s geometry to place a subplot. We need to set the number of rows and columns of the grid.}
\sortitem{{\color{blue}{mpl$\_$toolkits.axes$\_$grid1}}: is a collection of helper classes to ease displaying (multiple) images with matplotlib. In matplotlib, the axes location (and size) is specified in the normalized figure coordinates, which may not be ideal for displaying images that needs to have a given aspect ratio.}
\sortitem{{\color{blue}{make$\_$axes$\_$locatable}}: takes an existing axes instance and creates a divider for it. It provides append\_axes() method that creates a new axes on the given side (“top”, “right”, “bottom” and “left”) of the original axes.}
\sortitem{{\color{blue}{SciPy optimize}}: provides functions for minimizing (or maximizing) objective functions, possibly subject to constraints. It includes solvers for nonlinear problems (with support for both local and global optimization algorithms), linear programing, constrained and nonlinear least-squares, root finding, and curve fitting.}
\sortitem{{\color{blue}{seaborn}}: Seaborn is a Python data visualization library based on matplotlib. It provides a high-level interface for drawing attractive and informative statistical graphics.}
\sortitem{{\color{blue}{ matplotlib.lines}}: This module contains all the 2D line class which can draw with a variety of line styles, markers and colors.}
\sortitem{{\color{blue}{np.random.seed()}}: generate random numbers based on “pseudo-random number generators” algorithms. The seed() is used to initialize the random number generator to be able to generate a random number.}
\sortitem{{\color{blue}{np.clip}}: function is used to Clip (limit) the values in an array. Given an interval, values outside the interval are clipped to the interval edges. For example, if an interval of $[0, 1]$ is specified, values smaller than 0 become 0, and values larger than 1 become 1.}
\sortitem{{\color{blue}{np.random.randn}}: Return a sample (or samples) from the “standard normal” distribution.}
\sortitem{{\color{blue}{random.uniform(low, high, size)}}: Draw samples from a uniform distribution. Samples are uniformly distributed over the half-open interval [low, high) (includes low, but excludes high). In other words, any value within the given interval is equally likely to be drawn by uniform.}
\sortitem{{\color{blue}{round}}: returns a floating point number that is a rounded version of the specified number.}
\sortitem{{\color{blue}{plt.hist}}: function in pyplot module of matplotlib library is used to plot a histogram.}
\sortitem{{\color{blue}{np.linspace}}: Return evenly spaced numbers over a specified interval.}
\sortitem{{\color{blue}{np.ravel}}: return 1D array with all the input-array elements and with the same type as it.}
\sortitem{{\color{blue}{np.matmul}}: is Matrix product of two arrays.}
\sortitem{{\color{blue}{sum()}}: to calculate the sum of an iterable like range and list.}
\sortitem{{\color{blue}{np.random.rand}}: Create an array of a given shape and populate it with random samples from a uniform distribution over $[0, 1)$.} 
\sortitem{{\color{blue}{len}}: Return the number of items in a list.}
\sortitem{{\color{blue}{tuple}}: Tuples are used to store multiple items in a single variable. Items are ordered, unchangeable, and allow duplicate values.}
\sortitem{{\color{blue}{SLSQP}}: Sequential Least-Squares Quadratic Programming algorithm for nonlinearly constrained, gradient-based optimization, supporting both equality and inequality constraints.}
\sortitem{{\color{blue}{tol}}: is the tolerance for the stopping criteria. This tells scikit to stop searching for a minimum (or maximum) once some tolerance is achieved.}
\end{sortedlist}

\chapter{Glossary}

\begin{center}
\textsf{$|\,$\hyperlink{ar}{\bl{A}}$\,|\,$\hyperlink{bk}{\bl{B}}$\,|\,$\hyperlink{cr}{\bl{C}}$\,|\,$D$\,|\,$E$\,|\,$\hyperlink{fr}{\bl{F}}$\,|\,$G$\,|\,$\hyperlink{hb}{\bl{H}}$\,|\,$\hyperlink{ie}{\bl{I}}$\,|\,$\hyperlink{jb}{\bl{J}}$\,|\,$K$\,|\,$L$\,|\,$M$\,|\,$N$\,|\,$O$\,|\,$\hyperlink{pv}{\bl{P}}$\,|\,$\hyperlink{kit}{\bl{Q}}$\,|\,$R$\,|\,$\hyperlink{sh}{\bl{S}}$\,|\,$T$\,|\,$U$\,|\,$V$\,|\,$W$\,|\,$X$\,|\,$Y$\,|\,$Z$\,|$}\\[1cm]
\end{center}

\begin{itemize}
\hypertarget{ar}{\sor{Aer}}:

Aer is an element of \hyperlink{kit}{\textbf{\bl{Qiskit}}}, which provides \emph{high-performance quantum computing \textbf{simulators}} with realistic \textbf{noise models}.

\hypertarget{bk}{\sor{Backend}}:

The term backend can refer to either a \textbf{quantum computer} or a \textit{high-performance} classical \textbf{simulator} of a quantum computer.

\hypertarget{cr}{\sor{Classical Register}}:

A classical register consists of bits that can be written to and read within the coherence time of the quantum circuit.

\hypertarget{fr}{\sor{Frequency}}:

In IBM Quantum, a \textit{frequency} (or \textit{counts}) of a \textbf{\hyperlink{jb}{\bl{job}}} is defined as the number of times a given \textit{basis state} is \textbf{observed} after measurement.

{\hypertarget{jb}{\sor{Job}}}:

A job ties together all of the relevant information about a computation on IBM Q: a \hyperlink{qc}{\textbf{quantum circuit}}, choice of \hyperlink{bk}{\textbf{backend}}, the choice of how many \hyperlink{sh}{\textbf{shots}} to execute on the backend, and the results upon executing the quantum circuit on the backend.

\sor{\hypertarget{jp}{Jupyter Notebook}}:

Jupyter Notebook is a \textit{web-based interactive} computational environment for creating notebook documents.

\hypertarget{hb}{\sor{Hub}}:

A hub is the top level of a given organization and contains within it one or more groups. These groups are in turn populated with projects.

\hypertarget{ie}{\sor{IBM \textsc{Quantum Experience}}}:

IBM Q Experience is an \textit{online platform} that allow public and premium access to \textbf{cloud-based quantum computing} services.  Users interact with a quantum processor through the quantum circuit model of computation. Circuits can be created either using a \textit{graphically} with the \textbf{Quantum Composer}, or \textit{programmatically} within the \hyperlink{jp}{\textbf{Jupyter notebooks}} of the \textbf{Quantum Lab}. 

\hypertarget{pv}{\sor{Provider}}:

Access to the various services offered by IBM Q is controlled by the providers to which you are assigned. A provider is defined by a \textit{hierarchical organization} of \hyperlink{hb}{\textbf{hub}}, \textbf{group}, and \textbf{project}. Users can belong to more than one provider at any given time.

\sor{Python}:

Python is an interpreted, multi-paradigm, cross-platform \textbf{programming language}.

\hypertarget{kit}{\sor{Qiskit}}:

Qiskit is an \textit{open-source software development kit} (SDK) for working with \textbf{quantum computers} at the level of \textbf{circuits}, \textbf{pulses}, and \textbf{algorithms}. It provides tools for creating and manipulating quantum programs and running them on prototype \textbf{quantum devices} on IBM \textsc{Quantum Experience} or on \textbf{simulators} on a local computer.

\hypertarget{qc}{\sor{Quantum Circuit}}:

A quantum circuit is an ordered sequence of \textbf{quantum gates}, \textbf{measurements}, and \textbf{resets}.

\sor{Quantum Gate}:

A quantum gate is a \textbf{unitary} operation that can act on a single or multiple qubits.

\sor{Qubit}:

A qubit is the \textbf{basic} unit of \textit{quantum information}. A qubit is defined as a \textit{two-level} quantum system.

\sor{Quantum Register}:

A quantum register is a collection of qubits on which gates and other operations act.

\hypertarget{sh}{\sor{Shot}}:

The number of shots defines how many times a quantum circuit is \textbf{executed} in a quantum computer.

\end{itemize}

\setlength{\baselineskip}{0pt} 

{\renewcommand*\MakeUppercase[1]{#1}%
\printbibliography[heading=bibintoc,title={\bibtitle}]}
\end{singlespace}

\end{document}